\documentclass[11pt,a4paper,oneside,dvipsnames]{article}
\usepackage{graphicx}
\usepackage{rotating}
\usepackage{wrapfig}
\usepackage{placeins}
\usepackage{todonotes}

\usepackage[top=1in]{geometry}

\usepackage{latexsym}
\usepackage{graphicx}
\usepackage{multicol,multirow}
\usepackage{amsmath,amssymb,amsfonts}
\usepackage{mathrsfs}
\usepackage{amsthm}
\usepackage{rotating}
\usepackage{appendix}
\usepackage{ifpdf}
\usepackage[T1]{fontenc}
\usepackage{times}
\usepackage{sourcesanspro}
\usepackage{newtxmath}
\usepackage{textcomp}%
\usepackage{xcolor}%
\usepackage{csvsimple}
\usepackage{adjustbox}
\usepackage{xr}
\usepackage{soul}

\usepackage{xr-hyper}
\usepackage{hyperref}
\usepackage{siunitx}
\usepackage{authblk}

\usepackage{subcaption}
\usepackage[utf8]{inputenc}

\usepackage{array}

\DeclareUnicodeCharacter{2080}{\textsubscript{0}}
\DeclareUnicodeCharacter{2081}{\textsubscript{1}}
\DeclareUnicodeCharacter{2082}{\textsubscript{2}}
\DeclareUnicodeCharacter{2083}{\textsubscript{3}}
\DeclareUnicodeCharacter{2084}{\textsubscript{4}}
\DeclareUnicodeCharacter{2085}{\textsubscript{5}}
\DeclareUnicodeCharacter{2093}{\textsubscript{x}}
\DeclareUnicodeCharacter{2024}{\textsubscript{.}}

\definecolor{reddish}{HTML}{d6002b}
\definecolor{blueish}{HTML}{B3CDE3}
\definecolor{magentish}{HTML}{FF00AA}
\definecolor{greenish}{HTML}{2a8d0c}
\definecolor{yellowish}{HTML}{fbce18}

\usepackage{geometry}

\newgeometry{vmargin={15mm}, hmargin={12mm,17mm}}   % set the margins

\usepackage{booktabs}
\usepackage{pgfplotstable} % Generates table from .csv

\definecolor{reddish}{HTML}{FBB4AE}
\definecolor{blueish}{HTML}{B3CDE3}
\definecolor{magentish}{HTML}{FF00AA}
\definecolor{greenish}{HTML}{a1d99b}

\makeatletter
\newcommand*{\addFileDependency}[1]{% argument=file name and extension
  \typeout{(#1)}
  \@addtofilelist{#1}
  \IfFileExists{#1}{}{\typeout{No file #1.}}
}
\makeatother

\newcommand*{\myexternaldocument}[2]{%
    \externaldocument[#2]{#1}%
    \addFileDependency{#2.tex}%
    \addFileDependency{#2.aux}%
}

\myexternaldocument{supplementary_data}{S-}

\begin{document}

\title{A Framework for Scalable Ambient Air Pollution Concentration Estimation}

% \author{Omitted For Review}

\author{
  Liam J. Berrisford$^{1,2,3,*}$\\
  \texttt{l.berrisford@exeter.ac.uk}
  \and
  Lucy S. Neal$^{4}$\\
  \texttt{lucy.neal@metoffice.gov.uk}
  \and
  Helen J. Buttery$^{4}$\\
  \texttt{helen.buttery@metoffice.gov.uk}
\and
  Benjamin R. Evans$^{4}$\\
  \texttt{benjamin.evans@metoffice.gov.uk}
 \and
  Ronaldo Menezes$^{1, 5}$\\
  \texttt{r.menezes@exeter.ac.uk}\\
  \texttt{rmenezes@biocomplexlab.org}
}

%\authormark{L. J. Berrisford and R. Menezes}
\date{%
    $^1$ BioComplex Laboratory, Department of Computer Science, University of Exeter, England\\%
    $^2$ Department of Mathematics, University of Exeter, England\\
    $^3$ UKRI Centre for Doctoral Training in Environmental Intelligence, University of Exeter, England\\%
    $^4$ Met Office, FitzRoy Road, Exeter, EX1 3PB, U.K.\\
    $^5$ Department of Computer Science, Federal University of Ceará, Fortaleza, Brazil\\%
    $^{*}$ Corresponding Author\\[2ex]%
    \today
}

\maketitle

\abstract{Ambient air pollution remains a critical issue in the United Kingdom, where data on air pollution concentrations form the foundation for interventions aimed at improving air quality. However, the current air pollution monitoring station network in the UK is characterized by spatial sparsity, heterogeneous placement, and frequent temporal data gaps, often due to issues such as power outages. 
We introduce a scalable data-driven supervised machine learning model framework designed to address temporal and spatial data gaps by filling missing measurements. This approach provides a comprehensive dataset for England throughout 2018 at a 1km$^2$ hourly resolution. Leveraging machine learning techniques and real-world data from the sparsely distributed monitoring stations, we generate 355,827 synthetic monitoring stations across the study area, yielding data valued at approximately \pounds70 billion. 
Validation was conducted to assess the model's performance in forecasting, estimating missing locations, and capturing peak concentrations. The resulting dataset is of particular interest to a diverse range of stakeholders engaged in downstream assessments supported by outdoor air pollution concentration data for NO$_2$, O$_3$, PM$_{10}$, PM$_{2.5}$, and SO$_2$. This resource empowers stakeholders to conduct studies at a higher resolution than was previously possible.

\textbf{Impact Statement}: The current high-quality air pollution monitoring station network in the UK is spatially sparse with heterogeneous placement and commonly suffers from missing data temporally from issues such as power outages. We present a scalable data-driven supervised machine learning model framework to fill missing measurements temporally and spatially, providing a complete dataset for England during 2018 at a 1kmx1km hourly resolution. The approach leverages machine learning and data from the sparse real-world monitoring stations to create 355,827 synthetic monitoring stations across the study area to acquire data worth approximately £70 billion. Validation was conducted regarding the model's performance in forecasting, estimating missing locations and capturing peak concentrations. The dataset provided is of interest to a range of stakeholders conducting downstream assessments underpinned by outdoor air pollution concentration data for NO2, O3, PM10, PM25, and SO2, empowering stakeholders to perform studies at a higher resolution than previously possible. }

\clearpage
\section{Introduction}
\label{sec:introduction}

Air pollution presents a significant health risk to individuals, with 28-36 thousand deaths per year in the UK associated with air pollution exposure \cite{PublicHealthEngland:2022:AirPollutionDeaths}. Estimating ambient air pollution concentration levels is an essential step in tackling the health burden of air pollution due to the high cost of individual monitoring stations. The potential cost for a single multi-pollutant monitoring station could be as high as £198,000 \cite{AEATechnology:2006:PurchasingAURNCost}. Even for a country such as the United Kingdom that has highlighted tackling ambient air pollution as a key priority \cite{Parliament:2021:Environment}, there are only 171 monitoring stations across the UK for all pollutants currently being monitored\footnote{https://uk-air.defra.gov.uk/networks/network-info?view=aurn}. Therefore, areas without a dedicated monitoring station need to have their ambient air pollution concentrations estimated through models, which go on to inform policy for interventions into air pollution, making the models used for this process of pivotal importance.

% Current datasets at the national level output estimates at the annual temporal scale and 1km\textsuperscript{2} spatial level \cite{DEFRA:2019:ModelledBackgroundPollutionData}. However, a wide range of health advice from organisations such as the World Health Organisation details limits not only the annual level mean of the air pollution concentrations within a given area but also the daily mean. For example, for NO$_{2}$, there is a 10\si{\micro\gram/\meter^3}  annual mean limit and a 25\si{\micro\gram/\meter^3}  24-hour limit \cite{WHO:2021:WHOGuidelinesTable}, with no mention of an hourly limit \cite{WHO:2021:GlobalAirQualityGuidelines}. UK legislation, the Air Quality Standard Regulations 2010 \cite{UKGOV:2010:AirQualityStandardsUK} sets out a range of limit values, which are legally binding parameters not to be exceeded, and target values, defined in the same manner as limit values to be achieved but without legal bindings. This legislation mentions hourly level means, such as NO$_2$, with a detailed limit of 200\si{\micro\gram/\meter^3}  not exceeding 18 times a year. 

Existing national-level datasets produce estimations at the annual temporal scale and 1 km\textsuperscript{2} spatial resolution \cite{DEFRA:2019:ModelledBackgroundPollutionData}. However, it is imperative to note that health advisories, as articulated by organizations like the World Health Organisation, delineate constraints not solely on the annual mean of air pollution concentrations within a specified region but also on the daily mean. To illustrate, for NO$_{2}$, the stipulated limits include a 10 \si{\micro\gram/\meter^3} annual mean and a 25 \si{\micro\gram/\meter^3} 24-hour limit \cite{WHO:2021:WHOGuidelinesTable}, with the absence of an explicitly defined hourly limit \cite{WHO:2021:GlobalAirQualityGuidelines}. The regulatory landscape in the United Kingdom, as governed by the Air Quality Standard Regulations 2010 \cite{UKGOV:2010:AirQualityStandardsUK}, delineates both limit values---legally binding parameters not to be surpassed---and target values, akin to limit values but lacking legal bindings. Notably, this legislation addresses hourly level means for pollutants like NO$_2$, with a meticulous limit of 200 \si{\micro\gram/\meter^3}, not to be exceeded more than 18 times in a year.

The prevailing methodology, limited to generating mean annual estimates at the national scale, introduces a challenge. This stems from the fact that only specific locales, equipped with monitoring stations, possess hourly data on air pollution. Consequently, areas devoid of such monitoring infrastructure are excluded from any analysis of air pollution levels at a more granular temporal resolution. This discrepancy in data availability raises concerns regarding health inequalities, underscoring the imperative need for a more equitable and comprehensive approach.

\begin{figure}[!htb]
  \begin{subfigure}{\textwidth}
    \includegraphics[width=\linewidth]{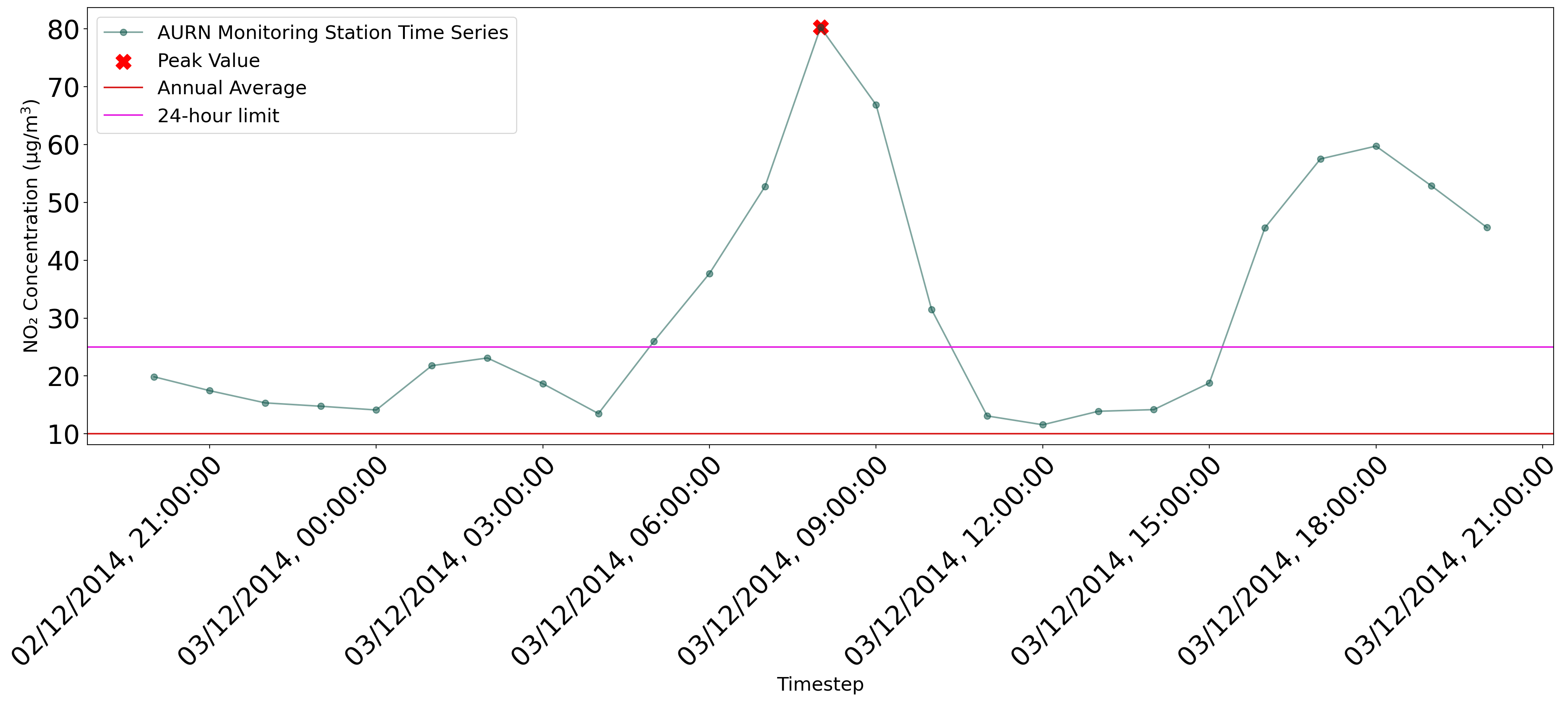}
    \caption{\textbf{Leominster day air pollution readings surrounding the 2014-2018 peak.} Shown is the peak value for the Leominster AURN monitoring station spanning 2014-2018, recorded at 08:00 on 03/12/2014. Presented alongside the peak value is the 24-hour window surrounding the peak, along with the annual and 24-hour limit averages.
} \label{fig:LeominsterAURNStationPeakDay}
  \end{subfigure}%
  \hspace*{\fill}\\   % maximize separation between the subfigures
  \begin{subfigure}{\textwidth}
    \includegraphics[width=\linewidth]{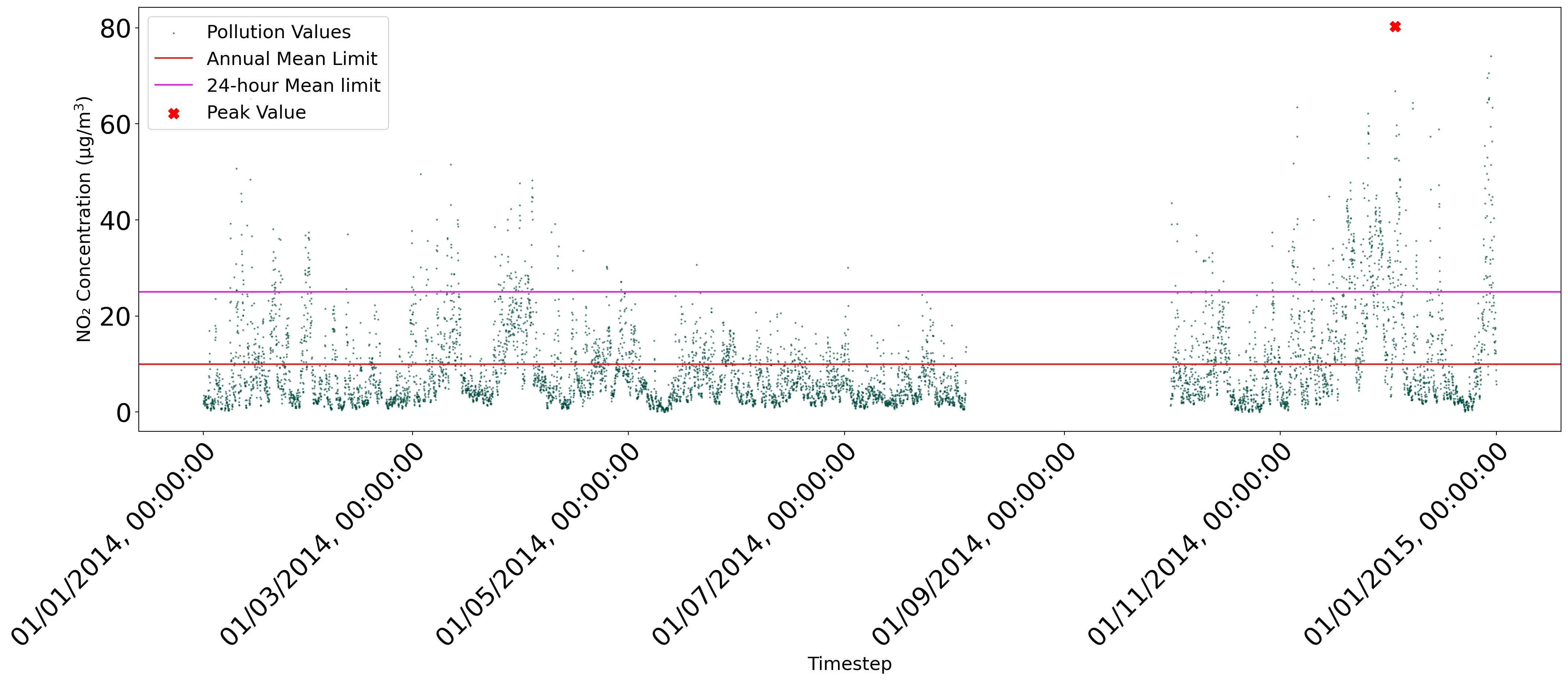}
    \caption{\textbf{Leominster 2014 air pollution readings.} Presented are all ambient air pollution measurements for NO$_2$ at the Leominster for the year 2014. Emphasized is the peak value, which is further examined in Figure \ref{fig:LeominsterAURNStationPeakDay}, along with the annual and 24-hour average limits.} \label{fig:LeominsterAURNStationPeakYear}
  \end{subfigure}%

\caption{{\bfseries Leominster AURN monitoring station NO$_{2}$ measurements.} Figure \ref{fig:LeominsterAURNStationPeakDay} shows how the peak air pollution reading for NO$_{2}$ at the Leominster station dramatically exceeds the 24-hour limit, even more so for the annual limit, showing how there can be periods of quite extreme pollution in the context of the annual limits. Figure \ref{fig:LeominsterAURNStationPeakYear} shows how there can be extended periods where the air pollution levels are below and exceed the designated limits and the relation of the monitoring station peak to all available data for the station in 2014.} \label{fig:LeominsterAURNStationPeak}
\end{figure}

The utilization of annual pollution levels provides a broad overview of the pollution within a designated study area. However, a notable challenge arises concerning information loss when transitioning from an hourly to a daily or annual temporal scale. This issue has manifested in the United Kingdom, where instances of divergent narratives emerge between the annual and daily means of specific locations. Take, for instance, Leominster\footnote{https://uk-air.defra.gov.uk/networks/site-info?site\_id=LEOM} on 03/12/2014 at 08:00, which recorded a peak pollution value for NO$_2$ of 80.2 \si{\micro\gram/\meter^3}. The 24-hour mean in the vicinity of this peak, spanning 12 hours on either side (from 02/12/2014 2000 to 03/12/2014 2000), stands at 31.5 \si{\micro\gram/\meter^3}, as illustrated in Figure \ref{fig:LeominsterAURNStationPeakDay}. This exceeds the WHO's daily mean guideline of 25 \si{\micro\gram/\meter^3}. The complexity deepens when examining Leominster's annual mean for 2014, registering a value of 9.5 \si{\micro\gram/\meter^3}, deemed safe by WHO guidelines and depicted in Figure \ref{fig:LeominsterAURNStationPeakYear}. Similar disparities are observable in other monitoring stations. For instance, London North Kensington exhibits unsafe levels at both the annual and daily scales, with a peak value of 209 \si{\micro\gram/\meter^3}, a daily mean of 122 \si{\micro\gram/\meter^3}, and an annual mean of 33 \si{\micro\gram/\meter^3} for the pollutant NO$_2$. Table \ref{S-tab:No2PeakValuesMeans} details the peak values for the five most polluted stations for NO$_2$ within the study, encompassing the Peak Value, Daily Mean surrounding the peak, and the annual mean for the year of the peak occurrence.

While there is evident importance in hourly air pollution concentration data for compliance and legislation purposes, the data serves a spectrum of other critical purposes. Researchers, policymakers, and public health officials routinely conduct human health assessments \cite{USEPA:1992:GuidelinesForExposureAssessment}, enabling informed decisions concerning interventions to protect vulnerable populations \cite{zou:2009:AirPollutionExposureMethodsEpidemiological}. Further, epidemiological studies assessments are routinely conducted \cite{atkinson:2016:UKEpidemiologicalStudy} and are of crucial importance when significant changes in air pollution are being observed, such as during the COVID-19 pandemic \cite{konstantinoudis:2021:UKEpidemiologicalStudyCovid19}. Beyond human health, ecosystem health can be significantly impacted by air pollution, leading to damage to plants, manifested as leaf injury and stunted growth \cite{molnar:2020:AirPollutionVegtation}. This has critical implications for concerns related to crop yields and food security \cite{tai:2014:AirPollutionOzoneFoodSecurity}.

As such, this work introduces a data-driven machine learning model designed to estimate the hourly concentration of air pollutants at the same spatial resolution (1 km\textsuperscript{2}) as the existing annual dataset available for the UK \footnote{\href{https://uk-air.defra.gov.uk/data/pcm-data}{UK-AIR Annual Modelled Background Air Pollution Data}.}. We argue that this work leads to a dataset of substantial value to various stakeholders.

%\clearpage
\section{Related Work}
\label{sec:existingWork}

\subsection{Measuring Air Pollution}
\label{sec:MeasuringAirPollution}

%% RMComment: I'm reading it now and will check if you're making a reference to your previous work. Alternativelly you can talk about your previous work at the end in the discussion.

Various methods exist to offer insight into air pollution concentrations in a given location. The most robust and straightforward method available involves specialized equipment designed to provide direct measurements of air pollution concentrations. In-situ measuring equipment can be broadly categorized into two groups: high-quality stationary monitoring stations discussed in Section \ref{sec:introduction} and more mobile low-cost air quality sensors \cite{kang:2022:LowCostAirPollutionSensor}.

While high-quality fixed monitoring stations provide a reliable method of obtaining air pollution concentration data, their deployment on a large scale is prohibitively expensive. In 2018, the UK had 165 high-quality monitoring stations online across the country within its premier monitoring network. Notably, a majority of these monitoring stations are situated in urban areas,  comprising 144 urban and 21 rural stations. The strategic decision of where to position these monitoring stations carries the potential to exacerbate inequality between urban and rural areas, potentially fostering a divide between rural and urban communities in terms of insight into air pollution where they live and work \cite{rosofsky:2018:UrbanRuralAirPollutionInequality}, particularly when considering that O$_3$ air pollution can often be worse in rural locations \cite{stasiuk:1974:RuralOzone,BIEA:2023:LargePointSources}. 

The emergence of low-cost sensors has made it possible to monitor air pollution concentrations over a larger geographic area. However, we see two critical problems with low-cost sensors. One such issue is the quality of the sensors themselves, which can be influenced by changes in atmospheric composition and meteorological conditions, or provide false signals if other air pollutants are present in high concentrations \cite{DEFRA:2023:LowCostSensors}. Another issue is the quality control that is conducted on the sensors, such as the calibration checks that go into ensuring that the measurement is made under the same conditions, such as the height at which the measurement is taken, affecting the reading that is produced, potentially making comparisons between different low-cost sensors and even the same sensor between locations more challenging \cite{concas:2021:LowCostSensorCalibration}. There is research being conducted to help combat the issues facing low-cost sensors; it is still an open challenge but rapidly improving \cite{Rai:2017:LowCostSensorReview}. For now, low-cost air pollution sensors are only suitable for raising awareness rather than applications requiring higher accuracy, such as epidemiological studies or compliance with air quality legislation \cite{Castell:2017:LowCostSensorEUComparison}.

An ex-situ indirect measurement of air pollution concentrations can be achieved with remote sensing. Sentinel 5P \cite{veefkind:2012:Sentinel5PDescription} is an ESA satellite platform that can provide insight into air pollution concentrations at a vast spatial extent. However, a major challenge associated with the use of Sentinel 5P is the issue of data completeness. Two primary drivers contribute to missing data from the Sentinel 5P platform. The first challenge arises from the platform's orbit, which follows a near-polar, sun-synchronous path \cite{ESA:2023:Sentinel5POrbit}. This orbit causes the platform to consistently pass over a region at a similar time each day. While this characteristic is advantageous for comparing locations, it complicates the provision of insight into air pollution concentrations across an entire day. As a result, questions such as the difference between rush hour and midnight air pollution concentrations become difficult to answer. Another factor contributing to data gaps is environmental conditions that may lead to a specific reading not passing quality control, resulting in missing measurements on certain days \cite{ESA:2017:ESAQualityAssurance}. Another remote sensing platform is the recently operational TEMPO \cite{zoogman:2017:NASATEMPORemoteSensing}, which provides hourly air pollution concentration measurements. However, TEMPO shares similar limitations with Sentinel 5P and only offers coverage over North America. While remote sensing is a valuable tool in certain circumstances, it cannot provide a complete picture of air pollution concentrations. This comprehensive understanding is crucial for designing effective interventions to tackle air pollution.

\subsection{Modelling Air Pollution}

As it is evident that all methods of measuring air pollution concentrations have drawbacks, models have been extensively utilized to complement real-world observations. Various types of model frameworks exist, each offering distinct benefits and drawbacks.

There are two widely used air pollution model frameworks: Lagrangian and Eulerian. Lagrangian models track individual air parcels (or particles) moving through the atmosphere. Each parcel is associated with a set of equations of motion, making the parcel the focal point of the model as it moves through space and time \cite{eliassen:1984:LagrangianAirModels}. On the other hand, Eulerian models do not concentrate on a single air parcel; instead, they divide the atmosphere into regions, using fixed points or cells to represent specific locations. The goal is to understand the concentrations of air pollutants at these specific locations at different times \cite{Byun:1984:EulerianDispersionModels}. Lagrangian models are particularly well-suited for studying problems where a specific pollution source is of interest, such as the ash emitted from a volcano \cite{Vitturi:2010:lagrangianVolcano}.

On the other hand, an Eulerian model is well-suited for studying the spatial distribution and long-term trends of air pollutants, albeit at the expense of not being able to provide specific information about a given source and pollutant. Consequently, this work will primarily focus on Eulerian models, as they provide the necessary data for conducting the analyses and assessments discussed in Section \ref{sec:introduction}.

Statistical Eulerian models, such as Land use regression (LUR), exist as a method for creating stochastic air pollution models. LUR incorporates a variety of predictors, including meteorological, terrain, land use, and road network data \cite{hoek:2008:review}. Another paradigm of Eulerian air pollution models is represented by mechanistic models, such as GEOS-Chem \cite{Henze:2007:GeosChem}. These models are open source and available for use, providing comprehensive spatial coverage of air pollution concentrations. However, they demand a high level of expertise in the domain field for interrogation due to their complexity. Additionally, these models come with extensive requirements for supporting infrastructure, with a GEOS-Chem \ang{4}x\ang{5} degree standard simulation requiring 15GB of RAM\footnote{\url{https://geos-chem.readthedocs.io/en/stable/gcc-guide/01-startup/memory.html}}.

A rapidly emerging area is the use of deterministic models to address the current gap within the existing suite of models, providing high-resolution air pollution concentration data both temporally and spatially; this empowers stakeholders to make informed decisions concerning air pollution. Several models in this category are based on data-driven supervised machine learning, where a target vector, typically representing air pollution concentrations, is estimated from a feature vector, such as meteorological variables (e.g., wind speed). The model's objective is to learn the relationship between the target and feature vectors in situations where both are available, enabling subsequent predictions of target vectors when only the feature vector is available. In the scientific literature, numerous studies utilize machine learning techniques to forecast air pollution concentrations \cite{freeman:2018:ForecastingAirPollutionConcentration, tao:2019:ForecastingAirPollutionConcentration2, harishkumar:2020:ForecastingAirPollutionConcentration3}. However, a limitation exists, as this approach requires air pollution concentration data from the location being predicted before the time predicted. Therefore, there is a need for historical air pollution data to be available. For example, a forecasting model will use air pollution concentration data from T-X to estimate air pollution at time T, where X is some defined time, such as 1/3/9 hours. If historical air pollution concentrations are used, it restricts the method's applicability to locations where an air pollution monitoring station exists.

Existing studies have tackled the problem of estimating air pollution concentrations in locations without monitoring stations. However, the studies focus either on small geographical areas, such as the Bay of Algeciras (Spain) with hourly temporal resolution \cite{van:2019:MissingLocationAirPollutionEstimationSmallArea} or a large geographical area with low temporal resolution, such as monthly \cite{chen:2021:MissingLocationAirPollutionMonthly}. Some work has been able to achieve higher spatial coverage with daily temporal resolution \cite{he:2023:AirPollutionLeaveOneOutValidationDaily, li:2020:AirPollutionLeaveOneOutValidationDaily2}. As such, this work presents a model combining these aspects, predicting hourly temporal resolution concentrations across England's large geographical area, a considerable challenge due to the variance of air pollution concentrations in locations covered. 

This manuscript aims to use machine learning to produce data similar to an Eulerian model framework. While traditionally, the concentration is resolved over an area in an Eulerian model, the model presented here can be considered an approach of using machine learning as a synthetic monitoring station. The process that is followed for the model is answering the question of the air pollution concentration reading of a monitoring station that experiences the environmental conditions described by the input data. The model takes the training data to learn the relationship between environmental conditions and air pollution, allowing us to use the environmental conditions that are known in all locations across the England and make predictions of air pollution concentrations that are not so readily available, providing a complete picture of air pollution concentrations in the England. Compared with other deterministic methods, such as mechanistic models, a key benefit of the approach is the improvement in computational speed. In contrast, more traditional Eulerian models involve spatial dependencies between grids, where, for example, two adjacent grids impact each other. The framework presented in this work, however, treats each synthetic monitoring station as independent from one another. This approach offers a significant speedup in computation through the parallelization of predictions, while also enabling more accessible exploration of data by predicting air pollution locations independently. This novel approach is a key contribution of this work, utilizing machine learning to underpin a scalable estimation of ambient air pollution concentrations. Importantly, this approach is linearly scalable concerning computational complexity, allowing stakeholders to employ a model capable of predicting air pollution concentrations at any spatial and temporal resolution.

\section{Data}
\label{sec:data}

The data-driven supervised machine learning model this paper proposes for air pollution concentration prediction is based on two primary sets of data: feature vectors and target vectors. In the case of air pollution concentration estimation, the target vector is the air pollution concentration itself, the data to be estimated, and the feature vector represents the data used to make predictions, for example, the wind speed. The model aims to understand the relationship between the feature and target vectors, e.g., what is the given NO$_x$ concentration at given wind speeds?

\subsection{Target Vector: Air Pollution Concentrations}
\label{sec:airPollutionData}

We obtained air pollution data for our study from the UK Automatic Urban and Rural Network (AURN) using the OpenAir package \cite{Carslaw:2012:OpenAir}. Our study focuses on seven pollutants: Nitrogen Oxide (NO), Nitrogen Dioxide (NO$_\text{2}$), Nitrogen Oxides (NO$_{x}$), Particles < 10\si{\micro\meter} (PM$_{10}$), Particles < 2.5\si{\micro\meter} (PM$_{2.5}$), Ozone (O$_\text{3}$), and Sulphur Dioxide (SO$_\text{2}$). All air pollutants are measured in micrograms per cubic meter (\si{\micro\gram/\meter^3}). All types of monitoring stations were included in the study. The number of station types per pollutant varied, resulting in different data point distributions, as shown in Table \ref{tab:AURNClassificationNumberOfStations}, with apparent gaps in some locations for certain air pollutants, such as Suburban Industrial for PM$_{10}$, PM$_{2.5}$, and SO$_{2}$. Clear spatial differences exist in the locations of monitoring stations. Figure \ref{S-fig:ukAURNNetwork} shows the spatial distribution of all AURN monitoring stations used in this study across three high-level environmental area classifications.

Each environmental area has a representative area over which the station's measurements relate \cite{UKAIR:BackgroundStations:2022}. Urban stations are defined as representative of a few square kilometres (km\textsuperscript{2}), suburban stations cover tens of square kilometres, and rural stations encompass at least 1,000 square kilometres. Each station within the network also has a location type that specifies the primary source of air pollution at the station. Background stations are strategically located to ensure that no single source or street significantly influences the readings at the station. Instead, the measurements reflect an integrated contribution from all sources upwind. Traffic stations are located so that the measurements represent a street segment of at least 100 meters, and industrial stations have a representative area of 250 square meters (m\textsuperscript{2}).

The monitoring station location was abstracted to the closest grid centroid for ease of creating the needed datasets. Consequently, there is some distance between the true location and the location where we created the feature vector for the monitoring station. This abstraction of location provided a common framework, reducing the computation required to build the associated feature and target vectors, and facilitating a more straightforward interpretation of the framework. While the AURN monitoring station guidelines specify a minimal representative sample area, and the maximum abstraction distance across the monitoring network locations was 399 meters for the London Hackney monitoring station, we deemed this to be a worthwhile tradeoff. Full details of the abstraction distance can be found in Supplementary Material Table \ref{S-tab:AURNStationDistanceFromCentroidOverall}. It is noteworthy that as the spatial resolution increases, the associated errors will decrease, leading to an improvement in the approach. Eventually, the error from abstracting the location of the monitoring station will be eliminated when the abstracted distance is below the monitoring station representative sample area. However, this comes at the cost of considerable additional computational expenses. Therefore, the experiments in this study represent a lower bound for the framework's performance, as any operational deployment could utilize increased spatial resolution for potential performance improvement.

For the study, we used the years 2014-2016 as the training set, 2017 as the validation set, and 2018 as the test set. To be included in the study, a station needed to have at least one measurement in each of these sets.

We conducted preprocessing on the collected air pollution concentration data. While UK-AIR performs some data validation \cite{UKAIR:2023:DataValidation}, we undertook additional preprocessing steps. The initial step involved removing negative values, which are possible in the UK-AIR dataset \cite{UKAIR:2023:AirPollutionNegativeValues}. The number of observations removed per air pollutant due to the presence of negative concentrations is detailed in Table \ref{S-tab:aurnNegativeValues}. The distribution of the positive air pollution concentration values can vary widely across air pollutants and exhibit apparent differences between different environmental locations of monitoring stations. This variability raises the question of how to identify outliers, alongside the issue of developing a model that can handle target vectors with stark differences in distribution. For example, O$_3$ is the only air pollutant with a non-zero-inflated distribution, with the skewing between distributions for different environmental classifications for each air pollutant showing varying degrees of skew, as depicted in KDE plots in Figure \ref{S-fig:airPollutionKDE}.

Points that are distant from the mean have the potential not to be genuine outliers, but rather data points generated by different phenomena compared with the other data points in the distribution. We aimed to identify and remove both outliers and anomalies within the dataset. The challenge in identifying outliers lies in the context of the dataset, where a single urban traffic data point within the context of rural background monitoring station data points might be flagged as an outlier using established methods like IQR \cite{crosby:1994:OutlierIQR}. We were also cognizant of the potential presence of anomalies in the dataset. We recognized that a single localized event could drive a high-value concentration data point. While this reading might be accurate, it does not align with the AURN monitoring stations' purpose, where concentrations are intended to represent a larger geographic area. Consequently, we considered identifying and removing these values beyond the scope of this work and proceeded with the study, acknowledging the presence of outliers and anomalous observations within the dataset that we could not explicitly identify. As such, we decided not to remove any such data points from the dataset.

\begin{table}[ht]
\centering
\resizebox{.7\linewidth}{!}{
\begin{filecontents*}{CSVFiles/Data/AURNClassificationNumberOfStations.csv}
,Pollutant,Urban Background,Urban Traffic,Rural Background,Suburban Background,Urban Industrial,Suburban Industrial
0,NOₓ,40,41,11,3,6,2
1,NO₂,40,41,11,3,6,2
2,NO,40,41,11,3,6,2
3,O₃,32,3,13,2,3,1
4,PM₁₀,17,22,2,0,5,0
5,PM₂₅,30,15,2,2,4,0
6,SO₂,9,1,5,0,3,0
\end{filecontents*}
\pgfplotstabletypeset[
    multicolumn names=l, 
    col sep=comma, 
    string type, 
    header=has colnames,
    columns={
        Pollutant,
        Urban Background,
        Urban Traffic,
        Rural Background,
        Suburban Background,
        Urban Industrial,
        Suburban Industrial
    },
    columns/Pollutant/.style={
        column type=l,
        column name={\shortstack{Pollutant\\Name}}
    },
    columns/Urban Background/.style={column type={S[round-precision=0, table-format=2.0, table-number-alignment=center]}, column name={\shortstack{Urban\\Background}}},
    columns/Urban Traffic/.style={column type={S[round-precision=0, table-format=2.0, table-number-alignment=center]}, column name={\shortstack{Urban\\Traffic}}},
    columns/Urban Industrial/.style={column type={S[round-precision=0, table-format=2.0, table-number-alignment=center]}, column name={\shortstack{Urban\\Industrial}}},
    columns/Suburban Background/.style={column type={S[round-precision=0, table-format=2.0, table-number-alignment=center]}, column name={\shortstack{Suburban\\Background}}},
    columns/Suburban Industrial/.style={column type={S[round-precision=0, table-format=2.0, table-number-alignment=center]}, column name={\shortstack{Suburban\\Industrial}}},
    columns/Rural Background/.style={column type={S[round-precision=0, table-format=2.0, table-number-alignment=center]}, column name={\shortstack{Rural\\Background}}},
    every head row/.style={before row=\toprule, after row=\midrule},
    every last row/.style={after row=\bottomrule}
]{CSVFiles/Data/AURNClassificationNumberOfStations.csv}}
\smallskip
\caption{{\bfseries AURN monitoring station counts by environmental classification per air pollutant. } The number of stations for each pollutant within the UK AURN network within England is shown. It can be seen that there is an unequal distribution across the different environment types, alongside some pollutants such as SO$_2$, missing some environmental types completely.} \label{tab:AURNClassificationNumberOfStations}
\end{table}

\subsection{Feature Vectors}
\label{sec:featureVectors}

\begin{figure}[ht]
\begin{center}
\includegraphics[width=.8\textwidth]{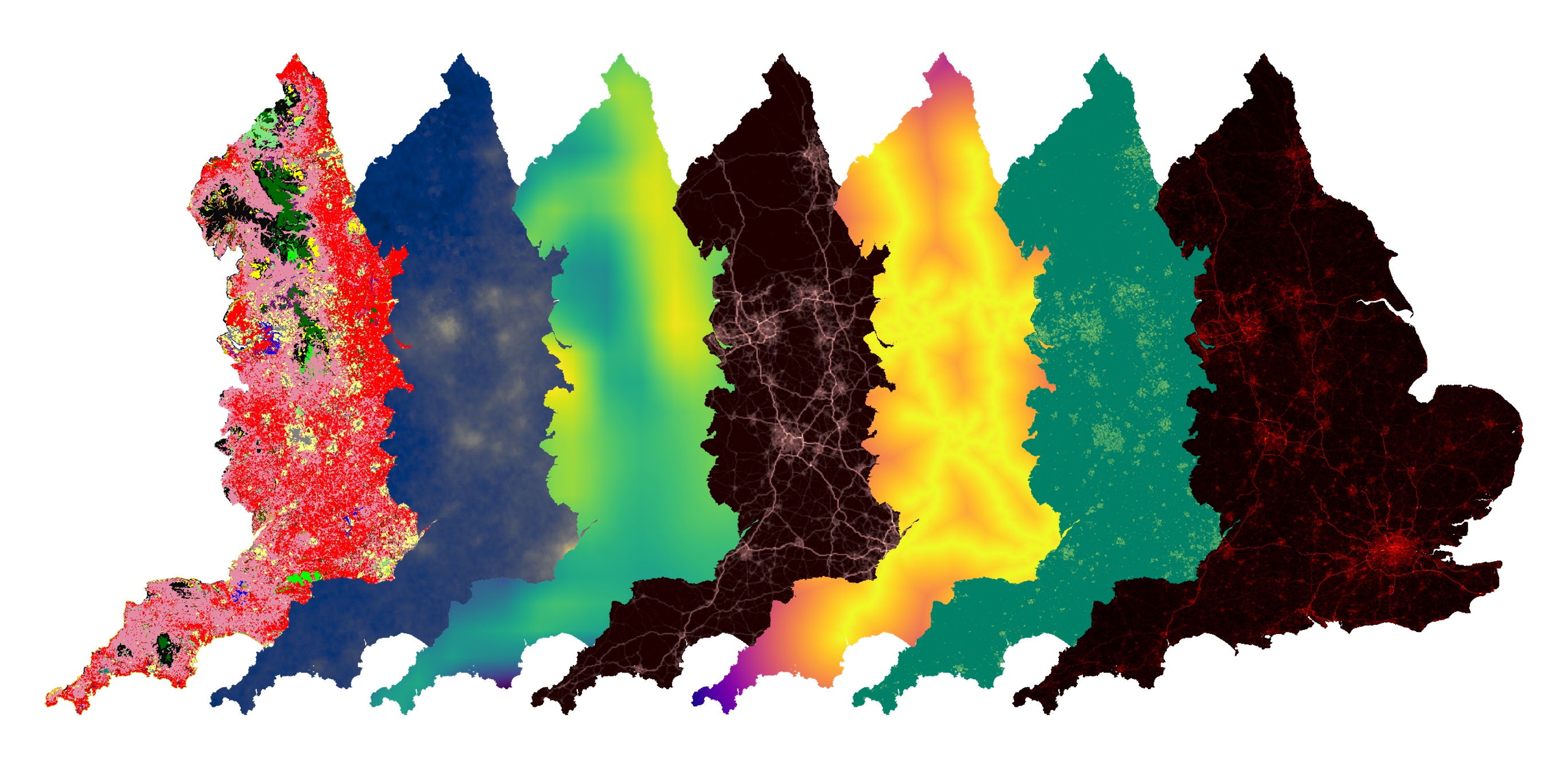}
\caption{\textbf{Example feature vector dataset from each dataset family.} From left to right, the example datasets are the majority land use classification for each grid (geographic family, discussed in Section \ref{S-sec:landUse}), Sentinel 5P NO$_2$ measurements (remote sensing family, discussed in Section \ref{S-sec:remoteSensingData}), 100m U component of wind (meteorological family, discussed in Section \ref{S-sec:DataDetails:metrological}), NAEI SNAP sector 7 (road transport) NO$_x$ emissions (emissions family, discussed in Section \ref{S-sec:Datadetails:emissions}), road infrastructure distance from the nearest motorway and total length of residential road per grid (transport infrastructure structural properties family, discussed in Section \ref{S-sec:Datadetails:TransportDataInfratsurtcureStructural}), and the car and taxis score (transport use family, discussed in Section \ref{S-sec:transportInfrastructureUseData}).}
\label{fig:englandLandGrids}
\end{center}
\end{figure}

The data considered in this study can be categorized into different dataset families, each containing a set of distinct but related datasets describing a phenomenon associated with air pollution concentrations. Addressing the temporal and spatial resolution differences between the datasets was a key challenge in creating a consistent feature vector to estimate the air pollution target vector. The common framework employed consisted of 355,827 1km\textsuperscript{2} grids covering the extent of the England land mass. England was chosen as the study area since it was a common geographical region in all the datasets examined during this study, as illustrated in Figure \ref{S-fig:englandLandGrids}. For the study, seven different dataset families were used, each providing a set of datasets describing a range of related phenomena that correlate with air pollution concentrations. Across all dataset families, there are 152 feature vector elements, with Figure \ref{fig:englandLandGrids} showing example feature vectors across England for each dataset family. 

\paragraph{Transport Infrastructure Structural Properties, 28 features.}
Transport Infrastructure has been shown to provide information concerning air pollution concentrations \cite{berrisford:2022:AirPollutionRoadInfrastructure}. We used Open Street Maps \cite{bennett:2010:OpenStreetMaps} to create annual snapshots of 14 transport infrastructure networks, from motorways\footnote{\href{https://wiki.openstreetmap.org/wiki/Tag:highway\%3Dmotorway\#:~:text=The\%20tag\%20highway\%20\%3D\%20motorway\%20is,local\%20context\%20and\%20prevailing\%20convention.}{Open Street Maps Motorway Highway Classification}.} to residential\footnote{\href{https://wiki.openstreetmap.org/wiki/Tag:highway\%3Dresidential\#:~:text=The\%20highway\%20\%3D\%20residential\%20tag\%20is,have\%20also\%20some\%20transit\%20traffic.}{Open Street Maps Motorway Residential Classification}.} roads. Using the road network, we calculated two sets of feature vectors. One detailing the distance to each road type from the grid centroid, and the second detailing the total length of the given road type within the grid. Further details on the process conducted can be seen in Section \ref{S-sec:Datadetails:TransportDataInfratsurtcureStructural}.

\paragraph{Transport Infrastructure Use, 5 features.}
Vehicles themselves are a primary driver of air pollution through multiple processes. Road vehicles exhaust gas air pollutants such as NO$_x$ and SO$_2$  \cite{watkins1991:RoadPollutionAirPollution} alongside causing PM air pollution \cite{yan:2011:RoadVechilesPMEmissions}. Further vehicles can cause air pollution through traffic resuspension \cite{amato:2010:TrafficResuspensionAirPollution} and fuel spillage and evaporation \cite{haagen:1959:FuelSpillageAirPollution}. Within England, only traffic counts from point locations detailing daily traffic flows across different road types for key vehicle types such as Cars and Heavy Goods vehicles (HGVs) were available from the Department of Transportation \cite{DepartmentOfTransport:RoadTrafficFlowData:2023}. Using OpenStreetMaps, we created a spatially complete dataset by providing the average daily traffic flow by road type per meter across the UK. We then used a spatial microsimulation using data from the UK Census (providing sociodemographic details of different regions of the UK) \cite{ons:2012:census} and the UK Time Use Survey (providing details of how different sociodemographic groups travel, both temporally and by which transportation mode) \cite{Sullivan:2023:UKTimeUse} to spatially distribute the daily traffic counts to produce an hourly spatially complete dataset of traffic counts. Complete details of the process are covered in Section \ref{S-sec:transportInfrastructureUseData}.

\paragraph{Meteorology, 11 features.}
Meteorological phenomena play a pivotal role in air pollution concentrations. Wind advects air pollution both to and from locations of interest through horizontal transport \cite{Jurado:2021:WindSpeedDirectionAirPollutionConcentrations, cichowicz:2017:OzoneWinterReduction}. Temperature can have a range of impacts on air pollution, impacting temperature inversions \cite{wallace:2010:temperatureInversionAirPollution}, the production of O$_3$ air pollution \cite{bloomer:2009:observedOzoneAndTemperatureRelationship} and the removal of air pollution by plants \cite{nowak:1998:VegetationTemperatureAirPollution}. UV radiation directly impacts O$_3$ production \cite{finlayson:1986atmosphericChemistryOzoneProduction}. The removal of air pollution from the atmosphere via deposition by precipitation is notable \cite{jolliet:2005:WetDeposition}, alongside wash-off from surfaces \cite{yuan:2017:RainfallRoadWashoff, xu:2019:RainfallLeafWashOff}. Pressure can also influence air pollution concentrations, either by vertical mixing in low-pressure systems \cite{ning:2018:LowPressureSystemAirPollution} or high-pressure systems, causing an accumulation of air pollution concentrations near the ground through a lack of vertical mixing and advection \cite{vukovich:1979:HighPressureSystemAirPollution}. Further, O$_3$ production is increased at higher pressures \cite{hippler:1990:OzoneFormationPressure}. The boundary layer also influences air pollution concentrations through vertical mixing within the layer. Larger boundary layer heights tend to produce less concentrated air pollution at the surface. For smaller boundary layer heights, the inverse is true. \cite{xiang:2019:BoundaryLayerHeightAirPollution, davies:2007:BoundaryLayerHeightAirPollution, xiang:2019:BoundaryLayerHeightAirPollution2}. We retrieved data from the ECMWF Re-Analysis Version 5 (ERA5) dataset \cite{hersbach:2016:era5}. ERA5 is a global dataset that details the environmental conditions at equal space points, at 0.25$^{\circ}$x0.25$^{\circ}$ hourly resolution. We choose the variables commonly associated with air pollution concentrations in the scientific literature. We interpolated across the study area to provide a value for each 1km$^2$ grid centroid; the details of the process of creating the dataset can be seen in Section \ref{S-sec:DataDetails:metrological}.

\paragraph{Remote Sensing, 5 features.}
While there are limitations to the data produced by remote sensing, as discussed in Section \ref{sec:MeasuringAirPollution}, they provide valuable insight into air pollution concentrations between locations. We used monthly averages of Sentinel 5P data \cite{veefkind:2012:Sentinel5PDescription} to ensure that all locations had a measurement value. Further details of the feature vector are discussed in Section \ref{S-sec:remoteSensingData}.

\paragraph{Emissions, 77 features.}

Emissions of air pollutants are the primary driver of a wide range of air pollutant concentrations. The emissions are classified into 11 SNAP (Selected Nomenclature for Air Pollutant) sectors denoting the emissions source, with particular details discussed in Section \ref{S-sec:Datadetails:emissions}. The first sector "Combustion Energy Production and Transformation" (SNAP 1) includes power generation which can produce air pollutants such as SO$_2$ \cite{chaaban:2004:NaturalGasSulfurEmissions, shi:2021:OilSulfurContent}. Road vehicles exhaust gas air pollutants such as CO, CO$_2$, NO$_x$, SO$_2$ \cite{watkins1991:RoadPollutionAirPollution} and are included in the "Road Transport" category (SNAP 7). SNAP 8, "Other Transport and Mobile Machinery" includes shipping which emits NO$_x$, PM, CO$_2$ and VOCs \cite{corbett:1997:ShipEmissions}, particularly SO$_x$ from the marine fuels which has a high sulfur content \cite{tao:2013:ShipHighSulfur}. Organic waste in landfills ("Waste Treatment and Disposal", SNAP 9) can produce VOCs, a precursor to O$_3$ \cite{nair:2019:LandfillVOCEmissions}. Agriculture emissions (part of SNAP 10, "Agriculture, Forestry and Land Use Change"), comprise a large source of air pollutants, for example 39\% of global PM$_{2.5}$ is caused by ammonia from livestock manure and urine and synthetic nitrogen fertilisers \cite{gu:2021:FarmingPM25Ammonia}. Other emission sectors are "Combustion in Commercial, Institutional, Residential and Agriculture" (SNAP 2),  "Combustion in Industry" (SNAP 3),  "Production Processes" (SNAP 4),  "Extraction and Distribution of Fossil Fuels" (SNAP 5),  "Solvent Use" (SNAP 6) and "Nature" (SNAP 11).

\paragraph{Land Use, 22 features.}
The land use composition of a given area is related to air pollution concentrations, such as throughout greenspace \cite{nowak:2002:UrbanTreesAirQuality, nowak:2006:UrbanTreeAirQualityPM} and urbanisation \cite{arnfield:1990:UrbanCanyonSolarRadiation, yassin:2011:WindTunnelAirPollution}. Land use composition profiles were created for each grid using the UKCEH 25m Land Cover Maps \cite{rowland:2017:UKCEHLandCoverMap}. Details of the process and the different land use types are discussed in Section \ref{S-sec:landUse}.

\paragraph{Temporal Aspects, 4 features.}
Air pollution displays various temporal cyclical elements, including diurnal cycles caused by rush hour for NO$_2$ \cite{goldberg:2021:NO2DailyCycles}, UV radiation for O$_3$ \cite{garland:1979:OzoneDailyCycle}, and boundary layer height for all pollutants \cite{su:2018:BoundaryLayerPollutantsRelationships}. Weekly trends also emerge due to the working week affecting transportation and industrial emissions for NO$_x$ \cite{beirle:2003:NOxWeekly}, with similar patterns observed for PM \cite{gietl:2009:PMWeeklyCycle}. Seasonal cycles for PM are evident due to winter residential heating \cite{feng:2014:WinterPMResidentialHeating}, with similar factors contributing to an increase in SO$_2$ \cite{meng:2018:SO2Increases}. Furthermore, winter has a higher probability of adverse meteorological conditions, which reduces vertical mixing \cite{Qianhui:2022:StableWinterAirPollutionIncrease}. Additionally, colder temperatures and reduced sunlight in winter months affect O$_3$ production \cite{cichowicz:2017:OzoneWinterReduction}. Consequently, the hour, day of the week, week number, and month were all included as elements in the feature vector.

The following section justifies the inclusion of such a broad range of different datasets by exploring the relationship between the feature vector and the air pollution concentrations at AURN monitoring stations. This analysis highlights the differences between types of air pollution and variations experienced at different environmental types for a single type of air pollution, ensuring robust estimation of air pollution concentrations across all environment types found in England.

% \clearpage
\section{Feature Selection}

\subsection{Air Pollution and Feature Vectors}
\label{sec:featureSelectionPollutants}

Air pollution can be attributed to various sources, with different processes influencing its concentration, as discussed in Section \ref{sec:featureVectors}. The issue of different drivers of air pollution is further complicated when considering that at different locations, the key phenomena driving the pollution concentrations are different, as detailed within the AURN environment classification discussed in Section \ref{sec:airPollutionData}. This tangled web of sources, such as road transport, and sinks, such as wind speed, makes it challenging to identify a common set of datasets that can provide insight into all the air pollutants of interest. Therefore, this section aims to untangle the relationship between the different air pollutants and monitoring station environment types to provide insight into the different feature vectors and their relationship with the air pollutant measurements. This provides insight into the benefit of each dataset, detailed in Section \ref{sec:featureVectors}.

\begin{figure}[!htb]
  \hspace*{\fill}   % maximize separation between the subfigures
  \begin{subfigure}{0.8\textwidth}
    \includegraphics[width=\linewidth]{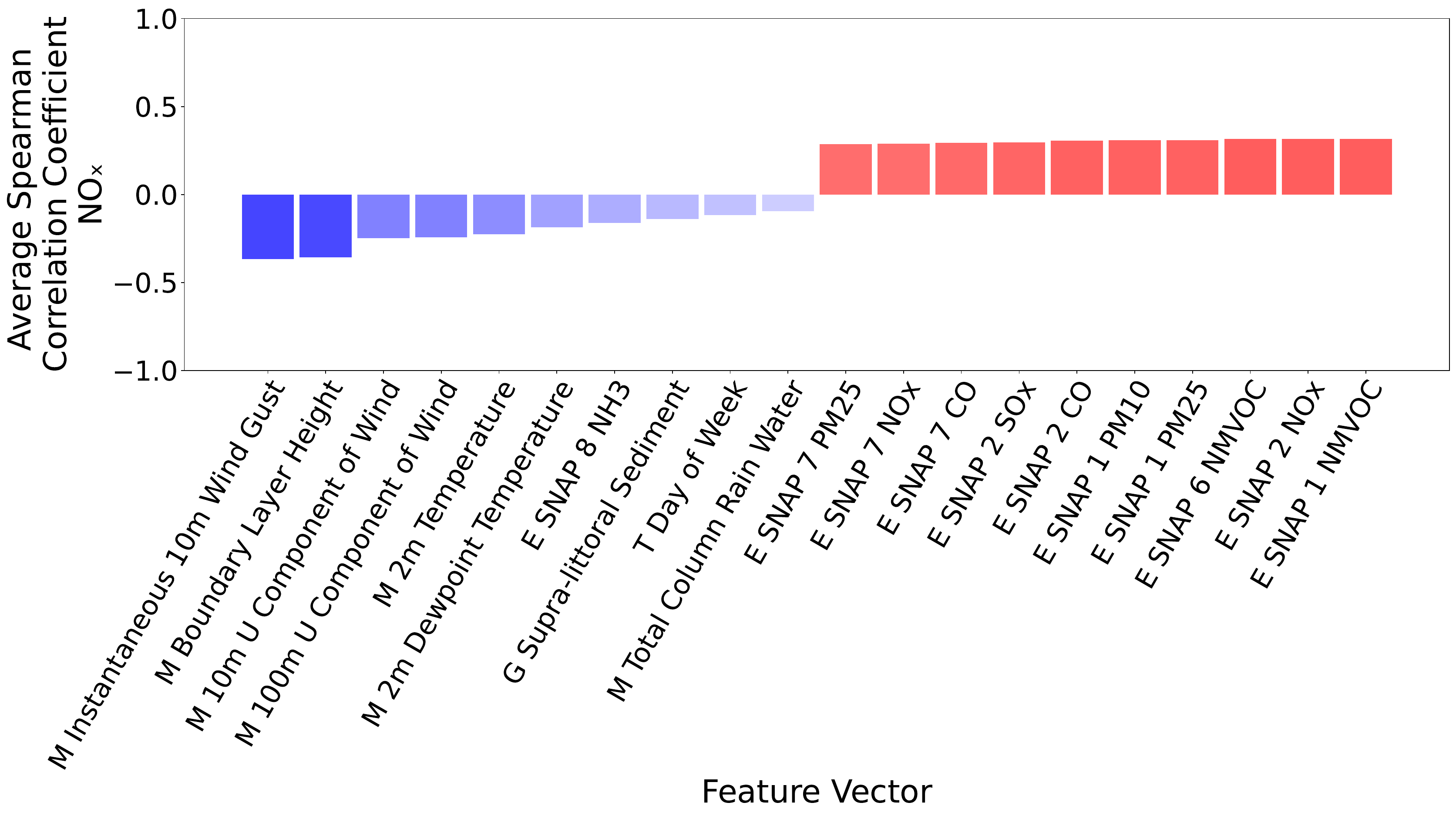}
  \end{subfigure}
  \hspace*{\fill}   % maximize separation between the subfigures
  \\
  \hspace*{\fill}   % maximize separation between the subfigures
  \begin{subfigure}{0.8\textwidth}
    \includegraphics[width=\linewidth]{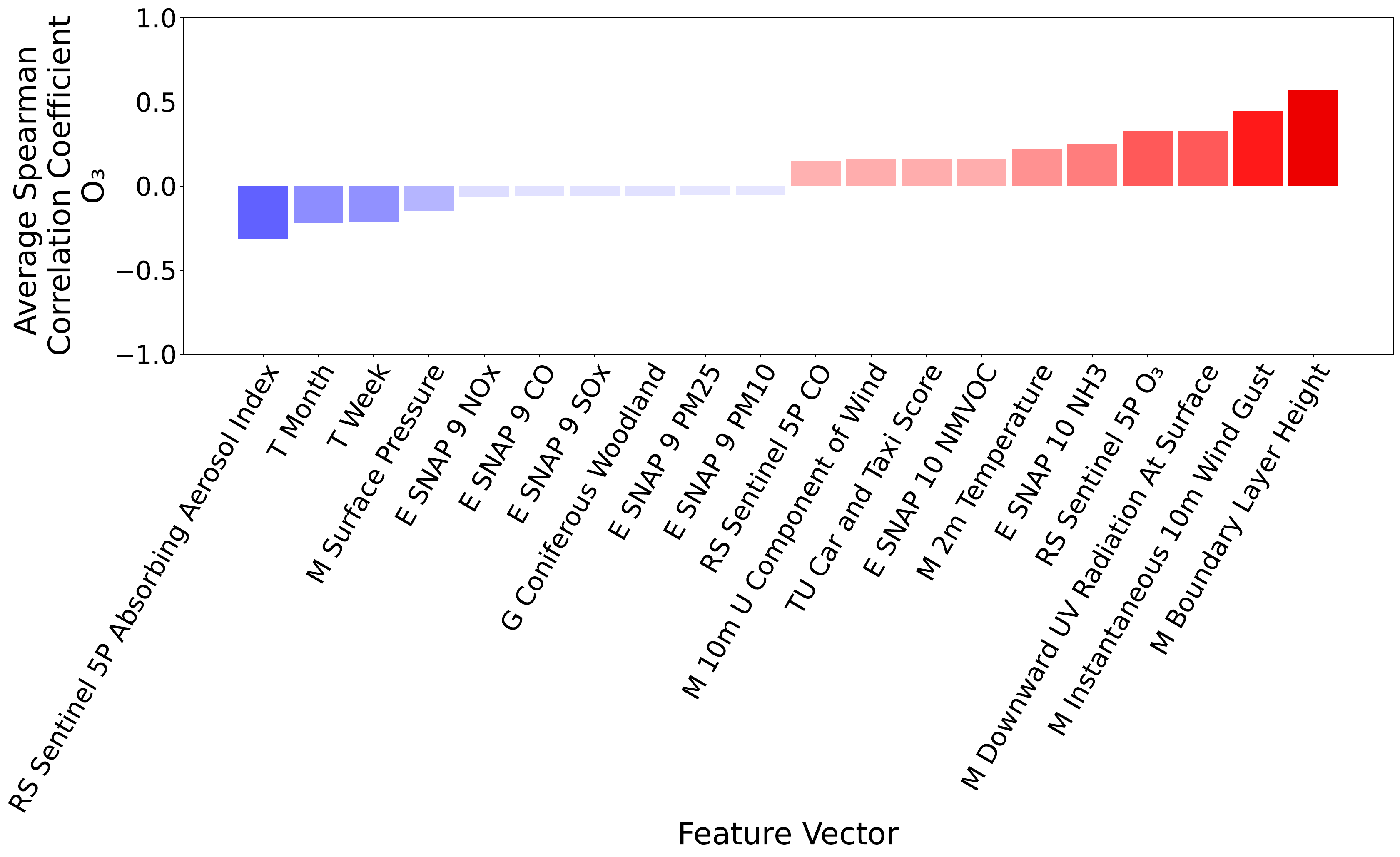}
  \end{subfigure}
  \hspace*{\fill}   % maximize separation between the subfigures
\caption{{\bfseries Spearman correlation coefficients overall mean for all pollutants.} The mean Spearman correlation coefficients for NO$_x$ and O$_3$ across all the environmental classifications of the AURN network for the ten most extreme, both positive and negative, for the feature vectors are shown. The sources and sinks of the air pollutants are different, aligning with the scientific literature (Section \ref{sec:featureVectors}), with NO$_x$ being highly positively correlated with emission features, whereas O$_3$ exhibits such a relationship mainly with meteorological features, such as wind gusts. Regarding negative correlations, the two air pollutants exhibit counter relationships, with NO$_x$ having a negative correlation with the meteorological. The analysis highlights how the relationships between a particular phenomenon and a given air pollutant can be widely different in strength.} \label{fig:spearmanCorrelationsFeatureVectorAirPollutants}
\end{figure}

Figure \ref{fig:spearmanCorrelationsFeatureVectorAirPollutants} shows the average Spearman correlation coefficient across all monitoring stations for NO$_x$ and O$_3$. These figures depict the ten highest magnitude feature vector elements in both directions. The differences in the contributing sources of air pollution for a given pollutant are evident and appear to support the scientific literature regarding the relationship between different air pollutants and their sources and sinks. 

For example, NO$_x$ has the highest positive Spearman correlation coefficient with the emissions dataset, particularly SNAP sectors 1 and 2, indicating a strong relationship between energy production and transformation emissions and the commercial, institutional, residential, and agriculture sectors. It is also notable that the highest magnitude negative Spearman correlation belongs to the sinks, namely wind speed and the boundary layer height, as expected.

In contrast, O$_3$ presents a very different situation compared to NO$_x$; the highest magnitude correlation is inverse. Both wind speed and boundary layer height have a high positive magnitude Spearman correlation, highlighting that the same phenomenon can have a completely opposite relationship on the concentrations of air pollutants. It also follows that the correlation between ``Downward UV Radiation At Surface'' and O$_3$ has a positive correlation, given that O$_3$ is produced under sunlight by the precursors of NO$_x$ and Volatile Organic Compounds (VOCs) \cite{EPA:2023:GroundLevelOzoneBasics}, highlighting that more sunlight results in more O$_3$ being produced.

\begin{figure}[!htb]
  \hspace*{\fill}   % maximize separation between the subfigures
  \begin{subfigure}{0.8\textwidth}
    \includegraphics[width=\linewidth]{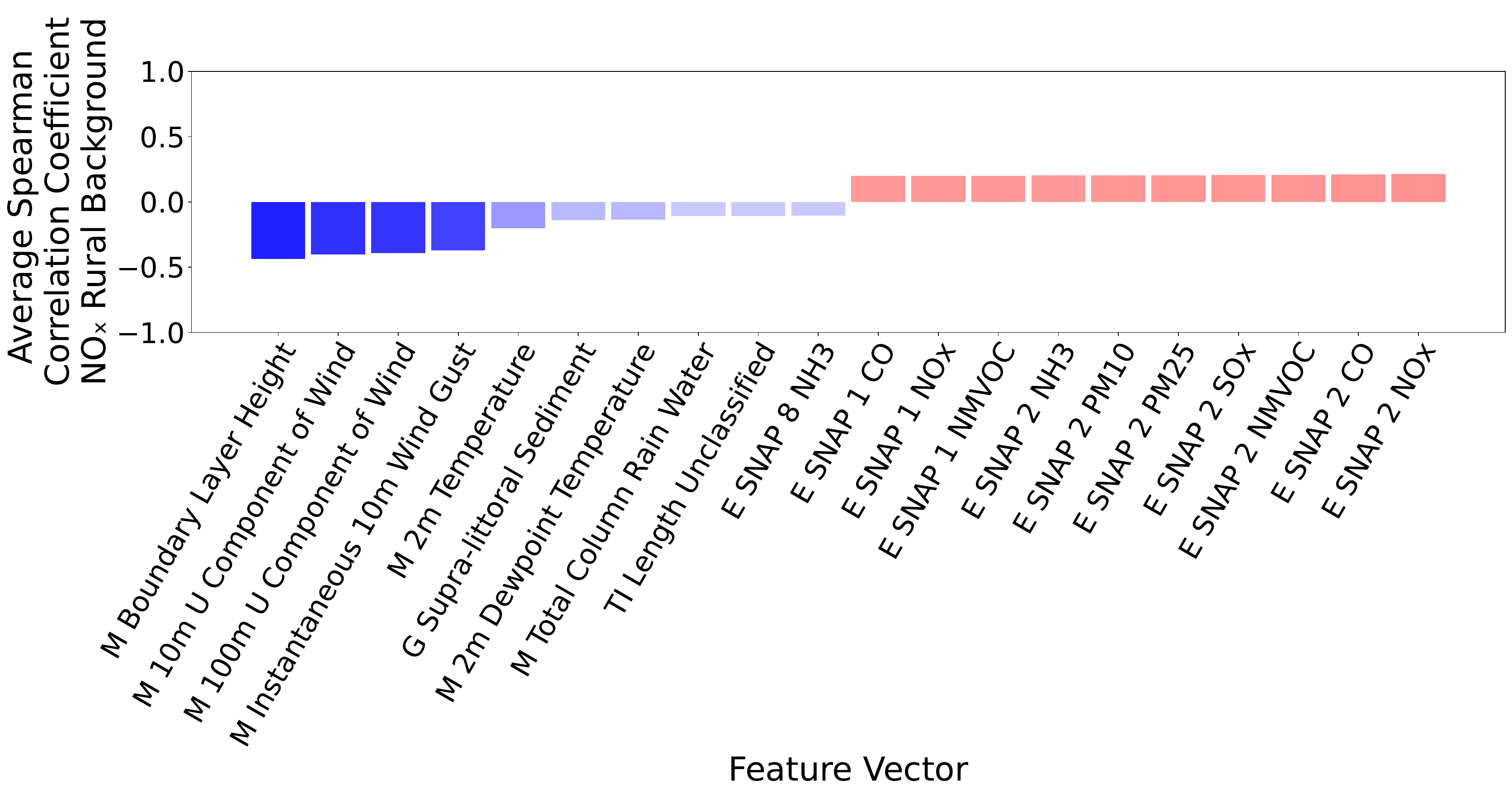}
  \end{subfigure}
  \hspace*{\fill}   % maximize separation between the subfigures
  \\
  \hspace*{\fill}   % maximize separation between the subfigures
  \begin{subfigure}{0.8\textwidth}
    \includegraphics[width=\linewidth]{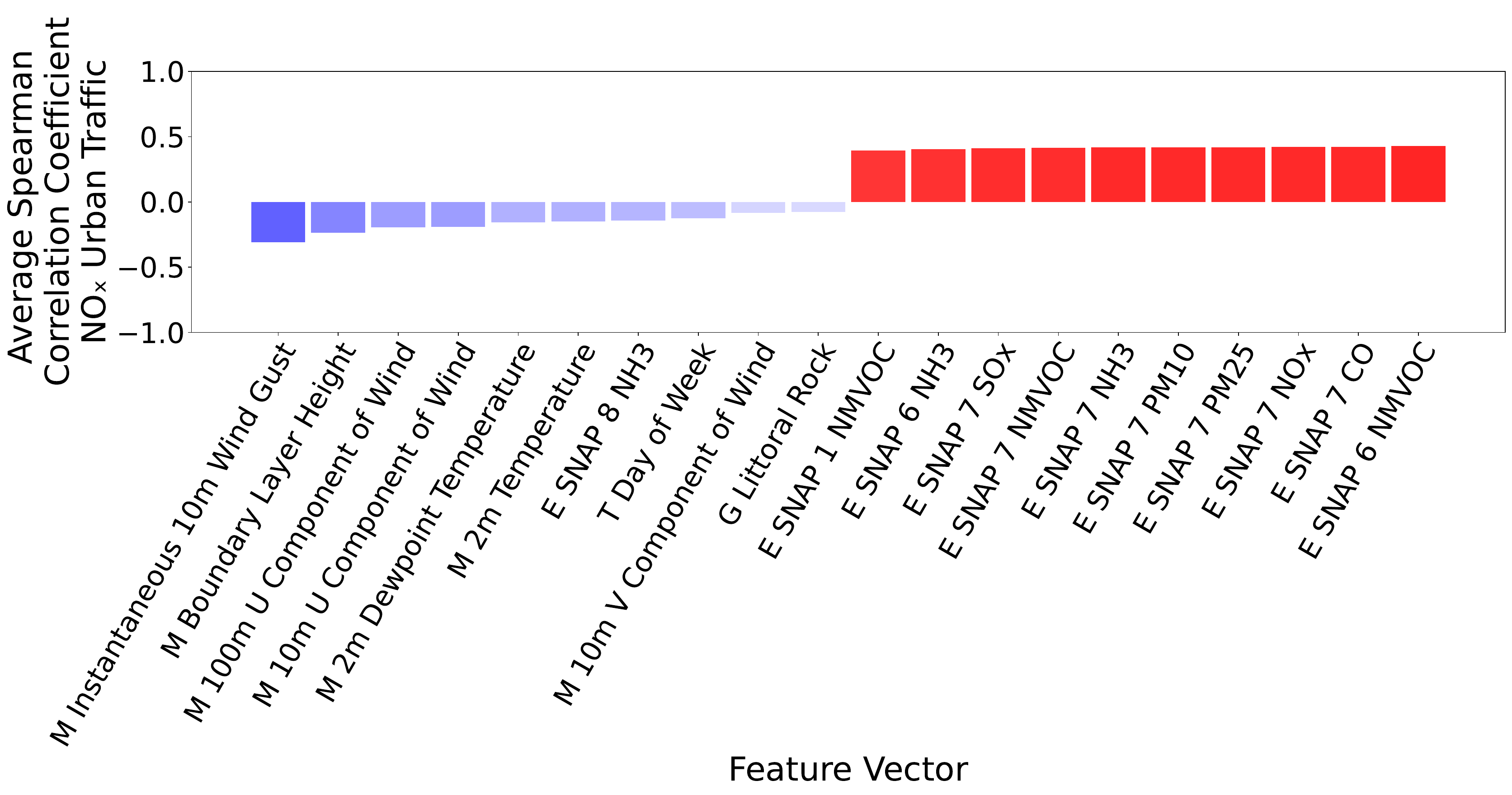}
  \end{subfigure}
  \hspace*{\fill}   % maximize separation between the subfigures
\caption{{\bfseries Spearman correlation coefficients for NO$_x$ monitoring station environmental subclassification locations, Rural Background and Urban Traffic.} While Figure \ref{fig:spearmanCorrelationsFeatureVectorAirPollutants} highlights the difference between phenomena and air pollutants, there exists a further difference between environmental subclassifications. For the Urban Traffic monitoring stations, it can be seen that the primary positive correlations are related to road transport as would be expected (the strong relationship with Solvent Use is likely an artefact of the scaling performed and discussed in Section \ref{sec:featureVectors} and \ref{S-sec:Datadetails:emissions}, alongside a limited sample size of 41 stations). In contrast, the Rural Background monitoring stations show a strong relationship with emissions from the residential sector, highlighting that the sources and sinks for an air pollutant depend on the air pollutant itself and the location of interest.  } \label{fig:spearmanCorrelationsFeatureVectorAirPollutantsSiteTypes}
\end{figure}

Figure \ref{fig:spearmanCorrelationsFeatureVectorAirPollutantsSiteTypes} illustrates the relationship between feature vectors for Rural Background and Urban Traffic monitoring station environment classifications (discussed in Section \ref{sec:airPollutionData}) for the air pollutant NO$_x$. Each subclassification of monitoring stations indicates the primary contributor to the measured air pollution. Background stations have no single primary source, while traffic stations are primarily driven by traffic.

For the Urban Traffic station, the strongest positive correlation across air pollutants is with SNAP Sector 7, denoting road transport emissions. Notably, there is a strong relationship with SNAP Sector 6 NMVOCs, indicating solvent use for NMVOCs. This might be explained by the small number of data points for monitoring stations (41 Urban Traffic stations), alongside emissions data being based on extensive scaling depending on the hour, week, and month of interest \cite{NAEI:2023:SNAPSector6Sources}. The potential for confounding variables, such as NMVOC emissions arising from vapor from petrol \cite{NAEI:2023:SNAPSector6Sources}, makes the 0.014 magnitude difference in relationship strength negligible. The current data quality is suitable for identifying general trends rather than pinpointing the most substantial relationship by sector.

The transport use dataset exhibits a positive but weaker Spearman correlation with the Urban Traffic site, with an average score of 0.36 across the five datasets. In contrast, the Rural Background sites for the transport use datasets have an average of 0.05. This aligns with the literature, showing a clear signal for increased traffic near an Urban Traffic monitoring site and increased NO$_x$ concentrations. It also agrees with the AURN environment classification, with a still positive but significantly reduced magnitude correlation. 

As the AURN environment classification is based on the primary emitters closest to the station, it then follows that both station types have the same feature vector for the highest negative magnitude Spearman correlation with the expected boundary layer height and wind, namely the meteorological variables that are present across all monitoring stations, and the sinks in the case of NO$_x$.

% \clearpage
\subsection{Inter Feature Vectors}
\label{sec:featureSelectonInterFeatureVectors}

While there is considerable existing literature about the relationship between different air pollutants and the phenomena covered in the datasets used in this study, considerably less literature covers the relationship between the phenomena comprising the feature vector. This section aims to understand the relationship between the different feature vectors to address the issue of multicollinearity, which can have significant implications for the machine learning approach implemented.

\begin{figure}[ht]
\begin{center}
\includegraphics[width=0.8\linewidth]{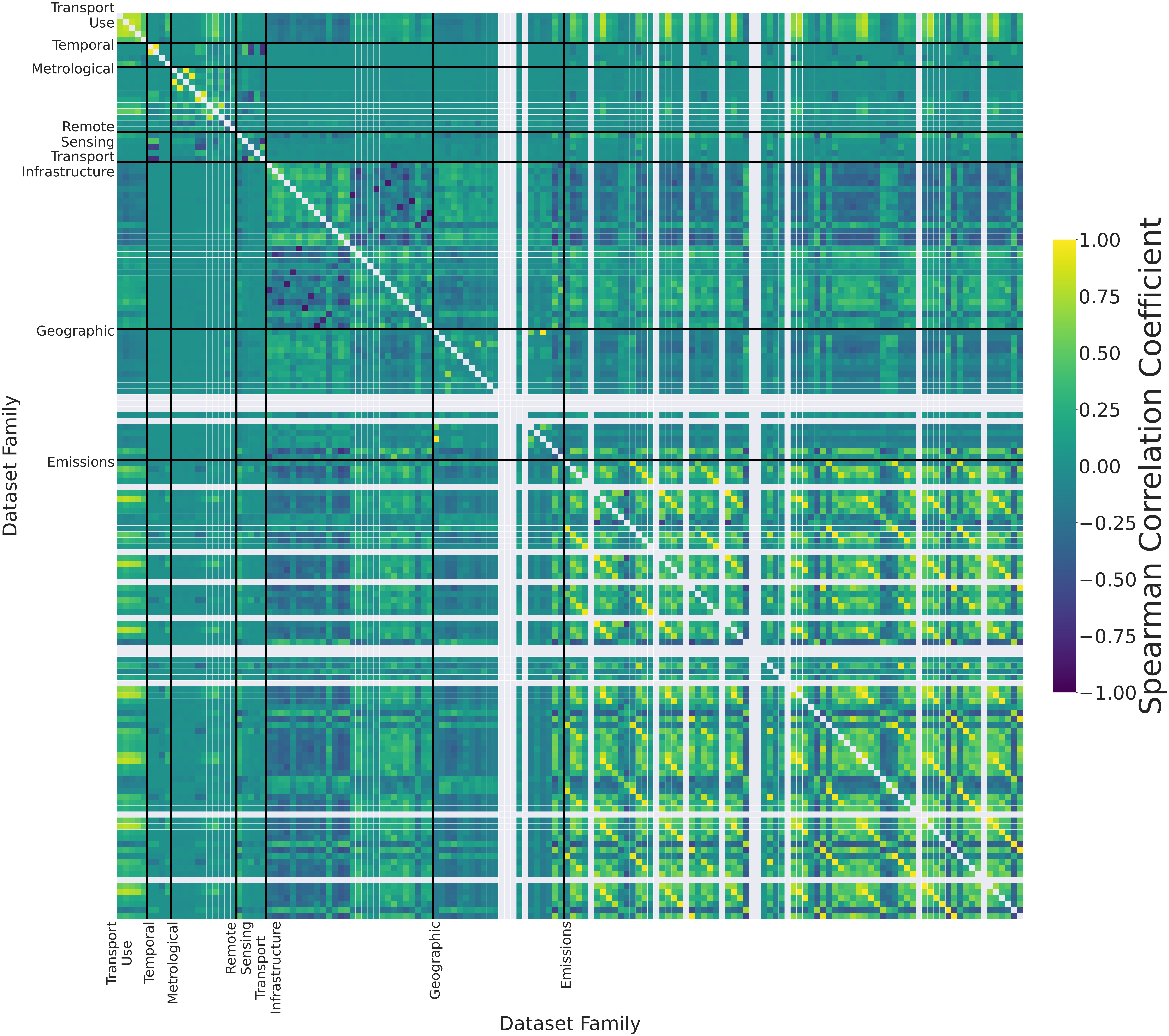}
\caption{{\bfseries Spearman correlation heatmap between all feature vectors.} The grey lines throughout the heatmap show the data points missing from the dataset, phenomena with no monitoring stations across all pollutants, including four geographic features and nine emissions features. } \label{fig:spearmanCorrelationHeatmap}
\end{center}
\end{figure}

The Spearman correlation coefficient was again used to calculate the relationship between each pair of feature vectors. Figure \ref{fig:spearmanCorrelationHeatmap} is a heatmap representing the Spearman correlation coefficient value for every pairing. There are no air pollution monitoring stations for some feature vectors—nine feature vectors for the emissions dataset family and four in the geographic dataset family. The lack of target vector data at some key feature vector locations presents the first significant problem with any model developed: not all environmental condition types experienced within England have observations. For example, there are no air pollution concentration data for the Land Use classification of Saltwater, representing a scenario in which the model has no exposure. 

From the heatmap in Figure \ref{fig:spearmanCorrelationHeatmap}, it is clear that multicollinearity is present between some features, for example between different species within the same emissions sector. Two main concerns were identified with the complete set of feature vectors: the implications on model interpretation depending on the model chosen and the redundant information between features. The model interpretation would impact future stakeholder engagement, and redundant features would make the hyperparameter search more complex. Therefore, we considered removing features entirely or creating a new set of features through dimensionality reduction. However, as stakeholder engagement with the model was an essential aspect of any operational model, and losing the semantic meaning of features would be detrimental to model interpretation, we decided against creating a new set of features.

% \noindent
% \begin{minipage}[c]{0.62\linewidth}
%   \centering
%   \begin{sideways}
%     \includegraphics[width=0.9\textheight]{Figures/Feature Selection/hierarchial_clustering_featureVectors_sideways.pdf}
%   \end{sideways}
%   \captionof{figure}{\textbf{Dendrogram depicting hierarchical clustering of feature vectors.} The lower the linkage distance between feature vectors, the more correlated the features are, indicating that they provide similar information. Table \ref{S-tab:linkageDistance} details the number of clusters for different linkage distances.}\label{fig:FeatureVectorHierarchialClustering}
% \end{minipage}%
% \hfill

\begin{figure}[p]
  \centering
  \begin{sideways}
    \includegraphics[width=0.9\textheight]{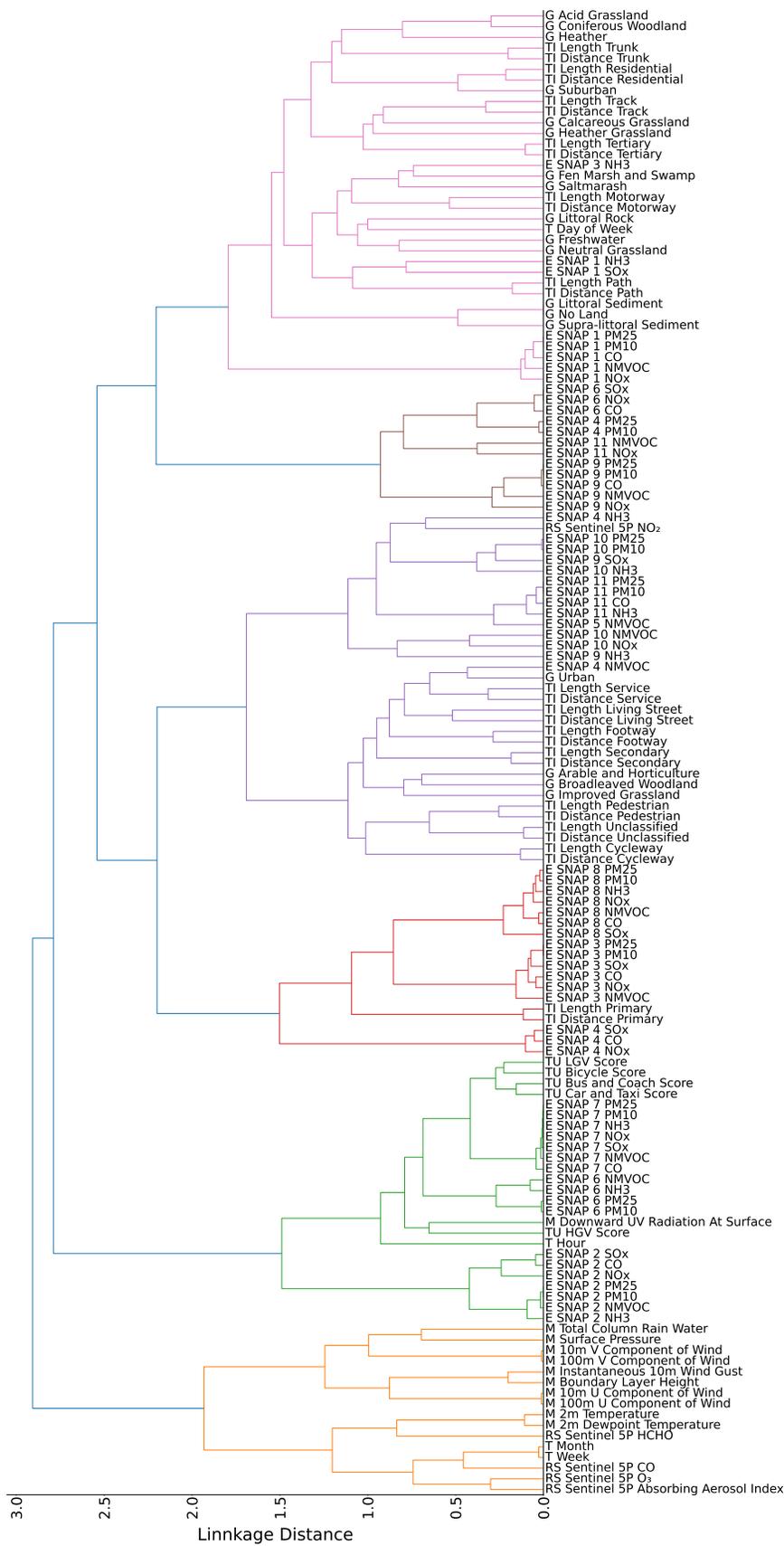}
  \end{sideways}
  \caption{\textbf{Dendrogram depicting hierarchical clustering of feature vectors.} The lower the linkage distance between feature vectors, the more correlated the features are, indicating that they provide similar information. Table \ref{S-tab:linkageDistance} details the number of clusters for different linkage distances.}
  \label{fig:FeatureVectorHierarchialClustering}
\end{figure}

% \begin{minipage}[c][\textheight]{0.33\linewidth}

Hierarchical clustering was performed between the Spearman correlation of the feature vectors, allowing us to create a more complex method of grouping together feature vectors. Figure \ref{fig:FeatureVectorHierarchialClustering} shows the clustering results. The linkage distance provides a consistent metric across all feature vectors to explore the similarity of features and provide clusters of features depending on the value of the linkage distance provided. Figure \ref{fig:FeatureVectorHierarchialClustering} shows how related some of the feature vectors are; for example, the 100m and 10m components of wind in both directions are highly correlated and therefore have a very low linkage distance. There are also more complex relations between the feature vectors, such as within the transport use datasets. Still, there are differences within the data set, such as car and taxi and bus and coach being highly related but not to the same degree as HGVs. The motivation for performing hierarchical clustering is to allow for a subset of features to be selected that provide the same information as one another, aiding model interpretation. In the case above, the idea is that including the 100m U component of wind provides the same information as the 10m U component of wind, so there isn't a need to include both.  

Table \ref{S-tab:linkageDistance} shows the number of clusters at varying linkage distances, where increasing the linkage distance results in fewer clusters as the information provided between datasets isn't required to be as strong. As including redundant feature vectors increases the computation costs of creating and using the model rather than impacting the performance of the predictions, we decided to keep all feature vectors when training the model. The intention is to provide a baseline performance of a machine learning model that utilises all the datasets covered while allowing for an understanding by individual stakeholders of the redundant feature through the hierarchical clustering performed, allowing them to subset the datasets as desired for their particular use case. Using all the feature vectors does mean, however, that a machine learning approach that is robust to multicollinearity needs to be chosen. The second issue of model interpretability implications is discussed in Section \ref{sec:modelDesign}. 

% \end{minipage}

% \clearpage
\section{Modelling}
\label{sec:modelling}

Section \ref{sec:modelling} starts by describing the reasoning behind different modelling choices. The model's performance in two critical scenarios is then explored: forecasting and estimating missing stations. Forecasting aims to answer the question of, given a location the model has already seen, how well the model performs when estimating a future year it has not. The estimating of a missing station then experiments with understanding model performance when predicting air pollution at a location it has never seen before. The model's performance on peak concentrations is also analysed, a critical situation for the model to perform well. Finally, a justification for including such a wide range of datasets is motivated by experimenting with a model that predicts based on one dataset family. 

\subsection{Model Design and Training}
\label{sec:modelDesign}

The first consideration when choosing the model framework was the need for the model to be robust to multiple uninformative and redundant features. As discussed in Sections \ref{sec:featureSelectionPollutants} and \ref{sec:featureSelectonInterFeatureVectors}, uninformative features exist for various reasons. Some features are uninformative for specific air pollutants, such as the transport use features with O$_3$. Features can also be uninformative for specific environmental locations of monitoring stations, as seen in Section \ref{sec:featureSelectionPollutants} for rural background NO$_x$ stations. Multicollinearity further compounds the issue, allowing multiple features to extract the same information about air pollution measurements, as discussed in Section \ref{sec:featureSelectonInterFeatureVectors}. The second consideration when choosing the model was for it to be robust to outliers and anomalies, as discussed in Section \ref{sec:airPollutionData}. The model we chose to use was LightGBM, a gradient-boosting algorithm based on decision trees.

LightGBM \cite{ke:2017:lightgbmDefinition} was identified as a machine learning approach that could address the concerns raised in the study through various techniques while also providing state-of-the-art performance on tabular prediction problems. LightGBM is a gradient-boosting decision tree (GBDT) algorithm where an ensemble of decision trees is trained in sequence, with the n+1 decision tree fitting the residuals of the first n decision trees, learning the difference between the actual target vector and the weighted sum of predictions of the first n decision trees.

LightGBM allows us to mitigate the impact of uninformative and redundant features on training time through the tree-building algorithm. The approach that LightGBM takes when building the decision trees is to split observations based on the feature vector values, looking for the best possible split regarding information gain and reducing the uncertainty regarding the target vector. This involves grouping homogenous instances of data points, such as instances where there is high transport use at a monitoring station that is measuring high concentration readings for NO$_x$.

One of the core issues with our air pollution training data is that many data points within the datasets repeat the same information due to the cyclical nature of air pollution measurements, causing a considerable amount of bloat in the datasets. The standard approach to identifying split points within a GBDT is the pre-sorted algorithm where all possible split points are explored, an approach which, in this use case, would be highly costly regarding computation and memory. LightGBM helps tackle this issue by using histograms when performing the splits, where continuous variables are put into discrete bins, changing the computational cost from being dependent on the number of data points to the number of discrete bins created.

The second concern identified when exploring the datasets was the presence of outliers and anomalies within the dataset. LightGBM inherently tackles this problem via decision trees being the underlying learner within the model. The decision tree's goal is to group homogenous data instances, and there is an ability to set a minimum number of data instances that comprise a valid leaf on the decision tree via the ``minimum data in leaf'' model parameter. 

The ``minimum data in leaf'' parameter allows for a minimum threshold of homogeneous data points for the LightGBM algorithm to view as a set of data points that should be learned from and used in predictions. Concerning the air pollution prediction problem as a notional example, say there is a high wind speed and low traffic count but a high air pollution concentration reading, which only occurs once in the dataset; LightGBM will not create a leaf for this data instance. The scenario described could plausibly happen if a single air pollution emitter passes by the station, causing an artificially high measurement that would not represent the geographic area intended for the station as outlined by the AURN documentation, as discussed in Section \ref{sec:airPollutionData}.

There are, however, some trade-offs to the LightGBM solutions presented above. The feature importances given for the feature vectors via the model will likely be misleading due to the multicollinearity present. For example, the most extreme case seen in the hierarchical clustering of multicollinearity is for the wind speeds at 10m and 100m, where both features exhibit a strong correlation, meaning that we can extract information about air pollution from either feature.
Therefore, during model building, in the split performed, the model would use only the 10m or 100m component of the wind direction, as they would present the same information gain about the target variable as each other. As the feature importance given by LightGBM is based on the number of times a feature vector is used, the total number of times the two feature vectors are used may be split across the two features, reducing the feature importance given to each one. The feature importances given must be analyzed considering the clusters presented in Table \ref{S-tab:linkageDistance}, treating each of the clusters' feature importances together, or a more sophisticated technique of model interrogation used, such as SHAP \cite{Lundberg:2017:SHAP}.

A key consideration during the model design was the choice of the loss function. The loss function represents the error of a given prediction, in this case, quantifying the difference between the prediction and the actual air pollution concentration measurement of a model, thereby allowing for comparisons between models and subsequent choice of the optimal model. The choice of the loss function in this situation was between the mean absolute error (MAE) and the mean squared error (MSE) \cite{hodson:2022:RMSEMAE}. The MAE would help reduce the influence of higher air pollution measurements on the model present due to the known presence of outliers and anomalies within the dataset. However, these high air pollution measurements are of vital interest within the context of air pollution, even if they are potentially erroneous. So, a tradeoff of potentially overfitting on these higher values was seen as a worthwhile tradeoff, and as such, the MSE was chosen as the loss function. The underlying premise is that 10\si{\micro\gram/\meter^3} is more than twice as bad for human health than 5\si{\micro\gram/\meter^3}, so using the MSE is more appropriate given the domain in which the model would be used, supporting the existing literature that there is a non-linear relationship between the detrimental effects of air pollution concentrations \cite{yang:2022:nonlinearAirPollution, zhao:2019:nonlinearAirPollution2}.

As an air pollution concentration can never measure less than zero, we trained the model on the log transformation of the target vector. We added a small constant of \(1 \times 10^{-7}\) to the target vector and then performed the log transformation due to the presence of 0 concentration measurements within the dataset. The log transformation and addition ensured that the model would never predict a negative value as the model output, as the reverse transformation of calculating the exponential and subtracting \(1 \times 10^{-7}\) was performed on the output. A further hyperparameter explored during model training was L2 regularization \cite{hoerl:1970:RidgeRegression}. Including L2 regularization helps distribute the weights within the decision tree, encouraging the weights to be closer to 0 but keeping all feature vectors, ensuring that no single feature vector drives predictions, which is key with the considerable number of feature vectors used.

The framework for choosing the model's hyperparameters was a randomised grid search of 40 parameter sets. The parameters we optimized during the randomised search were the L2 regularization and the min data in each leaf already discussed, alongside the number of leaves, the number of trees, and the max depth \cite{LightGBM:2023:LIghtGBMParameters}. The number of leaves search space was given the range of 1,000 to 4,095 \cite{Github:2017:ParameterGridSuggestions}, with the optimal values being chosen near the centre of this range, validating its choice. The number of trees was controlled via early stopping, where no additional tree would be added after 30 trees had been added without any improvement in the loss function performance. Similarly, the max depth was not limited and left to grow as needed until performance did not improve during training.

Some model parameters were kept constant throughout the search, such as the max bin, kept constant at 255. The max bin refers to the number of discrete bins created for a continuous feature vector. 255 was chosen to ensure that a range of different splits during model training could be created while also helping to reduce training time by allowing data to be stored optimally as an int8 data type. The boosting type used during training was Gradient-based One-Side Sampling (GOSS) \cite{ke:2017:lightgbmDefinition}. GOSS is a method of boosting that allows the n+1 decision tree discussed at the start of this section to be trained on a subsample of the data. The subsample of data chosen is the data that has a large gradient, i.e. the data has yet to learn well from in the model and a random sample of the small gradient data, helping to reduce the amount of data used drastically, and therefore training time. The tradeoff with GOSS is the potential for overfitting when the datasets are small; however, this was not a concern in the context of air pollution. 

The final consideration was the grouping and number of models to develop. One possible choice was creating a single unified model with all seven air pollutants comprising the target vector. However, due to the considerable imbalance in the number of data points and the issue of every monitoring station measuring a different subset of pollutants, the number of locations with every air pollutant measured at the same timestamp was minuscule. Another possibility was to create an individual model for each environment type covered in the AURN environment classification, such as Urban Traffic, Rural Background. However, this approach presented the problem of requiring a determination of the environment type of every grid in the study where there wasn't an existing monitoring station. Therefore, we created a single model for each of the different air pollutants mixed with all the different environment types, the benefit of which is simplifying the process of estimating a never before seen location while making use of all of the air pollution observations possible. 

During the hyperparameter grid search, data from 2014-2016 was used as the training set, with the validation set being 2017 and 2018 as the test set. We split the dataset temporally to ensure no data leakage and to give an intuitive sense of the performance metrics gathered. We chose the best parameter set based on the model's MSE on the validation set across the parameter sets. Subsequently, using the best parameter set to train a model with both the training and validation set, with performance evaluated using the test set, data the model has never seen. The R$^2$ score for each model on the different sets was calculated at each stage.

To allow flexibility in extending the model with new data, we deliberately excluded feature vector elements that would identify monitoring station details, such as names or locations. Additionally, we opted not to include lags of air pollution concentrations, like using the concentration at T-1 to estimate the concentration at time T. This decision enables us to make predictions even in the absence of a specific air pollution measurement, promoting the use of observations as independent entities. This structure facilitates the tabular format, leveraging the state-of-the-art performance of LightGBM. The temporal and spatial independence of observations supports the creation of a lightweight model, conducive to parallel computation for different locations and time points. Our experiments addressed two crucial scenarios, providing insights into the model's temporal and spatial performance. We assessed its ability to forecast air pollution concentrations into the future (discussed in Section \ref{sec:modelResultsTemporal}), and extended the analysis to evaluate the model's performance in estimating air pollution concentrations in spatial locations not previously encountered (discussed in Section \ref{sec:modelResultsSpatial}).

% \clearpage
\subsection{Filling Temporal Missing Data}
\label{sec:modelResultsTemporal}
The initial set of experiments focused on evaluating the models' capability to predict future air pollution concentrations at locations already included in the model. These experiments aimed to assess the model's performance in forecasting scenarios. The achieved performance serves as a conservative estimate for filling in missing data temporally in a given monitoring station's time series, considering that air pollution concentration readings at the estimated time would be available in an operational hindcast situation.

Table \ref{tab:datasubsetResultsAllDatasets} presents the R$^2$ score for the model developed at each stage, as discussed in Section \ref{sec:modelDesign}. The results illustrate a degradation in the model's performance as it moves temporally away from the data it was initially optimized for during the randomised parameter grid search. These experiments demonstrate that the model parameters identified during the randomised search remain consistent across the three datasets, with minimal performance loss observed between the validation and test sets. This observation supports the idea that the model is effectively learning the true relationship between the feature and target vectors.

\begin{table}[ht]
\centering
\resizebox{0.45\linewidth}{!}{
\begin{filecontents*}{CSVFiles/Model/temporal_experiment_datasetSubset_All_results_modelType_mean.csv}
Pollutant Name,Dataset Train Score,Dataset Validation Score,Dataset Test Score
NO₂,0.85,0.77,0.77
NOₓ,0.82,0.75,0.74
NO,0.73,0.67,0.65
O₃,0.8,0.7,0.67
SO₂,0.45,0.43,0.3
PM₁₀,0.51,0.38,0.32
PM₂₅,0.55,0.35,0.29
\end{filecontents*}
\pgfplotstabletypeset[
    multicolumn names=l, 
    col sep=comma, 
    string type, 
    header = has colnames, 
    columns={Pollutant Name, Dataset Train Score, Dataset Validation Score, Dataset Test Score},
    columns/Pollutant Name/.style={column type=l, column name=\shortstack{Pollutant\\ Name}},
    columns/Dataset Train Score/.style={column type={S[round-precision=2, table-format=-1.3, table-number-alignment=center]}, column name=\shortstack{Dataset\\ Train Score}},
    columns/Dataset Validation Score/.style={column type={S[round-precision=2, table-format=-1.3, table-number-alignment=center]}, column name=\shortstack{Dataset\\ Validation Score}},
    columns/Dataset Test Score/.style={column type={S[round-precision=2, table-format=-1.3, table-number-alignment=center]}, column name=\shortstack{Dataset\\ Test Score}},
    every head row/.style={before row=\toprule, after row=\midrule},
    every last row/.style={after row=\bottomrule}
    ]{CSVFiles/Model/temporal_experiment_datasetSubset_All_results_modelType_mean.csv}}
    \smallskip
    \caption{{\bfseries R$^2$ scores depicting forecasting performance (2014-2016 train score, 2017 validation socre, 2018 test score).}   The dataset train score shows the model's performance in capturing the relationship of the training data shown with the validation showing the performance in 2017 and the test score on 2018 data. The similar performance between the validation and test scores shows that the model optimised during the parameter search is learning the true relationship between the features and air pollution that is robust to data never seen before. }\label{tab:datasubsetResultsAllDatasets}
\end{table}

In the best-case scenario, NO$_2$ shows no drop in performance (rounded to 2 decimal places) between the validation and test sets. The most significant performance decrease is observed in SO$_2$; however, it's important to note that this may be influenced by a data issue, as SO$_2$ has significantly fewer stations (18) compared to NO$_2$ (103), as detailed in Table \ref{tab:AURNClassificationNumberOfStations}.

\begin{figure}[ht]
  \hspace*{\fill}   % maximize separation between the subfigures
  \begin{subfigure}{\textwidth}
    \includegraphics[width=\linewidth]{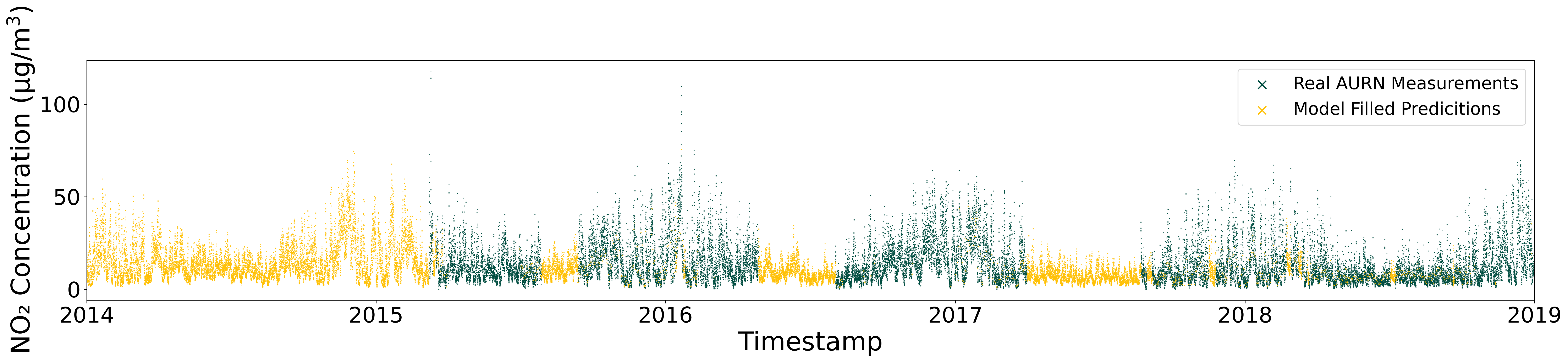}
  \end{subfigure}\\
  \hspace*{\fill}   % maximize separation between the subfigures
\caption{{\bfseries Chesterfield Loundsley Green NO$_2$ concentrations augmented dataset, with missing AURN measurements filled with model predictions.} Figure \ref{fig:aurnModelComparison} shows that the station's measurements (green) started in early 2015 with three clear periods of long-term missing data. The model predictions (yellow) can create a complete augmented time series using the model.} \label{fig:aurnModelComparison}
\end{figure}

While a benefit of the model presented is the ability to forecast air pollution concentrations into the future, answering the question of what air pollution concentrations at a station will look like in the next year, the adaptable temporal and spatial independence discussed in section \ref{sec:modelDesign} allows for the model to be used to estimate missing data. Figure \ref{fig:aurnModelComparison} shows the model used to estimate the missing data in the NO$_2$ observations for the Chesterfield Loundsley Green monitoring station from 2014-2018. There are two possible cases for the missing data being filled in. The first is to backdate or postdate the observations depending on when the station came online or was decommissioned. Chesterfield Loundsley Green came online on 01/03/2015 \cite{UKAIR:2023:ChesterfieldAURNBackground}, but the model can backdate the observations to 01/01/2014, extending the readings and helping to create a complete dataset. It is also possible to extend the life of a station if it was taken offline by filling in observations since the station was decommissioned \cite{UKAIR:2023:ClosedStationsAURN}. 

The second situation where the model can fill in missing data is when an issue at the monitoring station or associated infrastructure causes the station to go offline and measurements not to be reported \cite{UKAIR:2023:InteractiveMapDownStations}. However, one potential issue with this approach is if there is a particular reason that the site cannot report data, for example, when wind speeds are over a defined speed. This situation would indicate that no data is within the training set concerning this specific situation, indicating the model is extrapolating. However, this is not a concern in this situation as AURN reports the reasoning behind data not being reported, such as a communication issue or an instrument error. It is something, however, to be understood in the context of any future work that uses this framework where this situation could occur. In Figure \ref{fig:aurnModelComparison}, there are three prolonged periods in which Chesterfield was not reporting NO$_x$ measurements. The model presented can fill in these periods alongside the periods from before the station came online to create a time series for the station that has all available data as seen in Figure \ref{fig:aurnModelComparison} where the real measurements from the station have been augmented with the model output where real measurements are not available.

\subsection{Filling Spatial Missing Data}
\label{sec:modelResultsSpatial}

The second set of experiments that we conducted explored the ability of a model to be trained and predict the complete time series for another station, never seen before. We used 5-fold leave-one-out validation (LOOV) to experiment with this scenario. The results from this experiment provide an understanding of how the model performs when filling in missing air pollution concentration data spatially, a situation akin to using the model as synthetic stations across England at locations where no station has ever existed. 

The same experimental design as Section \ref{sec:modelResultsTemporal} was repeated alongside a final step that calculates the LOOV score for every station not included in the training, validation or test set. Table \ref{tab:spatialExperimentSpatialExperimentRepeatForTemporal} shows that the models trained during the 5-fold LOOV can retain their future predictive performance, with minor differences for the performance of ai pollutants across the different subsets of stations used, showing the results from section \ref{sec:modelResultsTemporal} to be robust to changing input datasets. Table \ref{tab:summaryStatisticsLOOVAll} shows the LOOV summary statistics for the experiments conducted, based on the R$^2$ retrieved from the model estimating the complete time series of a monitoring station's data. Four different summary statistics were considered from the set of LOOV results, namely the mean, median, min and max results. The max LOOV results are positive for all of the pollutants, indicating some merit to this approach across all pollutants. This is further supported by the majority of positive results in the mean and median LOOV for all pollutants apart from SO$_2$. Of central interest is the LOOV min, where for all pollutants other than PM$_10$ there is a negative R$^2$, indicating a prediction across the time series that is worse than simply predicting the average concentration. In this case, a potential hypothesis for the model performance is a lack of data available. The performance of SO$_2$ supports this hypothesis; the worst out of all pollutants and has the least available data.

\begin{table}[ht]
\centering
\resizebox{0.45\linewidth}{!}{
\begin{filecontents*}{CSVFiles/Model/spatial_experiment_overview_datasetSubset_All_results_modelType_mean_LOOV.csv}
Dataset Train Score,Dataset Validation Score,Dataset Test Score,Pollutant Name
0.85,0.77,0.77,NO₂
0.82,0.75,0.74,NOₓ
0.73,0.67,0.65,NO
0.81,0.7,0.67,O₃
0.46,0.42,0.29,SO₂
0.54,0.38,0.32,PM₁₀
0.58,0.34,0.29,PM₂₅
\end{filecontents*}
\pgfplotstabletypeset[
    multicolumn names=l, 
    col sep=comma, 
    string type, 
    header = has colnames, 
    columns={Pollutant Name, Dataset Train Score, Dataset Validation Score, Dataset Test Score},
    columns/Pollutant Name/.style={column type=l, column name=\shortstack{Pollutant\\ Name}},
    columns/Dataset Train Score/.style={column type={S[round-precision=2, table-format=-1.3, table-number-alignment=center]}, column name=\shortstack{Dataset\\ Train Score}},
    columns/Dataset Validation Score/.style={column type={S[round-precision=2, table-format=-1.3, table-number-alignment=center]}, column name=\shortstack{Dataset\\ Validation Score}},
    columns/Dataset Test Score/.style={column type={S[round-precision=2, table-format=-1.3, table-number-alignment=center]}, column name=\shortstack{Dataset\\ Test Score}},
    every head row/.style={before row=\toprule, after row=\midrule},
    every last row/.style={after row=\bottomrule}
    ]{CSVFiles/Model/spatial_experiment_overview_datasetSubset_All_results_modelType_mean_LOOV.csv}}
    \smallskip
    \caption{{\bfseries R$^2$ scores depicting forecasting performance for 5-fold leave-one-out-validation. } The experiment conducted aimed to ensure that with different subsets of monitoring stations, the forecasting performance of the model remains robust. Shown with Table \ref{tab:spatialExperimentSpatialExperimentRepeatForTemporal} having similar performance as the experiment result shown in Table \ref{tab:datasubsetResultsAllDatasets}, particularly the test score, data the models have never seen before. }\label{tab:spatialExperimentSpatialExperimentRepeatForTemporal}
\end{table}

Through the definition provided for the AURN station environment types, we know that air pollutant concentrations exhibit different signatures in different locations. There is the possibility of further subclassifications within these environment types. For example, taking the Urban Traffic environment type, there could be a distinction between the Urban Traffic stations within London and outside of London. Particularly for the approach used within this work, a data-driven model, this can have wide-ranging implications. Suppose it does exist that a subset or single station within the LOOV dataset is unique compared to others, resulting in no similar data present within the training data. In that case, the model presented will fail to replicate the time series measured. This hypothesis is supported as the temporal experiment framework performance is consistent across all the experiments conducted, shown in Figure \ref{tab:datasubsetResultsAllDatasets} and Figure \ref{tab:spatialExperimentSpatialExperimentRepeatForTemporal}, with the issue only appearing in the spatial experiments. 
As a more concrete example of this scenario, the Aston Hill AURN monitoring station is the only monitoring station in the NO$_x$ dataset with no roads nearby, a clearly unique monitoring station in the dataset. This issue is further complicated when considering the results from Section \ref{sec:featureSelectonInterFeatureVectors}, where not all feature vector elements have an air pollution monitoring station present, showing clear environment types that have no data available, denoting situations where the model is extrapolating and potentially widely wrong. 

\begin{figure}[!htb]
  \hspace*{\fill}   % maximize separation between the subfigures
  \begin{subfigure}{0.32\textwidth}
    \includegraphics[width=\linewidth]{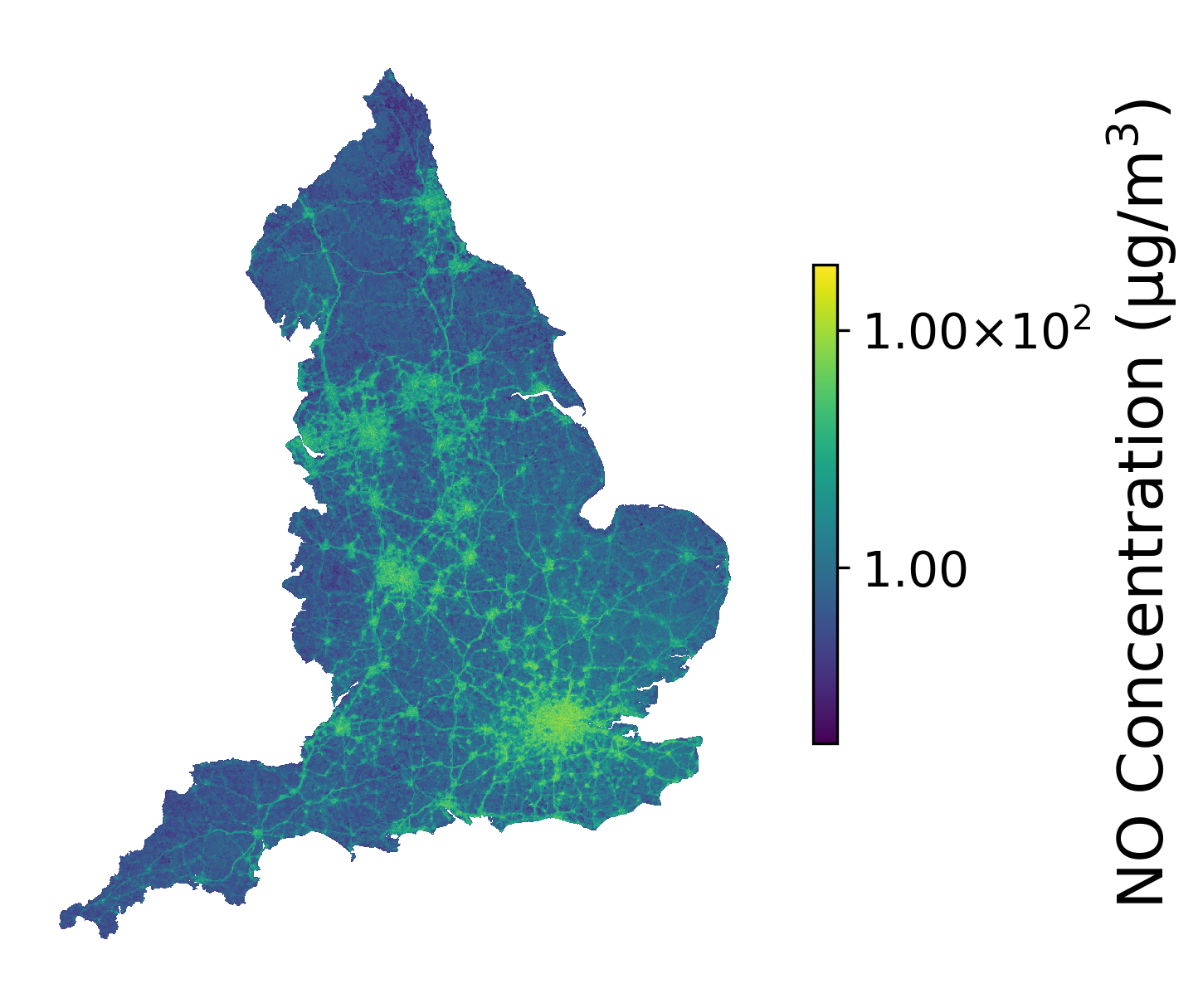}
  \end{subfigure}%
  \hspace*{\fill}   % maximize separation between the subfigures
  \hspace*{\fill}   % maximize separation between the subfigures
  \begin{subfigure}{0.32\textwidth}
    \includegraphics[width=\linewidth]{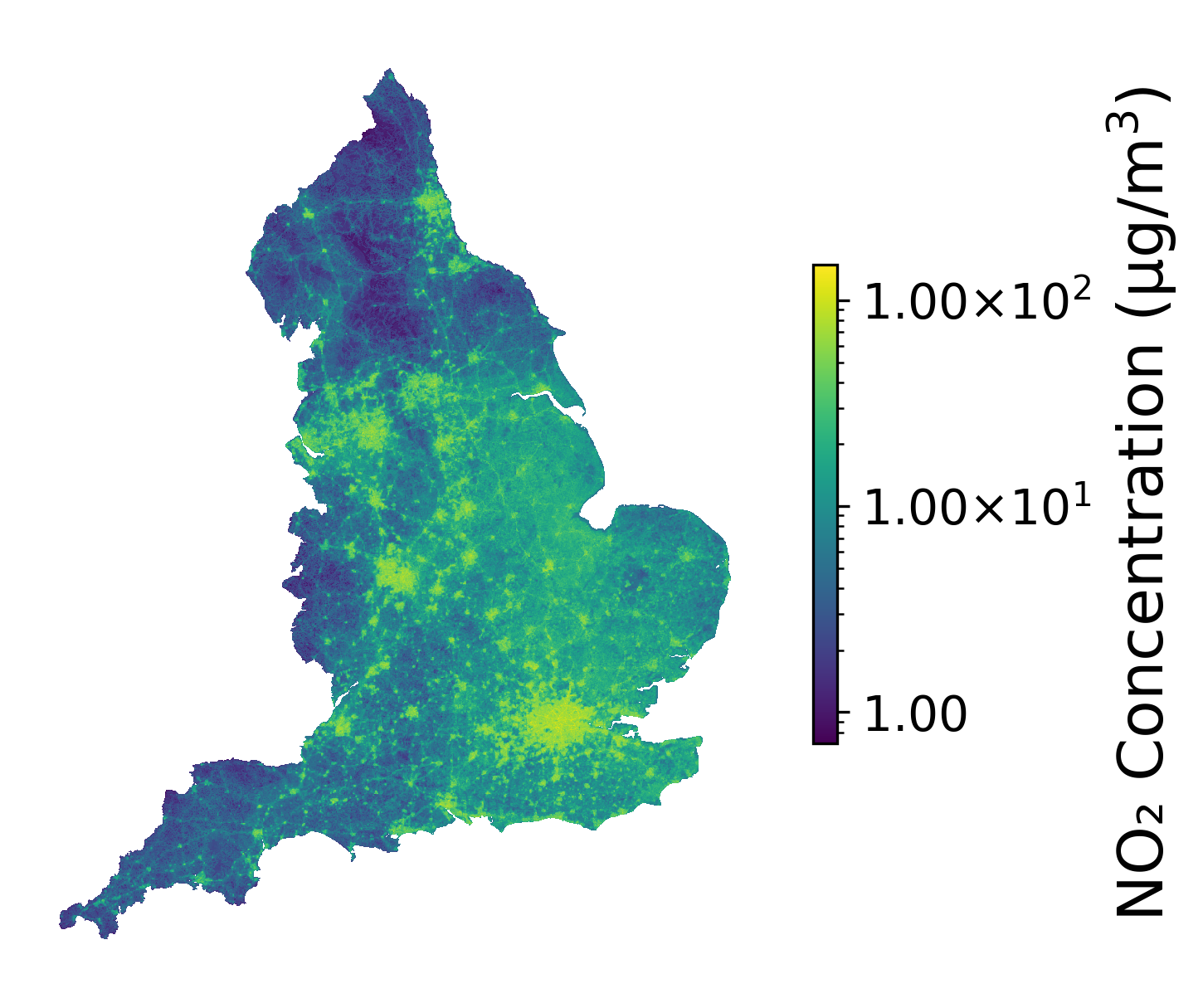}
  \end{subfigure}%
  \hspace*{\fill}   % maximize separation between the subfigures
  \hspace*{\fill}   % maximize separation between the subfigures
  \begin{subfigure}{0.32\textwidth}
    \includegraphics[width=\linewidth]{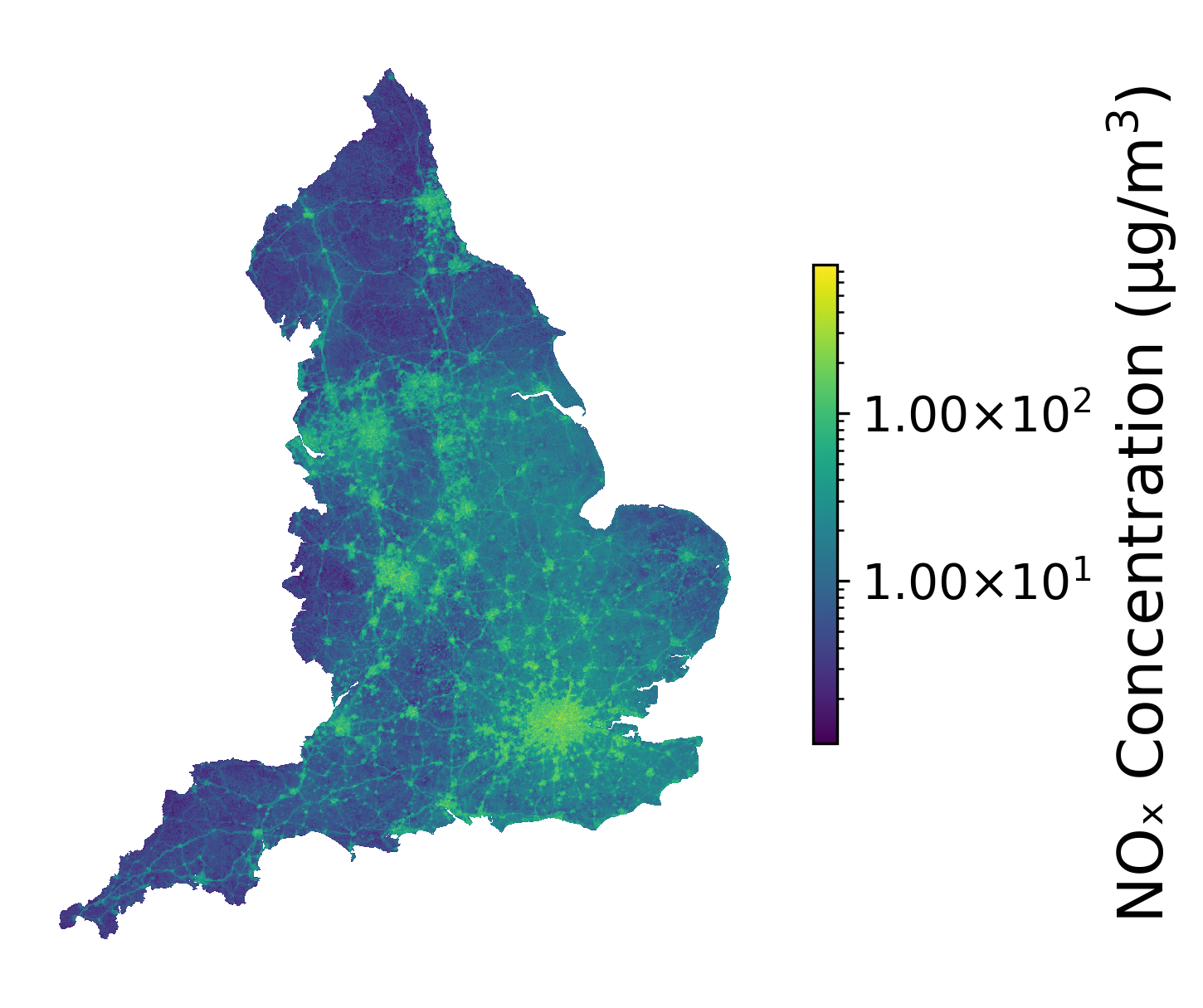}
  \end{subfigure}%
  \hspace*{\fill}   % maximize separation between the subfigures
  \\
  \hspace*{\fill}   % maximize separation between the subfigures
  \begin{subfigure}{0.32\textwidth}
    \includegraphics[width=\linewidth]{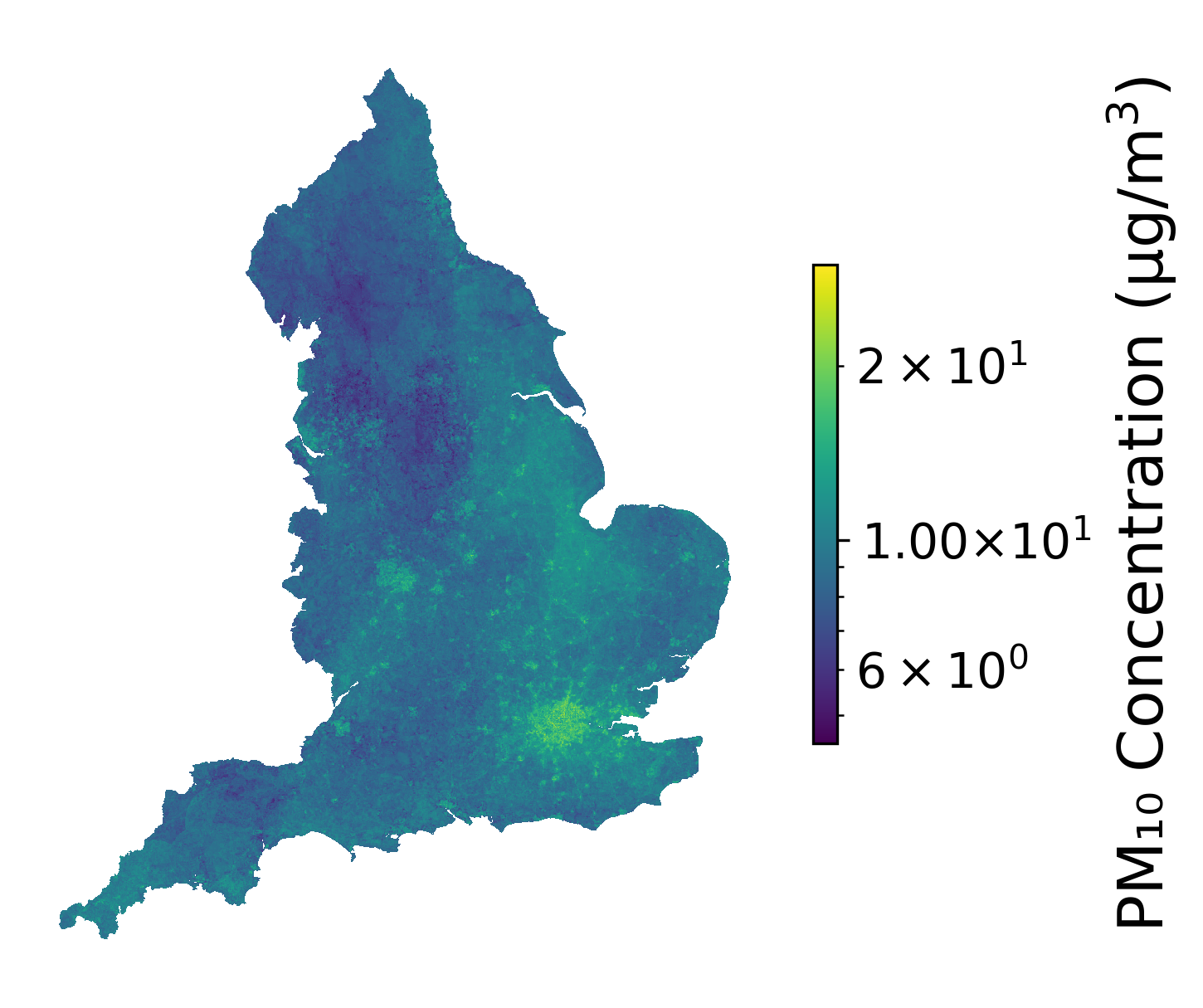}
  \end{subfigure}%
  \hspace*{\fill}   % maximize separation between the subfigures
  \hspace*{\fill}   % maximize separation between the subfigures
  \begin{subfigure}{0.32\textwidth}
    \includegraphics[width=\linewidth]{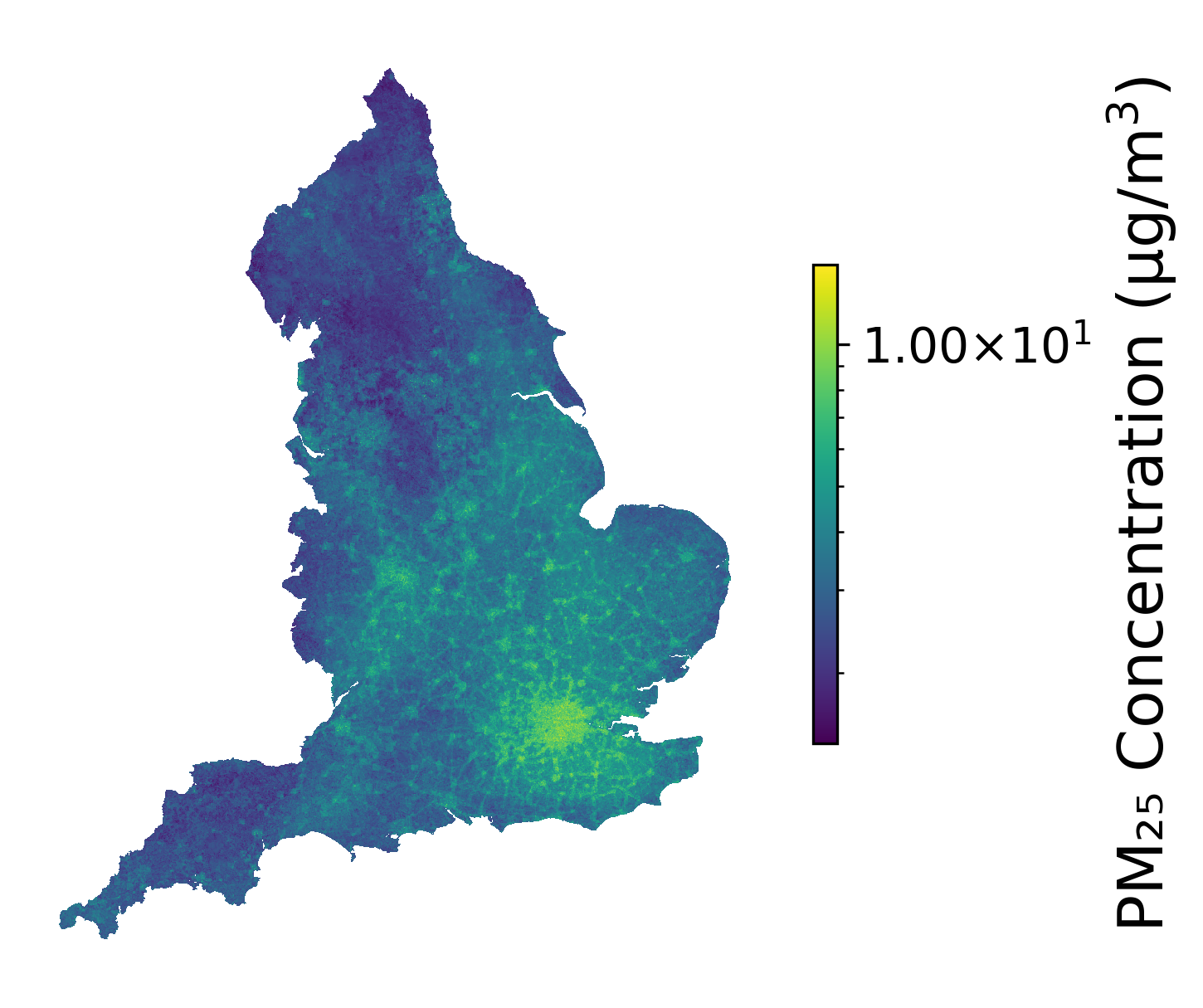}
  \end{subfigure}%
  \hspace*{\fill}   % maximize separation between the subfigures
  \\
  \hspace*{\fill}   % maximize separation between the subfigures
  \begin{subfigure}{0.32\textwidth}
    \includegraphics[width=\linewidth]{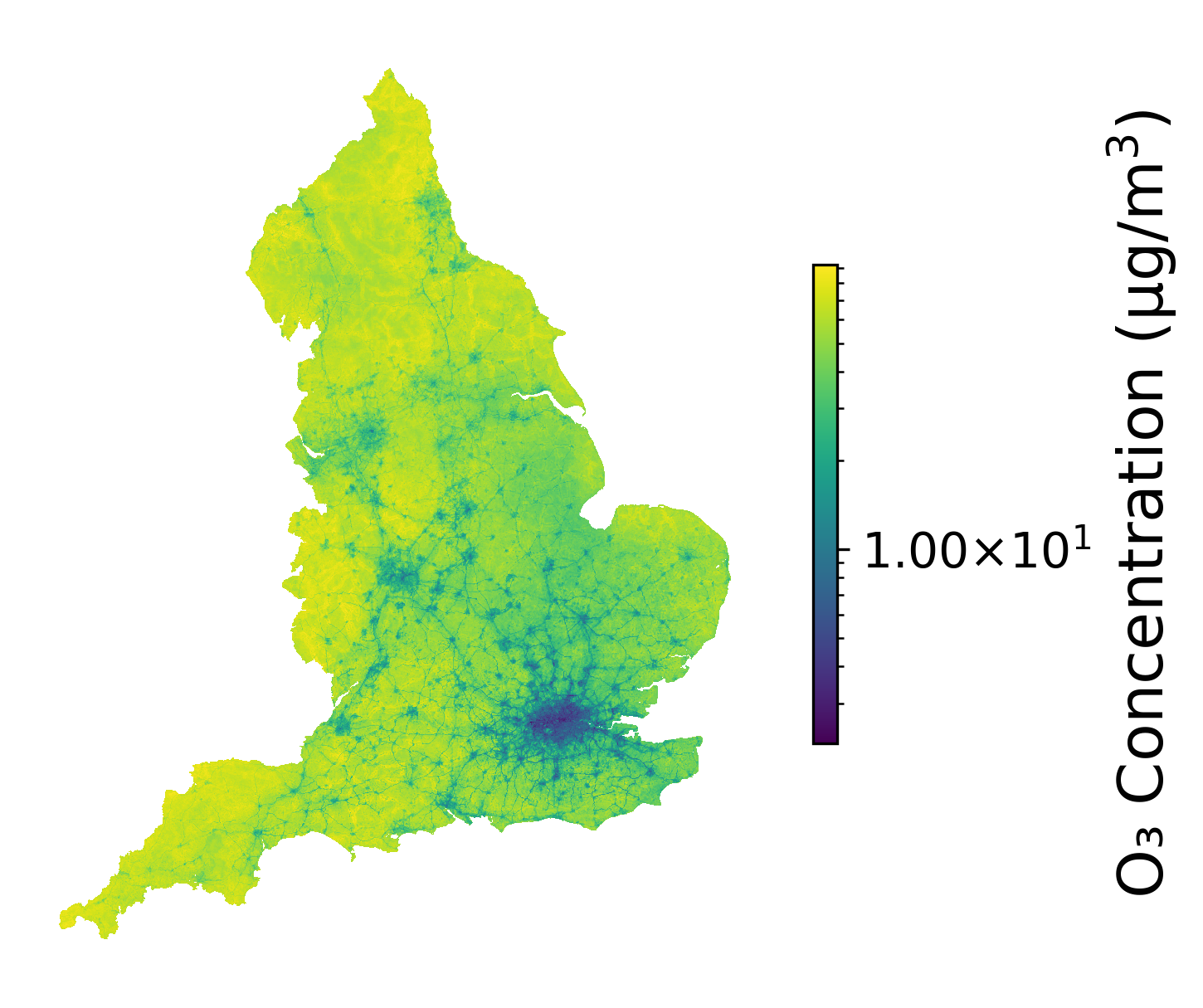}
  \end{subfigure}%
  \hspace*{\fill}   % maximize separation between the subfigures
  \hspace*{\fill}   % maximize separation between the subfigures
  \begin{subfigure}{0.32\textwidth}
    \includegraphics[width=\linewidth]{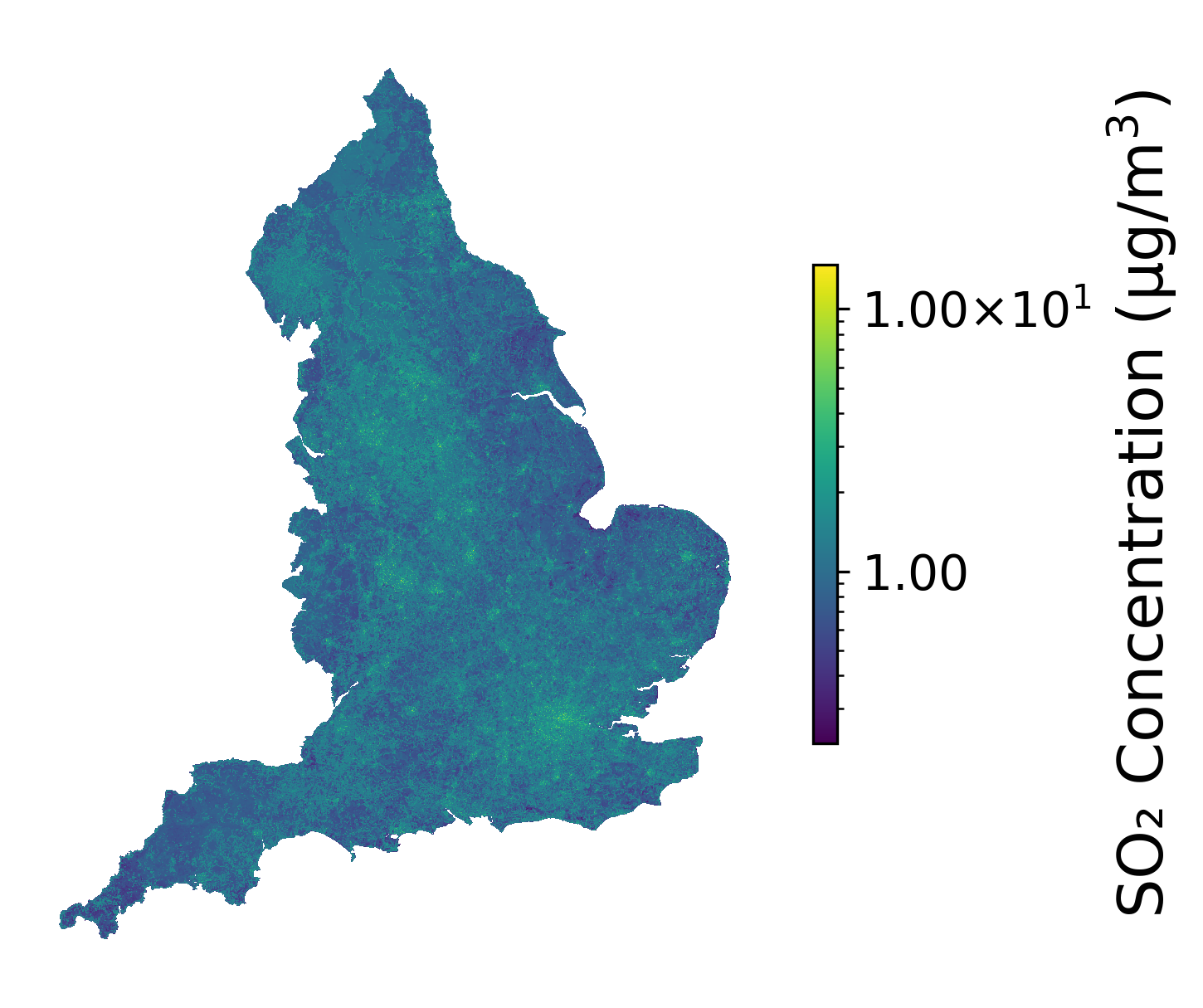}
  \end{subfigure}%
  \hspace*{\fill}   % maximize separation between the subfigures
\caption{{\bfseries Full spatial map of England for all pollutants for 8AM on 19/01/2018, chosen arbitrarily as a typical working day away from national holidays in England.} Plotted on a log scale to help highlight the differences within regions in the map. } \label{fig:fullSpatialPollutionMap}
\end{figure}

The experiment here provides the basis for using the model to create a complete spatial map of pollution across England and Wales with the framework of synthetic stations. In the case of the grid system used in this study, the framework acts as if 355,827 synthetic stations are present at the centroid of each grid, which gives a point sample measurement of the ambient air pollution concentration at a given time. Figure \ref{fig:fullSpatialPollutionMap} shows the resulting air pollution concentration map we can create from the model, predicting air pollution at every location across England. 

\begin{table}[ht]
\centering
\resizebox{0.55\linewidth}{!}{
\begin{filecontents*}{CSVFiles/Model/spatial_experiment_LOOV_datasetSubset_All_results_modelType_mean_LOOV.csv}
,Pollutant Name,Max,Min,Mean,Median
0,NO,0.62,-2.28,0.1,0.21
1,NO₂,0.7,-1.75,0.25,0.37
2,NOₓ,0.67,-1.46,0.2,0.32
3,O₃,0.78,-1.95,0.45,0.62
4,PM₁₀,0.59,-0.17,0.35,0.38
5,PM₂₅,0.59,0.23,0.45,0.46
6,SO₂,0.1,-1.65,-0.2,-0.02
\end{filecontents*}
\pgfplotstabletypeset[
    multicolumn names=l, 
    col sep=comma, 
    string type, 
    header = has colnames, 
    columns={Pollutant Name, Max, Min, Mean, Median},
    columns/Pollutant Name/.style={column type=l, column name=\shortstack{Pollutant\\ Name}},
    columns/Max/.style={column type={S[round-precision=2, table-format=-1.3, table-number-alignment=center]}, column name=\shortstack{Estimation \\ LOOV Max}},
    columns/Min/.style={column type={S[round-precision=2, table-format=-1.3, table-number-alignment=center]}, column name=\shortstack{Estimation \\ LOOV Min}},
    columns/Mean/.style={column type={S[round-precision=2, table-format=-1.3, table-number-alignment=center]}, column name=\shortstack{Estimation \\ LOOV Mean}},
    columns/Median/.style={column type={S[round-precision=2, table-format=-1.3, table-number-alignment=center]}, column name=\shortstack{Estimation \\ LOOV Median}},
    every head row/.style={before row=\toprule, after row=\midrule},
    every last row/.style={after row=\bottomrule}
    ]{CSVFiles/Model/spatial_experiment_LOOV_datasetSubset_All_results_modelType_mean_LOOV.csv}}
    \smallskip
    \caption{{\bfseries R$^2$ scores for missing monitoring stations performance summary statistics for 5-fold leave-one-out-validation.} The summary statistics show the R$^2$ for each monitoring station in the study for a model that has never seen the stations' data before. The approach has clear merit, with all air pollutants having a positive maximum score. However, some monitoring stations have a negative minimum score, driven by their unique nature concerning the feature vectors and phenomena driving the air pollution concentrations at a given location. The mean and median R$^2$ scores show that the approach works for most stations for most air pollutants in estimating air pollution concentrations at a missing location.   }\label{tab:summaryStatisticsLOOVAll}
\end{table}

% \clearpage
\subsection{Prediction of Peak Values}
\label{sec:predicitionExtremeValues}

While the R$^2$ score provides a metric for evaluating the overall performance of the model on the entire dataset, a critical consideration in the context of air pollution concentration estimation is the model's performance during peak concentrations. Given that peak concentrations have the most significant impact on human health and well-being and are the focus of policymakers when designing interventions, it is crucial to assess how well the model performs in these high-concentration scenarios.

\begin{figure}[!htb]
  \hspace*{\fill}   % maximize separation between the subfigures
  \begin{subfigure}{0.85\textwidth}
    \includegraphics[width=\linewidth]{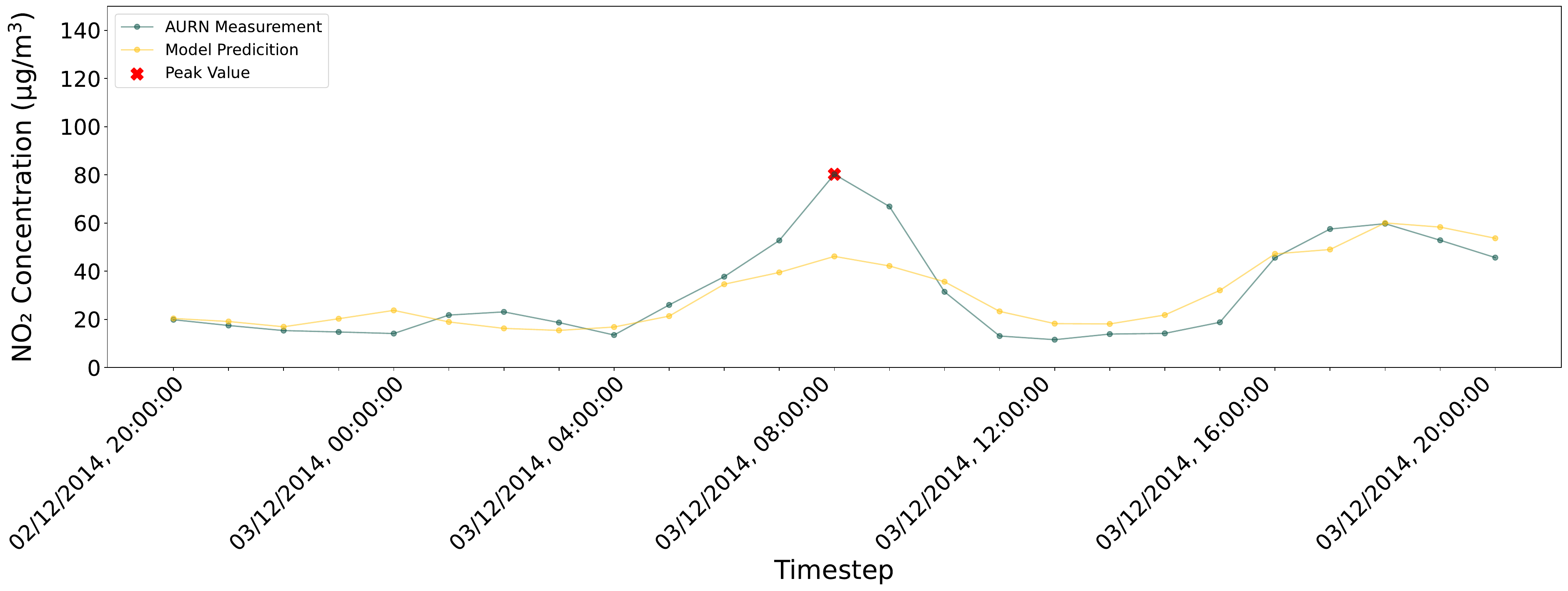}
    \caption{Leominster NO$_2$.}\label{fig:peakValueEstimationNO2Leominster}
  \end{subfigure}
  \hspace*{\fill}   % maximize separation between the subfigures
  \\
  \hspace*{\fill}   % maximize separation between the subfigures
  \begin{subfigure}{0.85\textwidth}
    \includegraphics[width=\linewidth]{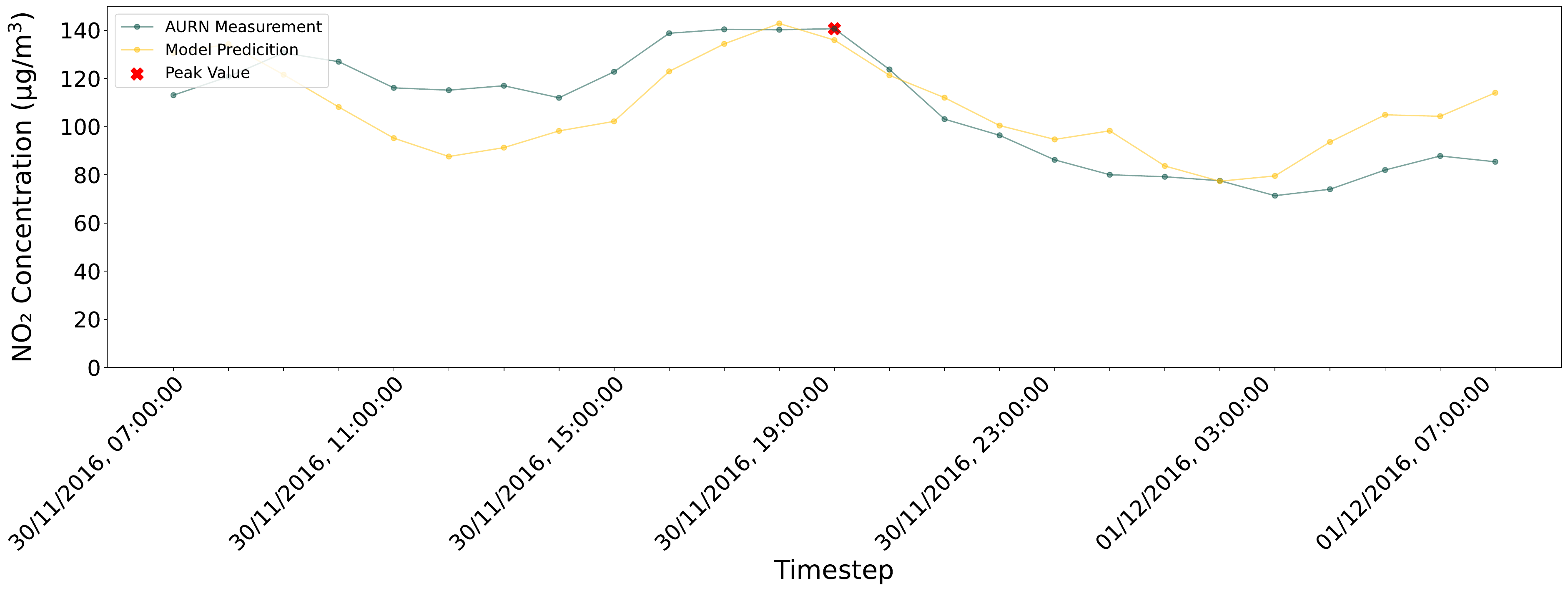}
    \caption{Stanford-le-Hope Roadside NO$_2$.}\label{fig:peakValueEstimationNO2StandfordLeHope}
  \end{subfigure}
  \hspace*{\fill}   % maximize separation between the subfigures
\caption{{\bfseries Prediction of peak values for NO$_2$ monitoring stations.} In Figure \ref{fig:peakValueEstimationNO2Leominster}, it is evident that the model failed to capture the peak concentration for the Leominster monitoring station. However, there is a noticeable uptick in the concentration prediction at the correct time, raising concerns about a consistent underestimation by the model. Conversely, Figure \ref{fig:peakValueEstimationNO2StandfordLeHope} illustrates the peak prediction for the Stanford-le-Hope monitoring station. The model not only captures the peak but also yields a magnitude considerably higher than that for Leominster, offering an initial indication that the model may not be systematically underpredicting concentrations.
} \label{fig:peakValueEstimation}
\end{figure}

We conducted an analysis of the model predictions during peak concentration events observed at each station. Figure \ref{fig:peakValueEstimation} illustrates the model's predictions compared to real-world measurements from AURN monitoring stations. Specifically, Figure \ref{fig:peakValueEstimationNO2Leominster} focuses on the Leominster monitoring station discussed in Section \ref{sec:introduction}. The visual comparison reveals that while the model did not capture the exact magnitude of the peak concentration at the station, it did exhibit an uptick at the correct time. This raises concerns about the model's ability to make high-magnitude predictions. However, further investigation indicates otherwise, as evidenced by the Stanford-le-Hope Roadside AURN station, which had a high-magnitude prediction of approximately ~140\si{\micro\gram/\meter^3} for its overall peak concentration reading between 2014-2018.

This prompts the question of whether the model's prediction for the Leominster peak value was an underprediction or if the peak value itself was an anomalous reading. The percentage difference for the Leominster monitoring station at 03/12/2014 8 AM was 42.5\%. In contrast, the mean peak percentage difference across all NO$_2$ monitoring stations was 22.1\% during the Leominster peak, considerably lower than the overall NO$_2$ average peak distance of 50.72\%. The considerable difference in expected peak distance suggests that the Leominster station differed from nearby stations at this time. The performance variation across different pollutants sheds light on a potential issue with the model stemming from the training data, as depicted in Table \ref{tab:peakPerformacneDifference}. Notably, O$_3$ exhibits the best performance in predicting peak concentrations. As outlined in Section \ref{sec:existingWork}, meteorological conditions predominantly drive O$_3$ concentrations. The ERA5 data used in this study stands out as the most robust dataset, featuring temporally and spatially unique data points.

In contrast, other datasets used in the model lack this level of uniqueness, relying on idealized values that may not accurately represent the true variability. For instance, the transport use dataset family follows a common temporal distribution, emissions are scaled, and datasets like transport infrastructure lack sufficient variability. These idealized values present an avenue for improving the model by seeking enhanced data representations of phenomena influencing air pollution. Additionally, addressing the potential impact of outlier and anomalous data points could further enhance model performance.
The peak distance percentage is defined in Equation  \ref{eq:peak_percentage}.
\begin{equation}
\left( \frac{(\text{Measured Peak} - \text{Model Prediction})}{\text{Peak Value}} \right) \times 100
\label{eq:peak_percentage}
\end{equation}

\begin{table}[ht]
\centering
\resizebox{0.3\linewidth}{!}{
\begin{filecontents*}{CSVFiles/Model/average_peak_difference_percentage_all_pollutants.csv}
,Pollutant Name,Average Peak Distance
3,O₃,32.06790304750911
0,NO₂,50.72448099040211
1,NOₓ,63.81598659908794
2,NO,72.31698591052208
5,PM₂₅,84.21265875483779
4,PM₁₀,87.3701142390662
6,SO₂,93.00075322598438
\end{filecontents*}
\pgfplotstabletypeset[
    multicolumn names=l, 
    col sep=comma, 
    string type, 
    header = has colnames, 
    columns={Pollutant Name, Average Peak Distance},
    columns/Pollutant Name/.style={column type=l, column name=\shortstack{Pollutant\\Name}},
    columns/Average Peak Distance/.style={column type={S[round-precision=2, table-format=-1.3, table-number-alignment=center ]}, column name=\shortstack{Average Peak Distance\\Percentage (\% of µg/m$^3$)}},
    every head row/.style={before row=\toprule, after row=\midrule},
    every last row/.style={after row=\bottomrule}
    ]{CSVFiles/Model/average_peak_difference_percentage_all_pollutants.csv}}
    \smallskip
    \caption{{\bfseries Average peak concentrations prediction difference.} The peak percentage difference is calculated according to formula \ref{eq:peak_percentage}. O$_3$ has the best performance for predicting the peak concentrations across all the monitoring stations, with SO$_2$ having the worst performance. This ordering presents further evidence that the likely explanation for the model not capturing the peak concentrations is not the model framework itself but rather the input data. SO$_2$ has by a considerable margin the least amount of data across the air pollutants (Table \ref{tab:AURNClassificationNumberOfStations}, alongside O$_3$ being most correlated (Section \ref{sec:featureSelectionPollutants}) and driven by meteorological phenomena according to the scientific literature (Section \ref{sec:featureVectors}), which given that ERA5 is the highest quality dataset, with unique points spatially and temporally indicates that the difference in data is likely driving the difference in peak concentration estimation performance. }\label{tab:peakPerformacneDifference}
\end{table}

\subsection{Data Subsetting}

The scalability of the framework has been considered primarily within temporal and spatial resolution dimensions. However, the framework's adaptability extends to different amounts of data, contingent on the availability of specific datasets. Variations in input datasets allow for the development of models tailored to different use cases. For instance, datasets that are only available after historical dates, such as remote sensing, may be excluded when creating models for forecasting purposes. Table \ref{S-tab:dataSubsetForecastingSpatial} provides experiment results for this type of model. Additionally, in situations where a location lacks certain datasets, such as traffic estimates for an entire country, models can be built using only the available datasets. Both meteorological and remote sensing datasets have global availability, making them suitable for creating baseline hindcast models for locations worldwide. Table \ref{S-tab:dataSubsetGlobalSpatial} showcases the performance of such a model.

Further experimentation delved into assessing the performance of each dataset family, as depicted in Table \ref{tab:dataSubsetResults}. These findings support the concepts presented in Section \ref{sec:featureSelectionPollutants}, emphasizing that no single dataset alone can achieve a positive mean LOOV score.

\begin{table}[ht]
\centering
\resizebox{.65\linewidth}{!}{
\begin{filecontents*}{CSVFiles/Model/datasubsetting_spatial_experiment_no2_mean.csv}
,Included Datasets,Dataset Train Score,Dataset Validation Score,Dataset Test Score,Dataset Leave One Out Validation Scores
0,Emissions,0.46,0.42,0.42,-0.23
1,Geographic,0.31,0.25,0.29,-0.46
2,Meteorological,0.15,0.17,0.14,-0.53
3,Remote Sensing,0.35,0.31,0.36,-0.38
4,Temporal,0,0.03,0.02,-0.64
5,Transport Infrastructure,0.32,0.29,0.29,-0.41
6,Transport Use,0.35,0.23,0.16,-0.47
\end{filecontents*}
\pgfplotstabletypeset[
    multicolumn names=l, 
    col sep=comma, 
    string type, 
    header=has colnames, 
    columns={
        Included Datasets,
        Dataset Train Score,
        Dataset Validation Score,
        Dataset Test Score,
        Dataset Leave One Out Validation Scores
    },
    columns/Included Datasets/.style={
        column type=l,
        column name=\shortstack{Dataset\\Family}
    },
    columns/Dataset Train Score/.style={
        column type={S[round-precision=2, table-format=-1.3, table-number-alignment=center]},
        column name=\shortstack{Dataset\\Train Score}
    },
    columns/Dataset Validation Score/.style={
        column type={S[round-precision=2, table-format=-1.3, table-number-alignment=center]},
        column name=\shortstack{Dataset\\Validation Score}
    },
    columns/Dataset Test Score/.style={
        column type={S[round-precision=2, table-format=-1.3, table-number-alignment=center]},
        column name=\shortstack{Dataset\\Test Score}
    },
    columns/Dataset Leave One Out Validation Scores/.style={
        column type={S[round-precision=2, table-format=-1.3, table-number-alignment=center]},
        column name=\shortstack{Mean\\LOOV}
    },
    every head row/.style={before row=\toprule, after row=\midrule},
    every last row/.style={after row=\bottomrule}
]{CSVFiles/Model/datasubsetting_spatial_experiment_no2_mean.csv}}
\smallskip
\caption{{\bfseries Repeat experiments results of Tables \ref{tab:datasubsetResultsAllDatasets} and \ref{tab:summaryStatisticsLOOVAll} for models trained on individual dataset families (Section \ref{sec:featureVectors}) for NO$_2$. } The framework presented can be used on varying amounts of input data, depending on available data, providing a basic understanding of limitations when moving the model between areas, such as being used to predict countries other than England. Table \ref{tab:dataSubsetResults} shows that in the case of England, while individual dataset families can forecast into the future, the performance of estimating missing monitoring stations is limited and requires datasets that cover a wide range of phenomena to achieve the same performance as \ref{tab:summaryStatisticsLOOVAll}. }\label{tab:dataSubsetResults}
\end{table}

\section{Research Data Output}
\label{sec:researchOutput}

As part of our ongoing efforts, we plan to release an open-source dataset consisting of two components. The first component is the augmented AURN dataset, as illustrated in Figure \ref{fig:aurnModelComparison}. This dataset includes model predictions for air pollution concentrations at all AURN monitoring stations for the period 2014-2018, which were utilized in this study. The second component is a comprehensive air pollution concentration map for England, encompassing each air pollutant for the year 2018. This dataset provides a spatial resolution of 1 km$^2$ and hourly temporal resolution. We anticipate that this dataset will be of significant interest to various stakeholders, as outlined in Section \ref{sec:introduction}. Moreover, it opens avenues for the research community to explore a diverse range of research topics that were previously constrained, given that current air pollution estimations at this spatial resolution typically operate at the annual temporal level.

\begin{figure}[ht]
  \hspace*{\fill}   % maximize separation between the subfigures
  \begin{subfigure}{0.2\textwidth}
    \includegraphics[width=\linewidth]{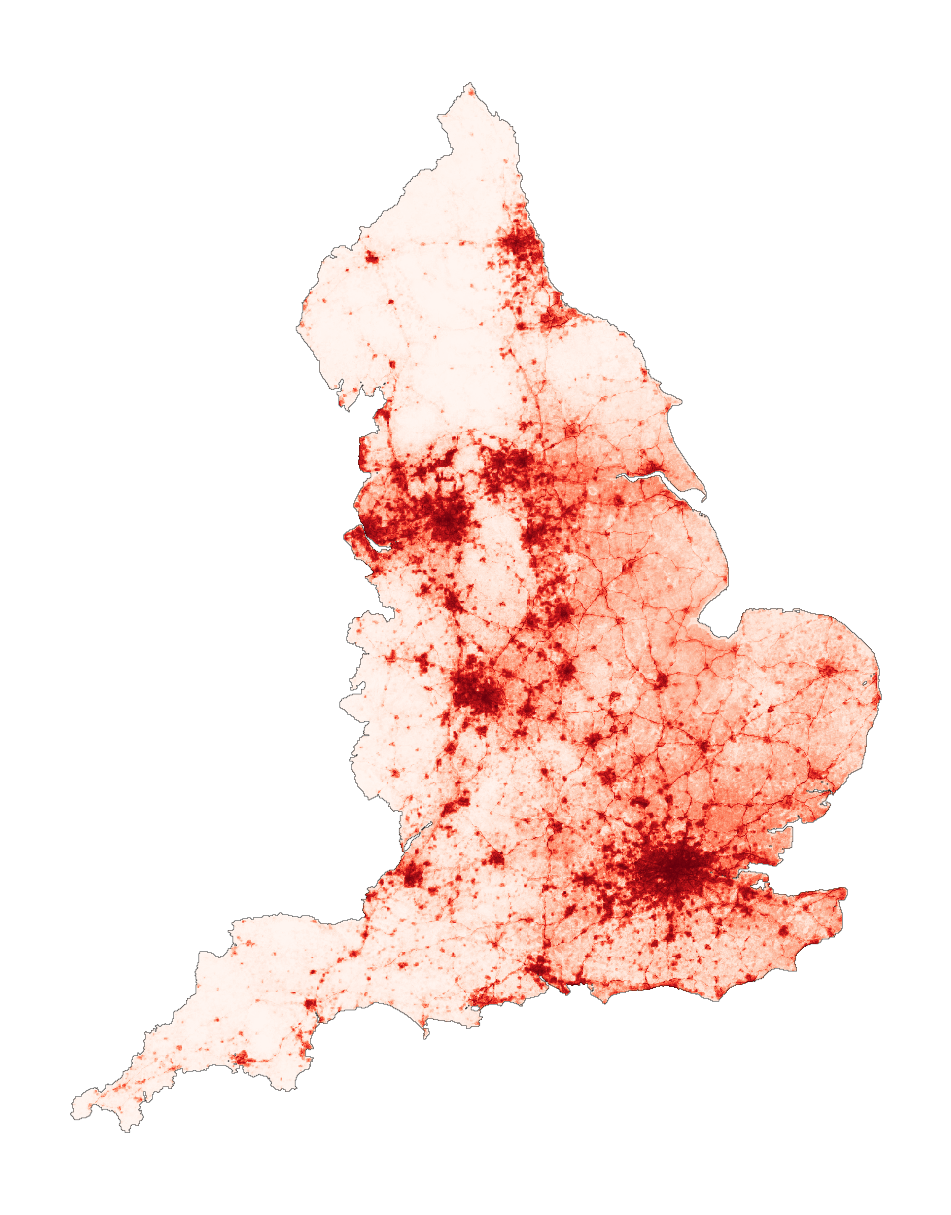}
    \caption{10\si{\micro\gram/\meter^3} threshold.}\label{fig:exceedanceLegislationCompliance10}
  \end{subfigure}
  \hspace*{\fill}   % maximize separation between the subfigures
  \begin{subfigure}{0.2\textwidth}
    \includegraphics[width=\linewidth]{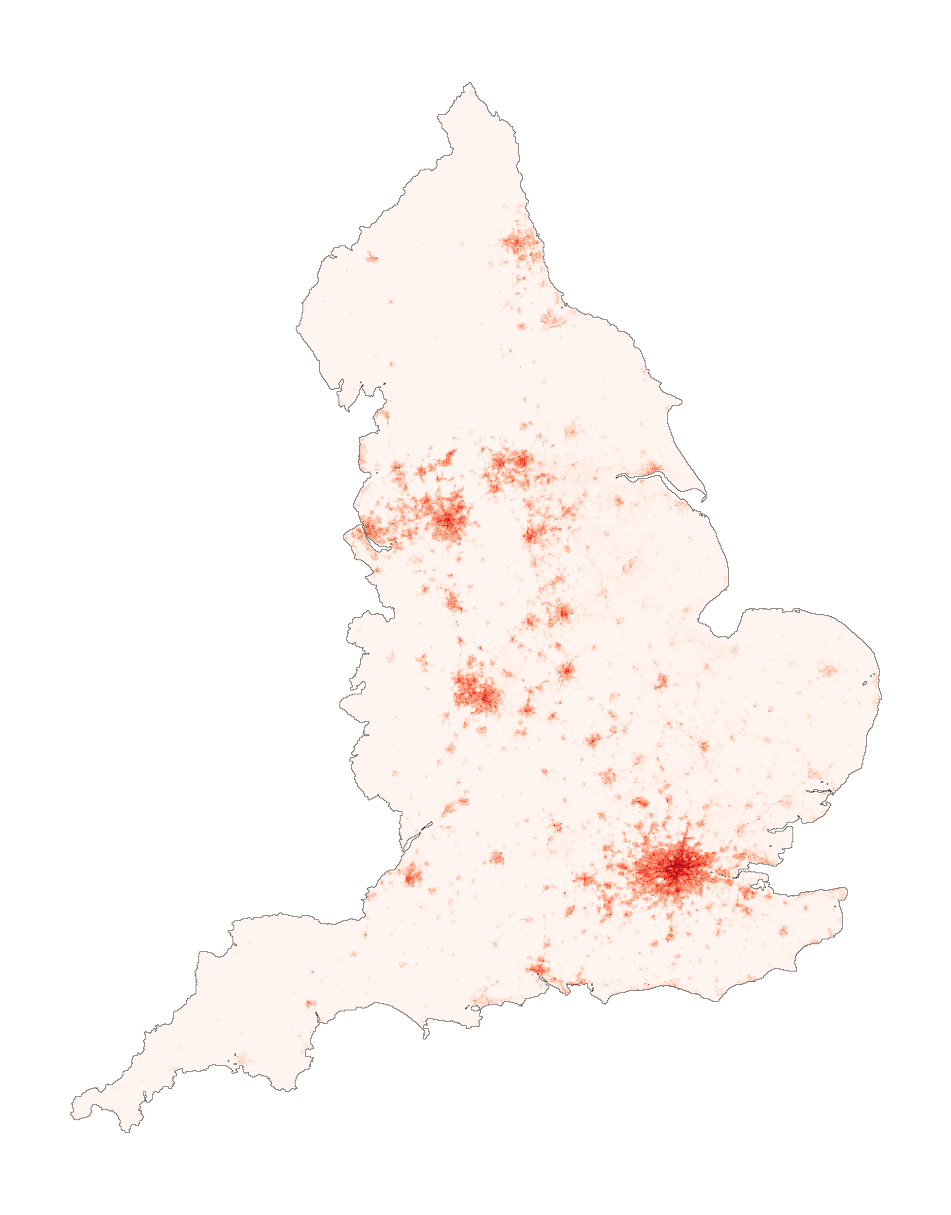}
    \caption{25\si{\micro\gram/\meter^3} threshold.}\label{fig:exceedanceLegislationCompliance25}
  \end{subfigure}
  \hspace*{\fill}   % maximize separation between the subfigures
  \begin{subfigure}{0.2\textwidth}
    \includegraphics[width=\linewidth]{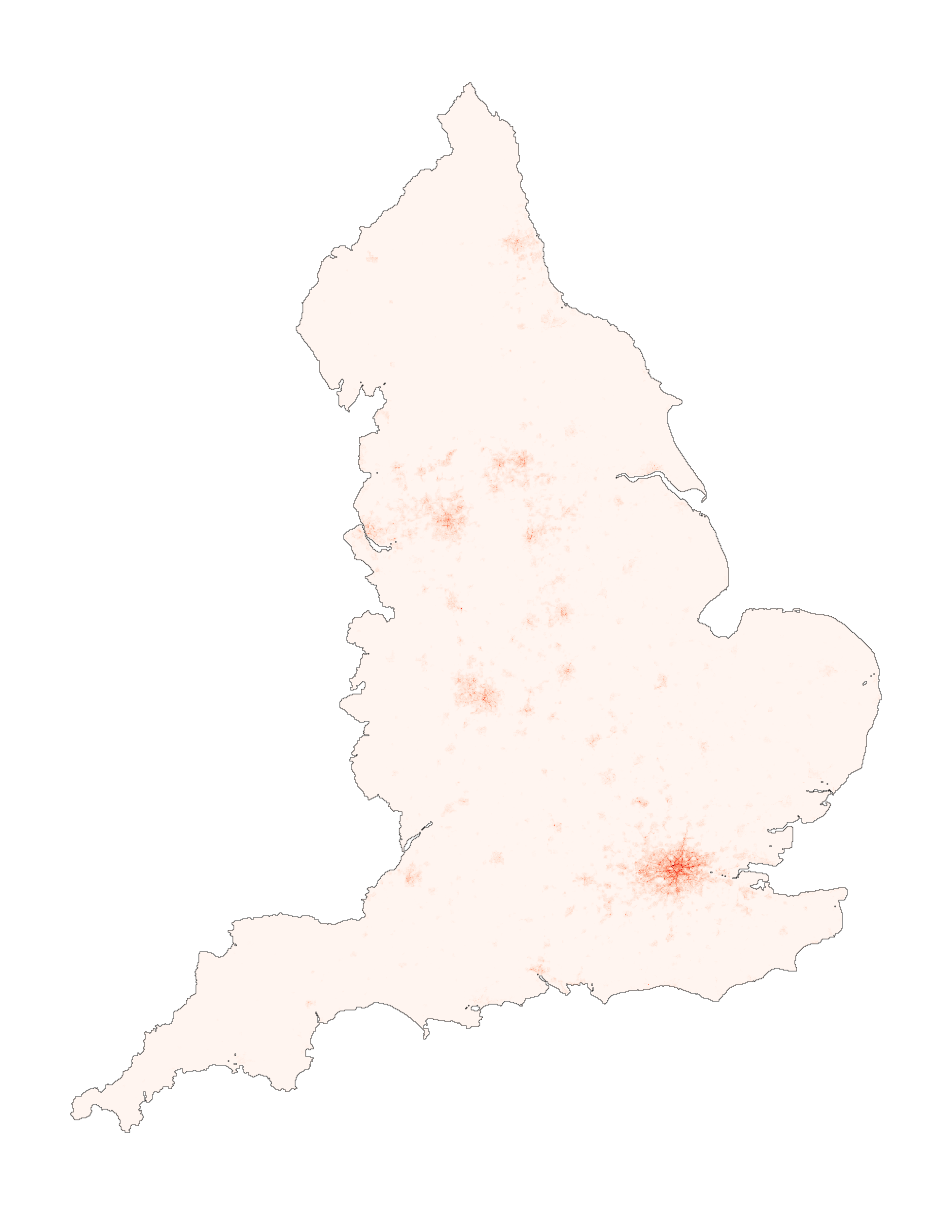}
    \caption{40\si{\micro\gram/\meter^3} threshold.}\label{fig:exceedanceLegislationCompliance40}
  \end{subfigure}
  \hspace*{\fill}   % maximize separation between the subfigures
  \begin{subfigure}{0.2\textwidth}
    \includegraphics[width=\linewidth]{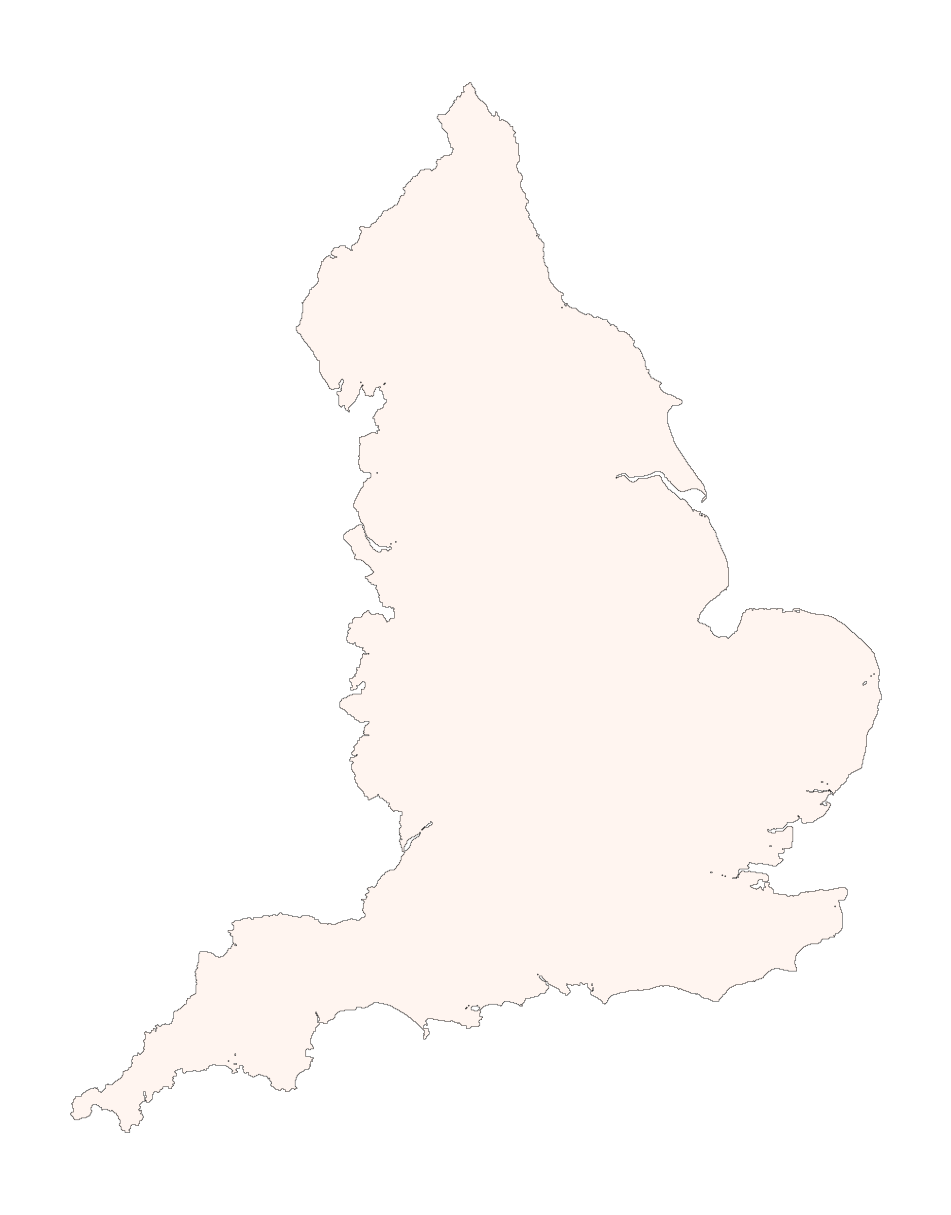}
    \caption{200\si{\micro\gram/\meter^3} threshold.}\label{fig:exceedanceLegislationCompliance200}
  \end{subfigure}
  \hspace*{\fill}   % maximize separation between the subfigures
  \raisebox{12mm}{
  \begin{subfigure}{0.1\textwidth}
    \includegraphics[width=\linewidth]{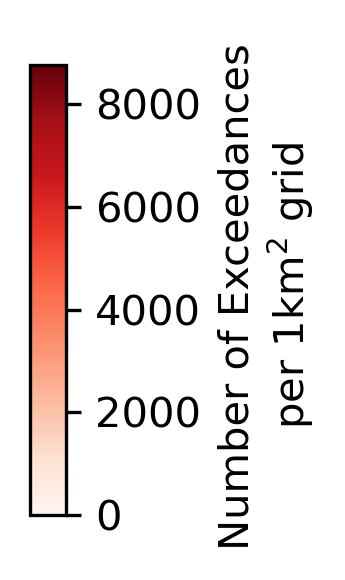}
  \end{subfigure}}
  \hspace*{\fill}   % maximize separation between the subfigures
\caption{{\bfseries Count of times that a grid exceeded the outlined thresholds for NO$_2$ in 2018.} Figure \ref{fig:exceedanceLegislationCompliance10} shows the 10\si{\micro\gram/\meter^3} threshold where one grid exceeds the threshold for every hour of the year, with 99.6\% of grids exceeding the threshold at least once. Figure \ref{fig:exceedanceLegislationCompliance25} depicts the counts for the 25\si{\micro\gram/\meter^3} where the max count was 8656 exceedances across the year, with 63\% of grids exceeding the threshold at least once. Figure \ref{fig:exceedanceLegislationCompliance40} uses a threshold of 40\si{\micro\gram/\meter^3} where the max count for exceedances was 8,086 across the year, with 26.2\% of grids exceeding the threshold at least once. Figure \ref{fig:exceedanceLegislationCompliance200} denotes a 200\si{\micro\gram/\meter^3} threshold, where only a single grid exceeded the threshold twice across the year. Latitude 51.5, longitude -0.15 was the location that exceeded the threshold, a central London location with the postcode W1G 6JA.} \label{fig:exceedanceLegislationCompliance}
\end{figure}

The presented dataset opens up diverse research possibilities, with one illustrative example being the examination of air pollution concentration variations across different locations concerning legislation compliance, as discussed in Section \ref{sec:introduction}. Figure \ref{fig:exceedanceLegislationCompliance} showcases a series of heatmaps representing the grids employed in this study. Each grid is colour-coded based on the number of times it exceeded specific concentration thresholds in 2018 at an hourly granularity. The maximum possible number of exceedances per grid is 8760, representing the total hours in 2018.

The thresholds considered include 10\si{\micro\gram/\meter^3}, aligning with the WHO NO$_2$ air quality guideline level for the annual temporal period. The second threshold is 25\si{\micro\gram/\meter^3}, corresponding to the WHO 24-hour aggregate air quality guideline \cite{WHO:2021:WHOGuidelinesTable}. The third threshold is 40\si{\micro\gram/\meter^3}, reflecting the UK National Air Quality Guideline annual limit. Lastly, 200\si{\micro\gram/\meter^3} represents the UK National Air Quality Guideline hourly target for NO$_2$ concentrations \cite{UKGOV:2010:AirQualityStandardsUK}. While not all these thresholds directly pertain to hourly concentration legislation, they provide a comprehensive set of benchmarks derived from actual legislation, offering insights into the distribution of air pollutants across England.

\begin{figure}[ht]
  \hspace*{\fill}   % maximize separation between the subfigures
  \begin{subfigure}{0.5\textwidth}
    \includegraphics[width=\linewidth]{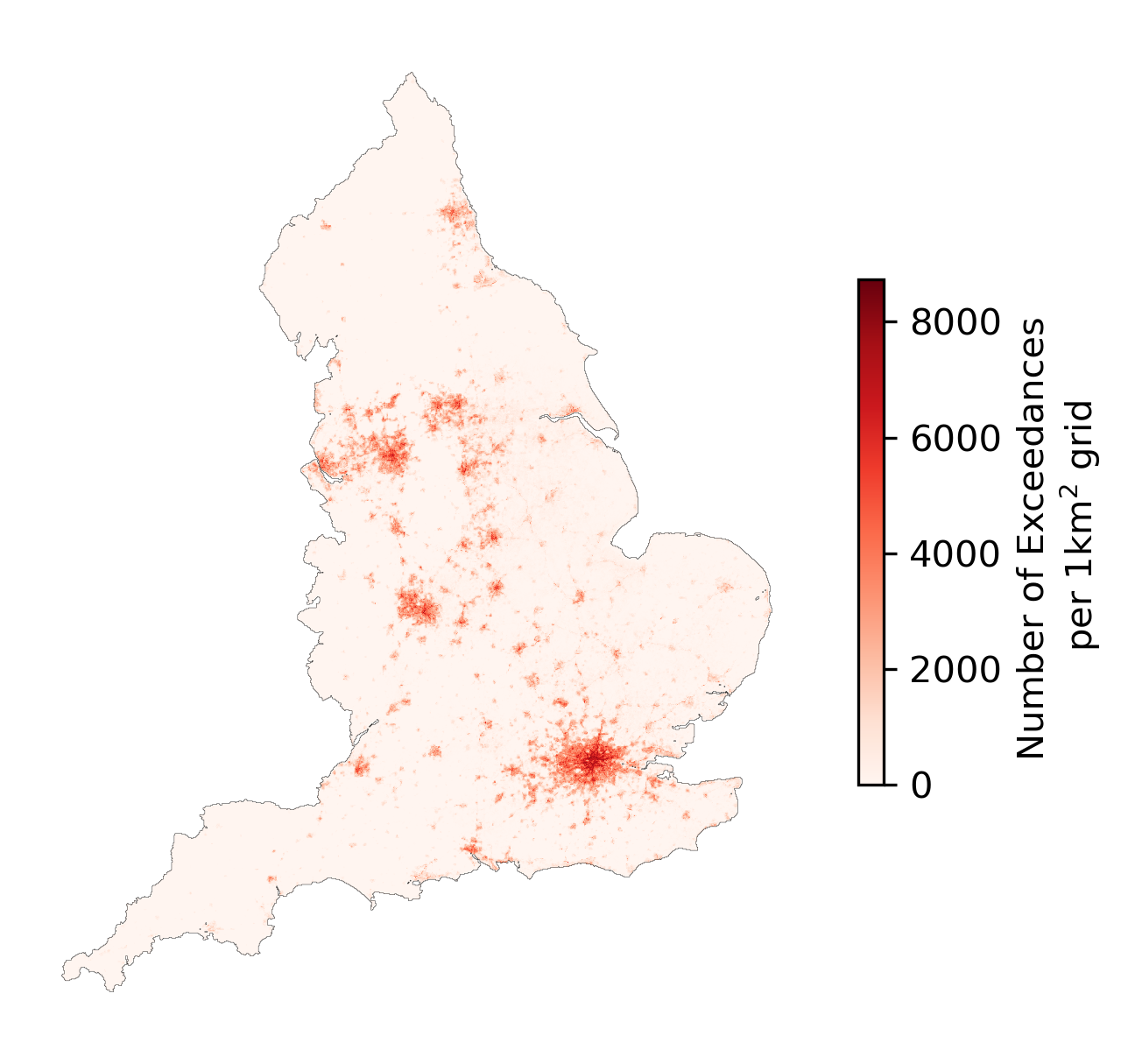}
  \end{subfigure}
  \hspace*{\fill}   % maximize separation between the subfigures
\caption{{\bfseries 24-hour mean \cite{UKAIR:2023:RunningMeanDefinition} exceedance counts example.} The threshold used is a mean of 25 \si{\micro\gram/\meter^3}. As the hourly level is the most common high-resolution temporal level mentioned in air quality legislation, pursuing data at this level allows for a more coarse temporal level to be calculated from the input data, resulting in the dataset providing complete legislation coverage no matter the resolution of interest. } \label{fig:exceedanceLimit24Hour}
\end{figure}

The analysis reveals compelling insights into air pollution concentration exceedances across various thresholds. It was found that 99.96\% of locations surpassed the 10\si{\micro\gram/\meter^3} threshold at least once, 63\% exceeded the 25\si{\micro\gram/\meter^3} threshold at least once, 26.2\% exceeded the 40\si{\micro\gram/\meter^3} threshold at least once, and only a single grid exceeded the 200\si{\micro\gram/\meter^3} threshold at least once. This analysis serves as a valuable tool to pinpoint locations demanding further investigation into air pollution concentration causes and potential interventions at the local level. For instance, the coordinates at latitude 51.5, longitude -0.15, representing a location exceeding the 200\si{\micro\gram/\meter^3} threshold with concentration predictions surpassing 10 for every hour in 2018, underscore the need for targeted attention. Moreover, leveraging the temporal precision of the predictions allows for flexible aggregation to various temporal levels stipulated in UK or EU legislation. For instance, aggregating to a 24-hour mean \cite{UKAIR:2023:RunningMeanDefinition} for each grid facilitates a comprehensive assessment of legislation compliance, as depicted in Figure \ref{fig:exceedanceLimit24Hour}.

% \clearpage
\section{Discussion}
\label{sec:discussion}

We have released two datasets that hold significant value for the scientific community, policymakers, and the public. The comprehensive spatial dataset offers valuable insights into locations where air pollution concentration data might be non-existent at the hourly temporal level. If one were to acquire the data produced by our model through real-world measurements, the cost would amount to £70B\footnote{Calculated based on 355827 monitoring stations at a cost of £198000 per station, as outlined in Section \ref{sec:introduction}.} through AURN monitoring stations.

The augmented AURN dataset, generated from this study, provides a complete temporal perspective of all monitoring station measurements across England. This is particularly crucial for compliance assessments related to absolute threshold exceedances, such as NO$_2$, where a detailed limit of 200\si{\micro\gram/\meter^3} should not be exceeded more than 18 times a year \cite{UKGOV:2010:AirQualityStandardsUK}. Ensuring a complete time series with measurements at each time step is essential when comparing two locations, especially in situations where missing data during peak pollution periods, like NO$_x$ during rush hour, could potentially mask crucial information. This consideration becomes paramount when creating higher temporal resolution statistics through UK AIR\footnote{\href{https://uk-air.defra.gov.uk/data/exceedance?f_exceedence_id=S3&f_year_start=2006&f_year_end=2007&f_group_id=4&f_region_reference_id=1&f_parameter_id=SO2&f_sub_region_id=1&f_output=screen&action=exceedance3&go=Submit}{UK AIR Statistics Using Incomplete Data, denoted by data capture rate.}}.

While the presented model has demonstrated its effectiveness in filling missing data from high-quality air quality monitoring stations, its most significant advantages lie in its potential application to low-cost monitoring sensors and citizen science initiatives. The symbiotic relationship between the model and low-cost sensors addresses a core issue present in both approaches. The model's performance is notably impacted by a lack of data. In locations where the installation of more expensive AURN-style stations might not be deemed a worthwhile investment, low-cost sensors can be strategically deployed to fill data gaps. These sensors, due to their minimal cost, can be implemented in various locations, enhancing spatial coverage.

Conversely, the model can contribute to overcoming the challenges associated with low-cost sensors, which are often less robust than AURN stations. By leveraging the model, missing data points can be backfilled temporally and spatially, ensuring the generation of a more complete dataset. This collaborative approach is particularly valuable in scenarios where the less frequent deployment of AURN stations results in gaps in the feature vector, as discussed in Section \ref{sec:featureSelectonInterFeatureVectors}. An additional advantage is the model's ability to handle temporally messy data commonly encountered in citizen science initiatives. Unlike AURN monitoring stations that provide data at regular intervals, citizen science datasets may exhibit irregular time stamps (e.g., 10:14, 10:47, and 11:46). The model enables a more sophisticated estimation of air pollution at specific times (e.g., 10:00, 11:00, 12:00), facilitating further analysis such as legislation compliance or integration with other forecasting systems.

Importantly, the model method offers substantial benefits compared to other approaches for filling missing data, such as interpolation, as it takes into account the nuanced patterns present in air pollution concentration time series datasets.

Although the proposed model demonstrates significant advantages, there is a clear pathway to extracting even more benefits from the model framework, given its inherent scalability in both spatial and temporal dimensions. The extension of the method to make predictions at a minute temporal-level is straightforward, and similarly, increasing the spatial resolution grid size to 100m$^2$ is feasible. This scalable approach empowers researchers by providing the desired data without being constrained by limitations in financial resources for monitoring station placement.

Furthermore, the encoding of the temporal aspect into a tabular format facilitates a substantial acceleration through parallelization. Each timestep and grid within the estimation is independent of one another, enabling the simultaneous calculation of all timesteps and grids. This approach yields a significant speedup over traditional forecasting methods, whether machine-learning or physics-based, that rely on lags from previous timesteps.

From a performance standpoint, the capacity to parallelize estimations becomes pivotal when combined with the scalability of the approach. This combination forms the basis for a computationally effective method of estimating air pollution concentrations at a global level. Future work could extend the experiment conducted in Section \ref{sec:modelResultsSpatial}, where air pollution concentrations at one station were estimated using data from other monitoring stations, to a study that analyzes the feasibility of estimating air pollution between countries and their respective air pollution monitoring networks. The potential benefit of this analysis is to help reduce inequalities between countries concerning monitoring stations, enabling the design of interventions based on air pollution without the need for high-cost, dedicated monitoring station networks to be implemented by a country's government.

While the datasets employed in this study successfully estimated air pollution concentrations under a variety of conditions, there remains room for improvement in the input feature vectors. Presently, the model does not consider variations associated with specific days, such as distinct travel patterns on bank holidays compared to regular weekdays. Incorporating local knowledge into the model, such as categorizing whether a day is a bank holiday or another national holiday, would enhance the model's understanding of unique circumstances on special days, such as Bonfire Night in the UK, known to have considerable impact on air pollution concentrations \cite{adams:2020:BonfireNightAirPollution}. Additionally, some feature vectors used in the model will improve over time as technology advances, enabling improved model performance. For example, remote sensing of trace gases over Europe will improve with the Sentinel-4 missions, which are currently scheduled for launch in 2024, on the MTG-s Satellite \cite{ESA:2024:Sentinel4SatellitePlatform}. Sentinel-4 will provide hourly temporal resolution, with a spatial resolution of ~8km for much of northern Europe for O$_3$, NO$_2$, SO$_2$, and aerosol optical depth \cite{EUMETSAT:2024:Sentinel4}. 

In addition to incorporating additional knowledge into the model, a thorough analysis of the training data used in the study is crucial to ensure comprehensive coverage of all scenarios, minimizing the need for extrapolation during model estimations. For instance, as discussed in Section \ref{sec:featureSelectonInterFeatureVectors}, there are environmental conditions where no air pollution concentration measurements are available. Future work could focus on analyzing missing scenarios in the training data, identifying locations where additional air pollution monitoring stations should be placed. When combined with low-cost sensors, this approach could form the basis for a dynamic mobile monitoring network to identify areas where the model predictions are most uncertain.

In summary, we believe this work holds significant importance for a broad audience, addressing critical challenges outlined in the United Nations (UN) Sustainable Development Goals (SDGs). The work presented empowers decision-makers with high-quality data for crucial indicators (UN SDG 3.9.1, 11.6.2) for essential goals such as Good Health and Well-being (SDG 3) and Sustainable Cities and Communities (SDG 11). The contribution to SDG 3 is evident in reallocating resources from monitoring air pollution to clean air initiatives, providing estimates in all regions, not just those with monitoring stations. Simultaneously, the research contributes to SDG 11 by advancing the understanding of the relationship between urban and rural air pollution. 

% \textbf{Competing Interests} The authors declare none. 
% \textbf{Data availability statement} All of the datasets used throughout this work are open source and available online at the locations described in the work. 
% \textbf{Ethics statement} The research meets all ethical guidelines, including adherence to the legal requirements of the study country
% \textbf{Funding Statement} 

\bibliographystyle{IEEEtran}
\bibliography{bibliography}
\end{document}

\maketitle

\clearpage
\section{Data Details}
\label{sec:Datadetails}

This section details the preprocessing and transformations to create a consistent dataset for training the data-driven supervised machine learning model.

\subsection{Common Data Format}

As the model framework is Eulerian \cite{Byun:1984:EulerianDispersionModels}, the first decision was the grid framework, taking into context the modifiable areal unit problem (MAUP) \cite{wong:2004:MAUP} for the grids in which the aggregation for predictors would be taken. The decision was made to use the same framework as existing air pollution concentration datasets, particularly the UK Modelled Background Annual dataset \cite{UKAIR:2023:ModelledBackgroundAirPollution}. 

\begin{figure}[!htb]
\begin{center}
\includegraphics[width=0.7\textwidth]{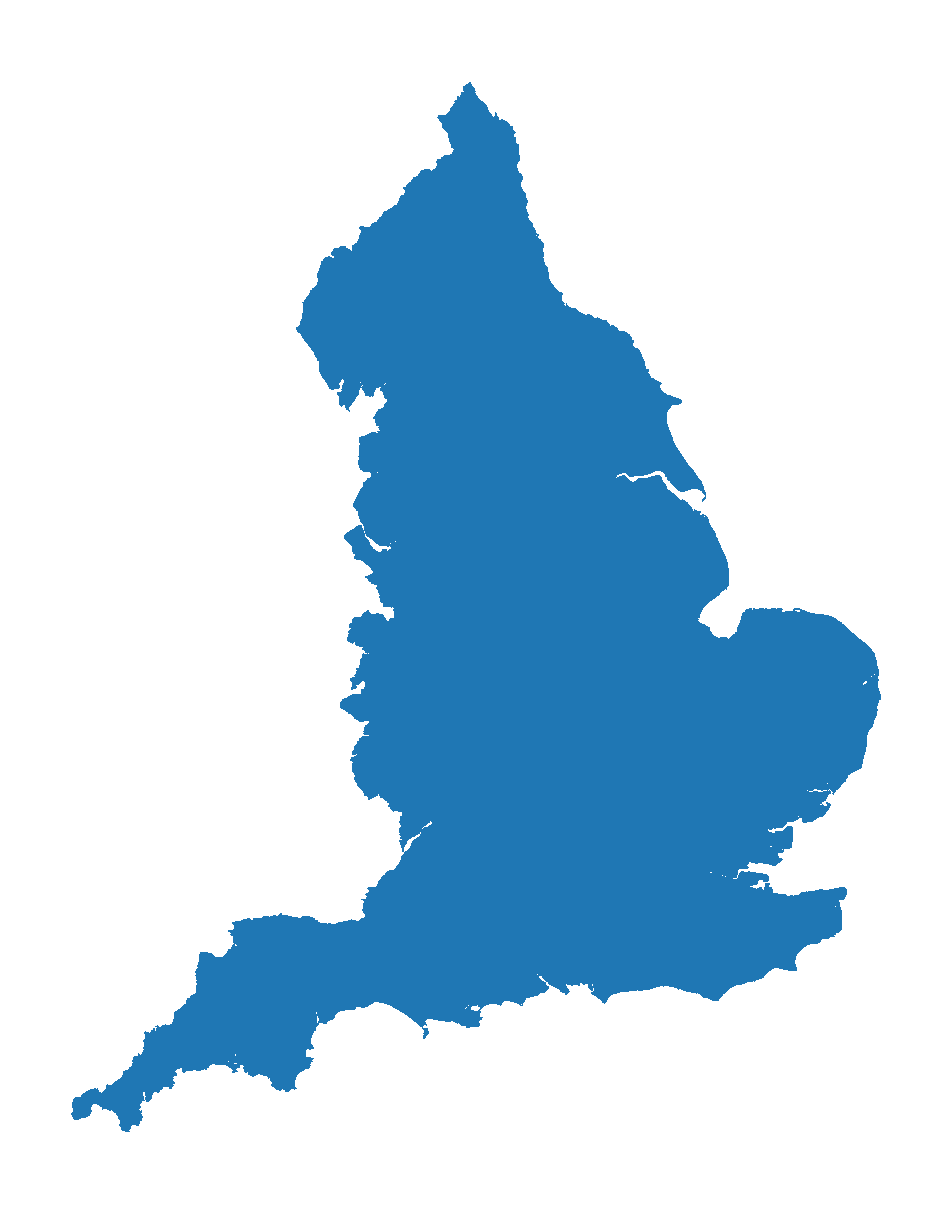}
\caption{{\bfseries 1km\textsuperscript{2} land grids For England.} The land grids covering the area of the England land mass provide the common framework for aggregating the datasets and providing estimates of ambient air pollution, a total of 355,827 point locations at the centroid of each grid for which measurements are sampled.}
\label{fig:englandLandGrids}
\end{center}
\end{figure}

\clearpage
\subsection{Air Pollution Concentrations}
\label{sec:Datadetails:airPollutionDataSupplementary}

\begin{figure}[!htb]
  \begin{subfigure}{0.25\textwidth}
    \includegraphics[width=\linewidth]{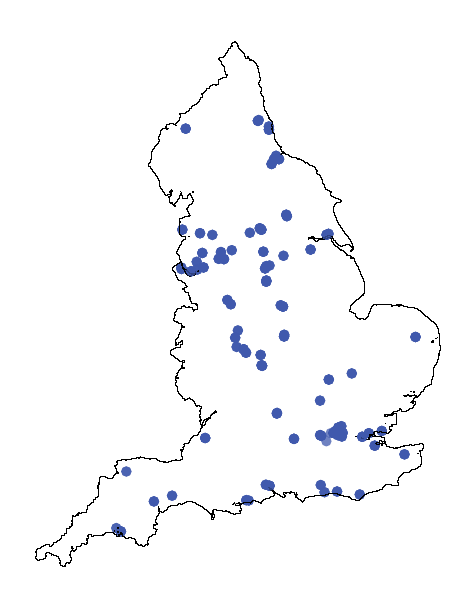}
    \caption{91 Unique Urban Station Locations} \label{fig:ukaurnUrban}
  \end{subfigure}%
  \hspace*{\fill}   % maximize separation between the subfigures
  \begin{subfigure}{0.25\textwidth}
    \includegraphics[width=\linewidth]{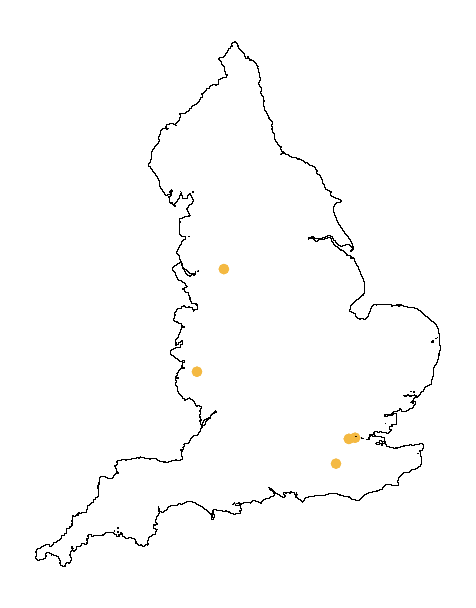}
    \caption{5 Unique Suburban Station Locations} \label{fig:ukaurnSuburban}
  \end{subfigure}%
  \hspace*{\fill}   % maximizeseparation between the subfigures
  \begin{subfigure}{0.25\textwidth}
    \includegraphics[width=\linewidth]{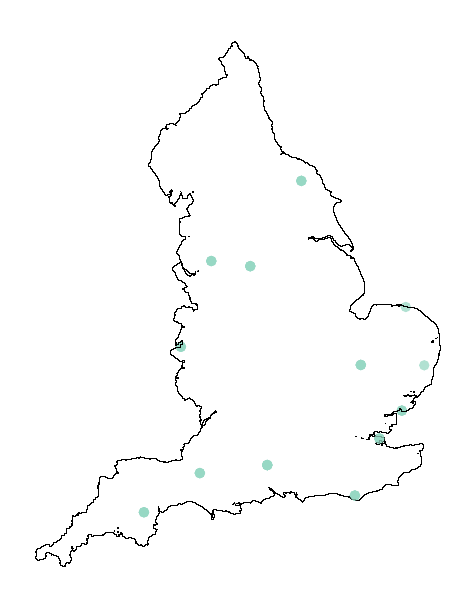}
    \caption{13 Unique Rural Station Locations} \label{fig:ukaurnrural}
  \end{subfigure}
\caption{{\bfseries Spatial distribution and classification of monitoring stations within England.} The Automatic Urban and Rural Network (AURN) stations are divided into 3 classes: urban, suburban, and rural. The urban stations are then further divided into background, traffic and industrial, suburban into background and industrial and rural background. Note the inequality of station numbers; most of the stations are in urban settings.} \label{fig:ukAURNNetwork}
\end{figure}

\begin{figure}[!htb]
  \hspace*{\fill}   % maximize separation between the subfigures
  \begin{subfigure}{0.32\textwidth}
    \includegraphics[width=\linewidth]{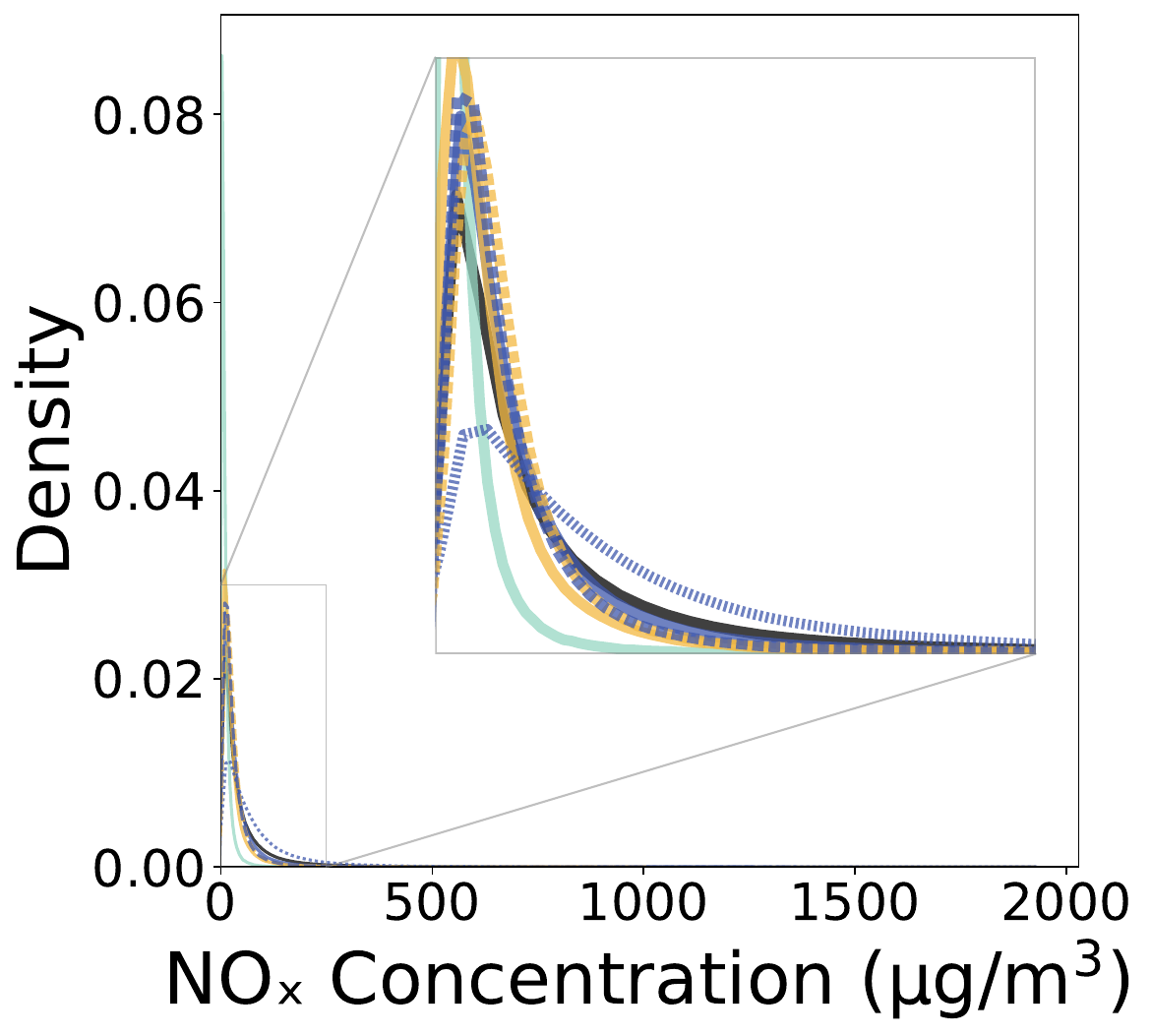}
    %\caption{NO$_X$}\label{fig:airPollutionKDENOX}
  \end{subfigure}
  \hspace*{\fill}   % maximize separation between the subfigures
  \begin{subfigure}{0.32\textwidth}
    \includegraphics[width=\linewidth]{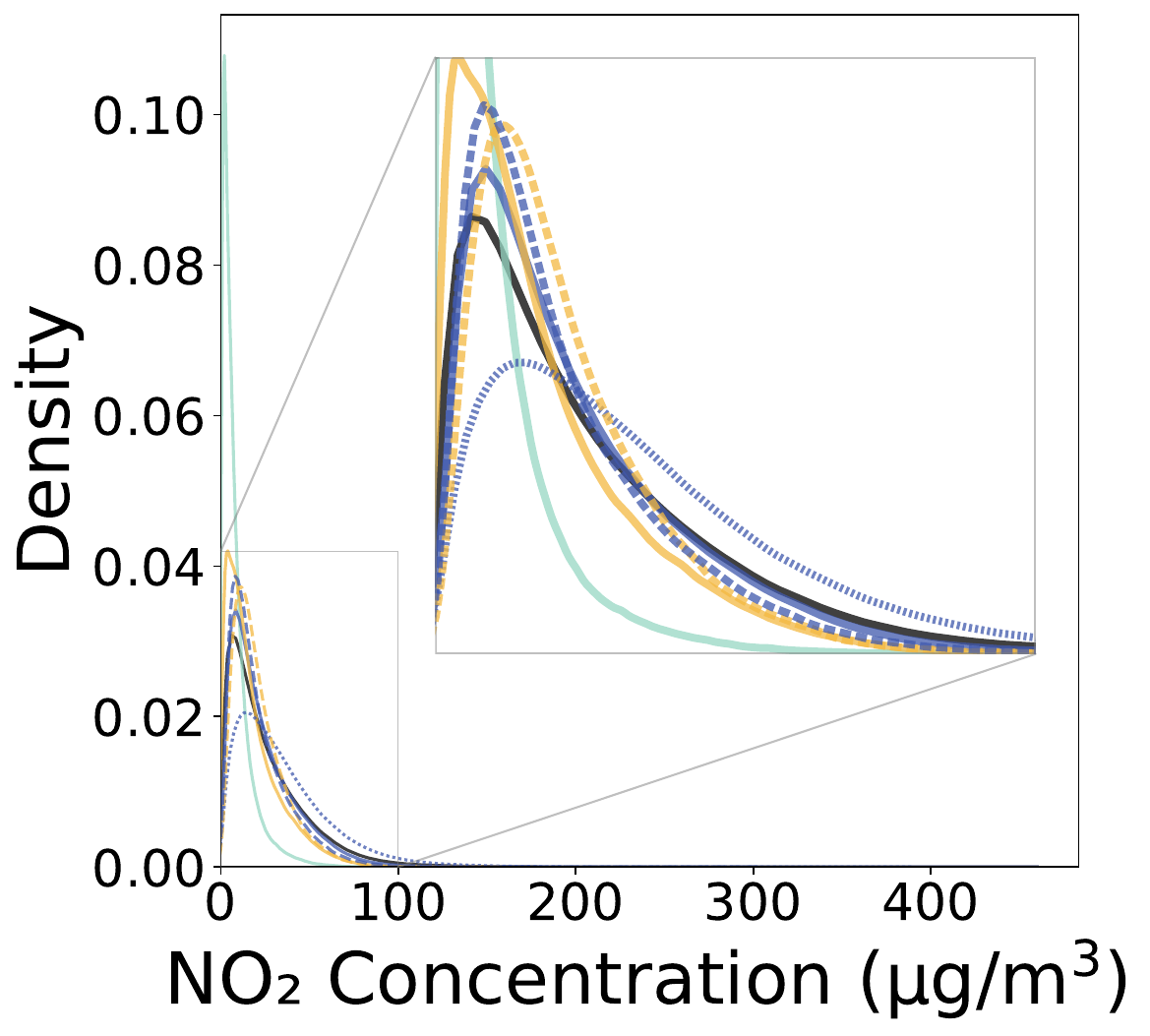}
    %\caption{NO$_2$}\label{fig:airPollutionKDENO2}
  \end{subfigure}
  \hspace*{\fill}   % maximize separation between the subfigures
  \begin{subfigure}{0.32\textwidth}
    \includegraphics[width=\linewidth]{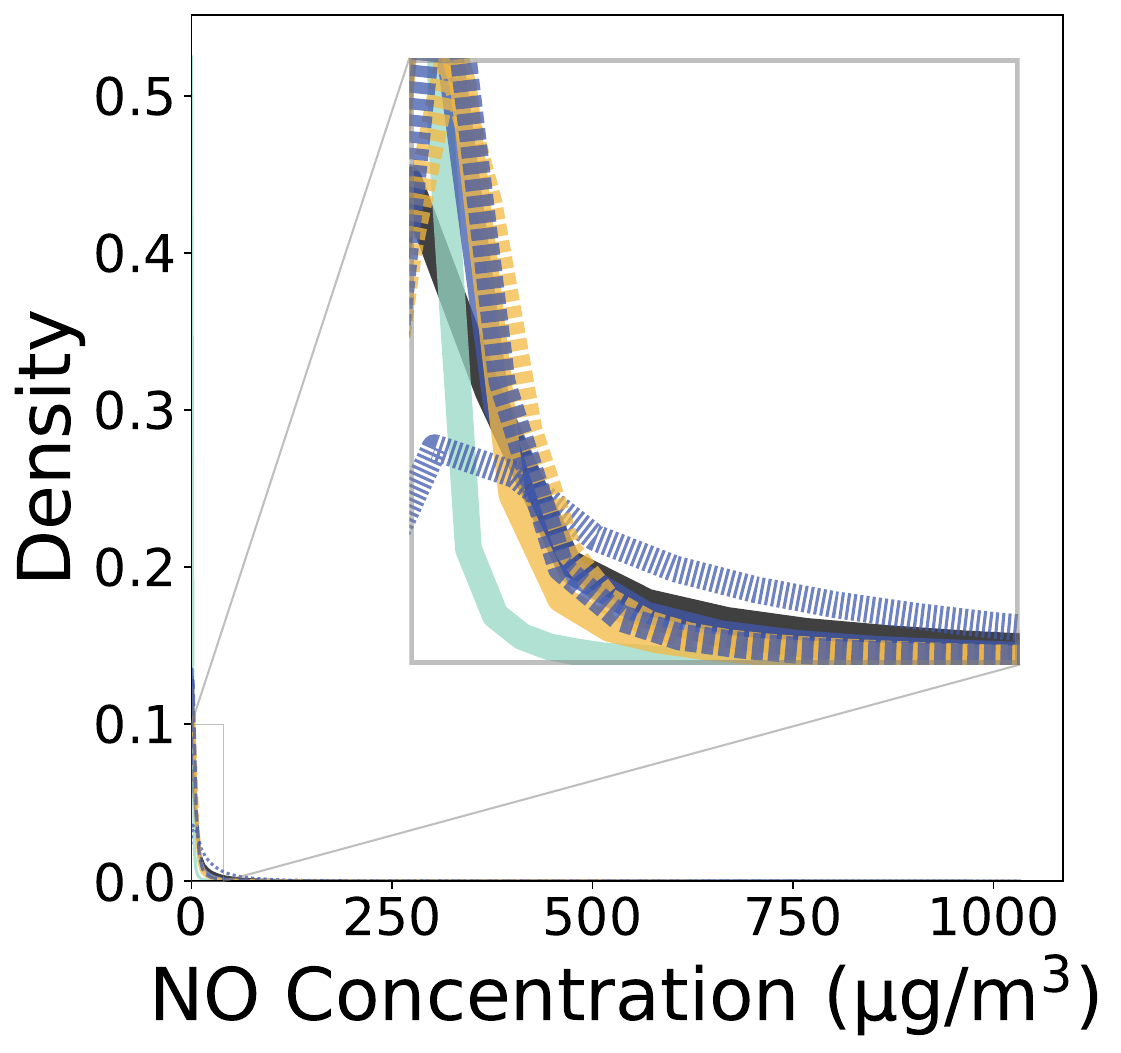}
    %\caption{NO}\label{fig:airPollutionKDENO}
  \end{subfigure}
  \hspace*{\fill}   % maximize separation between the subfigures
  \\
  \hspace*{\fill}   % maximize separation between the subfigures
  \begin{subfigure}{0.32\textwidth}
    \includegraphics[width=\linewidth]{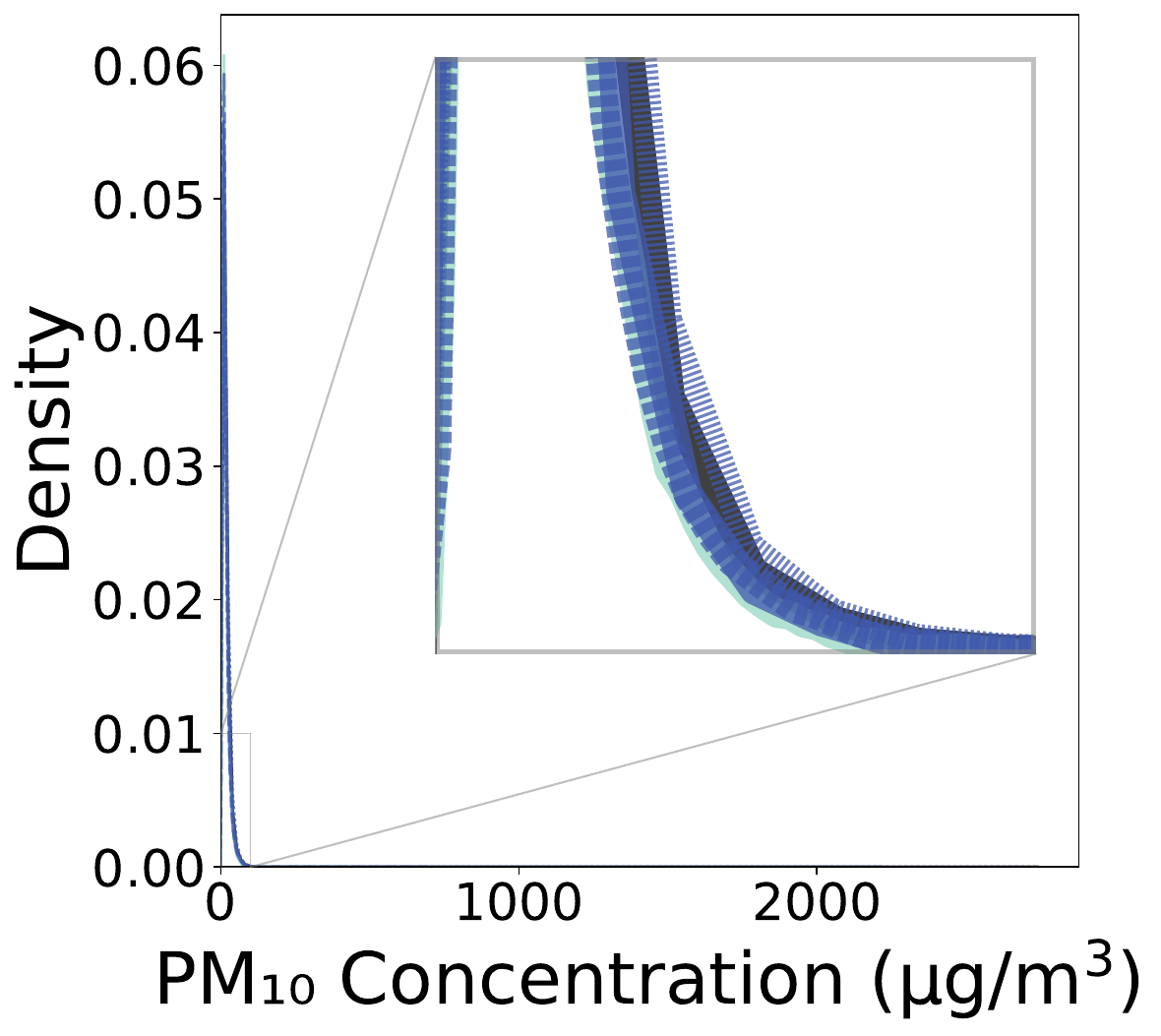}
    %\caption{PM$_{10}$}\label{fig:airPollutionKDEpm10}
  \end{subfigure}
  \hspace*{\fill}   % maximize separation between the subfigures
  \begin{subfigure}{0.32\textwidth}
    \includegraphics[width=\linewidth]{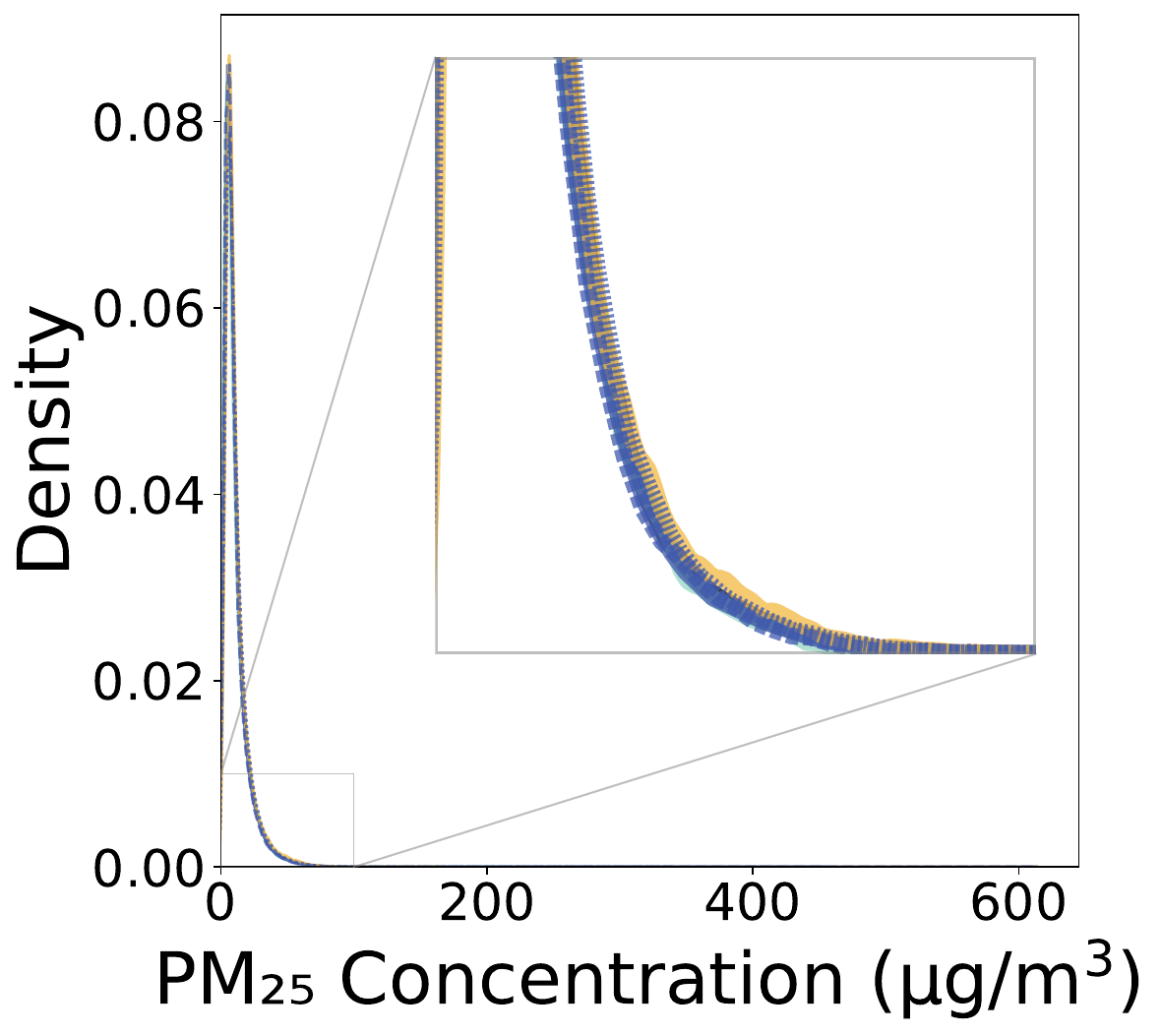}
    %\caption{PM$_{2.5}$}\label{fig:airPollutionKDEpm25}
  \end{subfigure}
  \hspace*{\fill}   % maximize separation between the subfigures
  \begin{subfigure}{0.32\textwidth}
    \includegraphics[width=\linewidth]{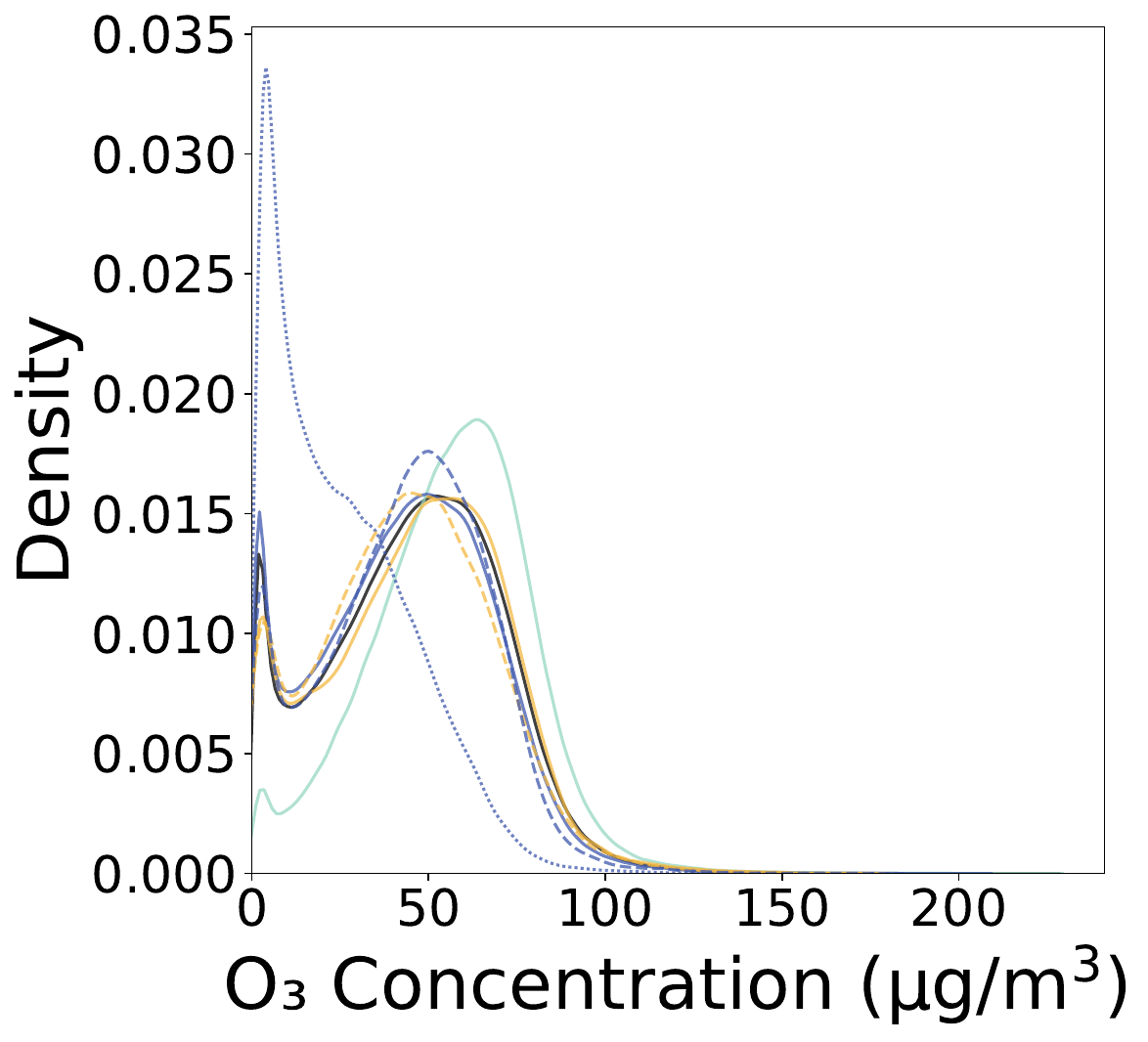}
    %\caption{O$_3$}\label{fig:airPollutionKDEo3}
  \end{subfigure}
  \hspace*{\fill}   % maximize separation between the subfigures
  \\
  \hspace*{\fill}   % maximize separation between the subfigures
  \begin{subfigure}{0.32\textwidth}
    \includegraphics[width=\linewidth]{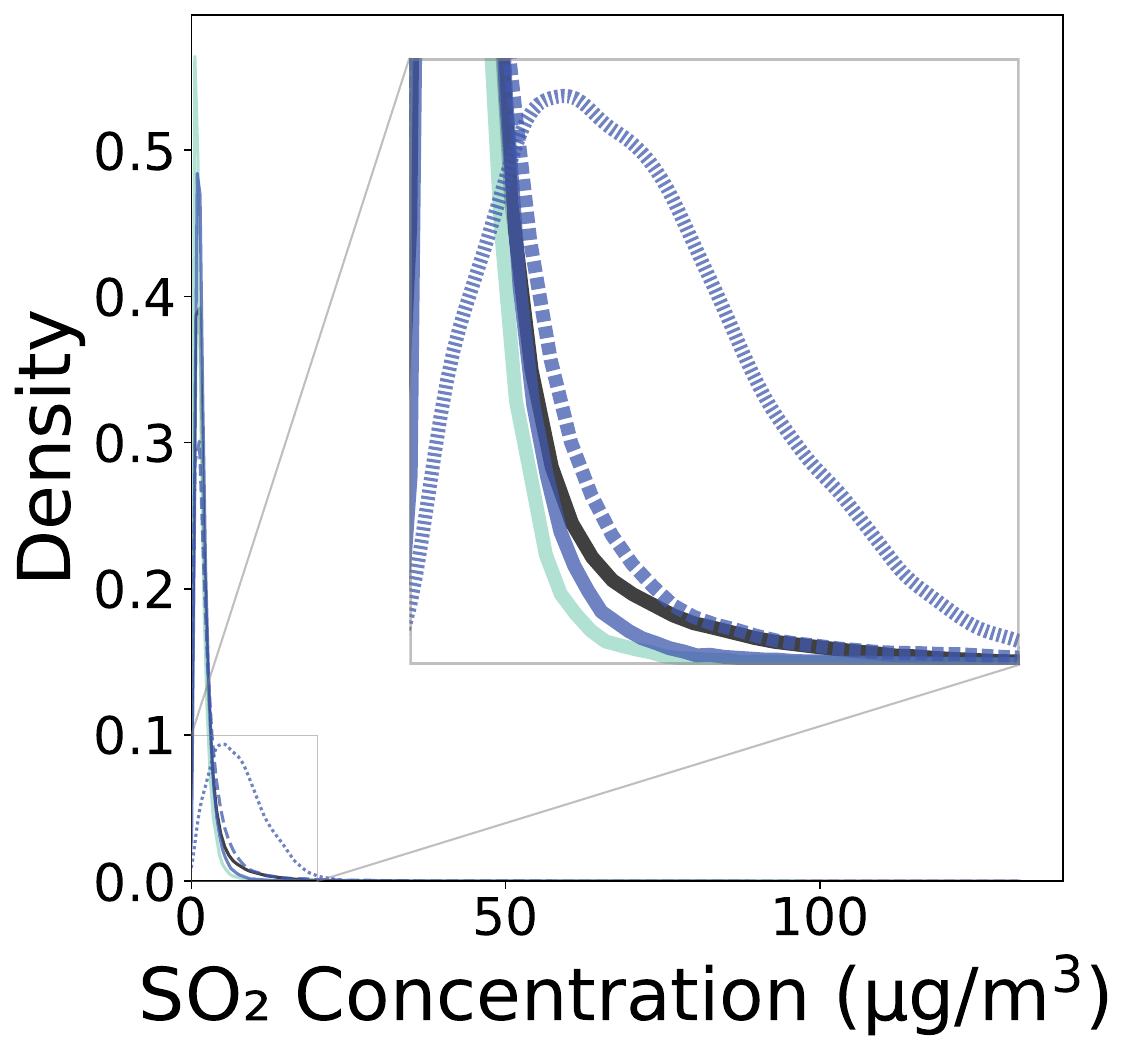}
    %\caption{SO$_2$}\label{fig:airPollutionKDEso2}
  \end{subfigure}
  \raisebox{8mm}{
  \begin{subfigure}{0.32\textwidth}
    \includegraphics[width=\linewidth]{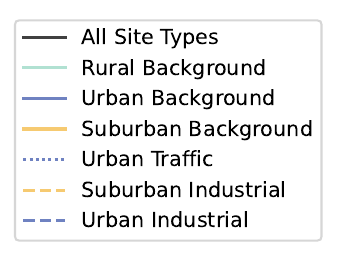}
  \end{subfigure}}
  \hspace*{\fill}   % maximize separation between the subfigures
\caption{{\bfseries KDE for air pollution measurements.} The kernel distribution estimation plots for each air pollutant show the distribution of concentration values. For PM$_{10}$ and PM$_{2.5}$, the distribution of values appears to remain constant across all sub classifications of the monitoring station, whereas NO$_x$, NO$_2$, NO and SO$_2$ appear to have different concentrations at different environmental locations. All air pollutants other than O$_3$ have a right skew, indicating a tendency for no air pollution to be the norm in the atmosphere, with pollution emitted and then subsequently dispersed, reducing the concentration measured. } \label{fig:airPollutionKDE}
\end{figure}

Detailed are the supporting analysis and figures for the air pollution concentration data. Included are the spatial distribution of the AURN network monitoring stations for each top-level environmental classification in Figure \ref{fig:ukAURNNetwork}, kernel distributions for each air pollutant concentration dataset Figure \ref{fig:airPollutionKDE}, and the abstracted distance of each AURN monitoring station from its real location in the model framework Table \ref{tab:AURNStationDistanceFromCentroidOverall}.

\begin{table}[!htb]
\resizebox{\linewidth}{!}{
\begin{filecontents*}{CSVFiles/Introduction/Peak_values_no2_subset_top_5.csv}
,Site Name,Peak Timestamp,Peak Value,8 Hour Mean,Peak Day Average,Peak Year Average
50,London Marylebone Road,10/06/2016 15:00:00,321.91104,192.41715555555555,139.6949472,89.12299093270228
71,Sandy Roadside,01/03/2018 16:00:00,336.72358,85.6664988888889,43.190186250000004,26.63246563116944
54,Luton A505 Roadside,19/01/2016 08:00:00,362.7214,255.08053777777775,176.85435080000002,49.75121235924008
14,Camden Kerbside,15/02/2016 18:00:00,385.34771,176.12021,124.422002,65.87905249485479
55,Manchester Piccadilly,13/09/2014 15:00:00,456.7772,111.93862444444444,69.9222112,40.48427679602461
\end{filecontents*}
\pgfplotstabletypeset[
    multicolumn names=l, 
    col sep=comma, 
    string type, 
    header = has colnames,
    columns={Site Name, Peak Timestamp, Peak Value, Peak Day Average, Peak Year Average},
    columns/Site Name/.style={column type=l, column name=AURN Site Name},
    columns/Peak Timestamp/.style={column type=l, column name=Peak Value Timestamp},
    columns/Peak Value/.style={column type={S[round-precision=1, table-format=5.0, table-number-alignment=left]}, column name=Peak Value (\si{\micro\gram/\meter^3})},
    columns/Peak Day Average/.style={column type={S[round-precision=1, table-format=5.0, table-number-alignment=left]}, column name=Peak Day Average (\si{\micro\gram/\meter^3})},
    columns/Peak Year Average/.style={column type={S[round-precision=1, table-format=5.0, table-number-alignment=left]}, column name=Peak Year Average (\si{\micro\gram/\meter^3})},
    every head row/.style={before row=\toprule, after row=\midrule},
    every last row/.style={after row=\bottomrule}
    ]{CSVFiles/Introduction/Peak_values_no2_subset_top_5.csv}}
    \smallskip
\caption{ { \bfseries AURN NO$_{2}$ monitoring station peak values between 2014-2018 with associated year daily mean peak and annual mean; for the year in which the peak measurement occurs. } These five monitoring stations show how there is not a simple relationship between the peak value, the peak daily average and the overall peak year average. London Marylebone Road station never has a peak as intense as Manchester Piccadilly but does experience consistently higher pollution across the year and similarly has a higher peak daily mean. However, Sandy Roadside has a higher overall peak than London Marylebone Road but a considerably lower peak day and year mean. The five stations have highlighted how multiple averages and a finer temporal scale are needed to uncover the intricacies of air pollution experienced at a single location.
} \label{tab:No2PeakValuesMeans}
\end{table}

\begin{table}[!htb]
\centering
\resizebox{\linewidth}{!}{
\begin{filecontents*}{CSVFiles/Data/air_pollution_negative_values.csv}
Pollutant,Number of Negative Values,Number Of Values,Negative Value Percentage
NOₓ,235,3855871,0.006094602231246845
NO₂,2697,3855760,0.06994729962445796
NO,2078,3859339,0.053843417227665155
O₃,659,2198186,0.029979264721001774
PM₁₀,3743,1646086,0.22738787645359962
PM₂₅,23515,2004487,1.173118109521289
SO₂,1016,683276,0.14869540273622958
\end{filecontents*}
\pgfplotstabletypeset[
    multicolumn names=l, 
    col sep=comma, 
    string type, 
    header = has colnames,
    columns={Pollutant, Number Of Values, Number of Negative Values, Negative Value Percentage},
    columns/Pollutant/.style={column type=l, column name=Pollutant Name},
    columns/Number of Values/.style={column type={S[round-precision=0, table-format=5, table-number-alignment=center]}, column name=Number of Values},
    columns/Number of Negative Values/.style={column type={S[round-precision=0, table-format=5, table-number-alignment=center]}, column name=Number of Negative Values},
    columns/Negative Value Percentage/.style={column type={S[round-precision=2, table-format=-1.3, table-number-alignment=center ]}, column name=Percentage of Negative Values (\%)},
    every head row/.style={before row=\toprule, after row=\midrule},
    every last row/.style={after row=\bottomrule}
    ]{CSVFiles/Data/air_pollution_negative_values.csv}}
    \smallskip
\caption{{\bfseries AURN negative data point summary.} Negative data points within the AURN air pollution concentrations were removed from the dataset as the only form of preprocessing performed on the dataset. Negative concentrations can't exist, and their presence in the dataset indicates a fault with the instruments at the monitoring station.} \label{tab:aurnNegativeValues}
\end{table}

\clearpage
\begin{table}[!htb]
\begin{subtable}[h]{\textwidth}
\centering
\resizebox{0.75\textwidth}{!}{
\begin{filecontents*}{CSVFiles/Data/aurnStationRealLocationFromGridCentroid_1.csv}
,site,site type,latitude,longitude,Grid ID,Grid Centroid Latitude,Grid Centroid Longitude,Station Distance From Grid Centroid
2000,Plymouth Tavistock Road,Urban Traffic,50.411058,-4.130288,97463,50.41111264,-4.130376807,8.747586046
37,Aston Hill,Rural Background,52.50385,-3.034178,256154,52.50390476,-3.03443216,18.24889882
2122,Salford Eccles,Urban Background,53.48481,-2.334139,373739,53.48488838,-2.333746238,27.40958474
858,Eastbourne,Urban Background,50.805778,0.271611,581778,50.80553743,0.271368086,31.73206074
1228,Leamington Spa Rugby Road,Urban Traffic,52.294884,-1.542911,431400,52.29510131,-1.543228788,32.4182673
187,Birkenhead Borough Road,Urban Traffic,53.388511,-3.025014,248483,53.3883334,-3.025449007,34.95977046
323,Blackpool Marton,Urban Background,53.80489,-3.007175,261012,53.80514847,-3.007482701,35.13175133
2563,Wharleycroft,Rural Background,54.616238,-2.468931,395280,54.61600683,-2.468493531,38.1335221
1432,London Brent,Urban Background,51.589769,-0.276223,562706,51.59003456,-0.276604238,39.5680024
1439,London Bridge Place,Urban Background,51.49521,-0.141655,594267,51.49480908,-0.141856945,46.72124789
1354,Liverpool Queen's Drive Roadside,Urban Traffic,53.446944,-2.9625,289455,53.44736541,-2.962566937,47.0682133
1716,Lullington Heath,Rural Background,50.7937,0.18125,578547,50.79415159,0.181536557,54.10382338
1050,Great Dun Fell,Rural Background,54.684233,-2.450799,397667,54.68372539,-2.450527225,59.08509074
193,Birmingham A4540 Roadside,Urban Traffic,52.47609,-1.875024,448581,52.47648735,-1.875605444,59.18539419
2184,Scunthorpe Town,Urban Industrial,53.58634,-0.636811,542696,53.58656851,-0.635930351,63.43940748
2116,Rotherham Centre,Urban Background,53.43186,-1.354444,470615,53.43127394,-1.354582579,65.81070386
1698,London Westminster,Urban Background,51.49467,-0.131931,593705,51.49480908,-0.132873792,67.07526151
2578,Wigan Centre,Urban Background,53.54914,-2.638139,334875,53.54913595,-2.639173435,68.34110453
2517,Tower Hamlets Roadside,Urban Traffic,51.52253,-0.042155,562130,51.52283737,-0.043042264,70.25950277
1783,Middlesbrough,Urban Industrial,54.569297,-1.220874,476604,54.56905833,-1.219835286,72.02519748
1190,Immingham Woodlands Avenue,Urban Background,53.619241,-0.213324,567550,53.61862707,-0.213722168,73.14270987
180,Billingham,Urban Industrial,54.60537,-1.275039,473328,54.60557853,-1.273734203,87.17512697
1445,London Bromley,Urban Traffic,51.40555,0.018869,570170,51.40500204,0.019839806,90.81321164
467,Bury Roadside,Urban Traffic,53.53911,-2.289611,402536,53.53843482,-2.288830474,91.08660253
498,Cambridge,Urban Traffic,51.995804,0.037756,595218,51.99667018,0.037806112,96.37567644
1652,London Teddington,Urban Background,51.42099,-0.339647,590245,51.42185439,-0.339486308,96.75967205
1460,London Cromwell Road 2,Urban Traffic,51.495483,-0.178709,569637,51.49480908,-0.177789557,98.3208982
2073,Reading London Road,Urban Traffic,51.454896,-0.940382,537621,51.45554035,-0.941357548,98.50132598
1267,Leeds Potternewton,Urban Background,53.82568,-1.535098,440658,53.82641232,-1.534245636,98.7944253
1524,London Haringey Priory Park South,Urban Background,51.584128,-0.125254,594839,51.58443861,-0.123890639,100.3300627
230,Birmingham East,Urban Background,52.49763,-1.831498,490673,52.49842266,-1.830689679,103.7433488
206,Birmingham Acocks Green,Urban Background,52.437165,-1.829999,490610,52.4380741,-1.830689679,111.4032562
603,Charlton Mackrell,Rural Background,51.05625,-2.68345,329202,51.05532073,-2.684089199,112.5746372
1875,Northampton,Urban Background,52.27349,-0.885933,517608,52.27306421,-0.887458631,114.0914292
2137,Saltash Callington Road,Urban Traffic,50.411463,-4.227678,138455,50.41111264,-4.229191488,114.1046487
351,Bottesford,Rural Background,52.93028,-0.814722,543944,52.9293956,-0.815593408,114.3783359
1213,Leamington Spa,Urban Background,52.28881,-1.533119,440884,52.28959307,-1.534245636,115.9898578
1154,Hove Roadside,Urban Traffic,50.82778,-0.170294,574609,50.82830075,-0.168806404,119.4558437
1958,Oxford Centre Roadside,Urban Traffic,51.751745,-1.257463,471747,51.75201602,-1.255767897,120.514992
2524,Walsall Alumwell,Urban Background,52.58167,-2.010483,450289,52.58058218,-2.010352736,121.2801918
2155,Sandwell Oldbury,Urban Background,52.50431,-2.017629,449270,52.50390476,-2.019335889,124.0071072
721,Coventry Allesley,Urban Background,52.411563,-1.560228,439265,52.41061544,-1.561195094,124.1144052
2459,Sunderland Wessington Way,Urban Traffic,54.91839,-1.408391,469692,54.91726161,-1.408481496,125.6047982
1771,Manchester Town Hall,Urban Background,53.4785,-2.2448,370238,53.47953,-2.24391471,128.6437712
2446,Sunderland,Urban Background,54.906106,-1.380081,497669,54.90691099,-1.381532037,128.9062828
2207,Sheffield Centre,Urban Background,53.37772,-1.473306,499731,53.37759146,-1.471363566,129.6360922
428,Bristol Old Market,Urban Traffic,51.45603,-2.583519,339705,51.45554035,-2.585274518,133.2655398
1080,Hartlepool St Abbs Walk,Urban Background,54.683242,-1.203838,505582,54.68372539,-1.20186898,137.5111205
2254,Southampton A33,Urban Traffic,50.920265,-1.463484,441951,50.91924248,-1.462380413,137.5211328
1360,Liverpool Speke,Urban Industrial,53.34633,-2.844333,307161,53.34534929,-2.84578595,145.5826251
365,Bradford Centre,Urban Background,53.79339,-1.748694,422171,53.79451248,-1.749841304,145.7984144
826,Doncaster A630 Cleveland Street,Urban Traffic,53.51824,-1.138057,487373,53.5170244,-1.138986911,148.4935494
1417,London Bloomsbury,Urban Background,51.52229,-0.125889,594828,51.52283737,-0.123890639,151.0631752
1642,London Southwark,Urban Background,51.49055,-0.096667,558953,51.48920134,-0.096941181,151.1610606
1347,Liverpool Centre,Urban Background,53.40845,-2.980249,257802,53.40980911,-2.980533243,152.2965142
2505,Thurrock,Urban Background,51.47707,0.317969,577591,51.47798378,0.31628385,154.7391421
646,Chesterfield Loundsley Green,Urban Background,53.244131,-1.454946,500438,53.24308732,-1.45339726,155.2039725
1885,Northampton Kingsthorpe,Urban Background,52.271886,-0.879898,516354,52.27306421,-0.878475478,162.8868138
785,Derby St Alkmund's Way,Urban Traffic,52.922983,-1.469507,499647,52.92396697,-1.471363566,165.7141957
378,Brentford Roadside,Urban Traffic,51.489448,-0.310121,557007,51.48920134,-0.312536849,169.4993944
262,Birmingham Ladywood,Urban Background,52.481346,-1.918235,460926,52.4819722,-1.920521208,169.7585246
2200,Sheffield Barnsley Road,Urban Traffic,53.40495,-1.455815,500366,53.40444121,-1.45339726,169.9642955
1775,Market Harborough,Rural Background,52.554444,-0.772222,544883,52.55321284,-0.770677644,172.1716854
1893,Northampton Spring Park,Urban Background,52.272257,-0.916605,512333,52.27306421,-0.914408089,174.3574288
523,Canterbury,Urban Background,51.27399,1.098061,628587,51.27559318,1.097818147,179.0648742
2095,Redcar,Suburban Background,54.610735,-1.0733,509787,54.61079301,-1.076104841,180.7362007
461,Burton-on-Trent Horninglow,Urban Background,52.82105,-1.635718,437408,52.82069286,-1.633060317,182.94819
397,Bristol Centre,Urban Background,51.45718,-2.585622,339705,51.45554035,-2.585274518,183.9037518
1205,Ladybower,Rural Background,53.40337,-1.752006,422098,53.40444121,-1.749841304,186.4944428
1911,Norwich Lakenfields,Urban Background,52.614193,1.301976,637538,52.61340273,1.304430662,187.5842592
2222,Sheffield Devonshire Green,Urban Background,53.378622,-1.478096,442278,53.37759146,-1.480346718,188.2010549
304,Blackburn Accrington Road,Urban Traffic,53.747751,-2.452724,397388,53.74661697,-2.450527225,191.7453582
1454,London Cromwell Road,Urban Traffic,51.49492,-0.180564,569637,51.49480908,-0.177789557,192.4657974
2598,Wirral Tranmere,Urban Background,53.37287,-3.022722,248480,53.37221947,-3.025449007,194.8355066
2242,Sibton,Rural Background,52.2944,1.463497,639159,52.29510131,1.466127413,195.1446942
435,Bristol St Paul's,Urban Background,51.462839,-2.584482,339706,51.46115225,-2.585274518,195.4293567
1338,Lincoln Canwick Road,Urban Traffic,53.221373,-0.534189,521770,53.22152713,-0.53711567,195.5957705
2559,Weybourne,Rural Background,52.95049,1.122017,628900,52.95110331,1.124767605,196.4913179
585,Carlisle Roadside,Urban Traffic,54.894834,-2.945307,293110,54.8965577,-2.944600632,196.9171037
310,Blackburn Darwen Roadside,Urban Traffic,53.715504,-2.483815,393928,53.71465613,-2.486459837,197.9391965
1627,London N. Kensington,Urban Background,51.52105,-0.213492,567947,51.52283737,-0.213722168,199.3833912
2326,Southwark Roadside,Urban Traffic,51.48199,-0.0623,558194,51.48359291,-0.06100857,199.4120407
2374,Stockport Shaw Heath,Urban Background,53.40306,-2.161111,389011,53.40444121,-2.163066334,200.9717503
2348,Stanford-le-Hope Roadside,Urban Traffic,51.518167,0.439548,603611,51.5172331,0.44204799,201.7602044
1131,High Muffles,Rural Background,54.334944,-0.80855,522973,54.33349953,-0.806610255,203.9943081
505,Camden Kerbside,Urban Traffic,51.54421,-0.175269,569646,51.54524752,-0.177789557,209.0242516
1326,Leominster,Suburban Background,52.22174,-2.736665,320311,52.22344045,-2.737988116,209.4628375
2659,York Fishergate,Urban Traffic,53.951889,-1.075861,509868,53.95376791,-1.076104841,209.5339765
1315,Leicester University,Urban Background,52.619823,-1.127311,478481,52.61887042,-1.130003758,210.3896015
1138,Honiton,Urban Background,50.792287,-3.196702,206235,50.79415159,-3.196128911,211.2104561
2549,West Bromwich Kenrick Park,Urban Background,52.508337,-1.986008,453337,52.50938619,-1.983403278,211.3900094
2161,Sandwell West Bromwich,Urban Background,52.52062,-1.995556,452345,52.52034697,-1.99238643,216.5905307
1159,Hull Centre,Urban Background,53.74479,-0.338322,589854,53.74661697,-0.339486308,217.0988207
1952,Oldbury Birmingham Road,Urban Traffic,52.502436,-2.003497,451352,52.50390476,-2.001369583,217.7347184
1255,Leeds Headingley Kerbside,Urban Traffic,53.819972,-1.576361,433438,53.82109737,-1.5791614,222.3698015
1928,Nottingham Centre,Urban Background,52.95473,-1.146447,488697,52.95652852,-1.147970063,224.508635
2591,Wigan Leigh,Urban Background,53.49422,-2.506899,375064,53.49560311,-2.504426142,224.5227697
1344,Lincoln Roadside,Urban Traffic,53.22889,-0.537895,521769,53.2269182,-0.53711567,225.3079295
2302,Southend-on-Sea,Urban Background,51.544206,0.678408,613245,51.54524752,0.675609963,225.502508
634,Chesterfield,Urban Background,53.230583,-1.433611,502019,53.23230859,-1.435430954,226.9152842
747,Coventry Memorial Park,Urban Background,52.394399,-1.519612,498208,52.394132,-1.51627933,228.0755116
1985,Plymouth Centre,Urban Background,50.37167,-4.142361,95959,50.37091269,-4.139359959,228.8912081
372,Bradford Mayo Avenue,Urban Traffic,53.771245,-1.759774,421447,53.77323236,-1.758824457,229.6262474
2493,Swindon Walcot,Urban Background,51.558061,-1.765678,493158,51.55644844,-1.767807609,232.0078135
2121,Rugeley,Urban Background,52.753297,-1.93773,458271,52.75533975,-1.938487513,232.7947923
2537,Warrington,Urban Industrial,53.38928,-2.615358,341984,53.3883334,-2.612223976,232.9661101
1531,London Harlington,Urban Industrial,51.48879,-0.441614,534875,51.48920134,-0.438300989,233.8995815
457,Bromley Roadside,Urban Traffic,51.4071,0.020128,570170,51.40500204,0.019839806,234.1372117
\end{filecontents*}
\pgfplotstabletypeset[
    multicolumn names=l, 
    col sep=comma, 
    string type, 
    header = has colnames,
    columns={site, site type, latitude, longitude, Grid Centroid Latitude, Grid Centroid Longitude, Station Distance From Grid Centroid},
    columns/site/.style={column type=l, column name=AURN Station Name},
    columns/site type/.style={column type=l, column name=Station Environment Type},
    columns/latitude/.style={column type={S[round-precision=1, table-format=2.1, table-number-alignment=center]}, column name=Station Latitude},
    columns/longitude/.style={column type={S[round-precision=1, table-format=2.1, table-number-alignment=center]}, column name=Station Longitude},
    columns/Grid Centroid Latitude/.style={column type={S[round-precision=1, table-format=2.1, table-number-alignment=center]}, column name=Grid Centroid Latitude},
    columns/Grid Centroid Longitude/.style={column type={S[round-precision=1, table-format=2.1, table-number-alignment=center]}, column name=Grid Centroid Longitude},
    columns/Station Distance From Grid Centroid/.style={column type={S[round-precision=1, table-format=2.1, table-number-alignment=center]}, column name=Station Distance From Grid Centroid (m)},
    every head row/.style={before row=\toprule, after row=\midrule},
    every last row/.style={after row=\bottomrule}
    ]{CSVFiles/Data/aurnStationRealLocationFromGridCentroid_1.csv}}
    \smallskip
\caption{ {\bfseries AURN station real distance from the centroid distance.} } \label{tab:AURNStationDistanceFromCentroid}
%\end{table}
\end{subtable}
\end{table}
\begin{table}\ContinuedFloat
\begin{subtable}[h]{\textwidth}
%\begin{table}[!htb]
\centering
\resizebox{0.75\textwidth}{!}{
\begin{filecontents*}{CSVFiles/Data/aurnStationRealLocationFromGridCentroid_2.csv}
,site,site type,latitude,longitude,Grid ID,Grid Centroid Latitude,Grid Centroid Longitude,Station Distance From Grid Centroid
1567,London Honor Oak Park,Urban Background,51.449674,-0.037418,563299,51.44992776,-0.034059111,234.464254
2146,Saltash Roadside,Urban Traffic,50.4131,-4.2303,138455,50.41111264,-4.229191488,234.5295023
851,Ealing Horn Lane,Urban Traffic,51.51895,-0.265617,563914,51.5172331,-0.267621085,235.9579734
338,Bolton,Urban Background,53.57232,-2.439583,398833,53.57053006,-2.441544072,237.4475015
1181,Hull Holderness Road,Urban Traffic,53.758971,-0.305749,557225,53.75726516,-0.303553696,238.3390362
2235,Sheffield Tinsley,Urban Background,53.41058,-1.396139,468592,53.40980911,-1.399498343,238.5918105
2572,Widnes Milton Road,Urban Traffic,53.365391,-2.73168,322999,53.36684679,-2.729004963,240.2216584
1404,London Bexley,Suburban Background,51.46603,0.184806,578412,51.46676345,0.181536557,240.7164972
1856,Newcastle Cradlewell Roadside,Urban Traffic,54.986405,-1.595362,435639,54.98447548,-1.597127705,242.3311184
499,Cambridge Roadside,Urban Traffic,52.20237,0.124456,596129,52.20136753,0.12763764,243.8017749
597,Central London,Urban Background,51.494722,-0.138333,594267,51.49480908,-0.141856945,244.1495652
658,Chesterfield Roadside,Urban Traffic,53.231722,-1.456944,500436,53.23230859,-1.45339726,244.9117569
1763,Manchester South,Suburban Industrial,53.369026,-2.24328,370217,53.36684679,-2.24391471,245.9490496
1068,Haringey Roadside,Urban Traffic,51.5993,-0.068218,555783,51.60122439,-0.069991722,246.5702288
2411,Stoke-on-Trent Centre,Urban Background,53.02821,-2.175133,405064,53.02699411,-2.172049487,246.5829354
1724,Luton A505 Roadside,Urban Traffic,51.892293,-0.46211,530803,51.89118818,-0.465250447,248.0658574
713,Chilworth,Suburban Background,50.962706,-1.424677,503213,50.96464646,-1.426447801,248.8688058
1901,Norwich Centre,Urban Background,52.63203,1.295019,637350,52.62980374,1.295447509,249.2322285
1150,Hounslow Roadside,Urban Traffic,51.48965,-0.308975,557007,51.48920134,-0.312536849,251.6060103
1517,London Haringey,Urban Background,51.58603,-0.126486,594839,51.58443861,-0.123890639,251.9273746
938,Featherstone,Urban Background,53.670222,-1.352146,470660,53.67200369,-1.354582579,254.9754632
2647,York Bootham,Urban Background,53.967513,-1.086514,485602,53.96965996,-1.085087993,256.304871
2389,Stockton-on-Tees Eaglescliffe,Urban Traffic,54.516667,-1.358547,470820,54.51682955,-1.354582579,256.5205626
1545,London Harrow,Suburban Background,51.57389,-0.352051,589648,51.57324462,-0.348469461,257.7092216
2564,Wicken Fen,Rural Background,52.2985,0.290917,598014,52.30060886,0.289334391,258.010261
276,Birmingham Tyburn,Urban Background,52.511722,-1.830583,490671,52.50938619,-1.830689679,259.8312343
404,Bristol East,Urban Background,51.45365,-2.578496,339665,51.45554035,-2.576291365,259.8424108
2182,Scunthorpe,Urban Industrial,53.58499,-0.633015,542696,53.58656851,-0.635930351,260.4601336
1240,Leeds Centre,Urban Background,53.80378,-1.546472,431169,53.80514847,-1.543228788,261.7434184
2263,Southampton Centre,Urban Background,50.90814,-1.395778,468143,50.90788452,-1.399498343,262.3978374
2367,Stewartby,Urban Industrial,52.071944,-0.511111,526788,52.07423427,-0.510166212,262.7262509
1581,London Marylebone Road,Urban Traffic,51.52253,-0.154611,593152,51.52283737,-0.150840098,263.1231847
382,Brighton Preston Park,Urban Background,50.840836,-0.147572,593235,50.83967822,-0.150840098,263.1246327
1300,Leicester Centre,Urban Background,52.631348,-1.133006,478483,52.62980374,-1.130003758,265.5956346
1166,Hull Freetown,Urban Background,53.74878,-0.341222,589854,53.74661697,-0.339486308,266.2222414
345,Borehamwood Meadow Park,Urban Background,51.661229,-0.270671,563838,51.66271894,-0.267621085,267.7716324
820,Dewsbury Ashworth Grove,Urban Background,53.693104,-1.637111,437264,53.69333534,-1.633060317,267.9329184
2613,Wolverhampton Centre,Urban Background,52.58818,-2.129008,384794,52.58605399,-2.127133723,268.176798
2397,Stockton-on-Tees Yarm,Urban Traffic,54.50918,-1.354319,470819,54.51160298,-1.354582579,269.9602927
2368,Stockport,Urban Background,53.40994,-2.1582,387785,53.40980911,-2.154083182,273.2583488
261,Birmingham Kerbside,Urban Traffic,52.328057,-1.907512,461940,52.32813631,-1.911538055,273.7352538
2194,Shaw Crompton Way,Urban Traffic,53.579283,-2.093786,380839,53.58122304,-2.091201112,275.0562199
2499,Telford Hollinswood,Urban Background,52.673471,-2.436692,355125,52.67350959,-2.43256092,278.5663781
482,Bury Whitefield Roadside,Urban Traffic,53.559029,-2.293772,367950,53.55983437,-2.297813627,281.5647985
2527,Walsall Willenhall,Urban Background,52.60821,-2.033144,446476,52.60793436,-2.037302195,282.4490408
2380,Stockton-on-Tees A1305 Roadside,Urban Traffic,54.565819,-1.3159,467059,54.56383847,-1.318649967,282.718434
670,Chilbolton Observatory,Rural Background,51.149617,-1.438228,502356,51.15146695,-1.435430954,283.5072305
2629,Wray,Rural Background,54.104666,-2.584182,344747,54.10713619,-2.585274518,283.7571797
450,Bristol Temple Way,Urban Traffic,51.457968,-2.583975,339705,51.45554035,-2.585274518,284.5628752
2465,Sutton Roadside,Urban Traffic,51.36636,-0.182789,592030,51.36565562,-0.186772709,287.4415033
316,Blackpool,Urban Background,53.79046,-3.029283,248538,53.78919346,-3.025449007,288.5508514
2640,Yarner Wood,Rural Background,50.5976,-3.71651,225196,50.60016537,-3.717151776,288.8305168
929,Exeter Roadside,Urban Traffic,50.725083,-3.532465,173560,50.72577797,-3.528505566,289.2221415
1670,London UCL,Urban Background,51.52378,-0.128958,593710,51.52283737,-0.132873792,290.4844281
1086,Harwell,Rural Background,51.571078,-1.325283,474700,51.57324462,-1.32763312,290.552984
732,Coventry Binley Road,Urban Traffic,52.407708,-1.490082,440076,52.40512165,-1.489329871,292.0803539
2331,St Helens Linkway,Urban Traffic,53.451826,-2.742134,320537,53.45272788,-2.737988116,292.2666709
1144,Horley,Suburban Industrial,51.165865,-0.167734,574529,51.16841306,-0.168806404,293.032741
110,Barnstaple A39,Urban Traffic,51.074793,-4.041924,149244,51.07230232,-4.040545278,293.2248969
1741,Manchester Piccadilly,Urban Background,53.48152,-2.237881,362117,53.47953,-2.234931557,295.0517086
2530,Walsall Woodlands,Urban Background,52.605621,-2.030523,447198,52.60793436,-2.028319042,297.1842023
2317,Southwark A2 Old Kent Road,Urban Traffic,51.480499,-0.05955,558193,51.47798378,-0.06100857,297.3609087
291,Birmingham Tyburn Roadside,Urban Traffic,52.512194,-1.830861,490670,52.51486692,-1.830689679,297.441849
1572,London Islington,Urban Background,51.531362,-0.096954,558961,51.53404383,-0.096941181,298.2076209
2102,Rochester Stoke,Rural Background,51.45617,0.634889,608517,51.45554035,0.630694199,298.9615515
1551,London Harrow Stanmore,Urban Background,51.617333,-0.298777,560392,51.61800394,-0.294570543,299.8504959
1756,Manchester Sharston,Suburban Industrial,53.371306,-2.239218,362097,53.37221947,-2.234931557,301.9639369
2426,Storrington Roadside,Urban Traffic,50.916932,-0.449548,529624,50.91924248,-0.447284142,301.9765492
2361,Stevenage,Suburban Background,51.886944,-0.200833,572175,51.88562959,-0.204739015,305.3320746
2447,Sunderland Silksworth,Urban Background,54.88361,-1.406878,469652,54.88102275,-1.408481496,305.4273117
1841,Newcastle Centre,Urban Background,54.97825,-1.610528,436623,54.97930919,-1.606110858,305.4859326
1576,London Lewisham,Urban Background,51.44541,-0.020139,569047,51.44431447,-0.016092805,305.7352658
2402,Stoke-on-Trent A50 Roadside,Urban Traffic,52.980436,-2.111898,382516,52.97822255,-2.109167417,306.5922594
1399,London A3 Roadside,Urban Traffic,51.37348,-0.291853,560450,51.37127862,-0.294570543,309.0344859
1908,Norwich Forum Roadside,Urban Traffic,52.62817,1.291714,637350,52.62980374,1.295447509,310.6409886
1663,London Teddington Bushy Park,Urban Background,51.425286,-0.345606,589622,51.42747045,-0.348469461,313.711409
1291,Leicester A594 Roadside,Urban Traffic,52.638677,-1.124228,477368,52.64073432,-1.121020605,314.915095
97,Barnsley 12,Urban Background,53.55593,-1.485153,439880,53.5544855,-1.489329871,319.2527162
610,Chatham Roadside,Urban Traffic,51.374264,0.54797,602245,51.37690092,0.549845824,320.8204623
2555,West London,Urban Background,51.4938,-0.200361,572309,51.49480908,-0.204739015,323.1892689
1023,Glazebury,Rural Background,53.46008,-2.472056,395010,53.45808966,-2.468493531,323.4312872
96,Barnsley,Urban Background,53.580045,-1.475865,442274,53.58122304,-1.480346718,323.5643603
101,Barnsley Gawber,Urban Background,53.56292,-1.510436,430106,53.56518255,-1.507296177,326.0252288
390,Brighton Roadside,Urban Traffic,50.82354,-0.137281,593830,50.82261096,-0.132873792,326.3602872
762,Crewe Coppenhall,Urban Background,53.115941,-2.453492,397065,53.11356246,-2.450527225,330.3087217
2080,Reading New Town,Urban Background,51.45309,-0.944067,537621,51.45554035,-0.941357548,330.8831654
186,Bircotes,Urban Background,53.422855,-1.054946,486454,53.42054289,-1.058138535,332.9428078
2622,Worthing A27 Roadside,Urban Traffic,50.832947,-0.379916,588506,50.83398983,-0.384402072,335.7110767
517,Cannock A5190 Roadside,Urban Traffic,52.687298,-1.980821,453322,52.68988796,-1.983403278,336.4984584
125,Bath Roadside,Urban Traffic,51.391127,-2.354155,396903,51.39376368,-2.351712544,338.639683
1648,London Sutton,Suburban Background,51.36789,-0.165489,574514,51.36565562,-0.168806404,338.7745518
2340,St Osyth,Rural Background,51.77798,1.049031,627989,51.77988502,1.052902383,340.3036346
218,Birmingham Centre,Urban Background,52.479724,-1.908078,461924,52.4819722,-1.911538055,342.6358801
1694,London Wandsworth,Urban Background,51.45696,-0.191164,592046,51.45554035,-0.186772709,342.7723792
714,Christchurch Barrack Road,Urban Traffic,50.735454,-1.780888,419274,50.73718055,-1.776790762,346.4071515
1557,London Hillingdon,Urban Background,51.49633,-0.460861,530834,51.49480908,-0.465250447,347.7613663
2250,Somerton,Rural Background,51.03572,-2.735253,321475,51.03833288,-2.737988116,347.8319171
1943,Nottingham Western Boulevard,Urban Traffic,52.969377,-1.188851,482224,52.9673769,-1.192885828,349.9604151
2051,Preston,Urban Background,53.76559,-2.680353,330736,53.76791064,-2.684089199,356.2110998
1925,Norwich Roadside,Urban Traffic,52.622,1.299064,637349,52.62433743,1.295447509,356.5770734
1966,Oxford St Ebbes,Urban Background,51.744806,-1.260278,472077,51.74644014,-1.26475105,357.5674789
1468,London Eltham,Suburban Background,51.45258,0.070766,575603,51.44992776,0.073738723,359.7345759
741,Coventry Centre,Urban Background,52.41345,-1.522133,443337,52.41610855,-1.525262483,363.9222008
2066,Reading,Urban Background,51.45352,-0.95518,540330,51.45554035,-0.959323854,364.5686571
2170,Sandy Roadside,Urban Traffic,52.132417,-0.300306,557538,52.13508264,-0.303553696,370.1256114
355,Bournemouth,Urban Background,50.73957,-1.826744,491348,50.73718055,-1.830689679,384.3046435
119,Bath A4 Roadside,Urban Traffic,51.390922,-2.35503,396904,51.38814345,-2.351712544,385.2858821
1512,London Hackney,Urban Background,51.55877,-0.056592,558159,51.55644844,-0.06100857,399.8329662
\end{filecontents*}
\pgfplotstabletypeset[
    multicolumn names=l, 
    col sep=comma, 
    string type, 
    header = has colnames,
    columns={site, site type, latitude, longitude, Grid Centroid Latitude, Grid Centroid Longitude, Station Distance From Grid Centroid},
    columns/site/.style={column type=l, column name=AURN Station Name},
    columns/site type/.style={column type=l, column name=Station Environment Type},
    columns/latitude/.style={column type={S[round-precision=1, table-format=2.1, table-number-alignment=center]}, column name=Station Latitude},
    columns/longitude/.style={column type={S[round-precision=1, table-format=2.1, table-number-alignment=center]}, column name=Station Longitude},
    columns/Grid Centroid Latitude/.style={column type={S[round-precision=1, table-format=2.1, table-number-alignment=center]}, column name=Grid Centroid Latitude},
    columns/Grid Centroid Longitude/.style={column type={S[round-precision=1, table-format=2.1, table-number-alignment=center]}, column name=Grid Centroid Longitude},
    columns/Station Distance From Grid Centroid/.style={column type={S[round-precision=1, table-format=2.1, table-number-alignment=center]}, column name=Station Distance From Grid Centroid (m)},
    every head row/.style={before row=\toprule, after row=\midrule},
    every last row/.style={after row=\bottomrule}
    ]{CSVFiles/Data/aurnStationRealLocationFromGridCentroid_2.csv}}
    \smallskip
\caption{ {\bfseries AURN station real distance from the centroid distance (cont.).} For each AURN monitoring station used within the study, the station's latitude and longitude are given, alongside the abstracted location of the station within the study, denoted by the grid centroids latitude and longitude. The station distance then gives the difference between the stations true location and the location used within the study.   } \label{tab:AURNStationDistanceFromCentroidCont}
\end{subtable}\caption{}\label{tab:AURNStationDistanceFromCentroidOverall}
\end{table}

\clearpage
\subsection{Transport Infrastructure Structural Properties}
\label{sec:Datadetails:TransportDataInfratsurtcureStructural}

Open Street Maps was used as the data set to build the transport infrastructure feature vector. Open Street Maps provides a high level of detail on the road location and the type of road, alongside providing a historical dataset that allows for historical roads to be acquired across years. Due to the computational cost of retrieving the feature vector for the transport infrastructure in a grided format, and the minimal change to the road infrastructure itself on a fine temporal level, especially hourly, we decided to take yearly snapshots of the road infrastructure. A possible improvement to the method would be to take more frequency snapshots of the road network at the expense of additional computation if desired. The snapshot of the road network used was the road network structural on the first day of the year. We then used this snapshot of the road network to create a feature vector for the following year of timestamps within the feature vector. 

\begin{figure}[!htb]
  \begin{subfigure}{0.48\textwidth}
    \includegraphics[width=\linewidth]{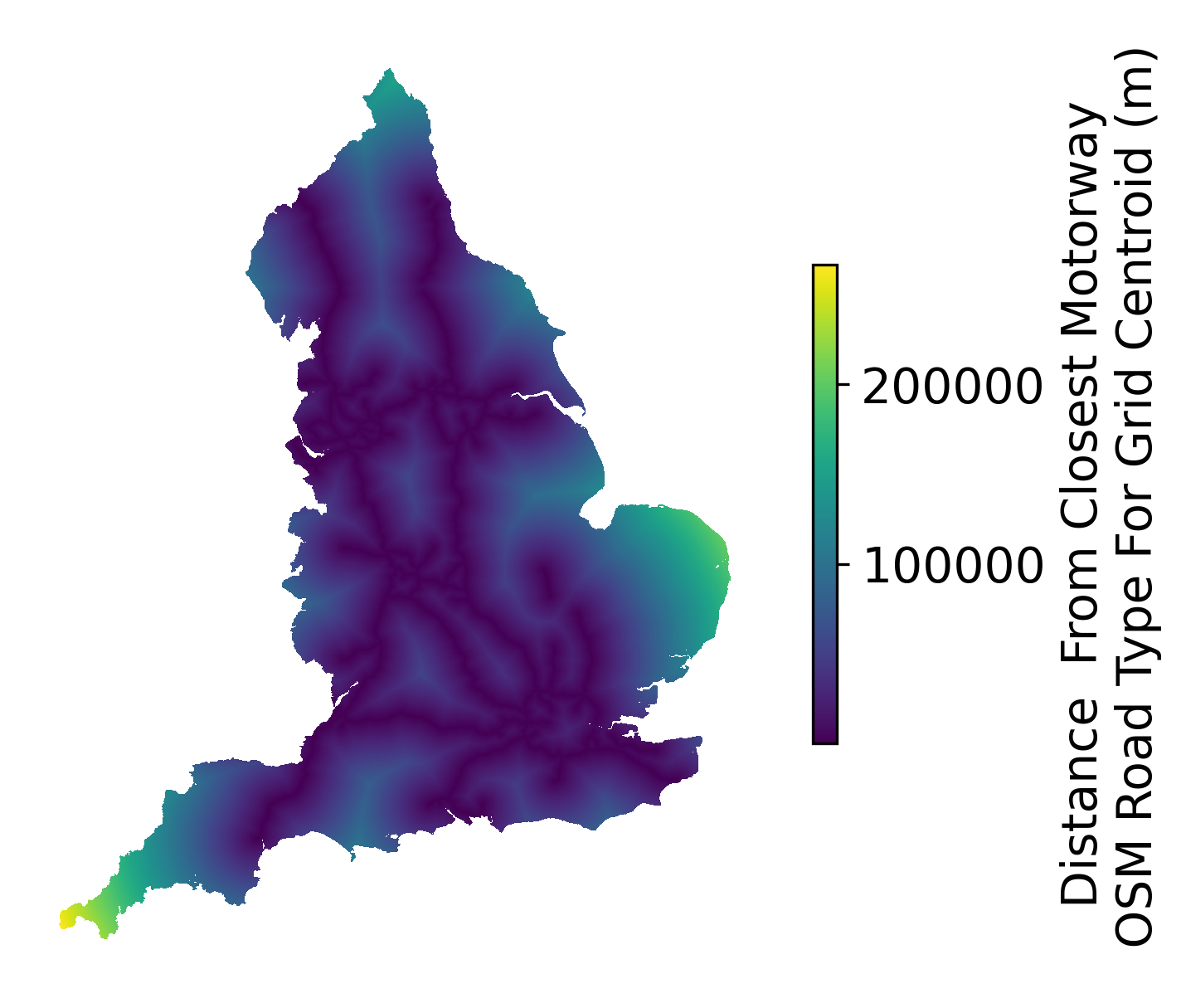}
    \caption{Grid Centroid Distance To Closest Motorway} \label{fig:transportInfrastructureDistanceMotorway}
  \end{subfigure}%
  \hspace*{\fill}   % maximize separation between the subfigures
  \begin{subfigure}{0.48\textwidth}
    \includegraphics[width=\linewidth]{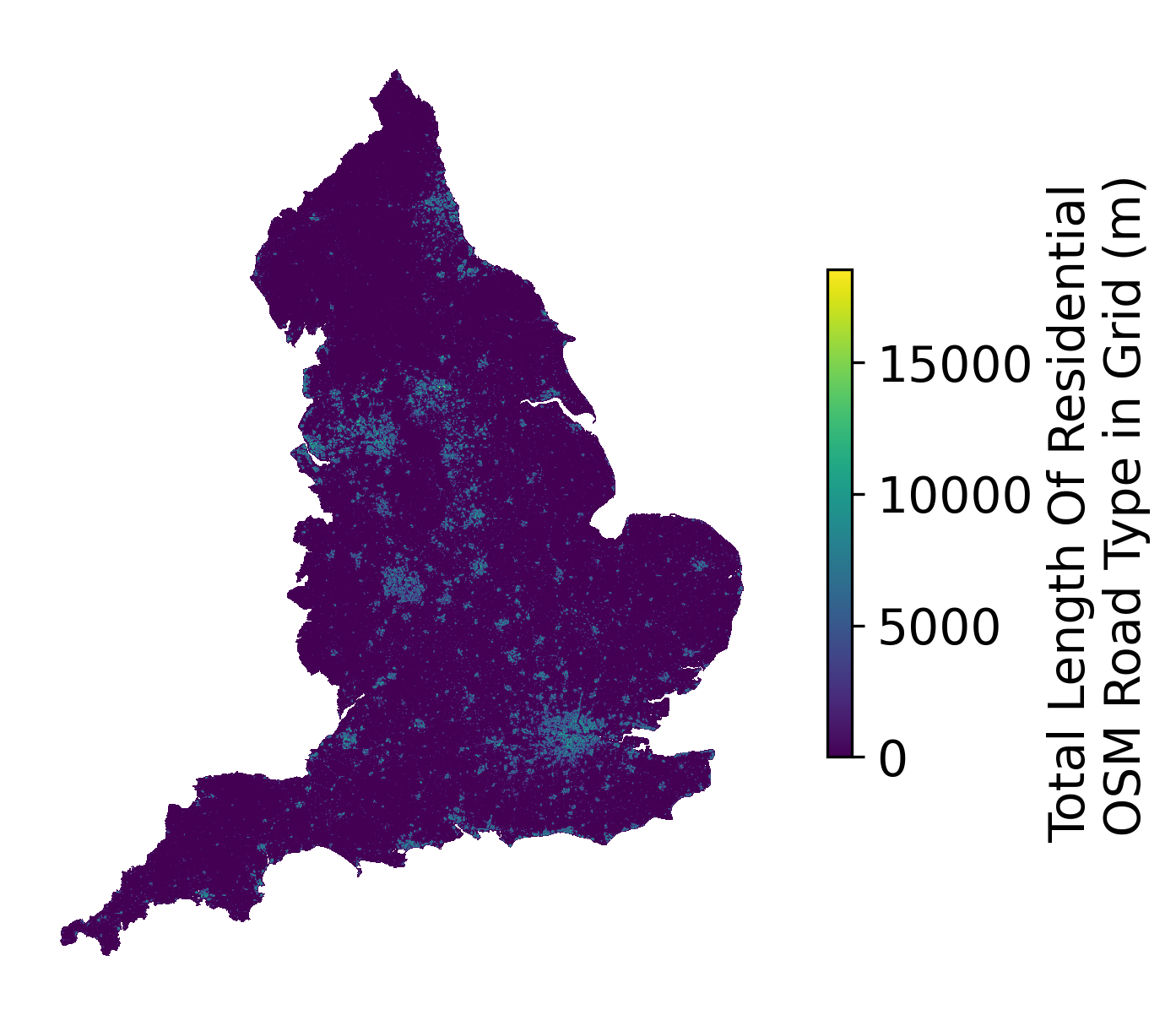}
    \caption{Total Length of Residential Road Within Grid} \label{fig:transportInfrastructureTotalResidential}
  \end{subfigure}%

\caption{{\bfseries Example complete England transport infrastructure structural properties datasets.} A feature vector element is created for the distance to the closest and the total length of the specific road type. 14 different road types are analysed, resulting in 28 feature vector elements contributed by the transport infrastructure structural properties dataset family. } \label{fig:transportInfrastructureFeatureVector}
\end{figure}

The first set of feature vectors concerning transport infrastructure structural properties detailed each grid's distance to the closest road type within the study in meters. Figure \ref{fig:transportInfrastructureDistanceMotorway} shows the feature vector for the distance to the closet motorway in 2018. The second set of feature vectors created concerning transport infrastructure structural properties details the total length in meters of each analysed road type for every grid within the study. Figure \ref{fig:transportInfrastructureTotalResidential} shows the total residential road length in meters in all grids for 2018.

The highway types analysed for creating the feature vector for the transport infrastructure structural properties dataset family included Residential, Footway, Service, Primary, Path, Cycleway, Tertiary, Secondary, Unclassified, Trunk, Track, Motorway, Pedestrian and Living Street. Figure \ref{fig:transportInfrastructureFeatureVector} provide an example of the full transport infrastructural properties dataset for the distance to the closet motorway and the total length of residential road for each grid.

\clearpage
\subsection{Transport Infrastructure Use}
\label{sec:transportInfrastructureUseData}

Traffic counts from point locations across England were used to estimate the daily traffic flow across given types of roads across different regions within England. The traffic counts used were part of the Department of Transports (DfT) Road usage data included in the annual average daily flow (AADF), major and minor roads dataset \cite{DepartmentOfTransport:RoadTrafficFlowData:2023}. The AADF dataset estimates a range of different transport methods, with the following aggregate transport types being used in the study:

\begin{enumerate}
    \item Bicycle Count 
    \item Car and Taxi Count 
    \item Bus and Coach Score
    \item Light Goods Van (LGV)
    \item Heavy Goods Vechile (HGV)
\end{enumerate}

The first step was to create a mean traffic flow per road type. The DfT AADF dataset gives traffic flow estimates on major and minor roads, from motorways to rural areas, including single-lane roads with passing bays \cite{DepartmentOfTransport:RoadTrafficFlowMetaData:2023}. In the transport infrastructural properties dataset family, we included 14 road types, with road types such as cycleways. As the AADF dataset only includes road types suitable for motor vehicles, there was a need to reduce the road types to only those related to the major and minor roads defined by DfT. Therefore, from the OpenStreetMap dataset, we included only the ten road types: motorway, trunk, primary, secondary, tertiary, unclassified, residential, living\_street, service, and track. We then matched the sample location from the AADF to the closest OpenStreetMaps road type to calculate a mean for the daily traffic flow for that road.

\begin{figure}[!htb]
\begin{center}
\includegraphics[width=0.75\textwidth]{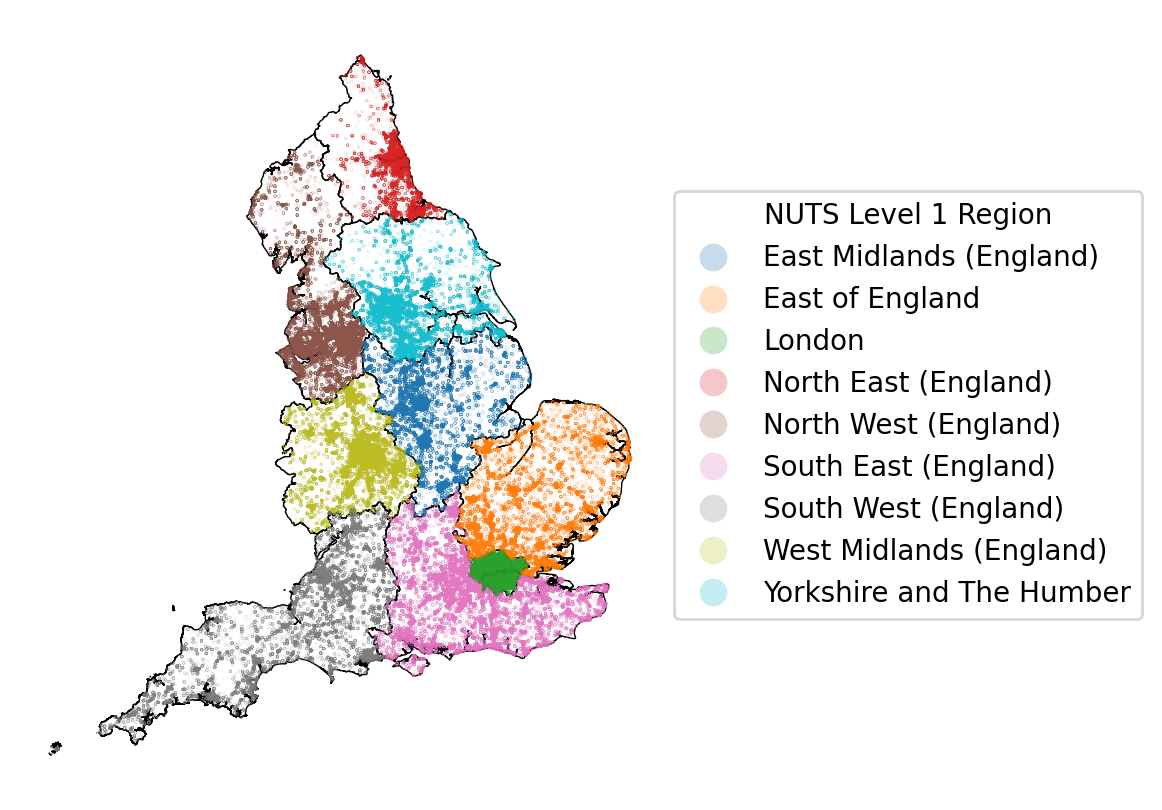}
\caption{{\bfseries Department for Transport point locations for Annual Average Daily Flow (AADF) of traffic.} Shown are the point sample locations for traffic counts along roads within England that the Department of Transport conducted, giving an annual average daily flow for various transport types. The points are slightly transparent to allow insight into where multiple points overlap, alongside being coloured by corresponding NUTS 1 region, indicating groups of point samples used to calculate an average flow for that region.}
\label{fig:DfTSampleLocations}
\end{center}
\end{figure}

As road type usage can be substantially different across England, such as a residential road in central London and a small town in the Midlands having widely different traffic flows, we created means for each road type within a set of defined geographic regions. The geographic boundary we chose for this aggregation of sample locations was the NUTS Level 1 Regions, seen in Figure \ref{fig:DfTSampleLocations}. Smaller region sizes of Local Authority Districts, with 371 geographic regions, were also trialled; however, some of the 10 OpenStreetMaps road types had no estimates for the mean traffic flow. Therefore, we used the coarser but more comprehensive aggregation of the NUTS boundary. 

\begin{figure}[!htb]
\begin{center}
\includegraphics[width=0.5\textwidth]{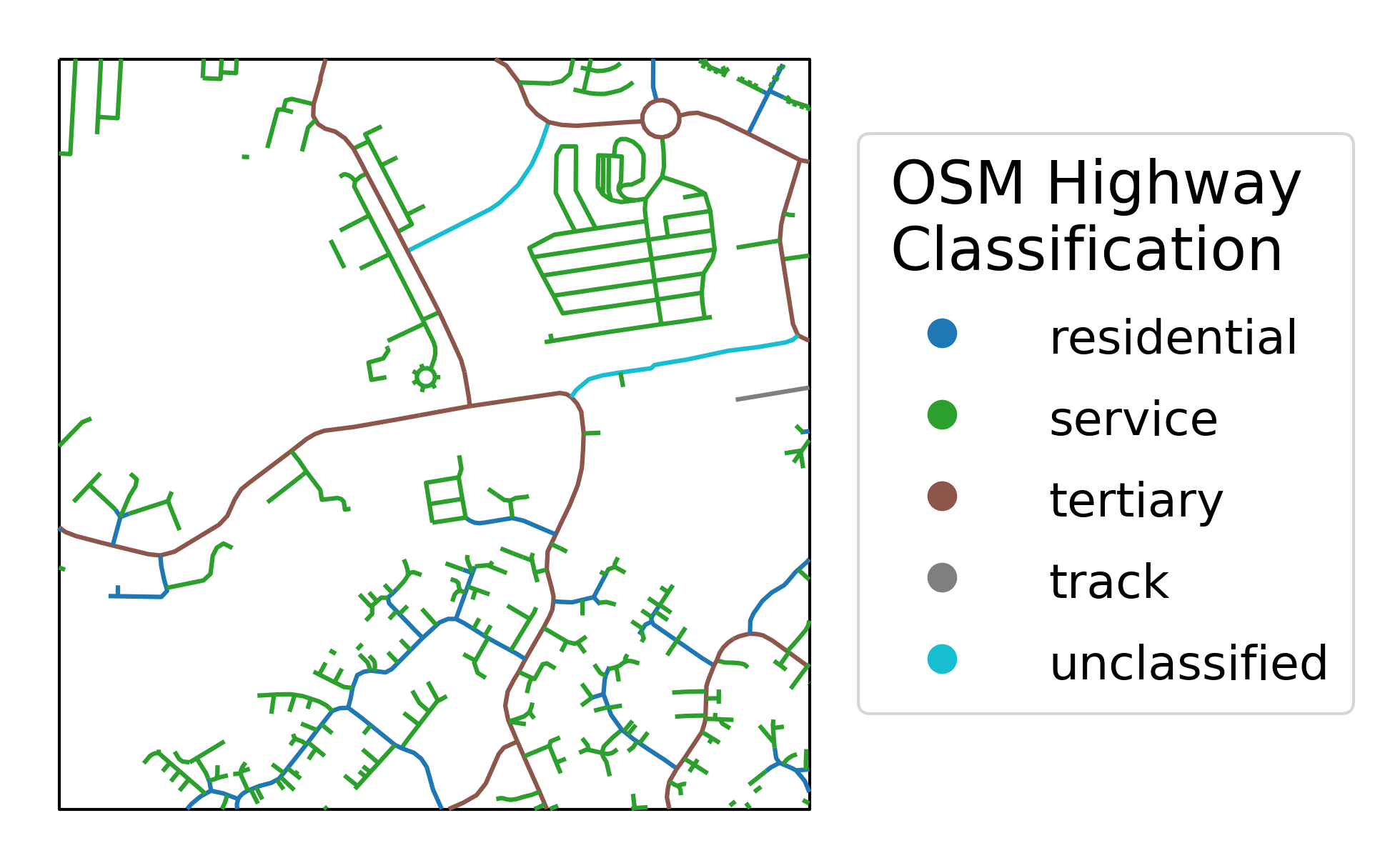}
\caption{{\bfseries Road network infrastructure for a single grid with centroid [52.218, -0.07] in South Cambridgeshire.} The sub-road networks are shown for each of the five road types within the 1km$^2$ grid. }
\label{fig:singleGridRoadInfrastructure}
\end{center}
\end{figure}

The next step was calculating the road network within each grid used within the study. Figure \ref{fig:singleGridRoadInfrastructure} shows the road infrastructure within a single 1km$^2$ grid in South Cambridgeshire at location Latitude 52.218, Longitude -0.07. We then calculated the total road length for each road type for each grid. Table \ref{tab:osmDFTTrafficCountRoadMeans} shows the total road length for each road type for the grid shown in Figure \ref{fig:singleGridRoadInfrastructure} alongside the number of traffic counts within the DfT dataset for that road type within that NUTS 1 region, in this case, the East of England region. Each road type's mean traffic flow per transport method was multiplied by the overall length of that road type to estimate the traffic flow for that transport method across that road type within the 1km$^2$ grid. Each of the road types multiplication was then summed to provide an overall estimate for the traffic of each transport method across the whole road network within that grid, with Table \ref{tab:signleGridOVerallTrafficScore} showing the overall traffic score for each transport method for the grid shown in Figure \ref{fig:singleGridRoadInfrastructure}. 

\begin{figure}[!htb]
  \hspace*{\fill}   % maximize separation between the subfigures
  \begin{subfigure}{0.9\textwidth}
    \includegraphics[width=\linewidth]{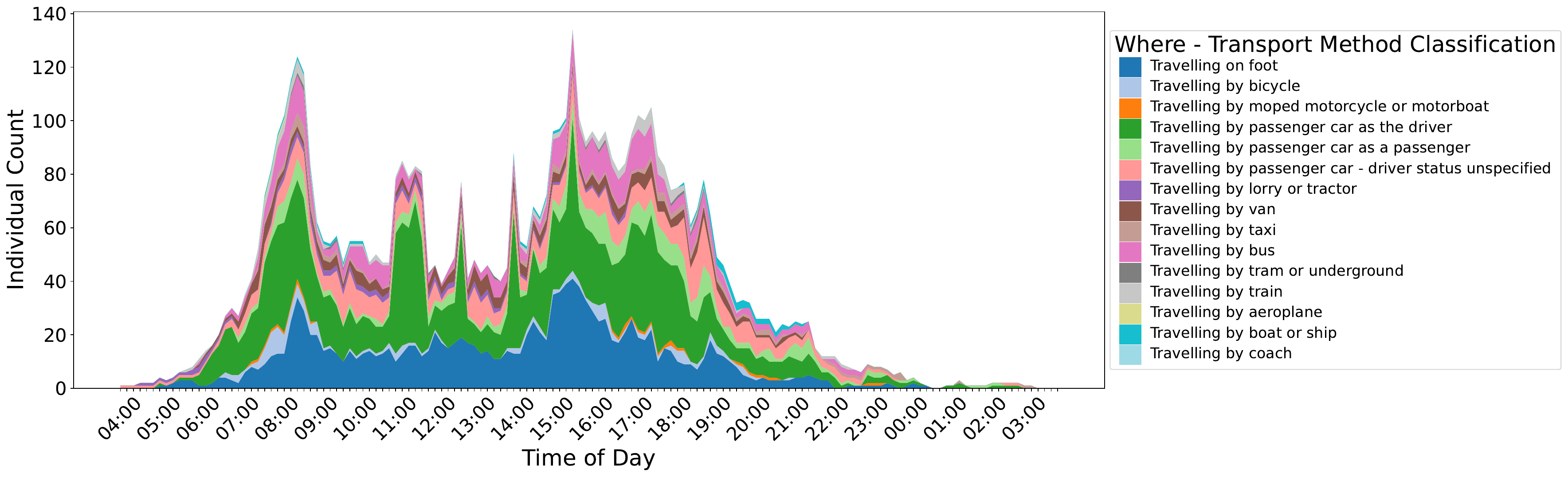}
    \caption{Weekday}\label{fig:stackplotForTravellingMethodWeekday}
  \end{subfigure}%
  \hspace*{\fill}   % maximize separation between the subfigures
  \\
  \hspace*{\fill}   % maximize separation between the subfigures
  \begin{subfigure}{0.9\textwidth}
    \includegraphics[width=\linewidth]{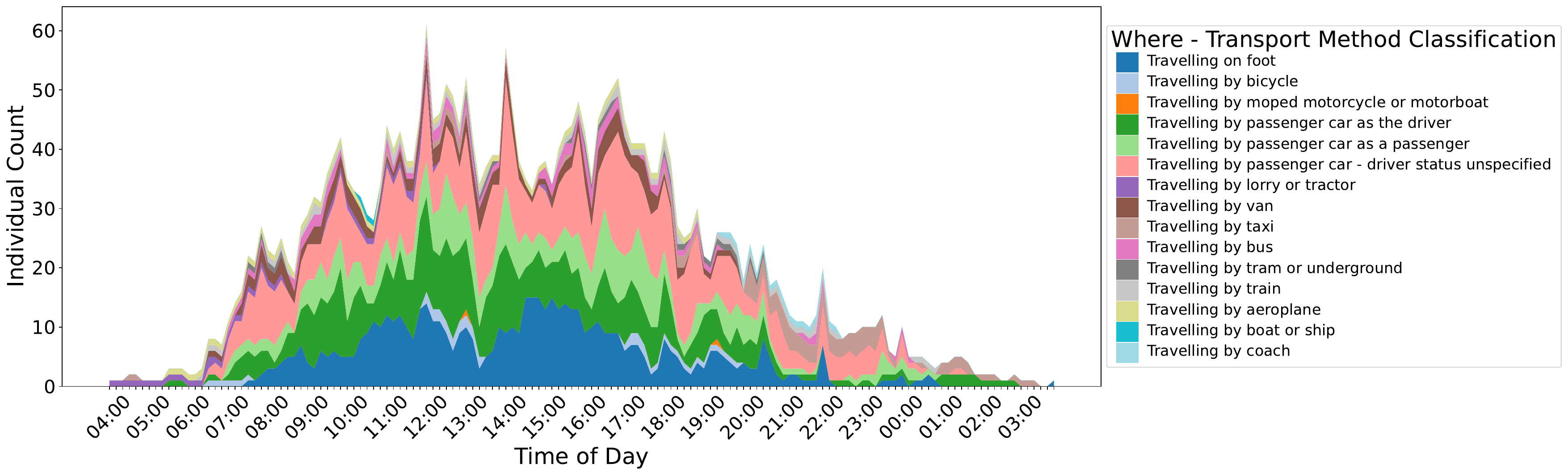}
    \caption{Saturday} \label{fig:stackplotForTravellingMethodSaturday}
  \end{subfigure}%
  \hspace*{\fill}   % maximize separation between the subfigures
  \\
  \hspace*{\fill}   % maximize separation between the subfigures
  \begin{subfigure}{0.9\textwidth}
    \includegraphics[width=\linewidth]{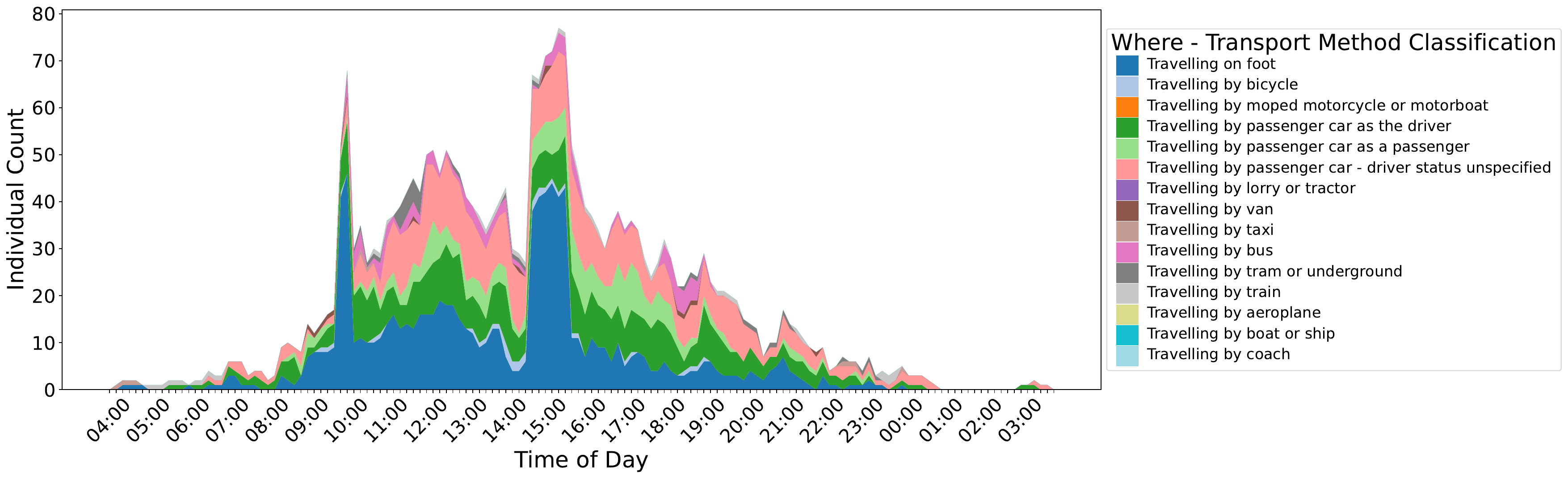}
    \caption{Sunday} \label{fig:stackplotForTravellingMethodSunday}
  \end{subfigure}%
  \hspace*{\fill}   % maximize separation between the subfigures

\caption{{\bfseries Spatial microsimulation for South Cambridgeshire 020 shows the proportion of different travel methods for all individuals travelling at the given time.} Of note are the different scales of the plots—total number of individuals taking a journey Weekday: 32622, Saturday: 16850, Sunday: 13857. Interestingly, even though Saturday has a higher overall number of individuals travelling, the peak on Sunday is higher.} \label{fig:spatialMicrosimulationTransportSouthCambridgeshire020MSOA}
\end{figure}

The traffic flow score in Table \ref{tab:signleGridOVerallTrafficScore} gives an estimate of traffic flow at the daily level. However, an estimate of traffic flow based on an hourly level was desired for the analysis of rush hour traffic. To achieve this, we temporally distributed the daily traffic flow based on a spatial microsimulation of the UK Time Use Survey \cite{gershuny:2017:UKTimeUseSurvey}. The UK Time Use Survey provides data on how 11421 individuals spent their time across the UK during weekdays, Saturdays and Sundays. One of the options for how they could specify how they were spending their time was for travelling, which included details of the transport method they were using. Profiling of travel habits was made possible as each participant has associated socio-demographic data. Using the UK census \cite{ons:2012:census}, we used a spatial microsimulation \cite{lovelace:2017:spatialMicrosimulation} to create a synthetic population of England. The input UK census data was the 7201 Middle Layer Super Output Area (MSOA) aggregate socio-demographic statistics. The spatial microsimulation provided a synthetic population for each MSOA that included data on when they would travel based on the UK Time Use Survey. We then created aggregate travel times for each MSOA region. Figure \ref{fig:spatialMicrosimulationTransportSouthCambridgeshire020MSOA} shows the travel profile for MSOA South Cambridgeshire 020, the MSOA that allowed an understanding of what times of day for a weekday, Saturday and Sunday individuals within a given MSOA travel by transport method. The travel profile seen in Figure \ref{fig:spatialMicrosimulationTransportSouthCambridgeshire020MSOA} was then used to temporally distribute the daily traffic flow score for each grid within that MSOA. The grid shown in Figure \ref{fig:singleGridRoadInfrastructure} is within the NUTS region East of England and MSOA South Cambridgeshire 020. As such, the travel profiles in Figure \ref{fig:spatialMicrosimulationTransportSouthCambridgeshire020MSOA} were used to distribute the daily traffic flows in Table \ref{tab:signleGridOVerallTrafficScore} temporally to produce the hourly traffic estimates.

\begin{figure}[!htb]
  \begin{subfigure}{0.32\textwidth}
    \includegraphics[width=\linewidth]{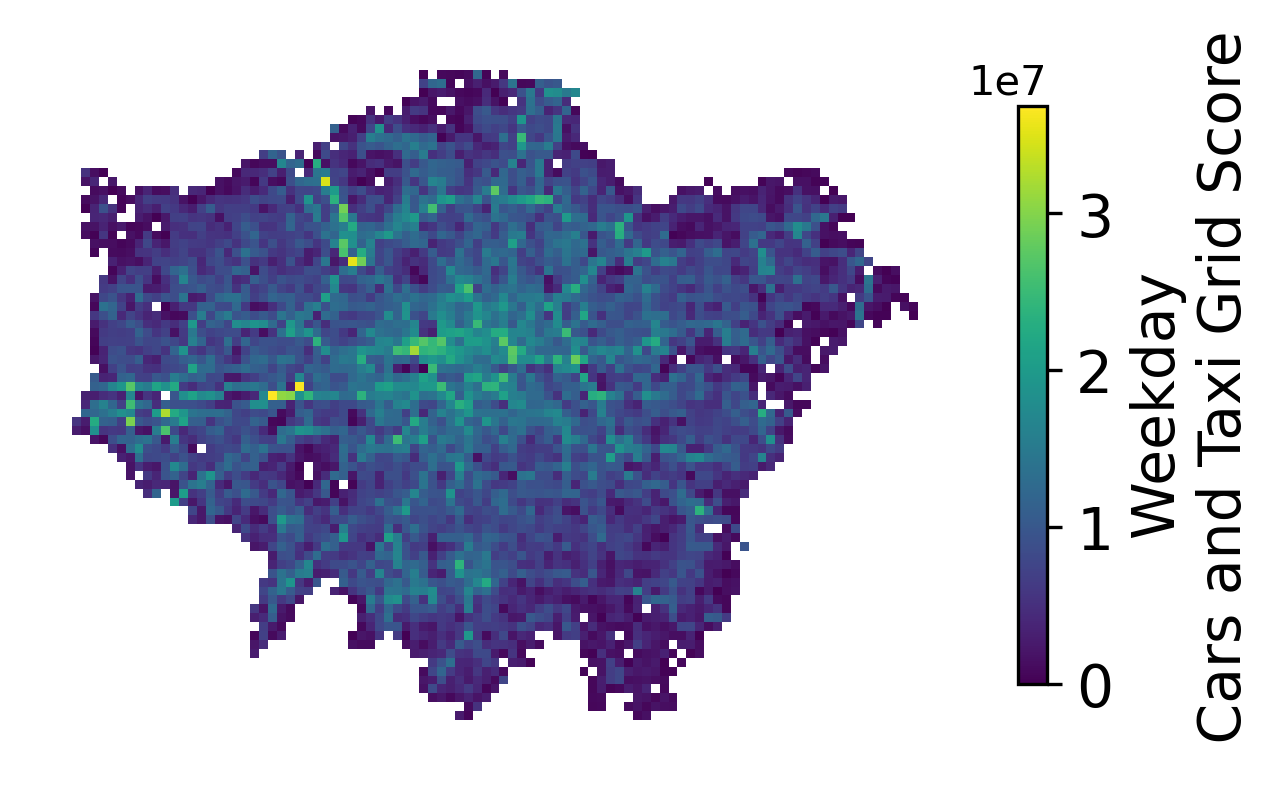}
    %\caption{Weekday}\label{fig:TemporalDistributionLondonWeekday}
  \end{subfigure}%
  \hspace*{\fill}   % maximize separation between the subfigures
  \begin{subfigure}{0.32\textwidth}
    \includegraphics[width=\linewidth]{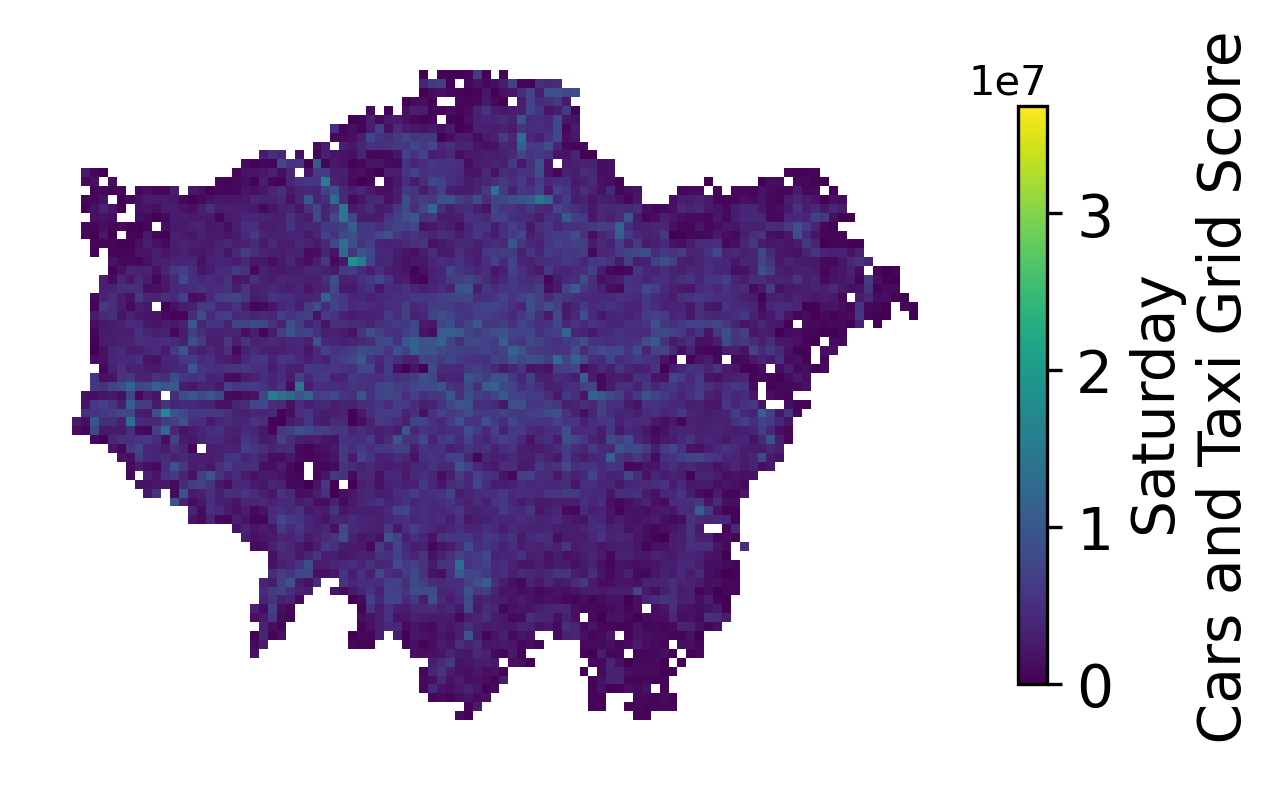}
    %\caption{Saturday} \label{fig:TemporalDistributionLondonSaturday}
  \end{subfigure}%
  \hspace*{\fill}   % maximize separation between the subfigures
  \begin{subfigure}{0.32\textwidth}
    \includegraphics[width=\linewidth]{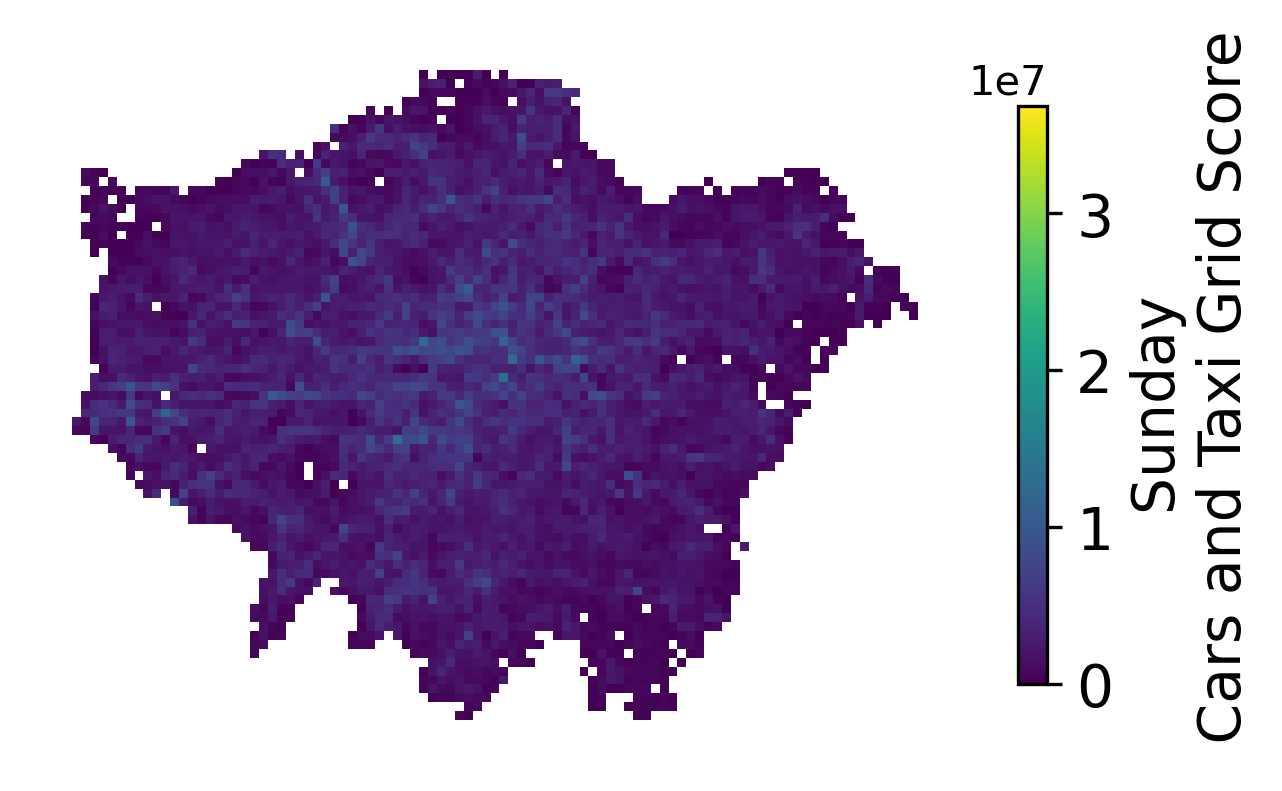}
    %\caption{Sunday} \label{fig:TemporalDistributionLondonSunday}
  \end{subfigure}%

\caption{{\bfseries Temporal distribution of traffic grid score within london at 08AM on 1st, 2nd and 3rd June 2018 for cars and taxis.} } \label{fig:TemporalDistributionLondon}
\end{figure}

\begin{figure}[!htb]
  \hspace*{\fill}   % maximize separation between the subfigures
  \begin{subfigure}{0.45\textwidth}
    \includegraphics[width=\linewidth]{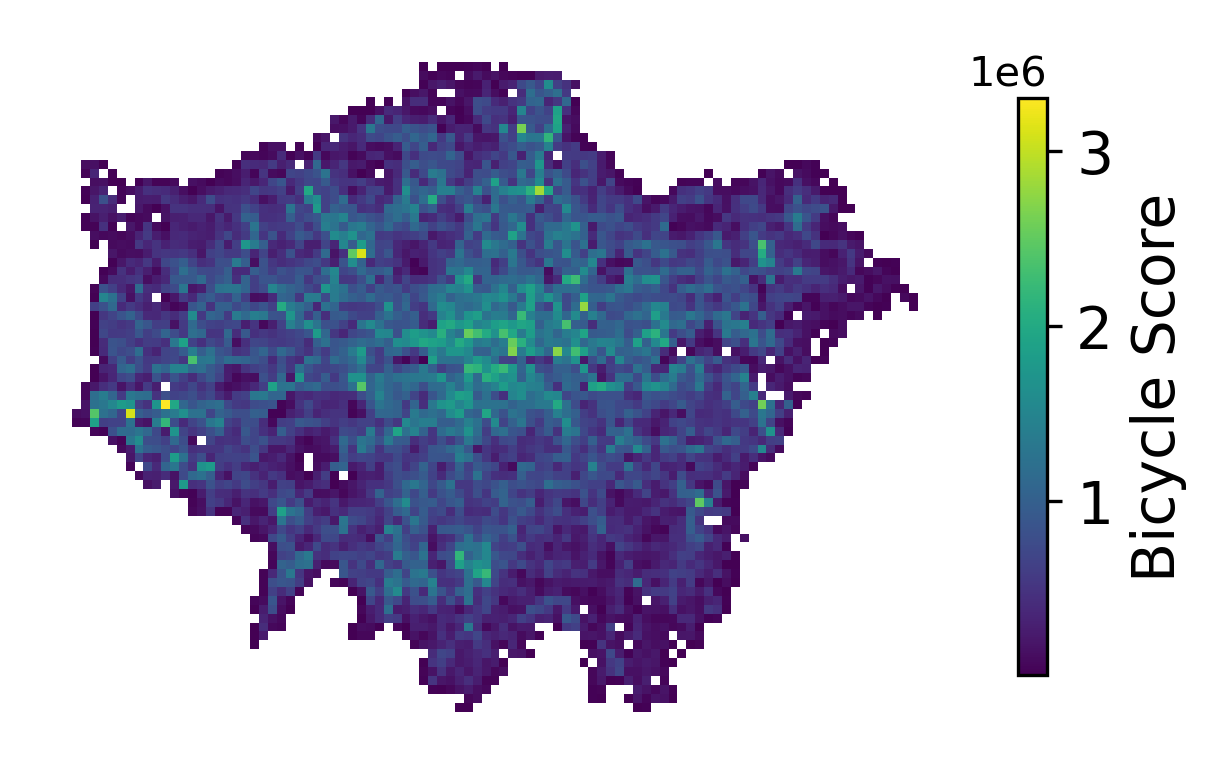}
    %\caption{Bicycle Score}\label{fig:transportUseDatasetBicycleScoreLondonSubset}
  \end{subfigure}%
  \hspace*{\fill}   % maximize separation between the subfigures
  \begin{subfigure}{0.45\textwidth}
    \includegraphics[width=\linewidth]{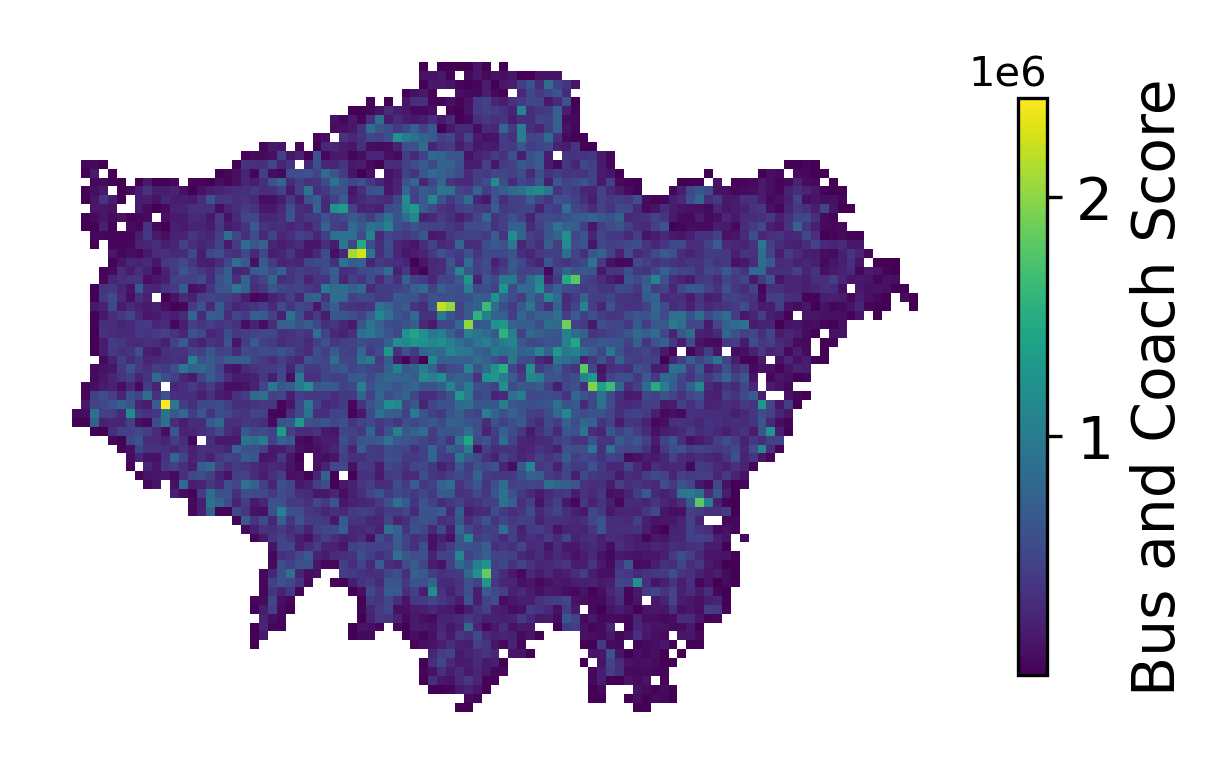}
    %\caption{Bus and Coach Score} \label{fig:transportUseDatasetBusAndCoachScoreLondonSubset}
  \end{subfigure}%
  \hspace*{\fill}   % maximize separation between the subfigures
  \\
  \hspace*{\fill}   % maximize separation between the subfigures
  \begin{subfigure}{0.45\textwidth}
    \includegraphics[width=\linewidth]{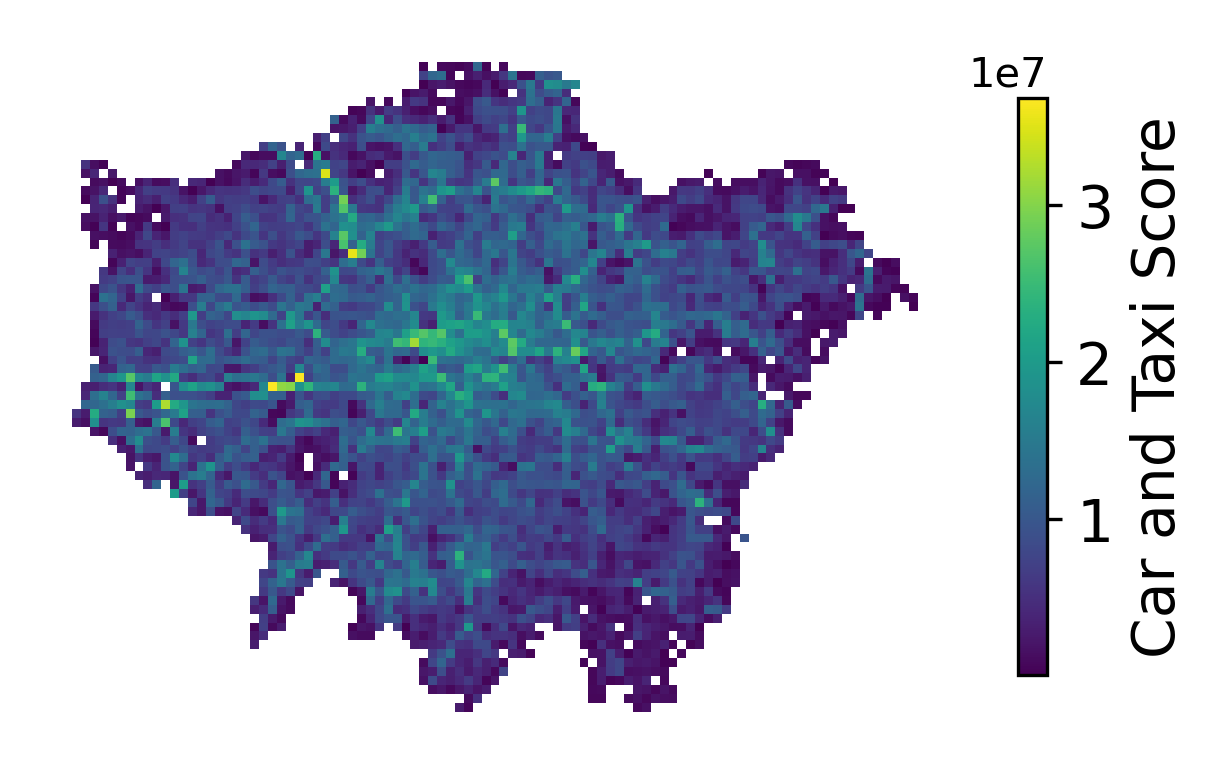}
    %\caption{Car and Taxi Score} \label{fig:transportUseDatasetCarAndTaxiScoreLondonSubset}
  \end{subfigure}%
  \hspace*{\fill}   % maximize separation between the subfigures
  \begin{subfigure}{0.45\textwidth}
    \includegraphics[width=\linewidth]{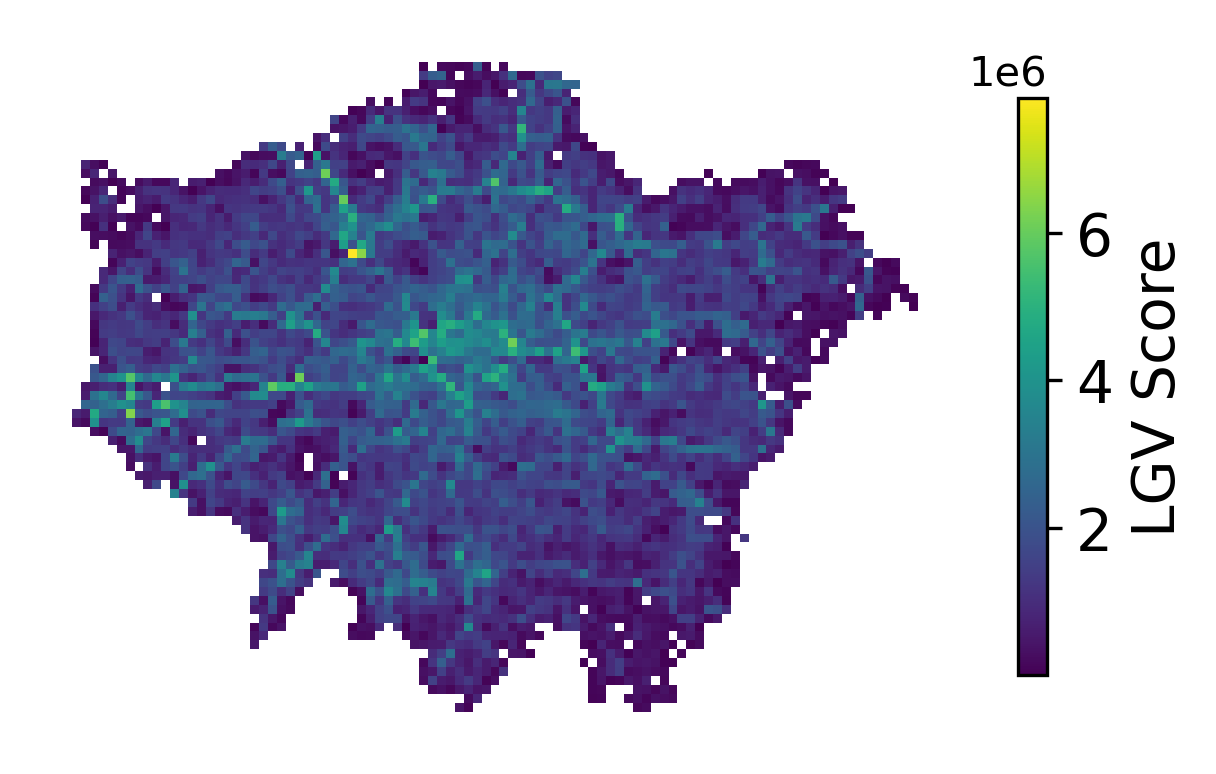}
    %\caption{Light Goods Vechile (LGV) Score} \label{fig:transportUseDatasetLGVScoreLondonSubset}
  \end{subfigure}%
  \hspace*{\fill}   % maximize separation between the subfigures
  \\
  \hspace*{\fill}   % maximize separation between the subfigures
  \begin{subfigure}{0.45\textwidth}
    \includegraphics[width=\linewidth]{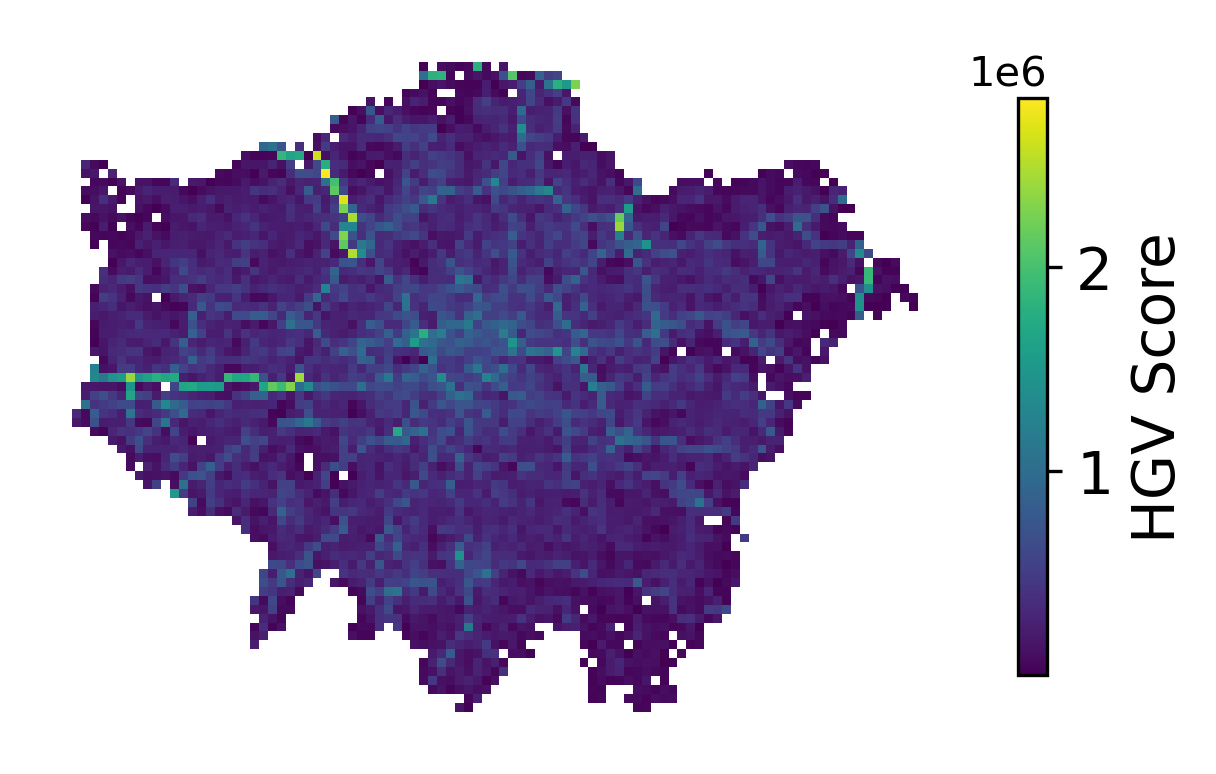}
    %\caption{Heavy Goods Vechile (HGV) Score} \label{fig:transportUseDatasetHGVScoreLondonSubset}
  \end{subfigure}%
  \hspace*{\fill}   % maximize separation between the subfigures

\caption{{\bfseries Transport use dataset for central London.} The transport use grids across central London help to highlight the difference in road usage across different road types across the five transport methods. Bicycle usage is more substantial in central London, with cars and taxis pervasive throughout and HGVs using the arterial main roads coming into the city.} \label{fig:transportUseDatasetLondonSubset}
\end{figure}

\begin{figure}
  \hspace*{\fill}   % maximize separation between the subfigures
  \begin{subfigure}{0.9\textwidth}
    \includegraphics[width=\linewidth]{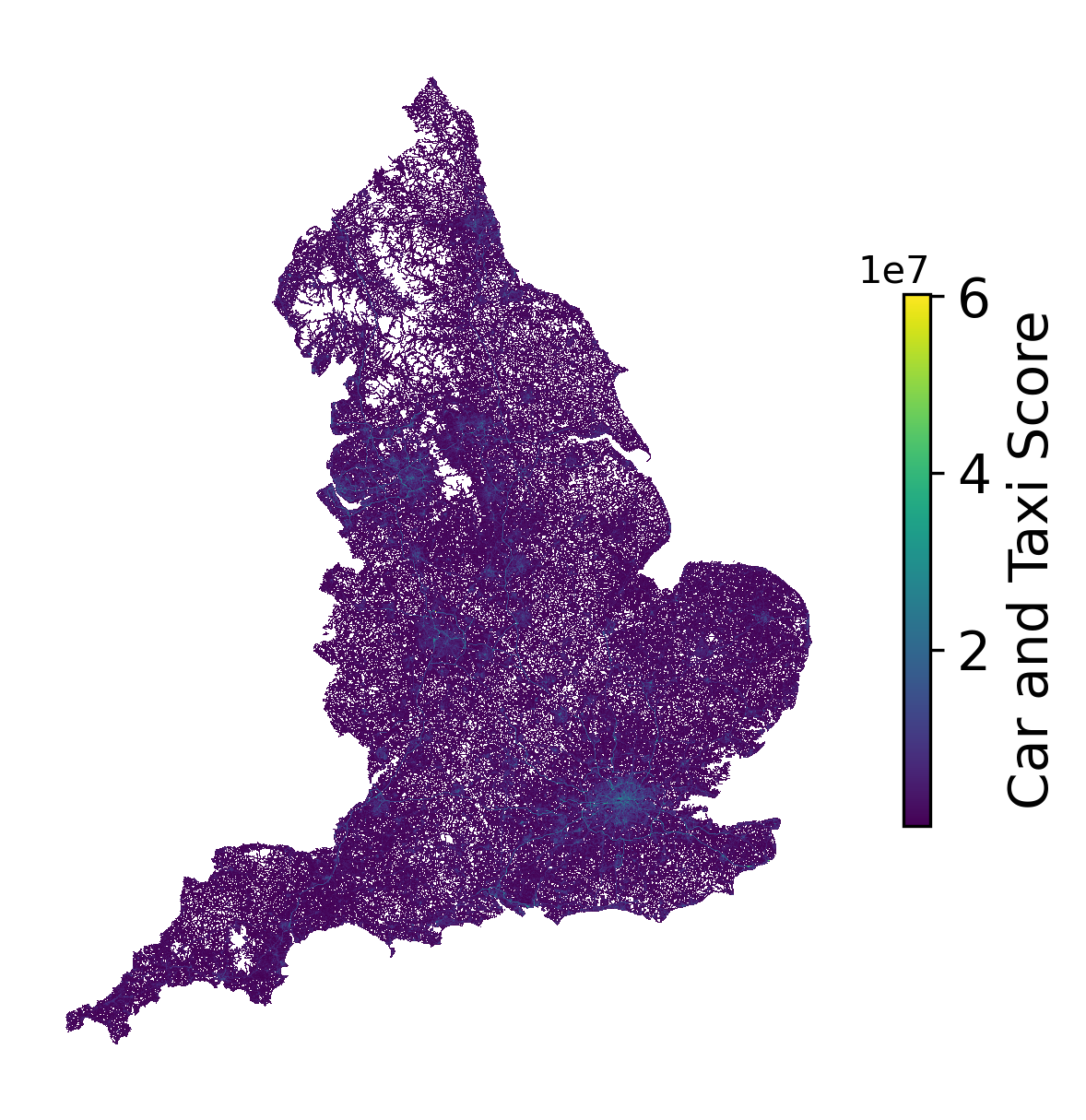}
    %\caption{Car and Taxi Score} \label{fig:transportUseDatasetCarAndTaxiScore}
  \end{subfigure}%
  \hspace*{\fill}   % maximize separation between the subfigures
\caption{{\bfseries Example complete England transport use dataset for Car and Taxi Score.} Of note is that not every grid is present within the figures, as not every 1km$^2$ grid within the study has any road infrastructure. The final feature vector for estimating the air pollution concentration is filled with zeros as those grids have no traffic. } \label{fig:transportUseDatasetAll}
\end{figure}

Figure \ref{fig:TemporalDistributionLondon} shows the difference in traffic score for grids within a London subset. As seen in the figure, each day exhibits a unique signal, with the weekday seeing the highest travel, followed by Saturday and finally Sunday. Figure \ref{fig:transportUseDatasetLondonSubset} shows how the traffic flow within a grid for each transport method differs depending on the road types present within the grid. The figure depicts the most significant traffic flow for HGVs on the arterial lines into London and the M25 ring road around central London. In contrast, bicycle use is most prominent in central London. Figure \ref{fig:transportUseDatasetAll} shows the full transport use dataset across all of the land grids, with the white land grids representing grids within the study area with no roads present. Hence, no traffic flow estimate has been made.

\begin{table}[!htb]
\resizebox{\linewidth}{!}{
\begin{filecontents*}{CSVFiles/Data/TrafficWithinGridPerHighwayType.csv}
Total Highway Type Road Length (m),OSM Highway Classification,Traffic Count Counts,Mean Pedal Cycles Traffic Per Meter,Mean Cars and Taxis Traffic Per Meter,Mean Bus and Coaches Traffic Per Meter,Mean LGV Traffic Per Meter,Mean HGV Traffic Per Meter,Pedal Cycles Traffic Total Traffic For Grid,Cars and Taxis Traffic Total Traffic For Grid,Bus and Coaches Traffic Total Traffic For Grid,LGV Traffic Total Traffic For Grid,HGV Traffic Total Traffic For Grid
2238.144897346576,residential,448,73.22544642857143,3954.8616071428573,38.870535714285715,659.7901785714286,129.26339285714286,163889.15928003218,8851553.325738665,86997.8911660564,1476706.0214890288,289310.2031369201
10138.152595112364,service,180,127.18333333333334,13983.855555555556,102.4888888888889,2701.8,1033.8555555555556,1289404.0408883742,141770461.49023202,1039047.9948590717,27391260.681474585,10481385.383526891
3220.237748555352,tertiary,366,60.54644808743169,4356.0,33.112021857923494,769.7213114754098,224.6120218579235,194973.95767209452,14027355.632707113,106628.58271787516,2478685.6230806466,723304.1115662251
99.49279473411143,track,58,15.810344827586206,14115.810344827587,64.22413793103448,3318.344827586207,2036.2068965517242,1573.0153926065548,1404421.4211435777,6389.838972147673,330151.4007880349,202587.91479480275
587.6030048249675,unclassified,305,25.619672131147542,3604.072131147541,18.934426229508198,742.6131147540983,342.6,15054.196326892774,2117763.6138682193,11125.925747095696,436361.69765193656,201312.78945303385
\end{filecontents*}
\pgfplotstabletypeset[
    multicolumn names=l, 
    col sep=comma, 
    string type, 
    header = has colnames,
    columns={Total Highway Type Road Length (m), OSM Highway Classification, Traffic Count Counts, Mean Pedal Cycles Traffic Per Meter, Mean Cars and Taxis Traffic Per Meter, Mean Bus and Coaches Traffic Per Meter, Mean LGV Traffic Per Meter, Mean HGV Traffic Per Meter},
    columns/OSM Highway Classification/.style={column type=l, column name=OSM Highway Classification},
    columns/Traffic Count Counts/.style={column type={S[round-precision=1, table-format=5.0, table-number-alignment=left]}, column name=Traffic Count Counts},
    columns/Total Highway Type Road Length (m)/.style={column type={S[round-precision=1, table-format=5.0, table-number-alignment=left]}, column name=Total Road Length (m)},
    columns/Mean Pedal Cycles Traffic Per Meter/.style={column type={S[round-precision=1, table-format=5.0, table-number-alignment=left]}, column name=Mean Pedal Cycles Traffic},
    columns/Mean Cars and Taxis Traffic Per Meter/.style={column type={S[round-precision=1, table-format=5.0, table-number-alignment=left]}, column name=Mean Cars and Taxis Traffic},
    columns/Mean Bus and Coaches Traffic Per Meter/.style={column type={S[round-precision=1, table-format=5.0, table-number-alignment=left]}, column name=Mean Bus and Coaches Traffic},
    columns/Mean LGV Traffic Per Meter/.style={column type={S[round-precision=1, table-format=5.0, table-number-alignment=left]}, column name=Mean LGV Traffic},
    columns/Mean HGV Traffic Per Meter/.style={column type={S[round-precision=1, table-format=5.0, table-number-alignment=left]}, column name=Mean HGV Traffic},
    every head row/.style={before row=\toprule, after row=\midrule},
    every last row/.style={after row=\bottomrule}
    ]{CSVFiles/Data/TrafficWithinGridPerHighwayType.csv}}
    \smallskip
\caption{{\bfseries Mean traffic count per road for East of England, with summary data road network length for Figure \ref{fig:singleGridRoadInfrastructure}} The summary data for the road network within a single grid within the East of England is shown, corresponding to the visualisation in Figure \ref{fig:singleGridRoadInfrastructure}. The table provides the data used to create the overall grid score for the grid shown in Table \ref{tab:signleGridOVerallTrafficScore}. To create the grid score, the sum of the multiplication of the length of the road type and the average traffic flow on that road type gives a single number indicating the traffic for that transport type across the road network within the grid. } \label{tab:osmDFTTrafficCountRoadMeans}
\end{table}

\begin{table}[!htb]
\centering
\resizebox{0.4\linewidth}{!}{
\begin{filecontents*}{CSVFiles/Data/TrafficWithinGridTotal.csv}
Transport Method,Grid Score
Pedal Cycles,1664894.3695600003
Cars and Taxis,168171555.4836896
Bus and Coaches,1250190.2334622468
LGV,32113165.424484234
HGV,11897900.402477873
\end{filecontents*}
\pgfplotstabletypeset[
    multicolumn names=l, 
    col sep=comma, 
    string type, 
    header = has colnames,
    columns={Transport Method, Grid Score},
    columns/Transport Method/.style={column type=l, column name=OSM Highway Classification},
    columns/Grid Score/.style={column type={S[round-precision=0, table-format=5.0, table-number-alignment=left]}, column name=Grid Score},
    every head row/.style={before row=\toprule, after row=\midrule},
    every last row/.style={after row=\bottomrule}
    ]{CSVFiles/Data/TrafficWithinGridTotal.csv}}
    \smallskip
\caption{{\bfseries Traffic flow average for road types for grid with centroid [52.218, -0.07] in South Cambridgeshire.} The overall grid score for each transport method for the grid is visualised in Figure \ref{fig:singleGridRoadInfrastructure}. The grid score provides an overall indication of the traffic across all road types within the grid.} \label{tab:signleGridOVerallTrafficScore}
\end{table}

\begin{figure}
  \hspace*{\fill}   % maximize separation between the subfigures
  \begin{subfigure}{0.4\textwidth}
    \includegraphics[width=\linewidth]{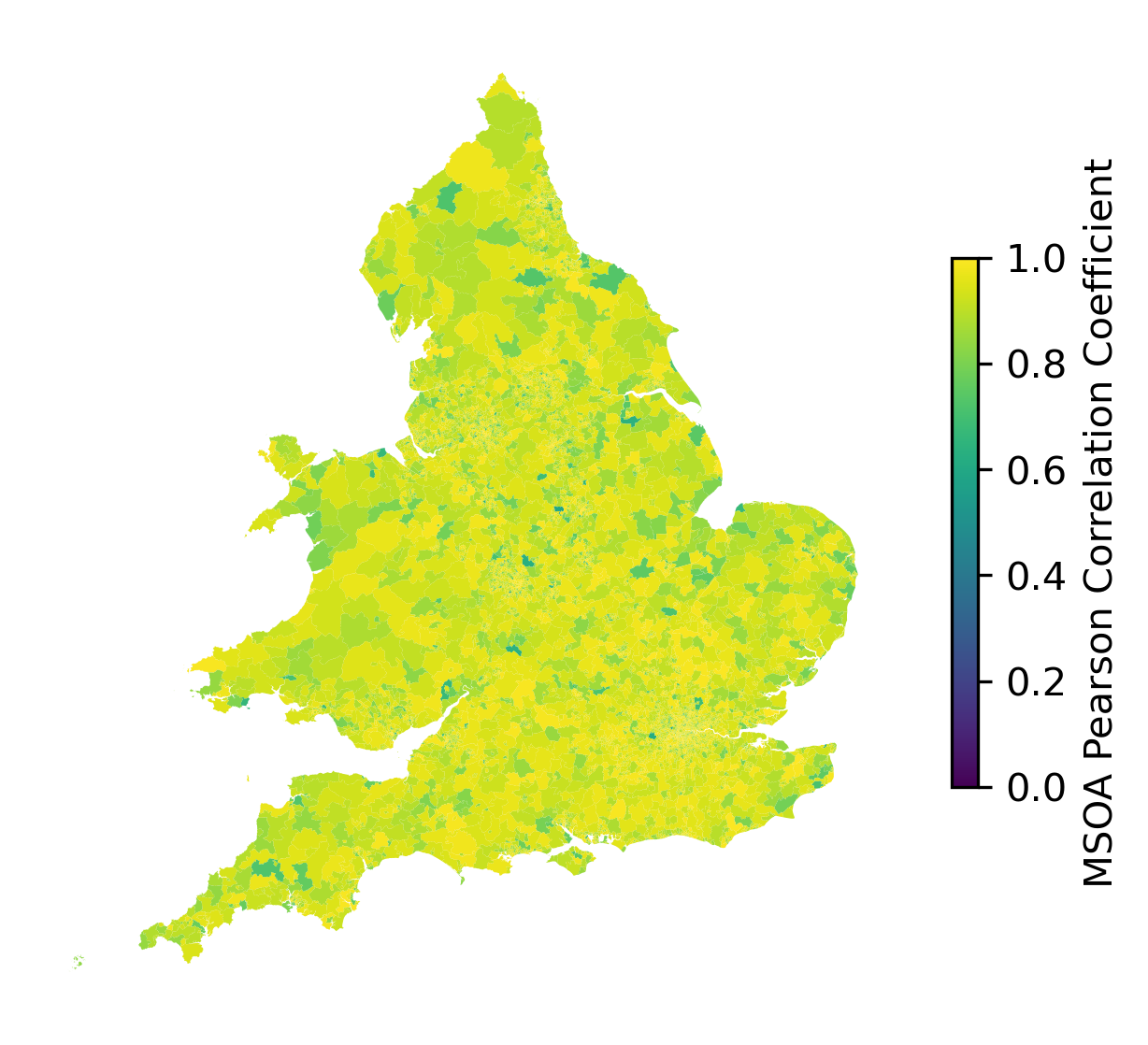}
    \caption{1 Million Individuals}\label{fig:spatialMicrosimulation1Million}
  \end{subfigure}%
  \hspace*{\fill}   % maximize separation between the subfigures
  \hspace*{\fill}   % maximize separation between the subfigures
  \begin{subfigure}{0.4\textwidth}
    \includegraphics[width=\linewidth]{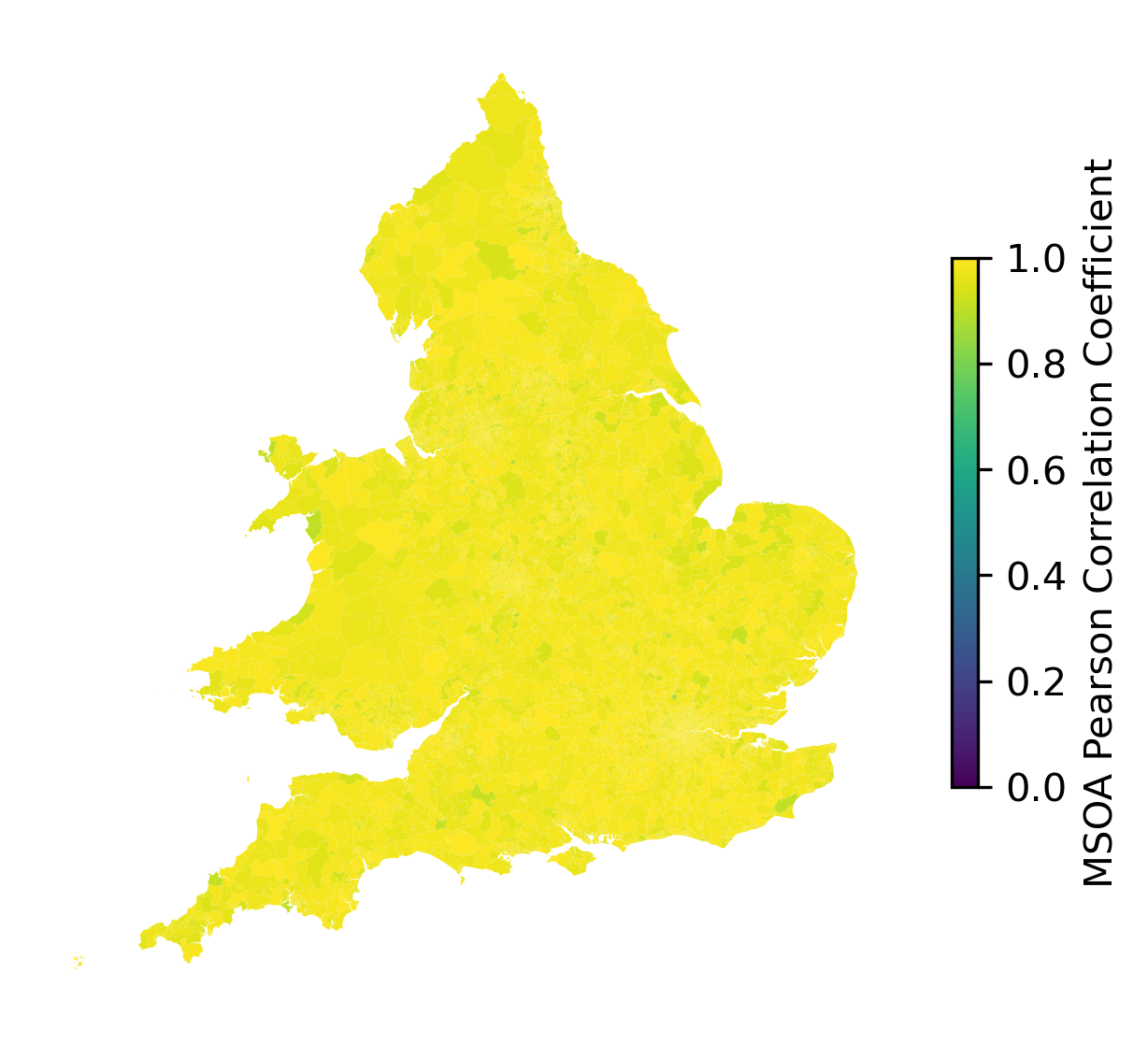}
    \caption{5 Million Individuals} \label{fig:spatialMicrosimulation5Million}
  \end{subfigure}%
  \hspace*{\fill}   % maximize separation between the subfigures
  \\
  \hspace*{\fill}   % maximize separation between the subfigures
  \begin{subfigure}{0.4\textwidth}
    \includegraphics[width=\linewidth]{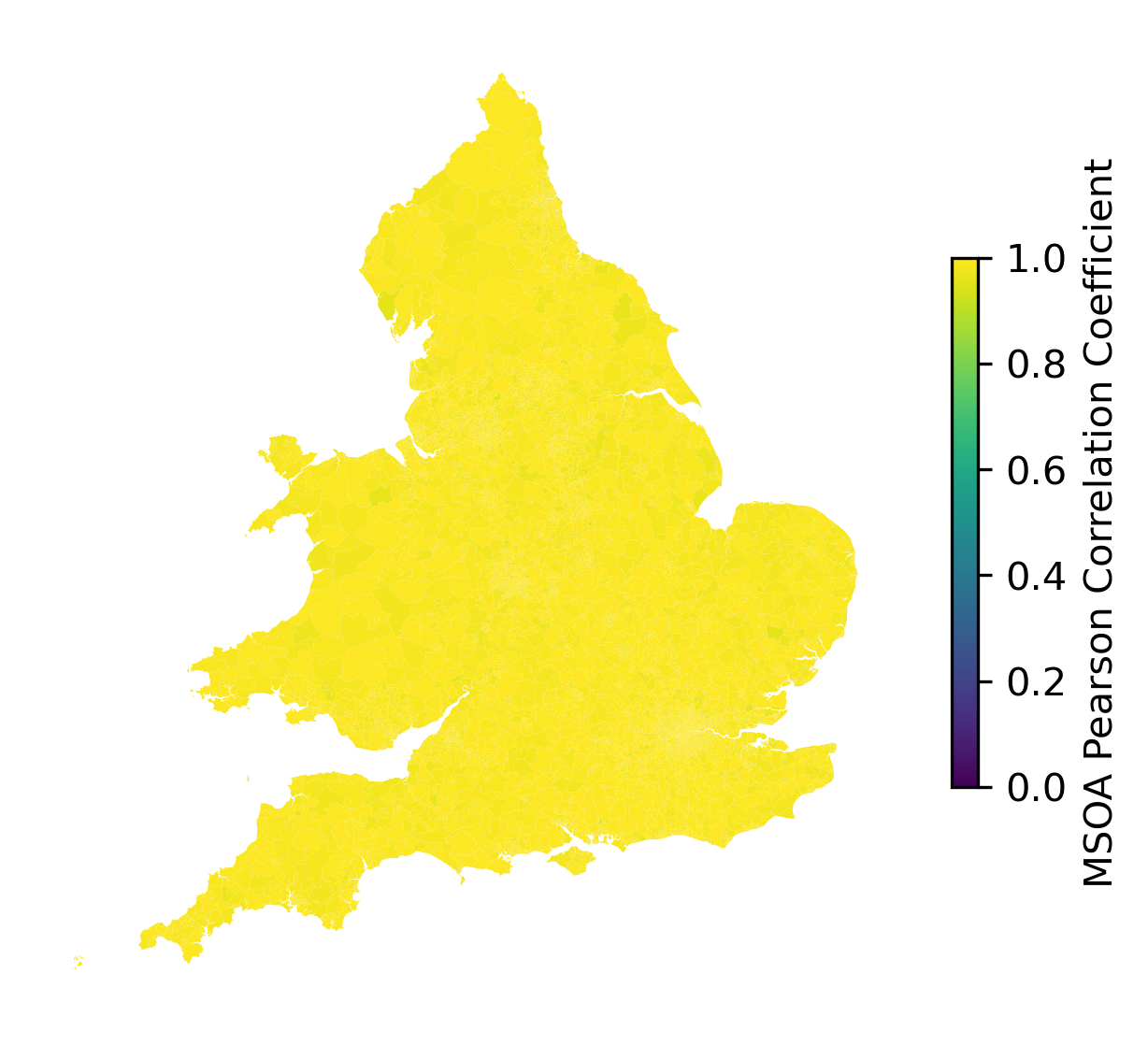}
    \caption{11 Million Individuals} \label{fig:spatialMicrosimulation11Million}
  \end{subfigure}%
  \hspace*{\fill}   % maximize separation between the subfigures
  \begin{subfigure}{0.4\textwidth}
    \includegraphics[width=\linewidth]{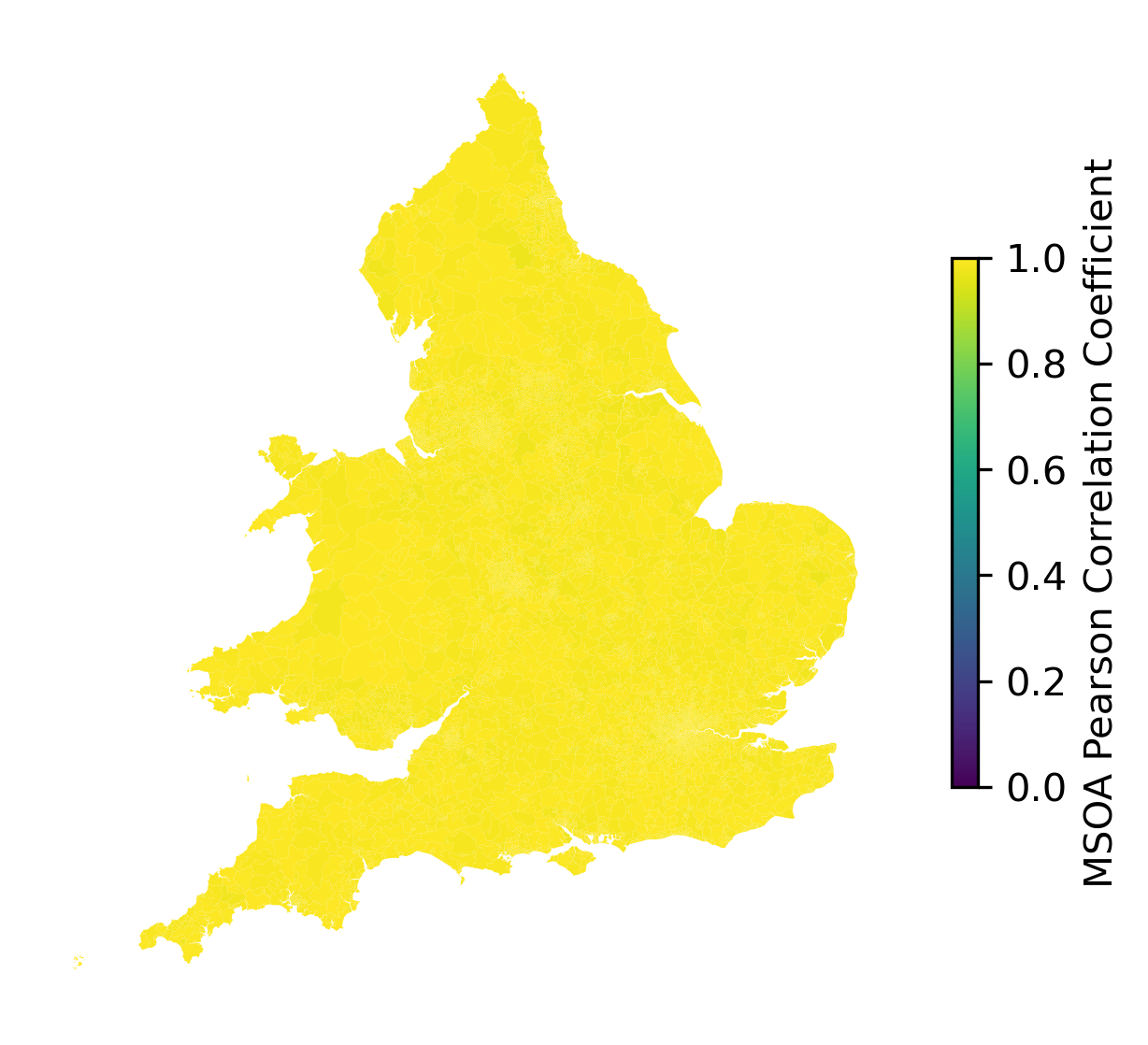}
    \caption{55 Million Individuals} \label{fig:spatialMicrosimulation55Million}
  \end{subfigure}%
  \hspace*{\fill}   % maximize separation between the subfigures

\caption{{\bfseries Pearson correlation coefficients across the MSOA regions used during the spatial microsimulation for different total numbers of simulated individuals.} During the spatial microsimulation, the number of individuals to be created was experimented with, intending to achieve a desirable Pearson Correlation coefficient across all MSOA regions ensuring model validity in population representation \cite{Dumont:2018:InternalModelValidation} while reducing the memory and computation burden associated with creating all individuals across the UK. While the simulation of all 55 million individuals in the UK resulted in a good pearson score, it was computationally expensive; as such, 11 million individuals were chosen to be simulated with the Pearson correlation being maintained at above 0.8 while significantly reducing computational costs.} \label{fig:spatialMicrosimulationCorrelation}
\end{figure}

\begin{figure}
  \hspace*{\fill}   % maximize separation between the subfigures
  \begin{subfigure}{\textwidth}
    \includegraphics[width=\linewidth]{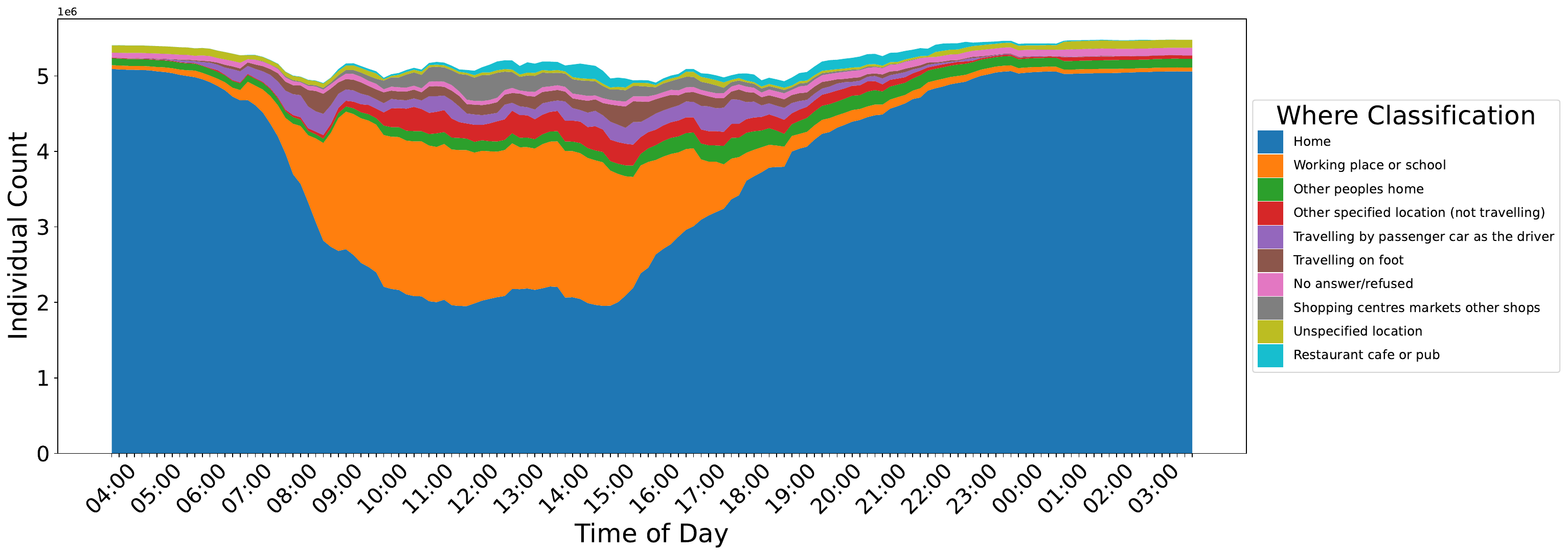}
    \caption{Weekday}\label{fig:uktimeUseSurveyWhereWeekday}
  \end{subfigure}%
  \hspace*{\fill}   % maximize separation between the subfigures
  \\
  \hspace*{\fill}   % maximize separation between the subfigures
  \begin{subfigure}{\textwidth}
    \includegraphics[width=\linewidth]{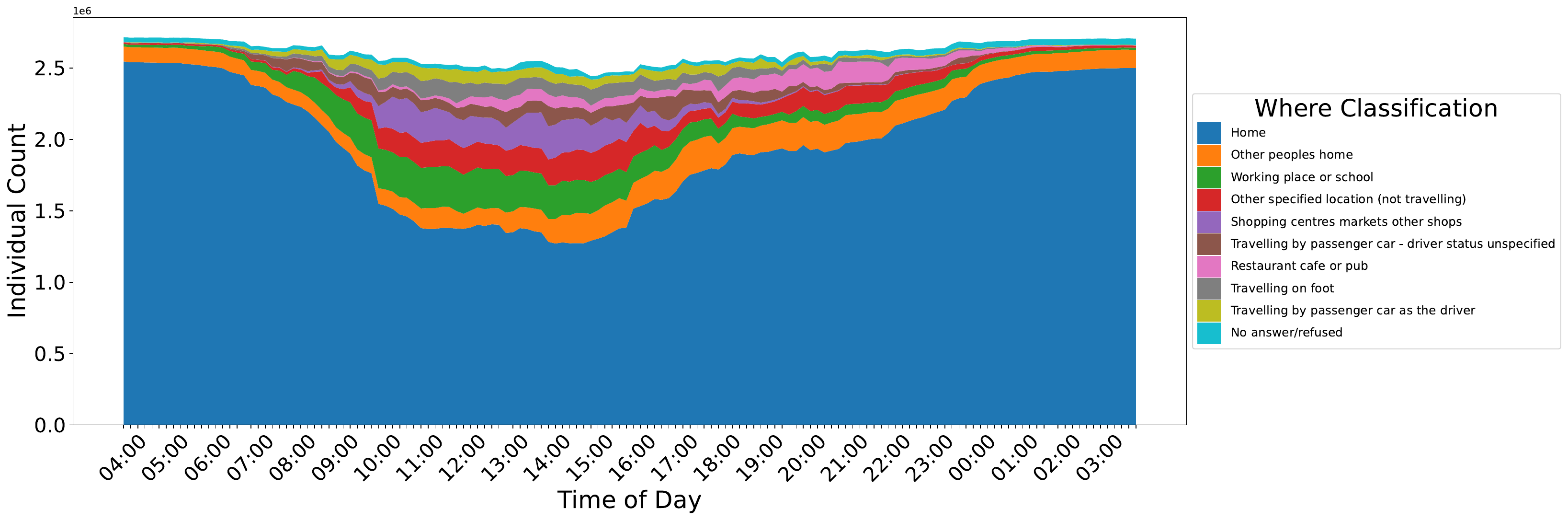}
    \caption{Saturday} \label{fig:uktimeUseSurveyWhereSaturday}
  \end{subfigure}%
  \hspace*{\fill}   % maximize separation between the subfigures
  \\
  \hspace*{\fill}   % maximize separation between the subfigures
  \begin{subfigure}{\textwidth}
    \includegraphics[width=\linewidth]{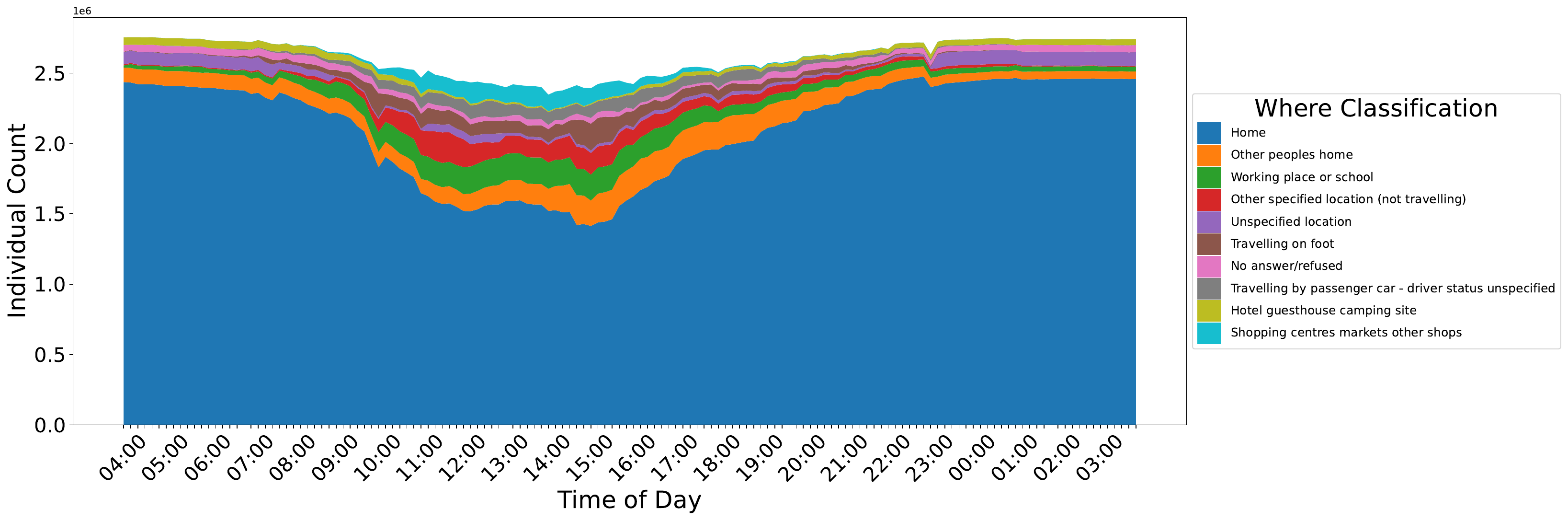}
    \caption{Sunday} \label{fig:uktimeUseSurveyWhereSunday}
  \end{subfigure}%
  \hspace*{\fill}   % maximize separation between the subfigures

\caption{{\bfseries UK Time Use stack plots showing where individuals are across all the major location categories included in the dataset. }} \label{fig:uktimeUseSurveyWhere}
\end{figure}

\clearpage
\subsection{Meteorological}
\label{sec:DataDetails:metrological}

\begin{figure}[!htb]
\begin{center}
\includegraphics[width=0.7\textwidth]{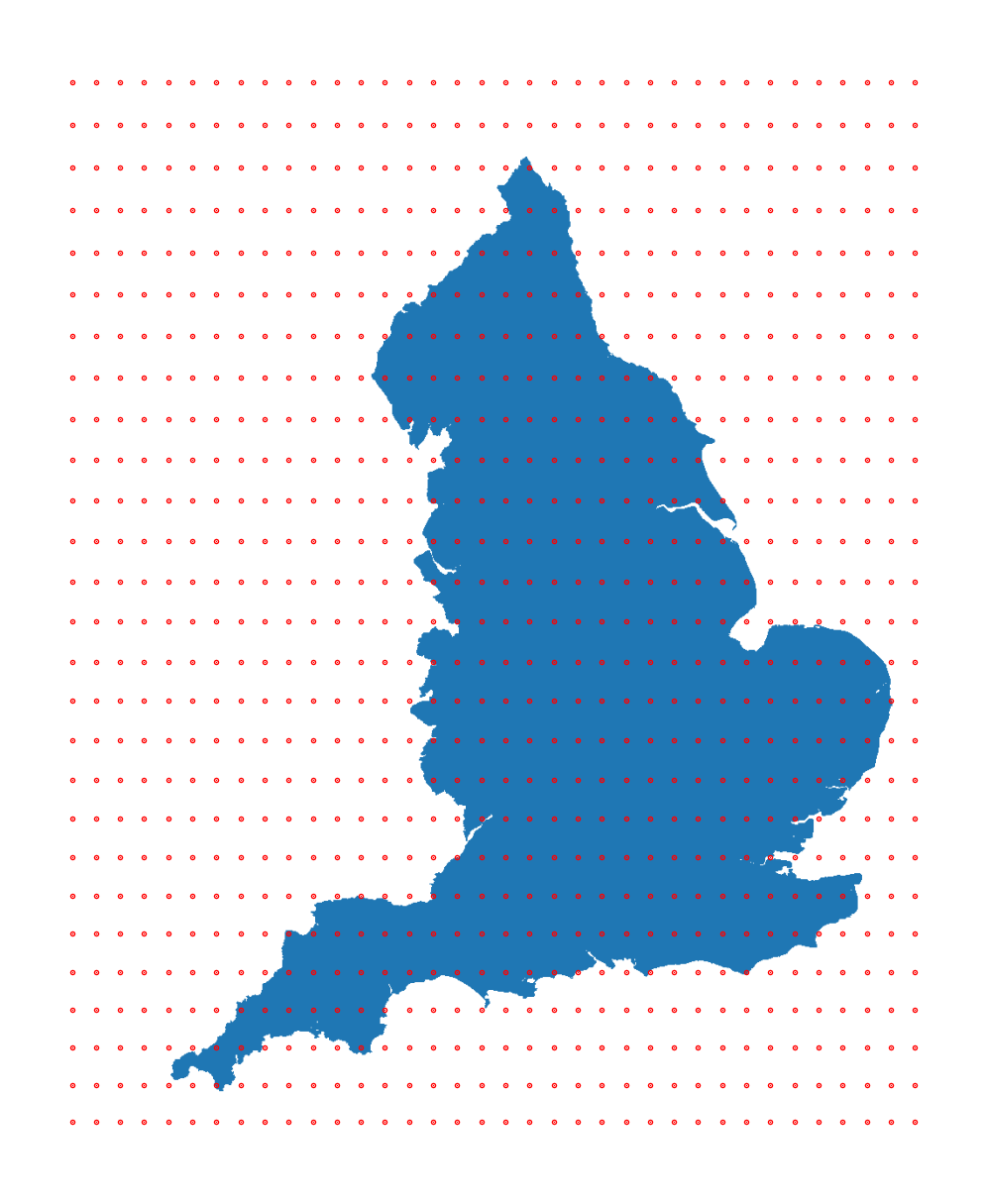}
\caption{\textbf{The blue region denotes the area of interest presented in Figure \ref{fig:englandLandGrids} with the red points showing the ERA5 sample locations across the UK.}}
\label{fig:era5SampleLocations}
\end{center}
\end{figure}

\begin{figure}
 \hspace*{\fill}
  \begin{subfigure}{0.9\textwidth}
    \includegraphics[width=\linewidth]{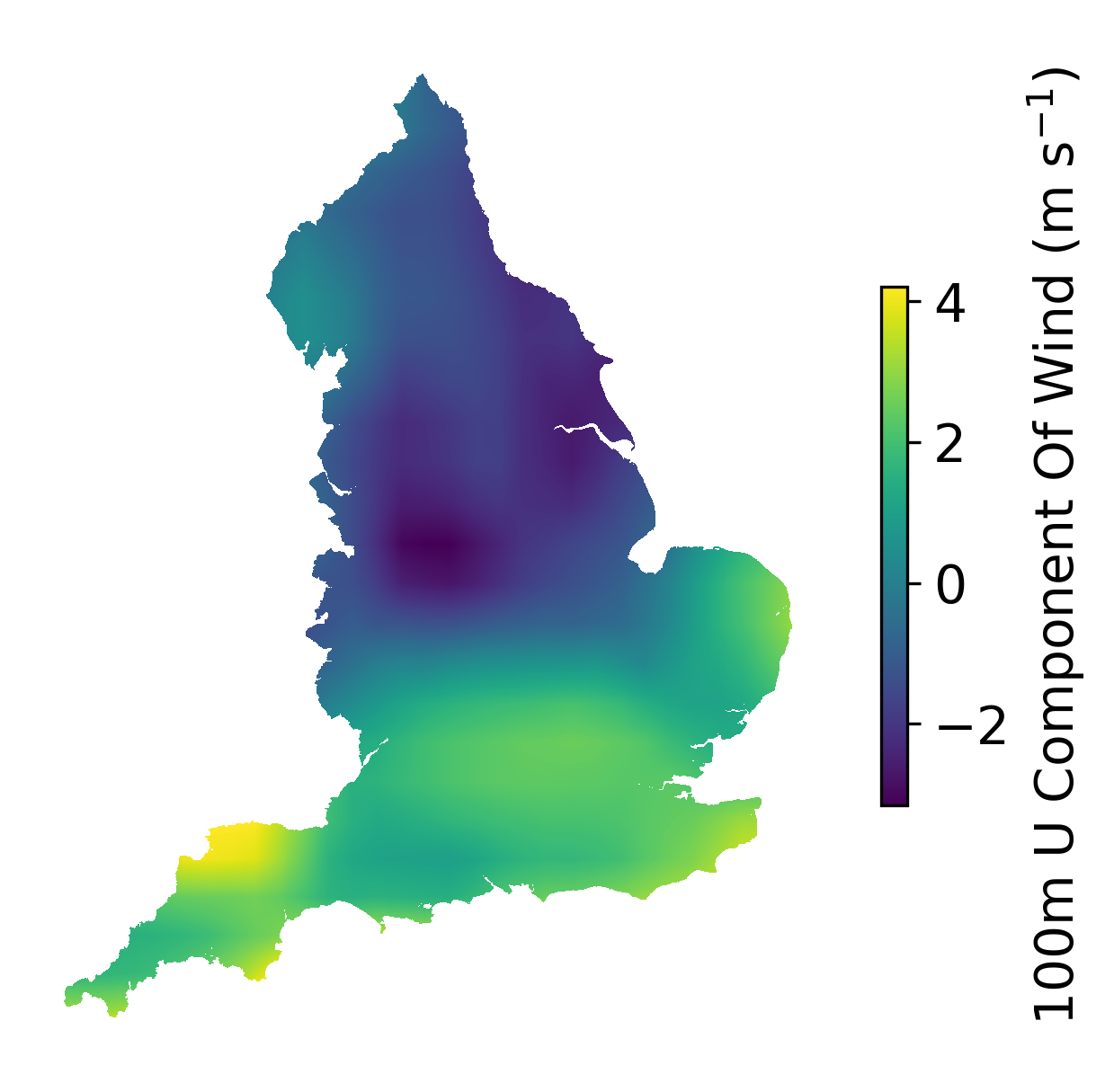}
    %\caption{100m U Component of Wind}\label{fig:metrologicalDatasets100mUWind}
  \end{subfigure}%
  \hspace*{\fill}   % maximize separation between the subfigures
\hspace*{\fill}   % maximize separation between the subfigures
\caption{{\bfseries Example complete England meteorological dataset from ERA5 for the 100m U Component of Wind feature vector.} } \label{fig:metrologicalDatasetsAll}
\end{figure}

We retrieved meteorological data from the ECMWF Re-analysis version 5 (ERA5) dataset \cite{hersbach:2016:era5}. ERA5 is a global dataset that details the environmental conditions at a range of point locations worldwide. There are 100s of variables available through the data set; we chose a subset of 11 based on meteorological variables detailed as being strongly associated with air pollution concentration in the existing literature. The subset of 11 variables to include was the 100m and 10m U component of wind, the 100m and 10m V component of wind, 2m dewpoint temperature, 2m temperature, boundary layer height, downwards UV radiation at the surface, the instantaneous 10m wind gust, surface pressure and total column rainwater. To create the feature vector, the point locations within the ERA5 dataset seen in Figure \ref{fig:era5SampleLocations} were interpolated across the study area to determine the variable value at each of the 1km$^2$ grid centroid. The resulting interpolated values at the grid centroids for a meteorological variable used, 100m U Component of Wind, are shown in Figure \ref{fig:metrologicalDatasetsAll}.

\clearpage
\subsection{Remote Sensing}
\label{sec:remoteSensingData}

Google Earth Engine \cite{gorelick:2013:GEE} derived datasets from Sentinel 5P \cite{veefkind:2012:Sentinel5PDescription} measurements comprised the remote sensing dataset family. The temporal period of datasets used was from 01-02-2019 to 01-03-2020, which allowed all available datasets from the Sentinel 5P platform to be studied. To ensure that all of the grids within the study area have a value for each timestamp, we aggregated the sentinel 5P datasets to the monthly mean temporal level. The grid would be interpolated from neighbouring values if any values were missing from the monthly aggregate, which was not the case for the variables used in the final study: NO$_2$, CO, HCHO, O$_3$, and the Absorbing Aerosol Index. Table \ref{tab:remoteSensingMissingData} shows the number of missing data points across the study area for each month's different variables of consideration from Sentinel 5P. Methane (CH$_4$) was missing many data points and was therefore excluded from the dataset.

\begin{figure}
  \hspace*{\fill}   % maximize separation between the subfigures
  \begin{subfigure}{0.9\textwidth}
    \includegraphics[width=\linewidth]{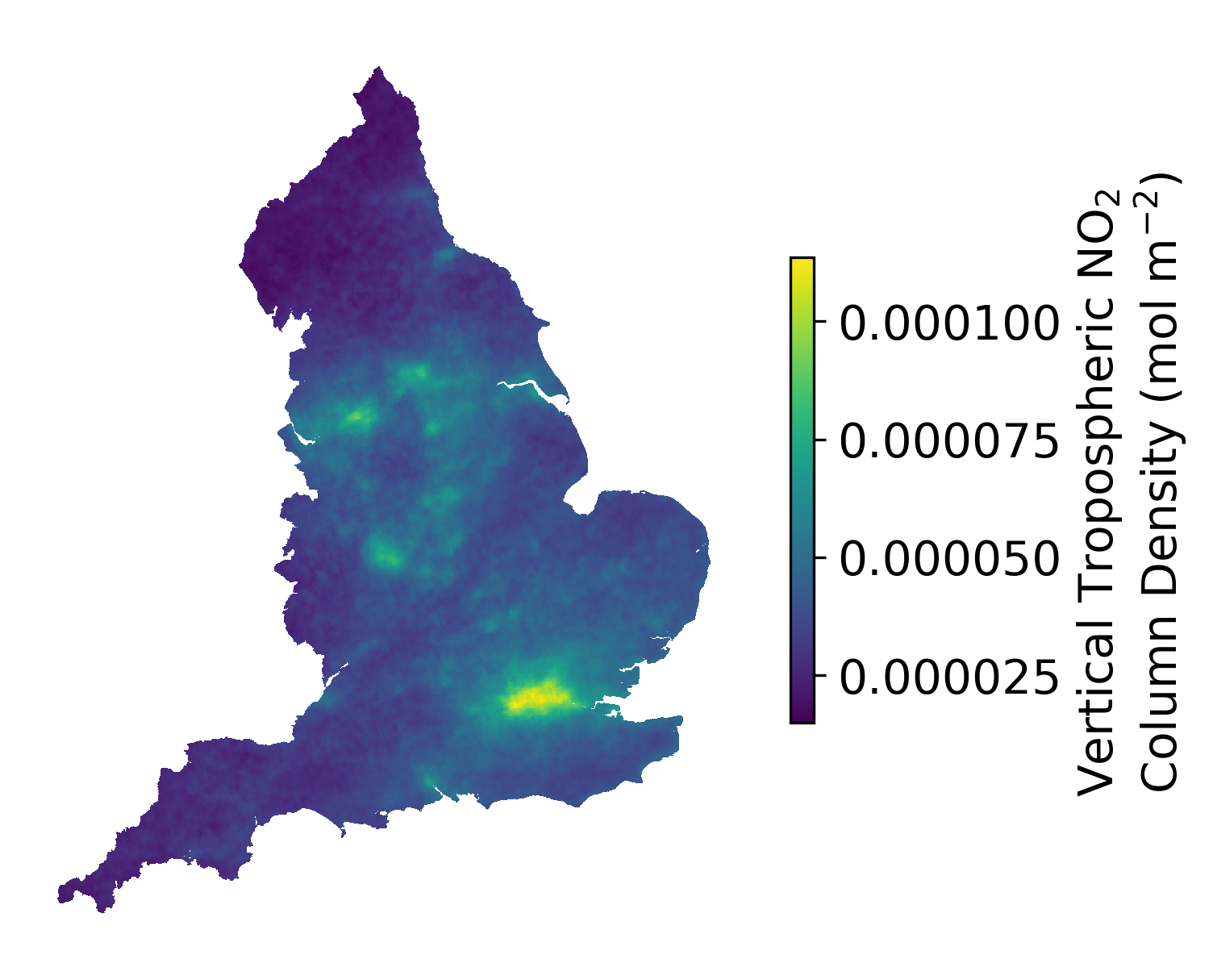}
    %\caption{NO2}\label{fig:remoteSensingNO2}
  \end{subfigure}
  \hspace*{\fill}   % maximize separation between the subfigures
\caption{{\bfseries Example complete England remote sensing dataset from Sentinel 5P Google Earth Engine for NO$_2$.} } \label{fig:remoteSensingDatasets}
\end{figure}

The process produced a spatially complete map of air pollution concentrations for each month of the year, which was then backfilled to other periods from before the Sentinel 5P platform came online to indicate typical air pollution concentrations during each month. Figure \ref{fig:remoteSensingDatasets} shows the complete spatial map of the remote sensing dataset produced for June for NO$_2$.

\begin{table}[!htb]
\hspace*{\fill}   % maximize separation between the subfigure
\begin{subtable}[!htb]{0.49\linewidth}
\centering
\resizebox{\textwidth}{!}{
\begin{filecontents*}{CSVFiles/Data/tropospheric_NO2_column_number_density_missing_data.csv}
Month,2019,2020,2021,2022,Monthly Median Overall,Monthly Median Overall Interpolated
1,-,100,99.6630385,100,100,100
2,100,100,100,-,100,100
3,100,100,100,-,100,100
4,100,100,100,-,100,100
5,100,100,100,-,100,100
6,100,100,100,-,100,100
7,100,99.91793765,100,-,100,100
8,100,100,100,-,100,100
9,100,100,100,-,100,100
10,100,99.8715668,100,-,100,100
11,100,100,100,-,100,100
12,100,99.98398098,100,-,100,100
\end{filecontents*}
\pgfplotstabletypeset[
    multicolumn names=l, 
    col sep=comma, 
    string type, 
    header = has colnames, 
    columns={Month, 2019, 2020, 2021, 2022, Monthly Median Overall, Monthly Median Overall Interpolated},
    columns/Month/.style={column type={S[round-precision=2, table-format=-1.3, table-number-alignment=center]}, column name=\shortstack{Month}},
    columns/2019/.style={column type={S[round-precision=2, table-format=-1.3, table-number-alignment=center]}, column name=\shortstack{2019}},
    columns/2020/.style={column type={S[round-precision=2, table-format=-1.3, table-number-alignment=center]}, column name=\shortstack{2020}},
    columns/2021/.style={column type={S[round-precision=2, table-format=-1.3, table-number-alignment=center]}, column name=\shortstack{2021}},
    columns/2022/.style={column type={S[round-precision=2, table-format=-1.3, table-number-alignment=center]}, column name=\shortstack{2022}},
    columns/Monthly Median Overall/.style={column type={S[round-precision=2, table-format=-1.3, table-number-alignment=center]}, column name=\shortstack{Monthly Median\\Overall}},
    columns/Monthly Median Overall Interpolated/.style={column type={S[round-precision=2, table-format=-1.3, table-number-alignment=center]}, column name=\shortstack{Monthly Median\\Overall Interpolated}},
    every head row/.style={before row=\toprule, after row=\midrule},
    every last row/.style={after row=\bottomrule}
    ]{CSVFiles/Data/tropospheric_NO2_column_number_density_missing_data.csv}}
    \smallskip
    \caption{{\bfseries  NO$_2$}  }\label{tab:remoteSensingMissingDataNO2}
%\end{table}
\end{subtable}
\hspace*{\fill}   % maximize separation between the subfigure
\begin{subtable}[h]{0.49\linewidth}
\centering
\resizebox{\textwidth}{!}{
\begin{filecontents*}{CSVFiles/Data/tropospheric_HCHO_column_number_density_missing_data.csv}
Month,2019,2020,2021,2022,Monthly Median Overall,Monthly Median Overall Interpolated
1,-,100,73.87859831,96.0865814,100,100
2,100,100,100,-,100,100
3,100,100,100,-,100,100
4,100,100,100,-,100,100
5,100,100,100,-,100,100
6,100,100,100,-,100,100
7,100,100,100,-,100,100
8,100,100,100,-,100,100
9,100,100,100,-,100,100
10,100,99.97976545,100,-,100,100
11,100,100,99.99718965,-,100,100
12,99.94716534,0.992898234,0.254618115,-,99.99381722,100
\end{filecontents*}
\pgfplotstabletypeset[
    multicolumn names=l, 
    col sep=comma, 
    string type, 
    header = has colnames, 
    columns={Month, 2019, 2020, 2021, 2022, Monthly Median Overall, Monthly Median Overall Interpolated},
    columns/Month/.style={column type={S[round-precision=2, table-format=-1.3, table-number-alignment=center]}, column name=\shortstack{Month}},
    columns/2019/.style={column type={S[round-precision=2, table-format=-1.3, table-number-alignment=center]}, column name=\shortstack{2019}},
    columns/2020/.style={column type={S[round-precision=2, table-format=-1.3, table-number-alignment=center]}, column name=\shortstack{2020}},
    columns/2021/.style={column type={S[round-precision=2, table-format=-1.3, table-number-alignment=center]}, column name=\shortstack{2021}},
    columns/2022/.style={column type={S[round-precision=2, table-format=-1.3, table-number-alignment=center]}, column name=\shortstack{2022}},
    columns/Monthly Median Overall/.style={column type={S[round-precision=2, table-format=-1.3, table-number-alignment=center]}, column name=\shortstack{Monthly Median\\Overall}},
    columns/Monthly Median Overall Interpolated/.style={column type={S[round-precision=2, table-format=-1.3, table-number-alignment=center]}, column name=\shortstack{Monthly Median\\Overall Interpolated}},
    every head row/.style={before row=\toprule, after row=\midrule},
    every last row/.style={after row=\bottomrule}
    ]{CSVFiles/Data/tropospheric_HCHO_column_number_density_missing_data.csv}}
    \smallskip
    \caption{{\bfseries HCHO}  }\label{tab:remoteSensingMissingDataHCHO}
\end{subtable}
\hspace*{\fill}   % maximize separation between the subfigure
\\
\hspace*{\fill}   % maximize separation between the subfigure
\begin{subtable}[h]{0.49\linewidth}
\centering
\resizebox{\textwidth}{!}{
\begin{filecontents*}{CSVFiles/Data/O3_column_number_density_missing_data.csv}
Month,2019,2020,2021,2022,Monthly Median Overall,Monthly Median Overall Interpolated
1,-,100,100,100,100,100
2,100,100,100,-,100,100
3,100,100,100,-,100,100
4,100,100,100,-,100,100
5,100,100,100,-,100,100
6,100,100,100,-,100,100
7,100,100,100,-,100,100
8,100,100,100,-,100,100
9,100,100,100,-,100,100
10,100,100,100,-,100,100
11,100,100,100,-,100,100
12,100,100,100,-,100,100
\end{filecontents*}
\pgfplotstabletypeset[
    multicolumn names=l, 
    col sep=comma, 
    string type, 
    header = has colnames, 
    columns={Month, 2019, 2020, 2021, 2022, Monthly Median Overall, Monthly Median Overall Interpolated},
    columns/Month/.style={column type={S[round-precision=2, table-format=-1.3, table-number-alignment=center]}, column name=\shortstack{Month}},
    columns/2019/.style={column type={S[round-precision=2, table-format=-1.3, table-number-alignment=center]}, column name=\shortstack{2019}},
    columns/2020/.style={column type={S[round-precision=2, table-format=-1.3, table-number-alignment=center]}, column name=\shortstack{2020}},
    columns/2021/.style={column type={S[round-precision=2, table-format=-1.3, table-number-alignment=center]}, column name=\shortstack{2021}},
    columns/2022/.style={column type={S[round-precision=2, table-format=-1.3, table-number-alignment=center]}, column name=\shortstack{2022}},
    columns/Monthly Median Overall/.style={column type={S[round-precision=2, table-format=-1.3, table-number-alignment=center]}, column name=\shortstack{Monthly Median\\Overall}},
    columns/Monthly Median Overall Interpolated/.style={column type={S[round-precision=2, table-format=-1.3, table-number-alignment=center]}, column name=\shortstack{Monthly Median\\Overall Interpolated}},
    every head row/.style={before row=\toprule, after row=\midrule},
    every last row/.style={after row=\bottomrule}
    ]{CSVFiles/Data/O3_column_number_density_missing_data.csv}}
    \smallskip
    \caption{{\bfseries O$_3$}  }\label{tab:remoteSensingMissingDataO3}
\end{subtable}
\hspace*{\fill}   % maximize separation between the subfigure
\begin{subtable}[h]{0.49\linewidth}
\centering
\resizebox{\textwidth}{!}{
\begin{filecontents*}{CSVFiles/Data/CO_column_number_density_missing_data.csv}
Month,2019,2020,2021,2022,Monthly Median Overall,Monthly Median Overall Interpolated
1,-,100,100,100,100,100
2,100,100,100,-,100,100
3,100,100,100,-,100,100
4,100,100,100,-,100,100
5,100,100,100,-,100,100
6,100,100,100,-,100,100
7,100,100,100,-,100,100
8,100,100,100,-,100,100
9,100,100,100,-,100,100
10,100,100,100,-,100,100
11,100,100,100,-,100,100
12,100,100,100,-,100,100
\end{filecontents*}
\pgfplotstabletypeset[
    multicolumn names=l, 
    col sep=comma, 
    string type, 
    header = has colnames, 
    columns={Month, 2019, 2020, 2021, 2022, Monthly Median Overall, Monthly Median Overall Interpolated},
    columns/Month/.style={column type={S[round-precision=2, table-format=-1.3, table-number-alignment=center]}, column name=\shortstack{Month}},
    columns/2019/.style={column type={S[round-precision=2, table-format=-1.3, table-number-alignment=center]}, column name=\shortstack{2019}},
    columns/2020/.style={column type={S[round-precision=2, table-format=-1.3, table-number-alignment=center]}, column name=\shortstack{2020}},
    columns/2021/.style={column type={S[round-precision=2, table-format=-1.3, table-number-alignment=center]}, column name=\shortstack{2021}},
    columns/2022/.style={column type={S[round-precision=2, table-format=-1.3, table-number-alignment=center]}, column name=\shortstack{2022}},
    columns/Monthly Median Overall/.style={column type={S[round-precision=2, table-format=-1.3, table-number-alignment=center]}, column name=\shortstack{Monthly Median\\Overall}},
    columns/Monthly Median Overall Interpolated/.style={column type={S[round-precision=2, table-format=-1.3, table-number-alignment=center]}, column name=\shortstack{Monthly Median\\Overall Interpolated}},
    every head row/.style={before row=\toprule, after row=\midrule},
    every last row/.style={after row=\bottomrule}
    ]{CSVFiles/Data/CO_column_number_density_missing_data.csv}}
    \smallskip
    \caption{{\bfseries CO}  }\label{tab:remoteSensingMissingDataCO}
\end{subtable}
\hspace*{\fill}   % maximize separation between the subfigure
\\
\hspace*{\fill}   % maximize separation between the subfigure
\begin{subtable}[h]{0.49\linewidth}
\centering
\resizebox{\textwidth}{!}{
\begin{filecontents*}{CSVFiles/Data/CH4_column_volume_mixing_ratio_dry_air_missing_data.csv}
Month,2019,2020,2021,2022,Monthly Median Overall,Monthly Median Overall Interpolated
1,-,0.159347098,1.236837002,12.62355021,13.52342571,100
2,77.06554028,46.45263007,63.80516375,-,82.08623854,100
3,37.2068449,47.65771007,56.0390302,-,76.84914298,100
4,63.74052559,72.79380148,83.10190064,-,85.99825196,100
5,22.89651994,69.35617589,60.84333117,-,80.16479919,100
6,49.86636764,55.90891079,60.14242876,-,77.18469931,100
7,23.53531351,32.61275845,78.60336624,-,82.90320858,100
8,23.20762618,47.09058053,16.11401046,-,64.43580729,100
9,53.4543472,81.4246811,79.15531986,-,87.12604721,100
10,39.47030439,3.464886026,49.57914942,-,64.93492624,100
11,3.685498852,31.5032305,16.62999154,-,38.5451357,100
\end{filecontents*}
\pgfplotstabletypeset[
    multicolumn names=l, 
    col sep=comma, 
    string type, 
    header = has colnames, 
    columns={Month, 2019, 2020, 2021, 2022, Monthly Median Overall, Monthly Median Overall Interpolated},
    columns/Month/.style={column type={S[round-precision=2, table-format=-1.3, table-number-alignment=center]}, column name=\shortstack{Month}},
    columns/2019/.style={column type={S[round-precision=2, table-format=-1.3, table-number-alignment=center]}, column name=\shortstack{2019}},
    columns/2020/.style={column type={S[round-precision=2, table-format=-1.3, table-number-alignment=center]}, column name=\shortstack{2020}},
    columns/2021/.style={column type={S[round-precision=2, table-format=-1.3, table-number-alignment=center]}, column name=\shortstack{2021}},
    columns/2022/.style={column type={S[round-precision=2, table-format=-1.3, table-number-alignment=center]}, column name=\shortstack{2022}},
    columns/Monthly Median Overall/.style={column type={S[round-precision=2, table-format=-1.3, table-number-alignment=center]}, column name=\shortstack{Monthly  Median\\Overall}},
    columns/Monthly Median Overall Interpolated/.style={column type={S[round-precision=2, table-format=-1.3, table-number-alignment=center]}, column name=\shortstack{Monthly Median\\Overall Interpolated}},
    every head row/.style={before row=\toprule, after row=\midrule},
    every last row/.style={after row=\bottomrule}
    ]{CSVFiles/Data/CH4_column_volume_mixing_ratio_dry_air_missing_data.csv}}
    \smallskip
    \caption{{\bfseries CH$_4$}  }\label{tab:remoteSensingMissingDataCH4}
\end{subtable}
\hspace*{\fill}   % maximize separation between the subfigure
\begin{subtable}[h]{0.49\linewidth}
\centering
\resizebox{\textwidth}{!}{
\begin{filecontents*}{CSVFiles/Data/absorbing_aerosol_index_missing_data.csv}
Month,2019,2020,2021,2022,Monthly Median Overall,Monthly Median Overall Interpolated
1,-,100,100,100,100,100
2,100,100,100,-,100,100
3,100,100,100,-,100,100
4,100,100,100,-,100,100
5,100,100,100,-,100,100
6,100,100,100,-,100,100
7,100,100,100,-,100,100
8,100,100,100,-,100,100
9,100,100,100,-,100,100
10,100,100,100,-,100,100
11,100,100,100,-,100,100
12,100,100,100,-,100,100
\end{filecontents*}
\pgfplotstabletypeset[
    multicolumn names=l, 
    col sep=comma, 
    string type, 
    header = has colnames, 
    columns={Month,2019,2020,2021,2022,Monthly Median Overall,Monthly Median Overall Interpolated},
    columns/Month/.style={column type={S[round-precision=2, table-format=-1.3, table-number-alignment=center]}, column name=\shortstack{Month}},
    columns/2019/.style={column type={S[round-precision=2, table-format=-1.3, table-number-alignment=center]}, column name=\shortstack{2019}},
    columns/2020/.style={column type={S[round-precision=2, table-format=-1.3, table-number-alignment=center]}, column name=\shortstack{2020}},
    columns/2021/.style={column type={S[round-precision=2, table-format=-1.3, table-number-alignment=center]}, column name=\shortstack{2021}},
    columns/2022/.style={column type={S[round-precision=2, table-format=-1.3, table-number-alignment=center]}, column name=\shortstack{2022}},
    columns/Monthly Median Overall/.style={column type={S[round-precision=2, table-format=-1.3, table-number-alignment=center]}, column name=\shortstack{Monthly Median\\Overall}},
    columns/Monthly Median Overall Interpolated/.style={column type={S[round-precision=2, table-format=-1.3, table-number-alignment=center]}, column name=\shortstack{Monthly Median\\Overall Interpolated}},
    every head row/.style={before row=\toprule, after row=\midrule},
    every last row/.style={after row=\bottomrule}
    ]{CSVFiles/Data/absorbing_aerosol_index_missing_data.csv}}
    \smallskip
    \caption{{\bfseries Absorbing Aerosol Index}  }\label{tab:remoteSensingMissingDataAAI}
\end{subtable}
\hspace*{\fill}   % maximize separation between the subfigure
\caption{{\bfseries Missing Data for each of the months for each of the variables considered in the study.} All of the variables considered other than CH$_4$ had a spatially complete dataset at the monthly aggregate level and, as such, were included in the study as shown in Figure \ref{fig:remoteSensingDatasets}. }
\label{tab:remoteSensingMissingData}
\end{table}

\clearpage
\subsection{Emissions}
\label{sec:Datadetails:emissions}

Emissions data is gathered from the UK National Atmospheric Emissions Inventory (NAEI) \cite{NAEI:2023:NAEIEmissionsData}. A set of seven air pollutants are included: PM$_{2.5}$, PM$_{10}$, (Non-methane volatile organic compounds (NMVOC), NH$_3$, SO$_x$, CO, NO$_x$ in the study. The emissions are classified into one of 11 sectors to denote the emission source, based on Selected Nomenclature for Air Pollutants (SNAP) sectors \cite{NAEI:2023:SNAPSectorDefinitions}:
\begin{itemize}
    \item SNAP Sector 1 (Combustion Energy Production and Transformation)
    \item SNAP Sector 2 (Combustion in Commercial, Institutional, Residential and Agriculture)
    \item SNAP Sector 3 (Combustion in Industry)
    \item SNAP Sector 4 (Production Processes)
    \item SNAP Sector 5 (Extraction and Distribution of Fossil Fuels)
    \item SNAP Sector 6 (Solvent Use)
    \item SNAP Sector 7 (Road Transport)
    \item SNAP Sector 8 (Other Transport and Mobile Machinery)
    \item SNAP Sector 9 (Waste Treatment and Disposal)
    \item SNAP Sector 10 (Agriculture, Forestry and Land Use Change)
    \item SNAP Sector 11 (Nature)
\end{itemize}

\begin{figure}[!htb]
\begin{center}
\includegraphics[width=0.9\textwidth]{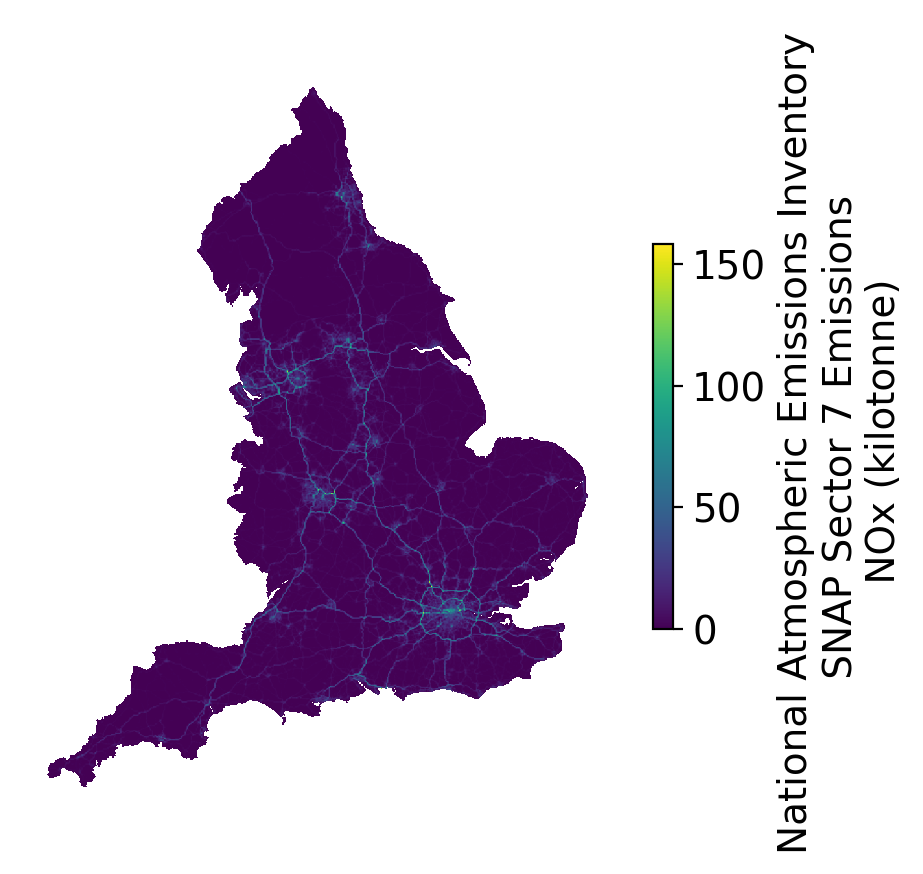}
\caption{{\bfseries Example complete England emission dataset for SNAP Sector 7 (Road Emissions) for NO$_x$ on 1st June 2018 at 0800AM.}}
\label{fig:emissionsDatasetSingle}
\end{center}
\end{figure}

We created the emissions feature vector by summing the point and area emissions data \cite{NAEI:2023:LargePointSources} from the NAEI for each year per SNAP sector per species to map the emissions across the study area. We then subsequently scaled the emissions map depending on the timestamp of interest, applying a scaling for the hour and day of the week and month of interest \cite{denier:2011:NAEIEmissionsScalingHourDay}. Figure \ref{fig:emissionsDatasetSingle} gives an example of the emission feature vector for NO$_x$ SNAP Sector 7 (Road) emissions across the study area.

\begin{figure}
    \hspace*{\fill}   % maximize separation between the subfigures
  \begin{subfigure}{0.49\textwidth}
    \includegraphics[width=\linewidth]{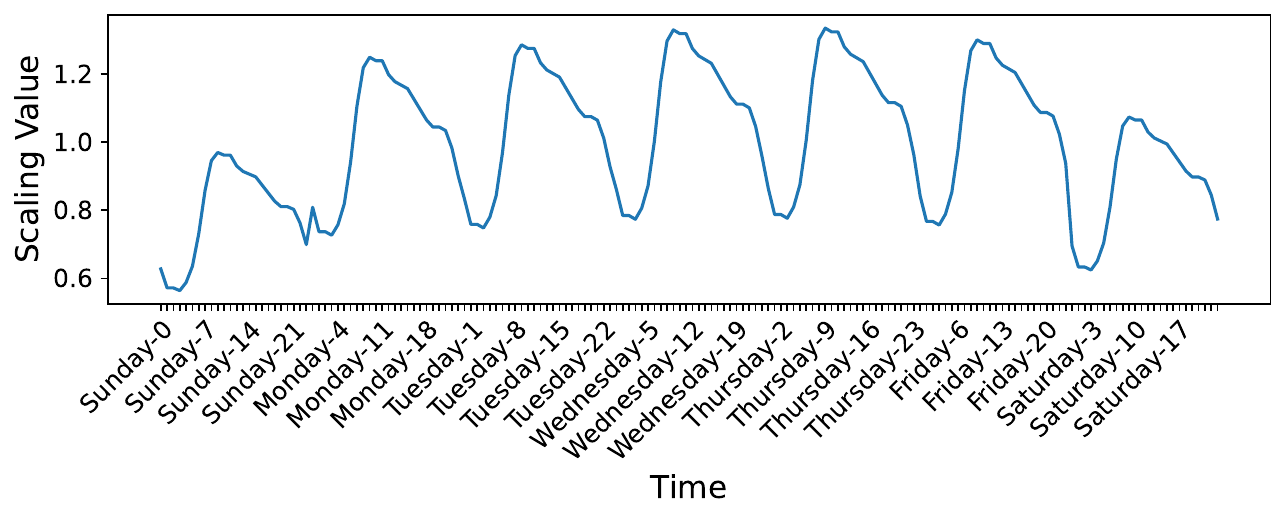}
    \caption{SNAP Sector 1}\label{fig:emissionsScalingWeeklySNAP1}
  \end{subfigure}%
  \hspace*{\fill}   % maximize separation between the subfigures
  \begin{subfigure}{0.49\textwidth}
    \includegraphics[width=\linewidth]{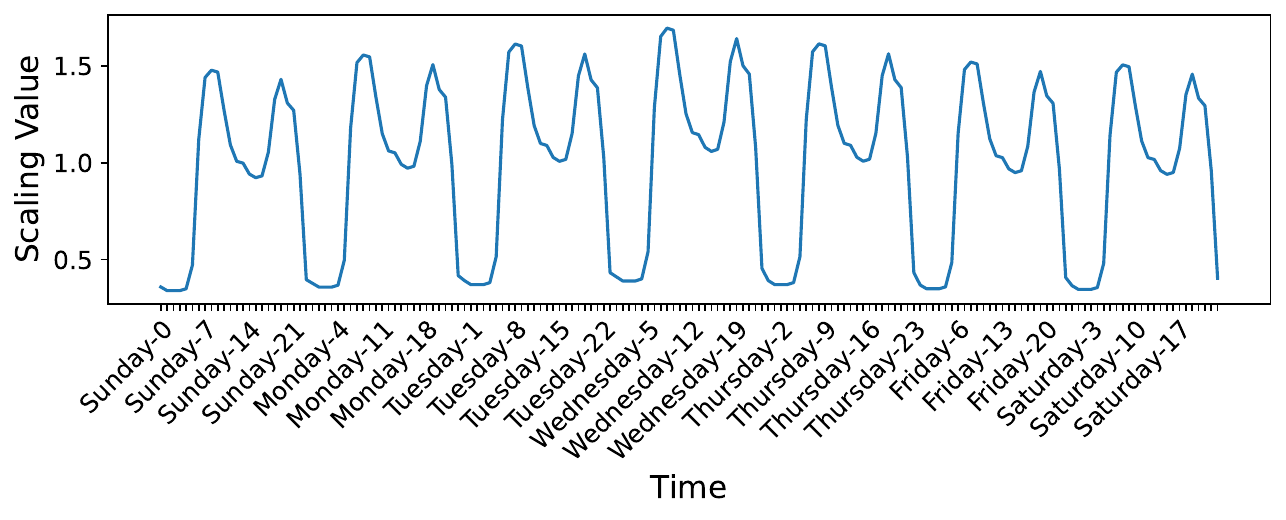}
    \caption{SNAP Sector 2}\label{fig:emissionsScalingWeeklySNAP2}
  \end{subfigure}%
  \hspace*{\fill}   % maximize separation between the subfigures
  \\
  \hspace*{\fill}   % maximize separation between the subfigures
  \begin{subfigure}{0.49\textwidth}
    \includegraphics[width=\linewidth]{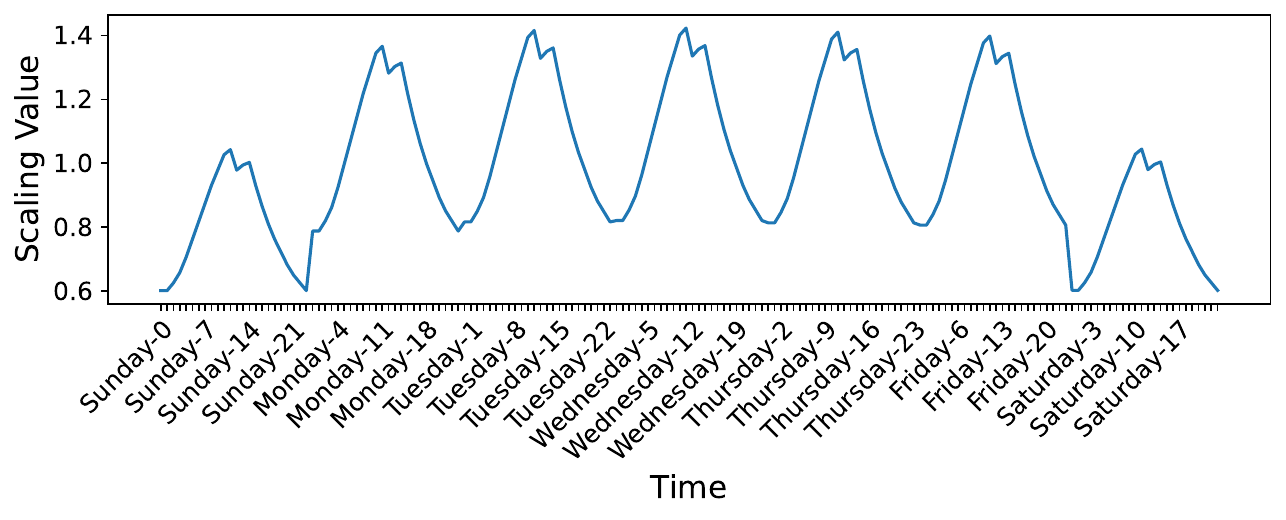}
    \caption{SNAP Sector 3}\label{fig:emissionsScalingWeeklySNAP3}
  \end{subfigure}%
  \hspace*{\fill}   % maximize separation between the subfigures
  \begin{subfigure}{0.49\textwidth}
    \includegraphics[width=\linewidth]{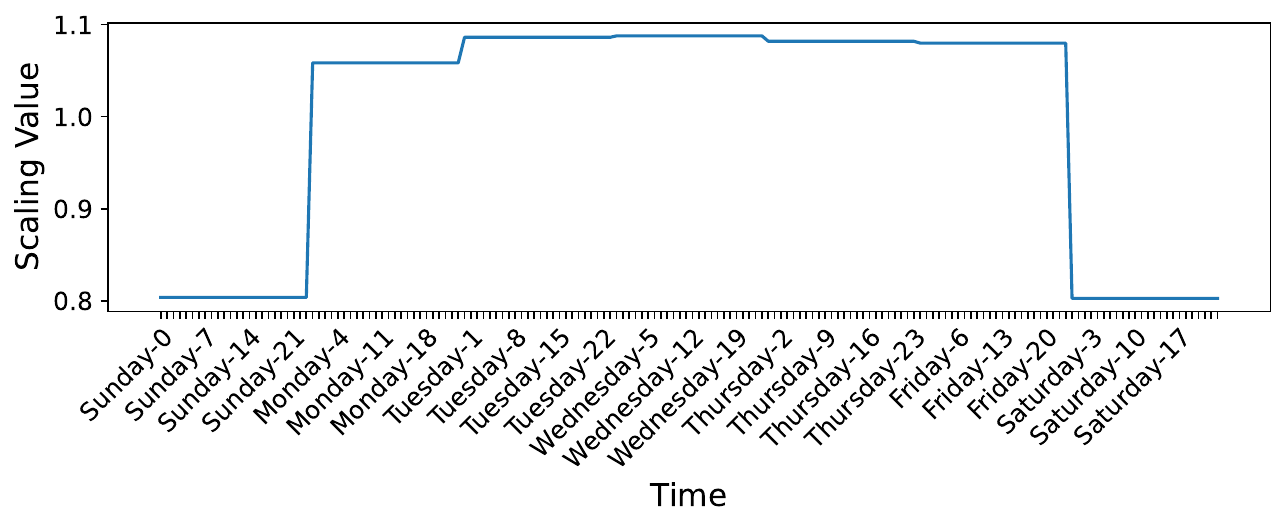}
    \caption{SNAP Sector 4}\label{fig:emissionsScalingWeeklySNAP4}
  \end{subfigure}%
  \hspace*{\fill}   % maximize separation between the subfigures
  \\
  \hspace*{\fill}   % maximize separation between the subfigures
  \begin{subfigure}{0.49\textwidth}
    \includegraphics[width=\linewidth]{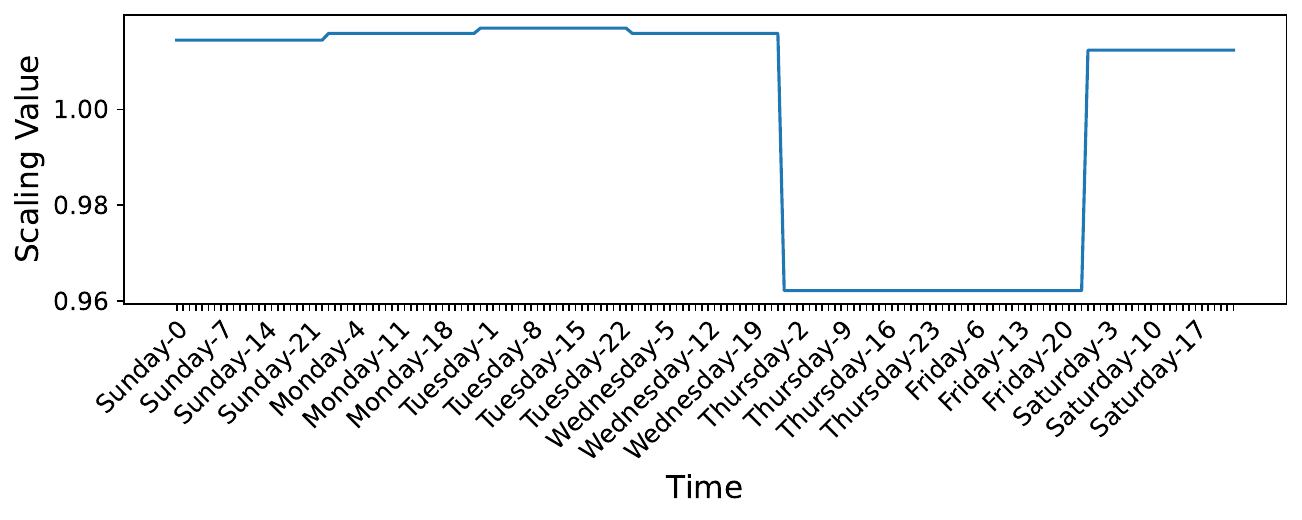}
    \caption{SNAP Sector 5}\label{fig:emissionsScalingWeeklySNAP5}
  \end{subfigure}%
  \hspace*{\fill}   % maximize separation between the subfigures
  \begin{subfigure}{0.49\textwidth}
    \includegraphics[width=\linewidth]{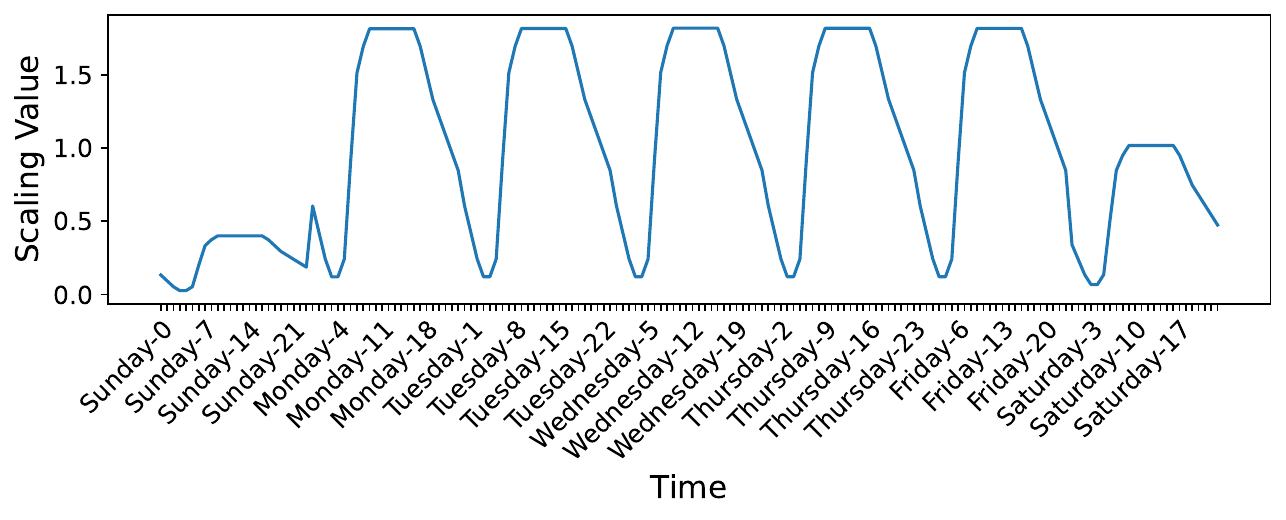}
    \caption{SNAP Sector 6}\label{fig:emissionsScalingWeeklySNAP6}
  \end{subfigure}%
  \hspace*{\fill}   % maximize separation between the subfigures
  \\
  \hspace*{\fill}   % maximize separation between the subfigures
  \begin{subfigure}{0.49\textwidth}
    \includegraphics[width=\linewidth]{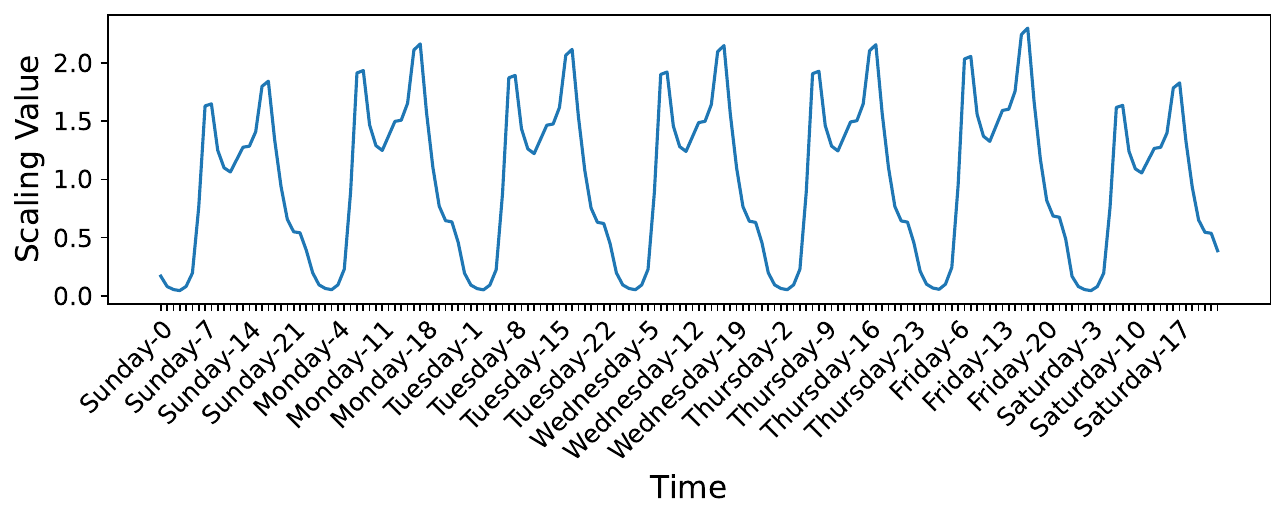}
    \caption{SNAP Sector 7}\label{fig:emissionsScalingWeeklySNAP7}
  \end{subfigure}%
  \hspace*{\fill}   % maximize separation between the subfigures
  \begin{subfigure}{0.49\textwidth}
    \includegraphics[width=\linewidth]{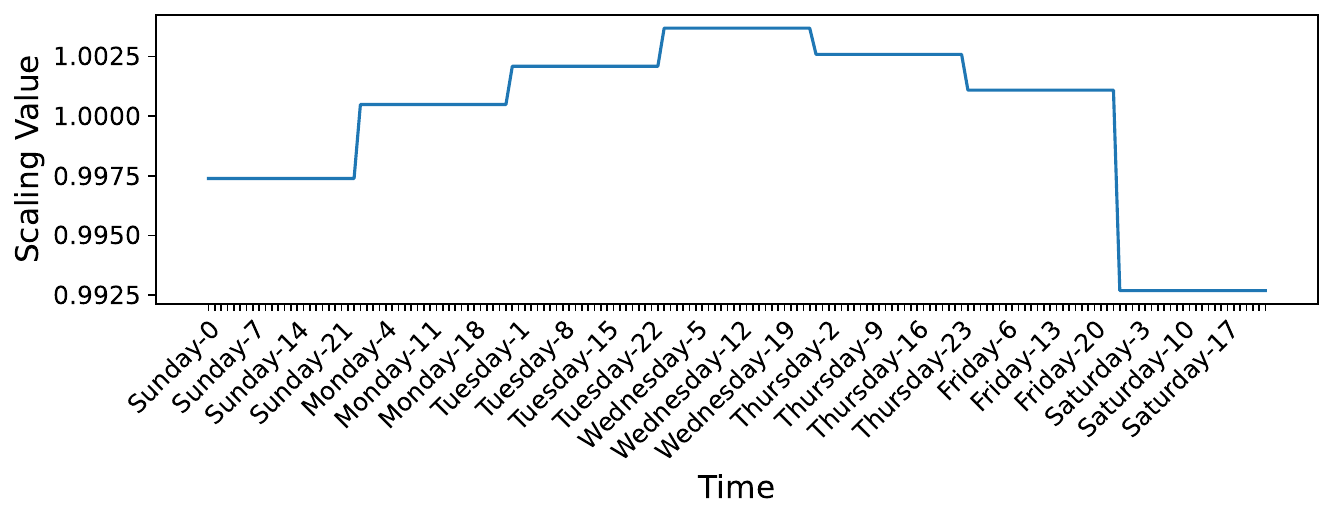}
    \caption{SNAP Sector 8}\label{fig:emissionsScalingWeeklySNAP8}
  \end{subfigure}%
  \hspace*{\fill}   % maximize separation between the subfigures
  \\
  \hspace*{\fill}   % maximize separation between the subfigures
  \begin{subfigure}{0.49\textwidth}
    \includegraphics[width=\linewidth]{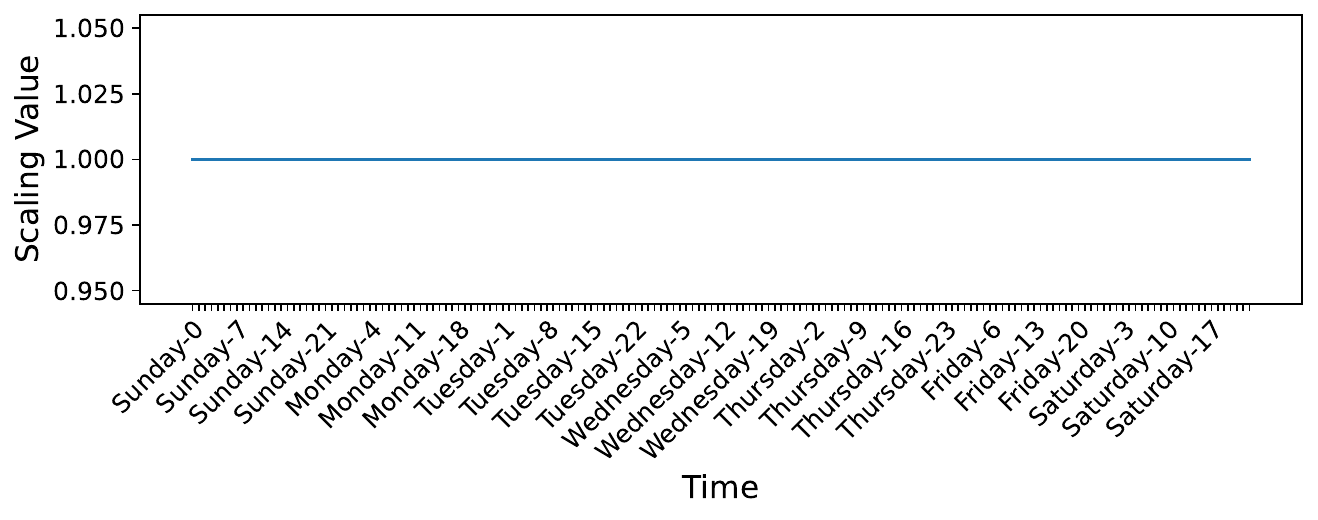}
    \caption{SNAP Sector 9}\label{fig:emissionsScalingWeeklySNAP9}
  \end{subfigure}%
  \hspace*{\fill}   % maximize separation between the subfigures
  \hspace*{\fill}   % maximize separation between the subfigures
  \begin{subfigure}{0.49\textwidth}
    \includegraphics[width=\linewidth]{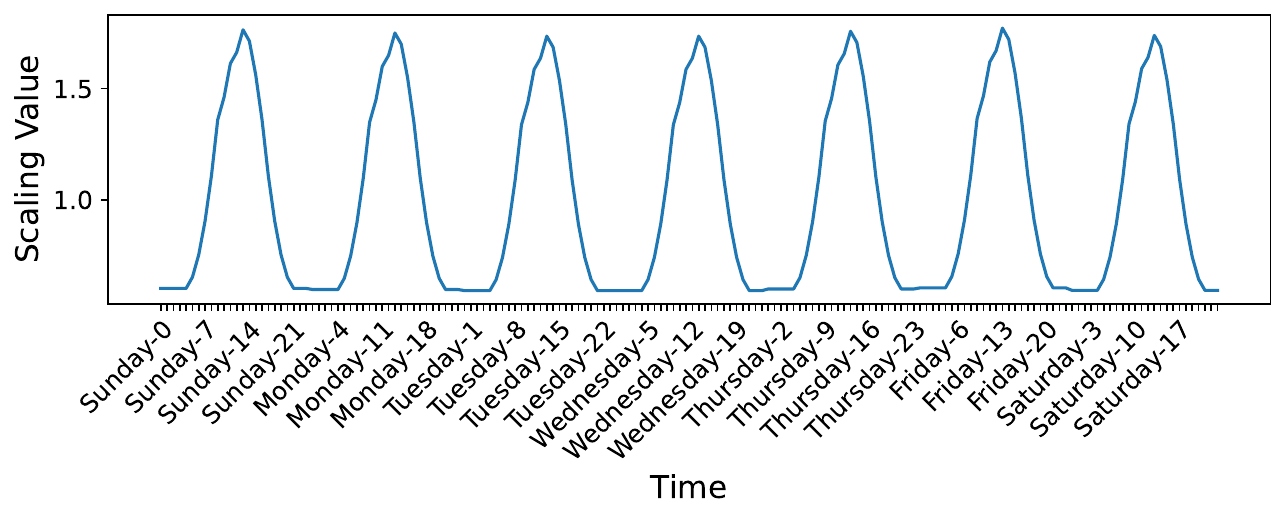}
    \caption{SNAP Sector 10}\label{fig:emissionsScalingWeeklySNAP10}
  \end{subfigure}%
  \hspace*{\fill}   % maximize separation between the subfigures
  \\
  \hspace*{\fill}   % maximize separation between the subfigures
  \begin{subfigure}{0.49\textwidth}
    \includegraphics[width=\linewidth]{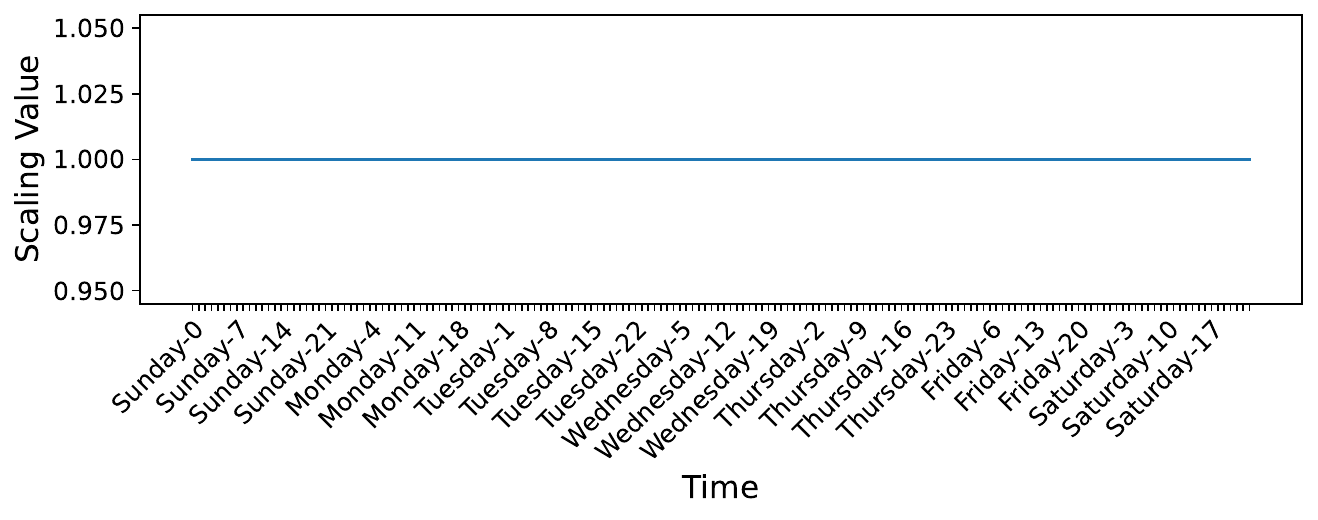}
    \caption{SNAP Sector 11}\label{fig:emissionsScalingWeeklySNAP11}
  \end{subfigure}%
  \hspace*{\fill}   % maximize separation between the subfigures
  
\caption{{\bfseries Hourly and daily emissions scaling per SNAP sector.} } \label{fig:emissionsScalingWeeklySNAP}
\end{figure}

\begin{figure}
    \hspace*{\fill}   % maximize separation between the subfigures
  \begin{subfigure}{0.49\textwidth}
    \includegraphics[width=\linewidth]{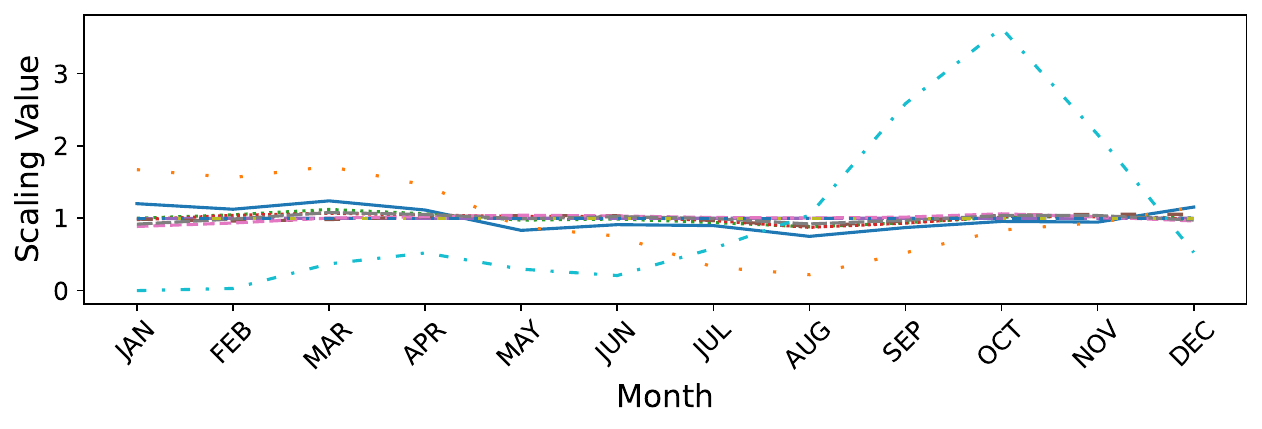}
    \caption{CO}\label{fig:emissionsScalingMonthlyCO}
  \end{subfigure}%
  \hspace*{\fill}   % maximize separation between the subfigures
  \begin{subfigure}{0.49\textwidth}
    \includegraphics[width=\linewidth]{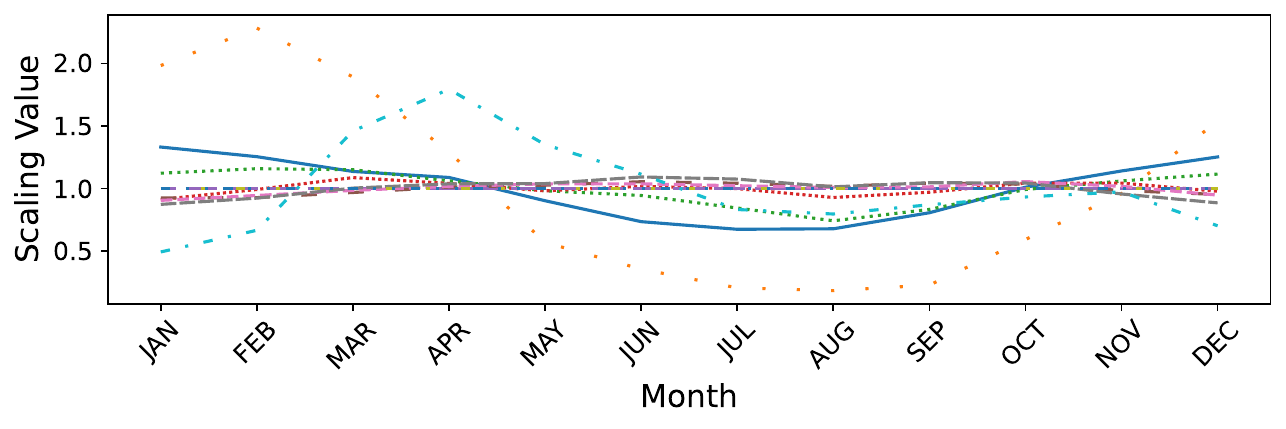}
    \caption{NH$_3$}\label{fig:emissionsScalingMonthlyNH3}
  \end{subfigure}%
  \hspace*{\fill}   % maximize separation between the subfigures
  \\
  \hspace*{\fill}   % maximize separation between the subfigures
  \begin{subfigure}{0.49\textwidth}
    \includegraphics[width=\linewidth]{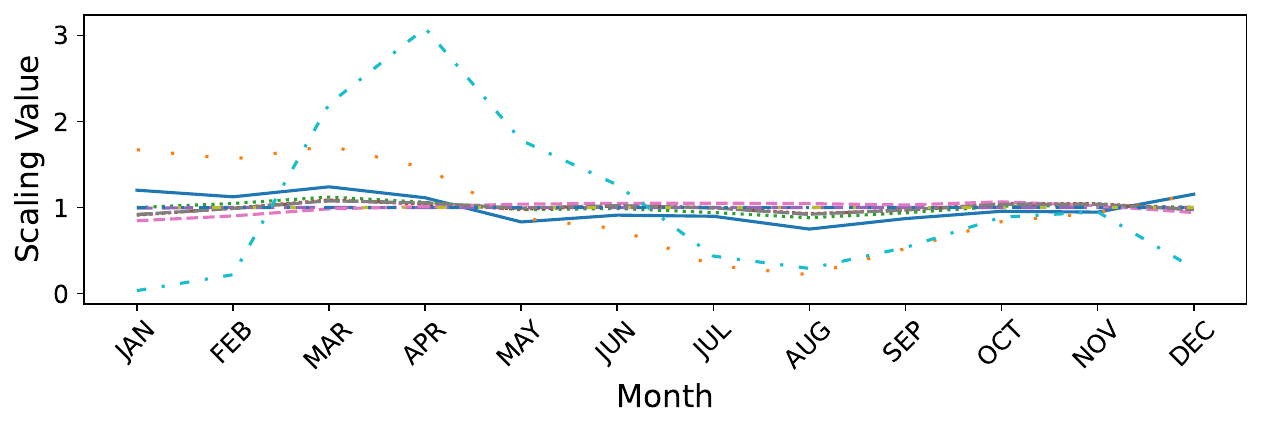}
    \caption{NMVOC}\label{fig:emissionsScalingMonthlyNMVOC}
  \end{subfigure}%
  \hspace*{\fill}   % maximize separation between the subfigures
  \begin{subfigure}{0.49\textwidth}
    \includegraphics[width=\linewidth]{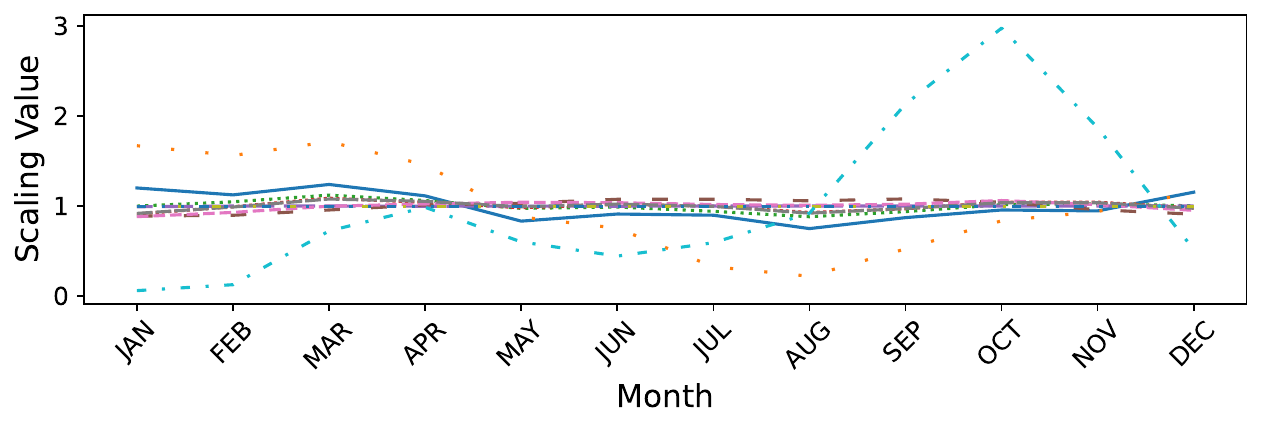}
    \caption{NO$_x$}\label{fig:emissionsScalingMonthlyNOX}
  \end{subfigure}%
  \hspace*{\fill}   % maximize separation between the subfigures
  \\
  \hspace*{\fill}   % maximize separation between the subfigures
  \begin{subfigure}{0.49\textwidth}
    \includegraphics[width=\linewidth]{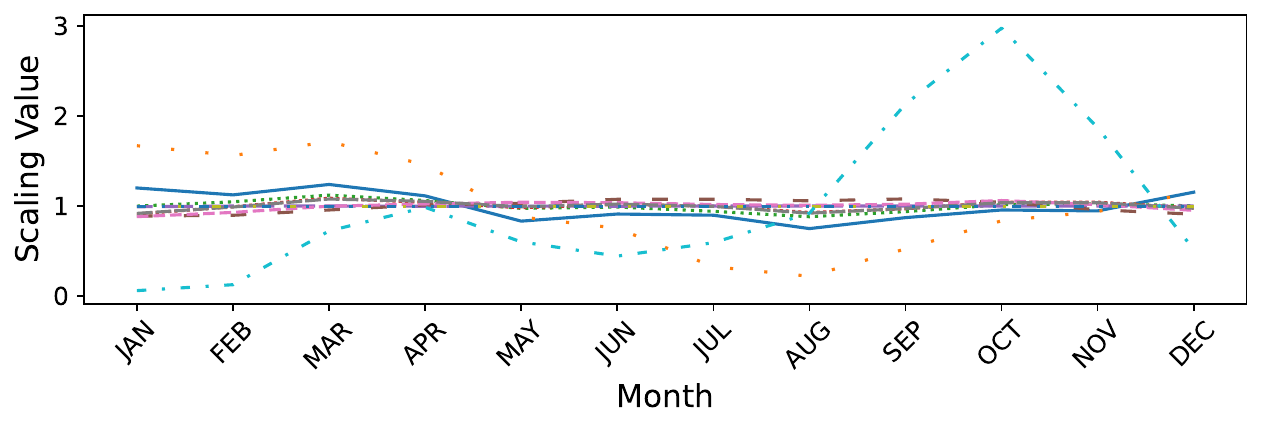}
    \caption{PM$_{10}$}\label{fig:emissionsScalingMonthlyPM10}
  \end{subfigure}%
  \hspace*{\fill}   % maximize separation between the subfigures
  \begin{subfigure}{0.49\textwidth}
    \includegraphics[width=\linewidth]{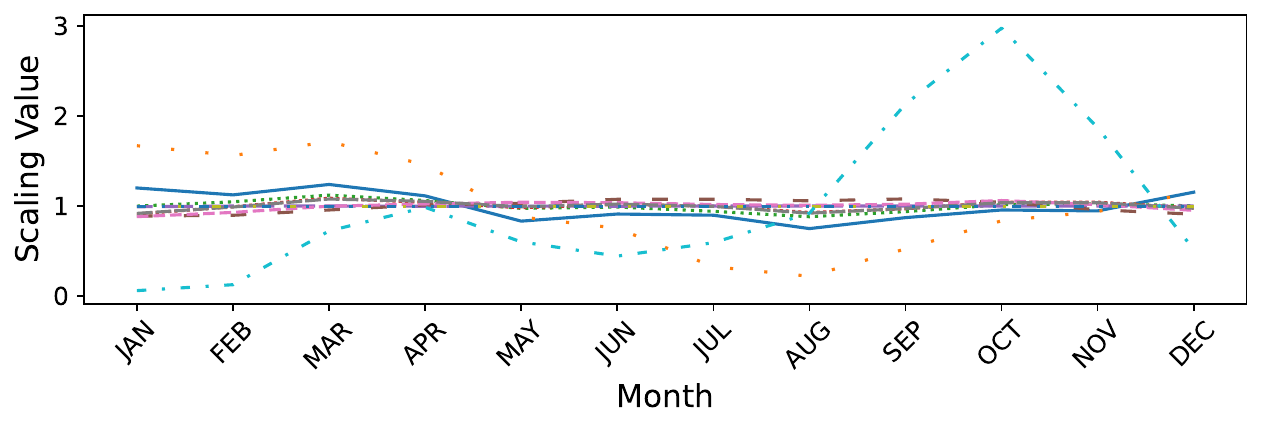}
    \caption{PM$_{2.5}$}\label{fig:emissionsScalingMonthlyPM25}
  \end{subfigure}%
  \hspace*{\fill}   % maximize separation between the subfigures
  \\
  %s\hspace*{\fill}   % maximize separation between the subfigures
  \begin{subfigure}{0.49\textwidth}
    \includegraphics[width=\linewidth]{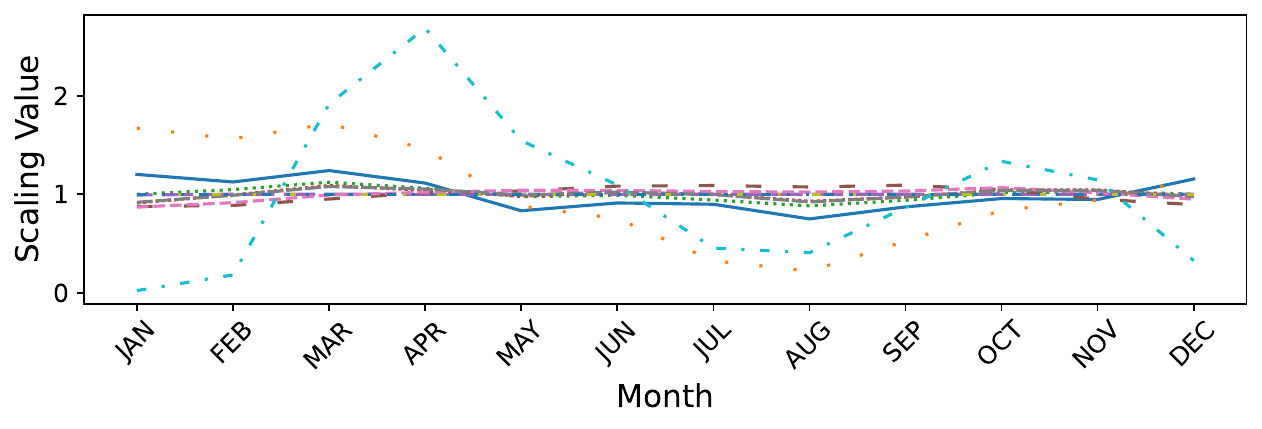}
    \caption{SO$_x$}\label{fig:emissionsScalingMonthlySOx}
  \end{subfigure}%
  \hspace*{\fill}   % maximize separation between the subfigure
\raisebox{11mm}{
  \begin{subfigure}{0.3\textwidth}
    \includegraphics[width=\linewidth]{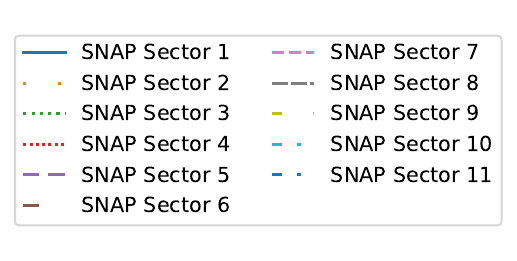}
  \end{subfigure}}%
  \hspace*{\fill}   % maximize separation between the subfigures
\caption{{\bfseries Emissions scaling for each emissions species by SNAP sector across the months.} } \label{fig:emissionsScalingMonthly}
\end{figure}

\clearpage
\subsection{Land Use}
\label{sec:landUse}

\begin{figure}[!htb]
\begin{center}
\includegraphics[width=0.9\textwidth]{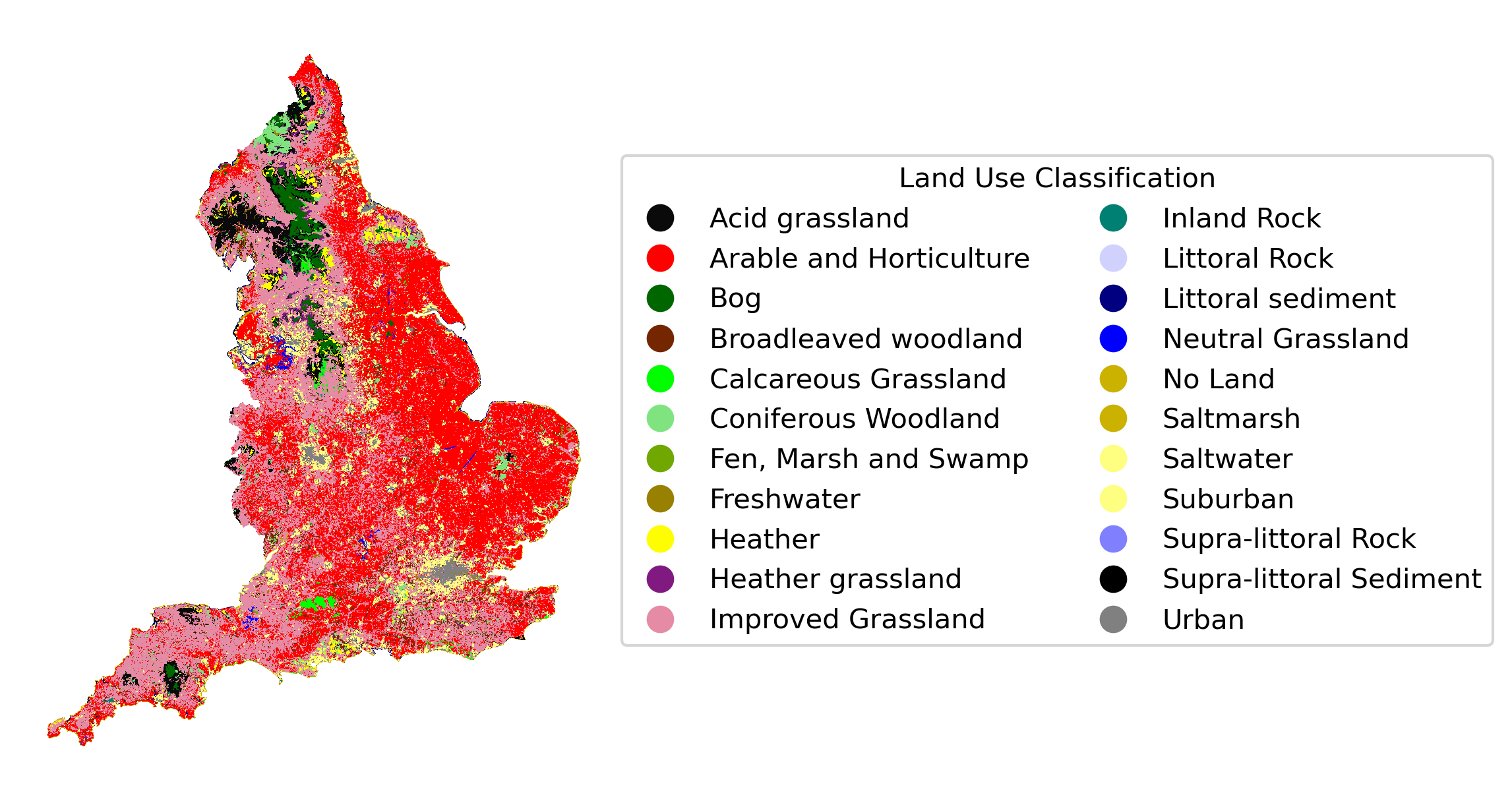}
\caption{{\bfseries Land use majority classification per grid.} } \label{fig:landUse}
\end{center}
\end{figure}

We created a geographic profile based on land use for each grid within the study. The 25m UKCEH Land Cover Maps \cite{rowland:2017:UKCEHLandCoverMap} were used to profile each grid's land use composition across 22 possible land use classifications. The feature vector elements represent the number of pixels with a given land cover classification in the raster. Figure \ref{fig:landUse} shows the majority classification for each grid within the study.  

\clearpage
\section{Feature Selection}

\subsection{Air Pollutants and Feature Vector}
\label{sec:featureSelection:AirPollutantsFeatureVector}

\begin{figure}[!htb]
  \hspace*{\fill}   % maximize separation between the subfigures
  \begin{subfigure}{0.45\linewidth}
    \includegraphics[width=\linewidth]{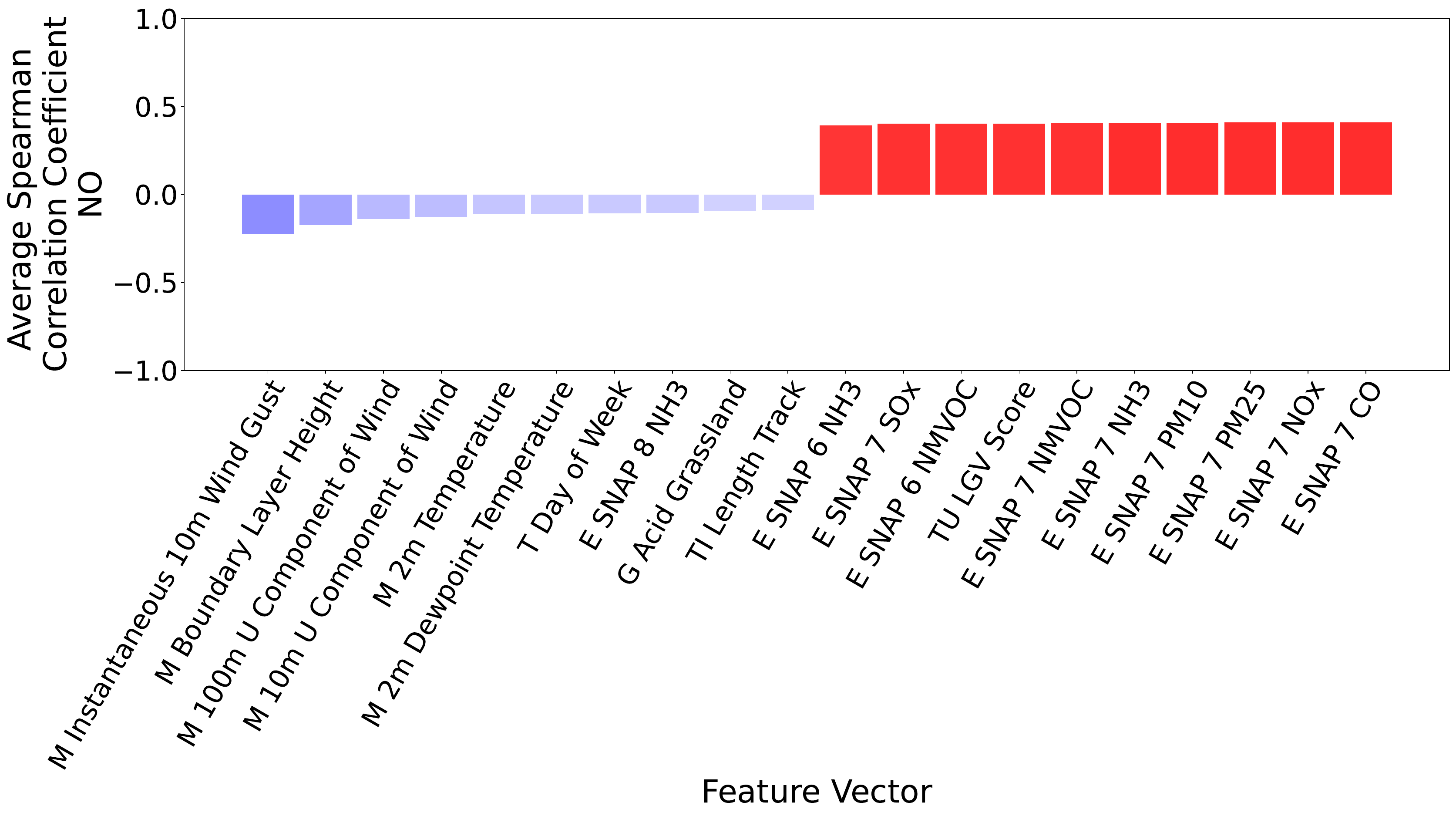}
    \caption{NO}\label{fig:spearmanCorrelationsFeatureVectorAirPollutantsNO}
  \end{subfigure}
  \hspace*{\fill}   % maximize separation between the subfigures
%  \\
  \hspace*{\fill}   % maximize separation between the subfigures
  \begin{subfigure}{0.45\linewidth}
    \includegraphics[width=\linewidth]{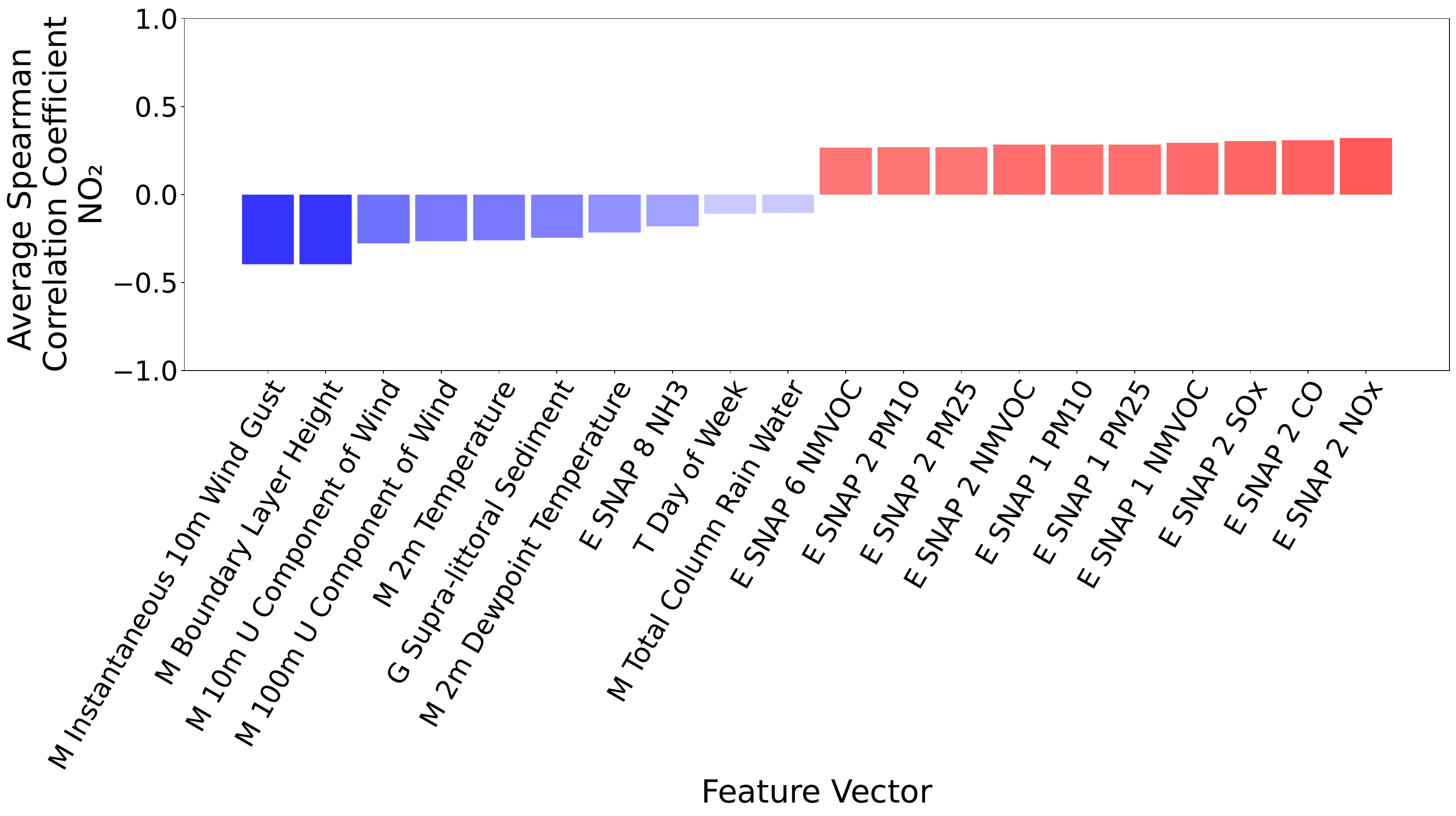}
    \caption{NO$_2$}\label{fig:spearmanCorrelationsFeatureVectorAirPollutantsNO2}
  \end{subfigure}
  \hspace*{\fill}   % maximize separation between the subfigures
  \\
  \hspace*{\fill}   % maximize separation between the subfigures
  \begin{subfigure}{0.45\textwidth}
    \includegraphics[width=\linewidth]{Figures/FeatureSelection/Spearman_Site_Type_All_Sites_nox_extreme_10.pdf}
    \caption{NO$_x$}\label{fig:spearmanCorrelationsFeatureVectorAirPollutantsNOX}
  \end{subfigure}
  \hspace*{\fill}   % maximize separation between the subfigures
%  \\
  \hspace*{\fill}   % maximize separation between the subfigures
  \begin{subfigure}{0.45\textwidth}
    \includegraphics[width=\linewidth]{Figures/FeatureSelection/Spearman_Site_Type_All_Sites_o3_extreme_10.pdf}
    \caption{O$_3$}\label{fig:spearmanCorrelationsFeatureVectorAirPollutantsO3}
  \end{subfigure}
  \hspace*{\fill}   % maximize separation between the subfigures
  \\
%  \caption{{\bfseries Spearman Correlation Coefficients Overall Mean For All Pollutants}}
%\end{figure}
%\begin{figure}[!htb]\ContinuedFloat
  \hspace*{\fill}   % maximize separation between the subfigures
  \begin{subfigure}{0.45\textwidth}
    \includegraphics[width=\linewidth]{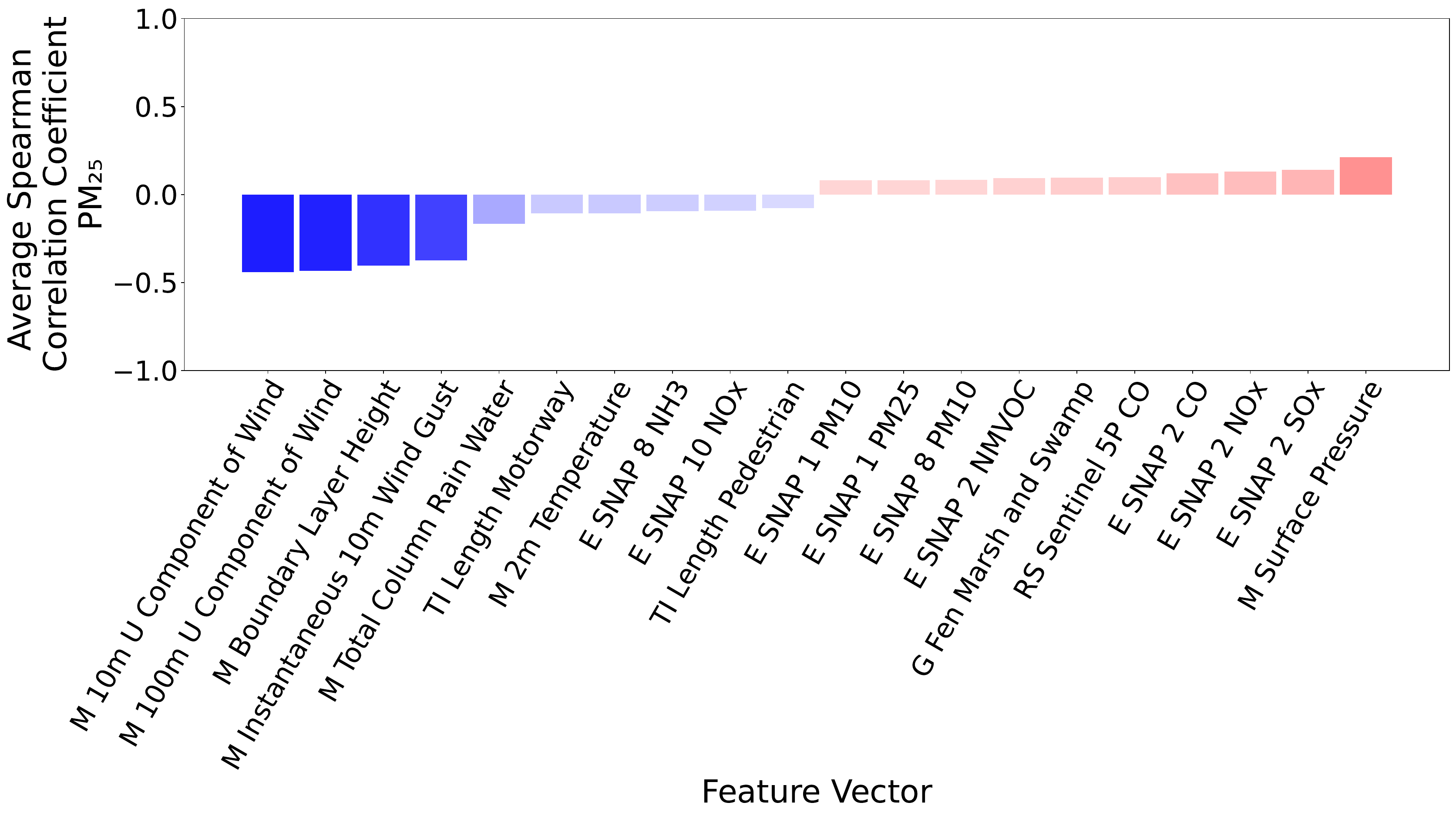}
    \caption{PM$_{2.5}$}\label{fig:spearmanCorrelationsFeatureVectorAirPollutantsPM25}
  \end{subfigure}
  \hspace*{\fill}   % maximize separation between the subfigures
%  \\
  \hspace*{\fill}   % maximize separation between the subfigures
  \begin{subfigure}{0.45\textwidth}
    \includegraphics[width=\linewidth]{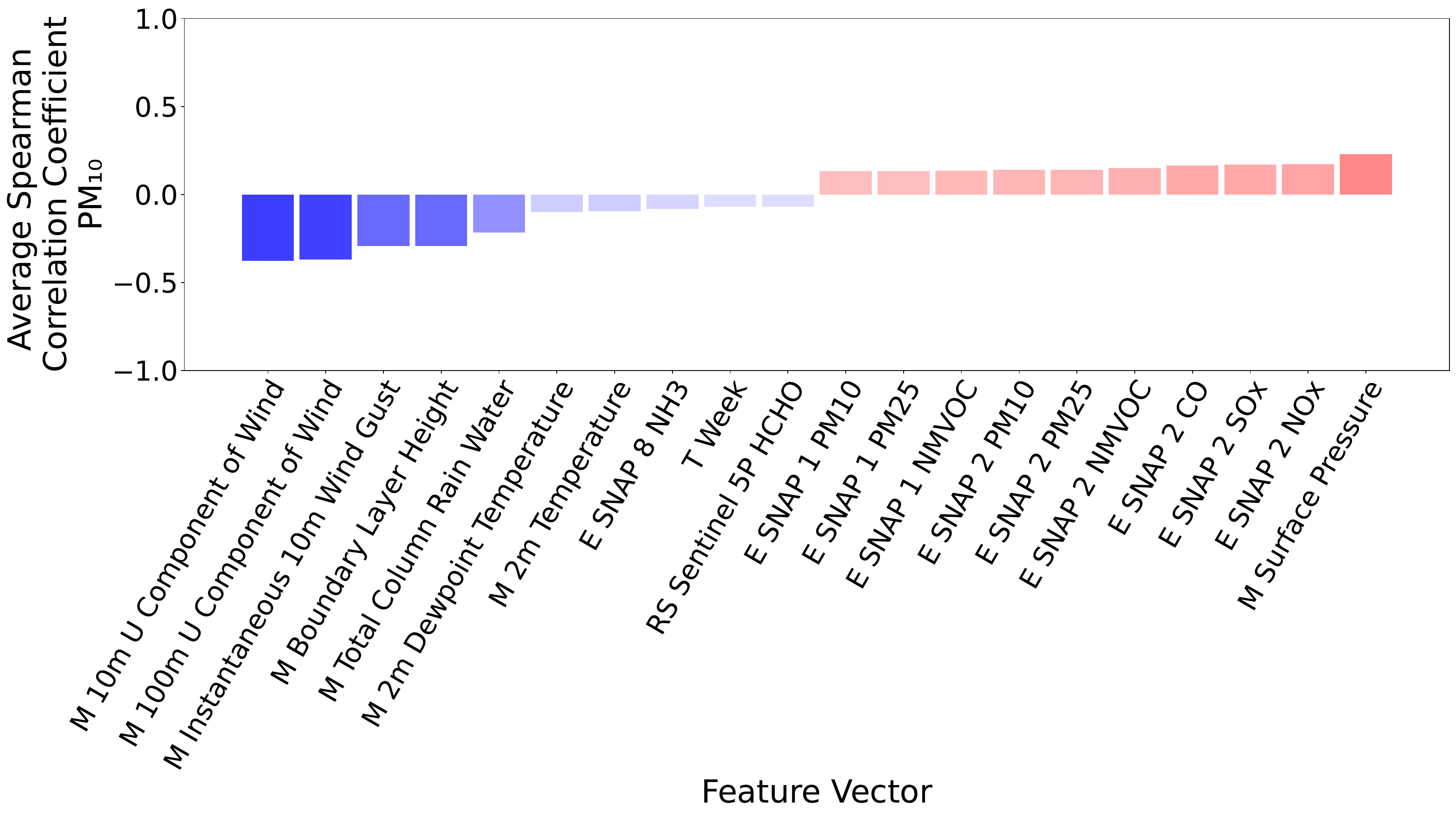}
    \caption{PM$_{10}$}\label{fig:spearmanCorrelationsFeatureVectorAirPollutantsPM10}
  \end{subfigure}
  \hspace*{\fill}   % maximize separation between the subfigures
  \\
  \hspace*{\fill}   % maximize separation between the subfigures
  \begin{subfigure}{0.45\textwidth}
    \includegraphics[width=\linewidth]{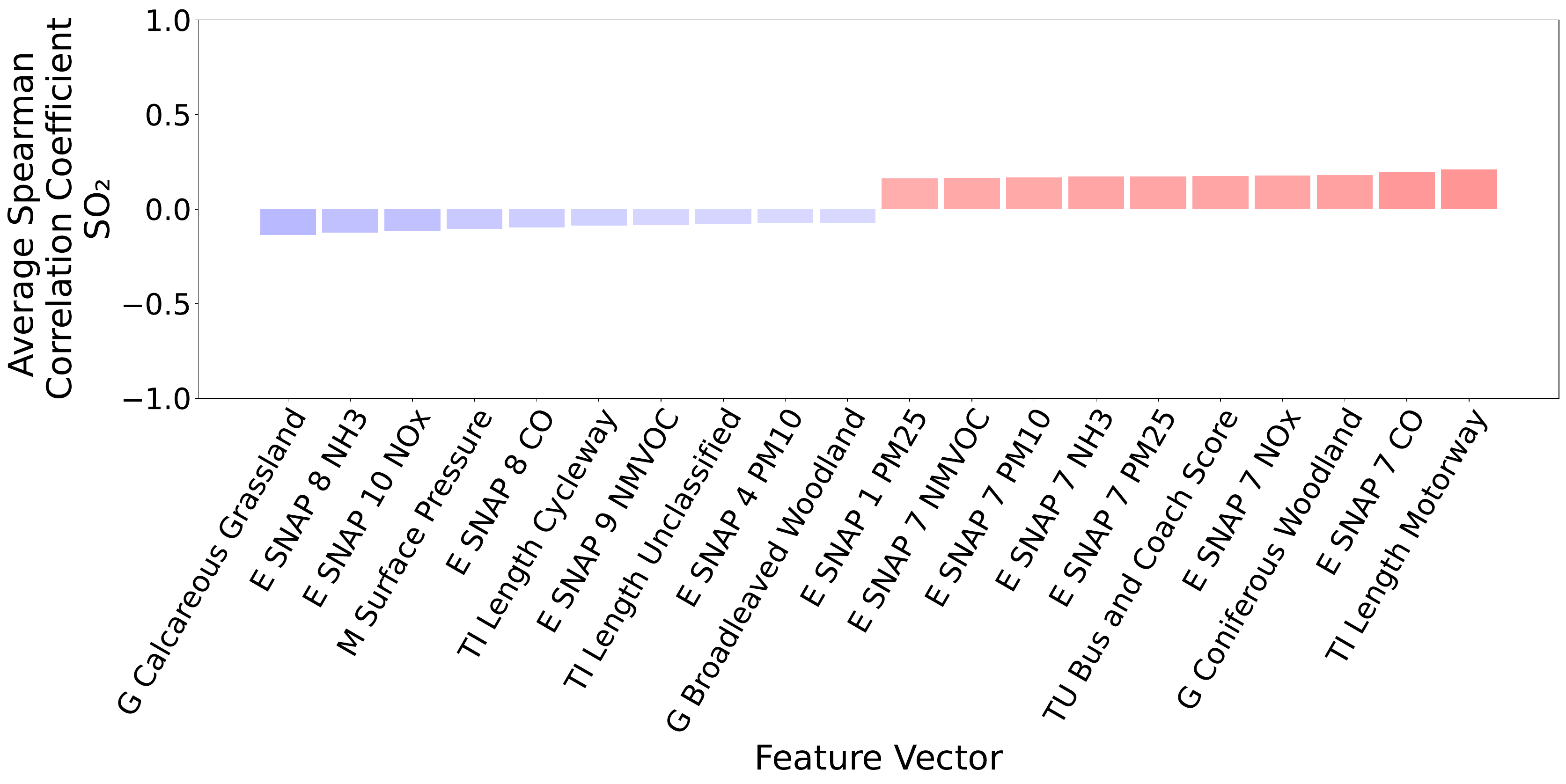}
    \caption{SO$_2$}\label{fig:spearmanCorrelationsFeatureVectorAirPollutantsSO2}
  \end{subfigure}
  \hspace*{\fill}   % maximize separation between the subfigures
\caption{{\bfseries Spearman correlation coefficients overall mean for all pollutants.} 
Of note is that the remote sensing dataset family does not have the highest correlation with the air pollutant that it measures directly, such as NO$_2$. The reason for this is the difference in temporal period, with the dataset being a monthly aggregate rather than the higher temporal resolution datasets included, such as the emissions datasets.} \label{fig:spearmanCorrelationsFeatureVectorAirPollutants}
\end{figure}

\clearpage
\subsection{Inter Feature Vectors}
\label{sec:featureSelection:InterFeatureVectors}

\begin{table}[!htb]
\centering
\resizebox{0.35\linewidth}{!}{
\begin{filecontents*}{CSVFiles/FeatureSelection/linkage_distance_table.csv}
,Linkage Distance,Number of Features
0,-1.0,139
1,0.05,112
2,0.1,103
3,0.25,83
4,0.5,62
5,0.75,51
6,1.0,29
7,1.25,15
8,1.5,11
9,1.75,8
10,2.0,6
11,2.5,4
12,3.0,1
\end{filecontents*}
\pgfplotstabletypeset[
    multicolumn names=l, 
    col sep=comma, 
    string type, 
    header = has colnames, 
    columns={Linkage Distance, Number of Features},
    columns/Linkage Distance/.style={column type={S[round-precision=2, table-format=-1.3, table-number-alignment=center]}, column name=\shortstack{Linkage Distance}},
    columns/Number of Features/.style={column type={S[round-precision=2, table-format=-1.3, table-number-alignment=center]}, column name=\shortstack{Number of Clusters}},
    every head row/.style={before row=\toprule, after row=\midrule},
    every last row/.style={after row=\bottomrule}
    ]{CSVFiles/FeatureSelection/linkage_distance_table.csv}}
    \smallskip
    \caption{{\bfseries Linkage distance for hierarchial clustering of the feature vectors. } The linkage distance column provides a set of thresholds and the associated number of clusters grouped when all features under the threshold are put into a cluster. The smaller the linkage, the more similar the feature vectors are—the -1.00 threshold indicates that all feature vectors should be individual clusters, resulting in 139 clusters, with 4 missing due to land use and 9 emissions feature vectors not being present in a location with a monitoring station; providing the 152 feature vectors elements considered in the study.  }\label{tab:linkageDistance}
\end{table}

\clearpage
\section{Modelling}

\subsection{Data Subsetting - Temporal}
\label{sec:modelResults:dataSubsetTemporal}

\begin{table}[!htb]
\hspace*{\fill}   % maximize separation between the subfigure
\begin{subtable}[!htb]{0.49\linewidth}
\centering
\resizebox{\textwidth}{!}{
\begin{filecontents*}{CSVFiles/Model/temporal_experiment_datasetSubset_temporal_results_modelType_mean.csv}
Pollutant Name,Dataset Train Score,Dataset Validation Score,Dataset Test Score
NO₂,-0.0,0.03,0.02
NOₓ,-0.02,-0.01,0.01
NO,-0.06,-0.06,-0.05
O₃,0.12,0.04,0.0
SO₂,-0.06,0.0,0.0
PM₁₀,-0.01,-0.01,-0.05
PM₂₅,-0.05,-0.03,-0.08
\end{filecontents*}
\pgfplotstabletypeset[
    multicolumn names=l, 
    col sep=comma, 
    string type, 
    header = has colnames, 
    columns={Pollutant Name, Dataset Train Score, Dataset Validation Score, Dataset Test Score},
    columns/Pollutant Name/.style={column type=l, column name=\shortstack{Pollutant\\ Name}},
    columns/Dataset Train Score Median/.style={column type={S[round-precision=2, table-format=-1.3, table-number-alignment=center]}, column name=\shortstack{Dataset Train Score}},
    columns/Dataset Validation Score Median/.style={column type={S[round-precision=2, table-format=-1.3, table-number-alignment=center]}, column name=\shortstack{Dataset Validation Score}},
    columns/Dataset Test Score Median/.style={column type={S[round-precision=2, table-format=-1.3, table-number-alignment=center]}, column name=\shortstack{Dataset Test Score}},
    every head row/.style={before row=\toprule, after row=\midrule},
    every last row/.style={after row=\bottomrule}
    ]{CSVFiles/Model/temporal_experiment_datasetSubset_temporal_results_modelType_mean.csv}}
    \smallskip
    \caption{{\bfseries Temporal}  }\label{tab:dataSubsettingResultsTemporalTemporal}
\end{subtable}
\hspace*{\fill}   % maximize separation between the subfigure
\begin{subtable}[!htb]{0.49\linewidth}
\centering
\resizebox{\textwidth}{!}{
\begin{filecontents*}{CSVFiles/Model/temporal_experiment_datasetSubset_Metrological_results_modelType_mean.csv}
Pollutant Name,Dataset Train Score,Dataset Validation Score,Dataset Test Score
NO₂,0.14,0.16,0.14
NOₓ,0.09,0.1,0.09
NO,0.03,0.02,0.01
O₃,0.43,0.41,0.34
SO₂,-0.02,0.04,0.03
PM₁₀,0.2,0.23,0.18
PM₂₅,0.25,0.26,0.24
\end{filecontents*}
\pgfplotstabletypeset[
    multicolumn names=l, 
    col sep=comma, 
    string type, 
    header = has colnames, 
    columns={Pollutant Name, Dataset Train Score, Dataset Validation Score, Dataset Test Score},
    columns/Pollutant Name/.style={column type=l, column name=\shortstack{Pollutant\\ Name}},
    columns/Dataset Train Score Median/.style={column type={S[round-precision=2, table-format=-1.3, table-number-alignment=center]}, column name=\shortstack{Dataset Train Score}},
    columns/Dataset Validation Score Median/.style={column type={S[round-precision=2, table-format=-1.3, table-number-alignment=center]}, column name=\shortstack{Dataset Validation Score}},
    columns/Dataset Test Score Median/.style={column type={S[round-precision=2, table-format=-1.3, table-number-alignment=center]}, column name=\shortstack{Dataset Test Score}},
    every head row/.style={before row=\toprule, after row=\midrule},
    every last row/.style={after row=\bottomrule}
    ]{CSVFiles/Model/temporal_experiment_datasetSubset_Metrological_results_modelType_mean.csv}}
    \smallskip
    \caption{{\bfseries Meteorology}  }\label{tab:dataSubsettingResultsMetrological}
\end{subtable}
\\
\hspace*{\fill}   % maximize separation between the subfigure
\begin{subtable}[!htb]{0.49\linewidth}
\centering
\resizebox{\textwidth}{!}{
\begin{filecontents*}{CSVFiles/Model/temporal_experiment_datasetSubset_Transport_Infrastructure_results_modelType_mean.csv}
Pollutant Name,Dataset Train Score,Dataset Validation Score,Dataset Test Score
NO₂,0.32,0.29,0.31
NOₓ,0.26,0.25,0.27
NO,0.19,0.18,0.2
O₃,-0.02,-0.04,-0.15
SO₂,0.16,0.26,0.12
PM₁₀,-0.01,0.02,-0.04
PM₂₅,-0.07,-0.02,-0.1
\end{filecontents*}
\pgfplotstabletypeset[
    multicolumn names=l, 
    col sep=comma, 
    string type, 
    header = has colnames, 
    columns={Pollutant Name, Dataset Train Score, Dataset Validation Score, Dataset Test Score},
    columns/Pollutant Name/.style={column type=l, column name=\shortstack{Pollutant\\ Name}},
    columns/Dataset Train Score Median/.style={column type={S[round-precision=2, table-format=-1.3, table-number-alignment=center]}, column name=\shortstack{Dataset Train Score}},
    columns/Dataset Validation Score Median/.style={column type={S[round-precision=2, table-format=-1.3, table-number-alignment=center]}, column name=\shortstack{Dataset Validation Score}},
    columns/Dataset Test Score Median/.style={column type={S[round-precision=2, table-format=-1.3, table-number-alignment=center]}, column name=\shortstack{Dataset Test Score}},
    every head row/.style={before row=\toprule, after row=\midrule},
    every last row/.style={after row=\bottomrule}
    ]{CSVFiles/Model/temporal_experiment_datasetSubset_Transport_Infrastructure_results_modelType_mean.csv}}
    \smallskip
    \caption{{\bfseries Transport Infrastructure}  }\label{tab:dataSubsettingResultsTransportInfrastructure}
\end{subtable}
\hspace*{\fill}   % maximize separation between the subfigure
\begin{subtable}[!htb]{0.49\linewidth}
\centering
\resizebox{\textwidth}{!}{
\begin{filecontents*}{CSVFiles/Model/temporal_experiment_datasetSubset_Transport_Use_results_modelType_mean.csv}
Pollutant Name,Dataset Train Score,Dataset Validation Score,Dataset Test Score
NO₂,0.36,0.23,0.16
NOₓ,0.28,0.18,0.14
NO,0.21,0.12,0.08
O₃,0.09,0.02,-0.09
SO₂,0.19,0.24,0.12
PM₁₀,0.02,0.03,-0.06
PM₂₅,-0.07,-0.03,-0.08
\end{filecontents*}
\pgfplotstabletypeset[
    multicolumn names=l, 
    col sep=comma, 
    string type, 
    header = has colnames, 
    columns={Pollutant Name, Dataset Train Score, Dataset Validation Score, Dataset Test Score},
    columns/Pollutant Name/.style={column type=l, column name=\shortstack{Pollutant\\ Name}},
    columns/Dataset Train Score Median/.style={column type={S[round-precision=2, table-format=-1.3, table-number-alignment=center]}, column name=\shortstack{Dataset Train Score}},
    columns/Dataset Validation Score Median/.style={column type={S[round-precision=2, table-format=-1.3, table-number-alignment=center]}, column name=\shortstack{Dataset Validation Score}},
    columns/Dataset Test Score Median/.style={column type={S[round-precision=2, table-format=-1.3, table-number-alignment=center]}, column name=\shortstack{Dataset Test Score}},
    every head row/.style={before row=\toprule, after row=\midrule},
    every last row/.style={after row=\bottomrule}
    ]{CSVFiles/Model/temporal_experiment_datasetSubset_Transport_Use_results_modelType_mean.csv}}
    \smallskip
    \caption{{\bfseries Transport Use}  }\label{tab:dataSubsettingResultsTransportUse}
\end{subtable}
\\
\hspace*{\fill}   % maximize separation between the subfigure
\begin{subtable}[!htb]{0.49\linewidth}
\centering
\resizebox{\textwidth}{!}{
\begin{filecontents*}{CSVFiles/Model/temporal_experiment_datasetSubset_Remote_Sensing_results_modelType_mean.csv}
Pollutant Name,Dataset Train Score,Dataset Validation Score,Dataset Test Score
NO₂,0.35,0.31,0.36
NOₓ,0.28,0.25,0.32
NO,0.2,0.18,0.24
O₃,0.14,0.11,0.05
SO₂,0.18,0.26,0.13
PM₁₀,0.01,0.01,-0.01
PM₂₅,-0.05,-0.03,-0.05
\end{filecontents*}
\pgfplotstabletypeset[
    multicolumn names=l, 
    col sep=comma, 
    string type, 
    header = has colnames, 
    columns={Pollutant Name, Dataset Train Score, Dataset Validation Score, Dataset Test Score},
    columns/Pollutant Name/.style={column type=l, column name=\shortstack{Pollutant\\ Name}},
    columns/Dataset Train Score Median/.style={column type={S[round-precision=2, table-format=-1.3, table-number-alignment=center]}, column name=\shortstack{Dataset Train Score}},
    columns/Dataset Validation Score Median/.style={column type={S[round-precision=2, table-format=-1.3, table-number-alignment=center]}, column name=\shortstack{Dataset Validation Score}},
    columns/Dataset Test Score Median/.style={column type={S[round-precision=2, table-format=-1.3, table-number-alignment=center]}, column name=\shortstack{Dataset Test Score}},
    every head row/.style={before row=\toprule, after row=\midrule},
    every last row/.style={after row=\bottomrule}
    ]{CSVFiles/Model/temporal_experiment_datasetSubset_Remote_Sensing_results_modelType_mean.csv}}
    \smallskip
    \caption{{\bfseries Remote Sensing}  }\label{tab:dataSubsettingResultsRemoteSensing}
\end{subtable}
\hspace*{\fill}   % maximize separation between the subfigure
\begin{subtable}[!htb]{0.49\linewidth}
\centering
\resizebox{\textwidth}{!}{
\begin{filecontents*}{CSVFiles/Model/temporal_experiment_datasetSubset_Geographic_results_modelType_mean.csv}
Pollutant Name,Dataset Train Score,Dataset Validation Score,Dataset Test Score
NO₂,0.31,0.23,0.29
NOₓ,0.25,0.2,0.26
NO,0.18,0.12,0.2
O₃,-0.04,-0.1,-0.2
SO₂,0.17,0.23,0.13
PM₁₀,-0.01,0.02,-0.07
PM₂₅,-0.08,-0.03,-0.11
\end{filecontents*}
\pgfplotstabletypeset[
    multicolumn names=l, 
    col sep=comma, 
    string type, 
    header = has colnames, 
    columns={Pollutant Name, Dataset Train Score, Dataset Validation Score, Dataset Test Score},
    columns/Pollutant Name/.style={column type=l, column name=\shortstack{Pollutant\\ Name}},
    columns/Dataset Train Score Median/.style={column type={S[round-precision=2, table-format=-1.3, table-number-alignment=center]}, column name=\shortstack{Dataset Train Score}},
    columns/Dataset Validation Score Median/.style={column type={S[round-precision=2, table-format=-1.3, table-number-alignment=center]}, column name=\shortstack{Dataset Validation Score}},
    columns/Dataset Test Score Median/.style={column type={S[round-precision=2, table-format=-1.3, table-number-alignment=center]}, column name=\shortstack{Dataset Test Score}},
    every head row/.style={before row=\toprule, after row=\midrule},
    every last row/.style={after row=\bottomrule}
    ]{CSVFiles/Model/temporal_experiment_datasetSubset_Geographic_results_modelType_mean.csv}}
    \smallskip
    \caption{{\bfseries Geographic}  }\label{tab:dataSubsettingResultsGeographic}
\end{subtable}
\\
\hspace*{\fill}   % maximize separation between the subfigure
\begin{subtable}[!htb]{0.49\linewidth}
\centering
\resizebox{\textwidth}{!}{
\begin{filecontents*}{CSVFiles/Model/temporal_experiment_datasetSubset_Emissions_results_modelType_mean.csv}
Pollutant Name,Dataset Train Score,Dataset Validation Score,Dataset Test Score
NO₂,0.47,0.42,0.43
NOₓ,0.39,0.34,0.39
NO,0.32,0.28,0.28
O₃,0.1,-0.02,-0.01
SO₂,0.23,0.29,0.13
PM₁₀,0.05,0.06,-0.02
PM₂₅,-0.02,-0.0,-0.08
\end{filecontents*}
\pgfplotstabletypeset[
    multicolumn names=l, 
    col sep=comma, 
    string type, 
    header = has colnames, 
    columns={Pollutant Name, Dataset Train Score, Dataset Validation Score, Dataset Test Score},
    columns/Pollutant Name/.style={column type=l, column name=\shortstack{Pollutant\\ Name}},
    columns/Dataset Train Score Median/.style={column type={S[round-precision=2, table-format=-1.3, table-number-alignment=center]}, column name=\shortstack{Dataset Train Score}},
    columns/Dataset Validation Score Median/.style={column type={S[round-precision=2, table-format=-1.3, table-number-alignment=center]}, column name=\shortstack{Dataset Validation Score}},
    columns/Dataset Test Score Median/.style={column type={S[round-precision=2, table-format=-1.3, table-number-alignment=center]}, column name=\shortstack{Dataset Test Score}},
    every head row/.style={before row=\toprule, after row=\midrule},
    every last row/.style={after row=\bottomrule}
    ]{CSVFiles/Model/temporal_experiment_datasetSubset_Emissions_results_modelType_mean.csv}}
    \smallskip
    \caption{{\bfseries Emissions}  }\label{tab:dataSubsettingResultsEmissions}
\end{subtable}
\hspace*{\fill}   % maximize separation between the subfigure
\caption{{\bfseries Temporal experiment results for different subsets of dataset families.} All dataset families provide some information about at least one of the air pollutants within the study. Still, none provide the complete picture of air pollution concentrations in the study, aligning with the scientific literature concerning air pollution sources and sinks with various phenomena from meteorological conditions, such as wind, and emissions sources, such as transportation and industry impacting seen air pollution concentrations. As such, it is clear that all of the datasets outlined are needed to predict air pollution concentrations in the future at an adequate level and no single dataset.}\label{tab:datasubsettingResultsTemporal}
\end{table}

\clearpage
\subsection{Data Subsetting - Spatial}
\label{sec:modelResults:dataSubsetSpatial}

\begin{table}[!htb]
\begin{subtable}[!htb]{0.49\linewidth}
\centering
\resizebox{\textwidth}{!}{
\begin{filecontents*}{CSVFiles/Model/spatial_experiment_LOOV_datasetSubset_Temporal_results_modelType_mean_LOOV.csv}
,Pollutant Name,Max,Min,Mean,Median
0,NO,0.11,-110.87,-2.16,-0.06
1,NO₂,0.29,-13.84,-0.63,-0.06
2,NOₓ,0.22,-30.3,-1.09,-0.04
3,O₃,0.37,-2.44,-0.11,0.11
4,PM₁₀,0.05,-0.6,-0.06,0.01
5,PM₂₅,0.08,-0.6,-0.06,-0.06
6,SO₂,0.01,-1.97,-0.28,-0.07
\end{filecontents*}
\pgfplotstabletypeset[
    multicolumn names=l, 
    col sep=comma, 
    string type, 
    header = has colnames, 
    columns={Pollutant Name, Max, Min, Mean, Median},
    columns/Pollutant Name/.style={column type=l, column name=\shortstack{Pollutant\\ Name}},
    columns/Max/.style={column type={S[round-precision=2, table-format=-1.3, table-number-alignment=center]}, column name=\shortstack{Estimation \\ LOOV Max}},
    columns/Min/.style={column type={S[round-precision=2, table-format=-1.3, table-number-alignment=center]}, column name=\shortstack{Estimation \\ LOOV Min}},
    columns/Mean/.style={column type={S[round-precision=2, table-format=-1.3, table-number-alignment=center]}, column name=\shortstack{Estimation \\ LOOV Mean}},
    columns/Median/.style={column type={S[round-precision=2, table-format=-1.3, table-number-alignment=center]}, column name=\shortstack{Estimation \\ LOOV Median}},
    every head row/.style={before row=\toprule, after row=\midrule},
    every last row/.style={after row=\bottomrule}
    ]{CSVFiles/Model/spatial_experiment_LOOV_datasetSubset_Temporal_results_modelType_mean_LOOV.csv}}
    \smallskip
    \caption{{\bfseries Temporal}  }\label{tab:summaryStatisticsLOOVSpatial}
\end{subtable}
\begin{subtable}[!htb]{0.49\linewidth}
\centering
\resizebox{\textwidth}{!}{
\begin{filecontents*}{CSVFiles/Model/spatial_experiment_LOOV_datasetSubset_Metrological_results_modelType_mean_LOOV.csv}
,Pollutant Name,Max,Min,Mean,Median
0,NO,0.35,-441.56,-5.97,0.08
1,NO₂,0.46,-14.89,-0.52,0.14
2,NOₓ,0.44,-61.68,-1.49,0.13
3,O₃,0.7,-2.96,0.25,0.47
4,PM₁₀,0.4,-0.34,0.2,0.24
5,PM₂₅,0.38,-0.2,0.25,0.27
6,SO₂,0.08,-3.15,-0.3,-0.03
\end{filecontents*}
\pgfplotstabletypeset[
    multicolumn names=l, 
    col sep=comma, 
    string type, 
    header = has colnames, 
    columns={Pollutant Name, Max, Min, Mean, Median},
    columns/Pollutant Name/.style={column type=l, column name=\shortstack{Pollutant\\ Name}},
    columns/Max/.style={column type={S[round-precision=2, table-format=-1.3, table-number-alignment=center]}, column name=\shortstack{Estimation \\ LOOV Max}},
    columns/Min/.style={column type={S[round-precision=2, table-format=-1.3, table-number-alignment=center]}, column name=\shortstack{Estimation \\ LOOV Min}},
    columns/Mean/.style={column type={S[round-precision=2, table-format=-1.3, table-number-alignment=center]}, column name=\shortstack{Estimation \\ LOOV Mean}},
    columns/Median/.style={column type={S[round-precision=2, table-format=-1.3, table-number-alignment=center]}, column name=\shortstack{Estimation \\ LOOV Median}},
    every head row/.style={before row=\toprule, after row=\midrule},
    every last row/.style={after row=\bottomrule}
    ]{CSVFiles/Model/spatial_experiment_LOOV_datasetSubset_Metrological_results_modelType_mean_LOOV.csv}}
    \smallskip
    \caption{{\bfseries Meteorology}  }\label{tab:summaryStatisticsLOOVMetrologicalSpatial}
\end{subtable}
\\
\begin{subtable}[!htb]{0.49\linewidth}
\centering
\resizebox{\textwidth}{!}{
\begin{filecontents*}{CSVFiles/Model/spatial_experiment_LOOV_datasetSubset_Transport_Infrastructure_results_modelType_mean_LOOV.csv}
,Pollutant Name,Max,Min,Mean,Median
0,NO,0.0,-1.33,-0.2,-0.07
1,NO₂,0.0,-3.02,-0.41,-0.23
2,NOₓ,0.0,-3.35,-0.31,-0.11
3,O₃,0.0,-1.39,-0.33,-0.2
4,PM₁₀,-0.0,-0.79,-0.15,-0.12
5,PM₂₅,0.0,-0.52,-0.11,-0.09
6,SO₂,0.0,-1.86,-0.22,-0.07
\end{filecontents*}
\pgfplotstabletypeset[
    multicolumn names=l, 
    col sep=comma, 
    string type, 
    header = has colnames, 
    columns={Pollutant Name, Max, Min, Mean, Median},
    columns/Pollutant Name/.style={column type=l, column name=\shortstack{Pollutant\\ Name}},
    columns/Max/.style={column type={S[round-precision=2, table-format=-1.3, table-number-alignment=center]}, column name=\shortstack{Estimation \\ LOOV Max}},
    columns/Min/.style={column type={S[round-precision=2, table-format=-1.3, table-number-alignment=center]}, column name=\shortstack{Estimation \\ LOOV Min}},
    columns/Mean/.style={column type={S[round-precision=2, table-format=-1.3, table-number-alignment=center]}, column name=\shortstack{Estimation \\ LOOV Mean}},
    columns/Median/.style={column type={S[round-precision=2, table-format=-1.3, table-number-alignment=center]}, column name=\shortstack{Estimation \\ LOOV Median}},
    every head row/.style={before row=\toprule, after row=\midrule},
    every last row/.style={after row=\bottomrule}
    ]{CSVFiles/Model/spatial_experiment_LOOV_datasetSubset_Transport_Infrastructure_results_modelType_mean_LOOV.csv}}
    \smallskip
    \caption{{\bfseries Transport Infrastructure}  }\label{tab:summaryStatisticsLOOVTransportInfrastructureSpatial}
\end{subtable}
\begin{subtable}[!htb]{0.49\linewidth}
\centering
\resizebox{\textwidth}{!}{
\begin{filecontents*}{CSVFiles/Model/spatial_experiment_LOOV_datasetSubset_Transport_Use_results_modelType_mean_LOOV.csv}
,Pollutant Name,Max,Min,Mean,Median
0,NO,0.03,-18.6,-0.59,-0.14
1,NO₂,0.01,-5.36,-0.46,-0.27
2,NOₓ,0.03,-11.08,-0.5,-0.18
3,O₃,0.04,-3.08,-0.3,-0.22
4,PM₁₀,-0.01,-0.58,-0.14,-0.09
5,PM₂₅,-0.01,-0.42,-0.12,-0.12
6,SO₂,-0.01,-1.73,-0.26,-0.11
\end{filecontents*}
\pgfplotstabletypeset[
    multicolumn names=l, 
    col sep=comma, 
    string type, 
    header = has colnames, 
    columns={Pollutant Name, Max, Min, Mean, Median},
    columns/Pollutant Name/.style={column type=l, column name=\shortstack{Pollutant\\ Name}},
    columns/Max/.style={column type={S[round-precision=2, table-format=-1.3, table-number-alignment=center]}, column name=\shortstack{Estimation \\ LOOV Max}},
    columns/Min/.style={column type={S[round-precision=2, table-format=-1.3, table-number-alignment=center]}, column name=\shortstack{Estimation \\ LOOV Min}},
    columns/Mean/.style={column type={S[round-precision=2, table-format=-1.3, table-number-alignment=center]}, column name=\shortstack{Estimation \\ LOOV Mean}},
    columns/Median/.style={column type={S[round-precision=2, table-format=-1.3, table-number-alignment=center]}, column name=\shortstack{Estimation \\ LOOV Median}},
    every head row/.style={before row=\toprule, after row=\midrule},
    every last row/.style={after row=\bottomrule}
    ]{CSVFiles/Model/spatial_experiment_LOOV_datasetSubset_Transport_Use_results_modelType_mean_LOOV.csv}}
    \smallskip
    \caption{{\bfseries Transport Use}  }\label{tab:summaryStatisticsLOOVTransportUseSpatial}
\end{subtable}
\\
\begin{subtable}[!htb]{0.49\linewidth}
\centering
\resizebox{\textwidth}{!}{
\begin{filecontents*}{CSVFiles/Model/spatial_experiment_LOOV_datasetSubset_Remote_Sensing_results_modelType_mean_LOOV.csv}
,Pollutant Name,Max,Min,Mean,Median
0,NO,0.01,-35.53,-0.8,-0.1
1,NO₂,0.05,-2.41,-0.38,-0.21
2,NOₓ,0.03,-6.96,-0.44,-0.21
3,O₃,0.2,-1.54,-0.16,-0.1
4,PM₁₀,0.03,-0.62,-0.1,-0.07
5,PM₂₅,0.02,-0.5,-0.1,-0.09
6,SO₂,-0.04,-1.79,-0.41,-0.21
\end{filecontents*}
\pgfplotstabletypeset[
    multicolumn names=l, 
    col sep=comma, 
    string type, 
    header = has colnames, 
    columns={Pollutant Name, Max, Min, Mean, Median},
    columns/Pollutant Name/.style={column type=l, column name=\shortstack{Pollutant\\ Name}},
    columns/Max/.style={column type={S[round-precision=2, table-format=-1.3, table-number-alignment=center]}, column name=\shortstack{Estimation \\ LOOV Max}},
    columns/Min/.style={column type={S[round-precision=2, table-format=-1.3, table-number-alignment=center]}, column name=\shortstack{Estimation \\ LOOV Min}},
    columns/Mean/.style={column type={S[round-precision=2, table-format=-1.3, table-number-alignment=center]}, column name=\shortstack{Estimation \\ LOOV Mean}},
    columns/Median/.style={column type={S[round-precision=2, table-format=-1.3, table-number-alignment=center]}, column name=\shortstack{Estimation \\ LOOV Median}},
    every head row/.style={before row=\toprule, after row=\midrule},
    every last row/.style={after row=\bottomrule}
    ]{CSVFiles/Model/spatial_experiment_LOOV_datasetSubset_Remote_Sensing_results_modelType_mean_LOOV.csv}}
    \smallskip
    \caption{{\bfseries Remote Sensing}  }\label{tab:summaryStatisticsLOOVTemporalSpatial}
\end{subtable}
\begin{subtable}[!htb]{0.49\linewidth}
\centering
\resizebox{\textwidth}{!}{
\begin{filecontents*}{CSVFiles/Model/spatial_experiment_LOOV_datasetSubset_Geographic_results_modelType_mean_LOOV.csv}
,Pollutant Name,Max,Min,Mean,Median
0,NO,0.0,-3.53,-0.31,-0.15
1,NO₂,0.04,-3.82,-0.46,-0.25
2,NOₓ,0.0,-4.98,-0.41,-0.22
3,O₃,-0.0,-1.28,-0.36,-0.23
4,PM₁₀,-0.0,-0.72,-0.13,-0.08
5,PM₂₅,-0.0,-0.53,-0.14,-0.11
6,SO₂,-0.0,-2.36,-0.28,-0.08
\end{filecontents*}
\pgfplotstabletypeset[
    multicolumn names=l, 
    col sep=comma, 
    string type, 
    header = has colnames, 
    columns={Pollutant Name, Max, Min, Mean, Median},
    columns/Pollutant Name/.style={column type=l, column name=\shortstack{Pollutant\\ Name}},
    columns/Max/.style={column type={S[round-precision=2, table-format=-1.3, table-number-alignment=center]}, column name=\shortstack{Estimation \\ LOOV Max}},
    columns/Min/.style={column type={S[round-precision=2, table-format=-1.3, table-number-alignment=center]}, column name=\shortstack{Estimation \\ LOOV Min}},
    columns/Mean/.style={column type={S[round-precision=2, table-format=-1.3, table-number-alignment=center]}, column name=\shortstack{Estimation \\ LOOV Mean}},
    columns/Median/.style={column type={S[round-precision=2, table-format=-1.3, table-number-alignment=center]}, column name=\shortstack{Estimation \\ LOOV Median}},
    every head row/.style={before row=\toprule, after row=\midrule},
    every last row/.style={after row=\bottomrule}
    ]{CSVFiles/Model/spatial_experiment_LOOV_datasetSubset_Geographic_results_modelType_mean_LOOV.csv}}
    \smallskip
    \caption{{\bfseries Geographic}  }\label{tab:summaryStatisticsLOOVGeographicSpatial}
\end{subtable}
\\
\hspace*{\fill}
\begin{subtable}[!htb]{0.49\linewidth}
\centering
\resizebox{\textwidth}{!}{
\begin{filecontents*}{CSVFiles/Model/spatial_experiment_LOOV_datasetSubset_Emissions_results_modelType_mean_LOOV.csv}
,Pollutant Name,Max,Min,Mean,Median
0,NO,0.13,-3.3,-0.23,-0.1
1,NO₂,0.25,-2.45,-0.23,-0.1
2,NOₓ,0.17,-3.29,-0.25,-0.11
3,O₃,0.15,-2.34,-0.26,-0.19
4,PM₁₀,0.04,-0.73,-0.1,-0.05
5,PM₂₅,0.02,-0.57,-0.09,-0.06
6,SO₂,0.02,-2.13,-0.31,-0.18
\end{filecontents*}
\pgfplotstabletypeset[
    multicolumn names=l, 
    col sep=comma, 
    string type, 
    header = has colnames, 
    columns={Pollutant Name, Max, Min, Mean, Median},
    columns/Pollutant Name/.style={column type=l, column name=\shortstack{Pollutant\\ Name}},
    columns/Max/.style={column type={S[round-precision=2, table-format=-1.3, table-number-alignment=center]}, column name=\shortstack{Estimation \\ LOOV Max}},
    columns/Min/.style={column type={S[round-precision=2, table-format=-1.3, table-number-alignment=center]}, column name=\shortstack{Estimation \\ LOOV Min}},
    columns/Mean/.style={column type={S[round-precision=2, table-format=-1.3, table-number-alignment=center]}, column name=\shortstack{Estimation \\ LOOV Mean}},
    columns/Median/.style={column type={S[round-precision=2, table-format=-1.3, table-number-alignment=center]}, column name=\shortstack{Estimation \\ LOOV Median}},
    every head row/.style={before row=\toprule, after row=\midrule},
    every last row/.style={after row=\bottomrule}
    ]{CSVFiles/Model/spatial_experiment_LOOV_datasetSubset_Emissions_results_modelType_mean_LOOV.csv}}
    \smallskip
    \caption{{\bfseries Emissions}  }\label{tab:summaryStatisticsLOOVEmissionsSpatial}
\end{subtable}
\hspace*{\fill}
\caption{{\bfseries Summary statistics for individual monitoring station leave-one-out-validation (LOOV) for the spatial experiment with different subsets of dataset families.} The spatial LOOV experiments echo the results seen in Table \ref{tab:datasubsettingResultsTemporal} to a more extreme degree, where the median for the majority of the LOOV is negative when only a single dataset family is included, highlighting that the prediction of the concentrations performs worse than simply predicting the average for the station's measurements, highlighting the importance of including a range of phenomena data to be able to accurately predict the air pollution concentrations of a monitoring stations concentration measurements.}
\end{table}

\subsection{Data Subsetting - Forecasting and Global Framework}
\begin{table}[!htb]
\resizebox{\linewidth}{!}{
\begin{filecontents*}{CSVFiles/Model/datasubsetting_spatial_experiment_all_pollutants_subset_Global_mean.csv}
,Included Datasets,Dataset Train Score,Dataset Validation Score,Dataset Test Score,Dataset Leave One Out Validation Scores,Pollutant Name
0,Global,0.7613014567828091,0.6809740716655945,0.7031201359145196,-0.0867588234597021,NO₂
1,Global,0.7256251996256257,0.6542075788461436,0.6750774499098579,-0.6890436250852998,NOₓ
2,Global,0.4562249015994306,0.42820108631548537,0.39624696768643874,-1.8707220086268024,NO
3,Global,0.765247229447086,0.6641111520313371,0.6222969887577285,0.4828477036236577,O₃
4,Global,0.2776242358141598,0.3670618725802447,0.24708901691147123,-0.47035347678081757,SO₂
5,Global,0.4919075572073234,0.3528519784525306,0.3074244188317806,0.37898576710046444,PM₁₀
6,Global,0.5344097808311706,0.3536770414397938,0.30114846876059553,0.46315451183908907,PM₂₅
\end{filecontents*}
\pgfplotstabletypeset[
    multicolumn names=l, 
    col sep=comma, 
    string type, 
    header = has colnames, 
    columns={Pollutant Name, Dataset Train Score, Dataset Validation Score, Dataset Test Score, Dataset Leave One Out Validation Scores},
    columns/Pollutant Name/.style={column type=l, column name=\shortstack{Pollutant Name}},
    columns/Dataset Train Score/.style={column type={S[round-precision=2, table-format=-1.3, table-number-alignment=center]}, column name=\shortstack{Dataset Train Score}},
    columns/Dataset Validation Score/.style={column type={S[round-precision=2, table-format=-1.3, table-number-alignment=center]}, column name=\shortstack{Dataset Validation Score}},
    columns/Dataset Test Score/.style={column type={S[round-precision=2, table-format=-1.3, table-number-alignment=center]}, column name=\shortstack{Dataset Test Score}},
    columns/Dataset Leave One Out Validation Scores/.style={column type={S[round-precision=2, table-format=-1.3, table-number-alignment=center]}, column name=\shortstack{Mean LOOV}},
    every head row/.style={before row=\toprule, after row=\midrule},
    every last row/.style={after row=\bottomrule}
    ]{CSVFiles/Model/datasubsetting_spatial_experiment_all_pollutants_subset_Global_mean.csv}}
    \smallskip
    \caption{{\bfseries Overview of forecasting (test score) and filling missing data spatial (LOOV) performance for a model trained with data that is only available globally.} When using the dataset families available globally, it can be seen that the model's performance in a forecasting situation is good. However, the LOOV is weaker, with some air pollutants having a negative performance overall, highlighting that the data that is available globally would be suitable for forecasting into the future in locations where monitoring stations are, but not when used to fill in missing monitoring station locations, without further improvements to the model or data used. The global datasets are the temporal, meteorological and remote sensing dataset families in this context.  }\label{tab:dataSubsetGlobalSpatial}
\end{table}

\begin{table}[!htb]
\resizebox{\linewidth}{!}{
\begin{filecontents*}{CSVFiles/Model/datasubsetting_spatial_experiment_all_pollutants_subset_Forecasting_mean.csv}
,Included Datasets,Dataset Train Score,Dataset Validation Score,Dataset Test Score,Dataset Leave One Out Validation Scores,Pollutant Name
0,Forecasting,0.8268633504747118,0.7654303949807763,0.7595315199045669,0.11335008064860175,NO₂
1,Forecasting,0.7981375001094417,0.7421061611597997,0.7301280683796756,0.028531569893517638,NOₓ
2,Forecasting,0.6867485705584139,0.6338816958556232,0.5899879368710039,-0.051698617554667,NO
3,Forecasting,0.7907058446544332,0.6911929651399203,0.6562141083845571,0.39482175168036593,O₃
4,Forecasting,0.37180965917990716,0.3812552594415236,0.2530927242272744,-0.10706867372427357,SO₂
5,Forecasting,0.5061386833653373,0.377493006014218,0.3344175149006915,0.3227371536915223,PM₁₀
6,Forecasting,0.5183471579203421,0.34566812064230323,0.30906282165827526,0.42069197228096816,PM₂₅
\end{filecontents*}
\pgfplotstabletypeset[
    multicolumn names=l, 
    col sep=comma, 
    string type, 
    header = has colnames, 
    columns={Pollutant Name, Dataset Train Score, Dataset Validation Score, Dataset Test Score, Dataset Leave One Out Validation Scores},
    columns/Pollutant Name/.style={column type=l, column name=\shortstack{Pollutant Name}},
    columns/Dataset Train Score/.style={column type={S[round-precision=2, table-format=-1.3, table-number-alignment=center]}, column name=\shortstack{Dataset Train Score}},
    columns/Dataset Validation Score/.style={column type={S[round-precision=2, table-format=-1.3, table-number-alignment=center]}, column name=\shortstack{Dataset Validation Score}},
    columns/Dataset Test Score/.style={column type={S[round-precision=2, table-format=-1.3, table-number-alignment=center]}, column name=\shortstack{Dataset Test Score}},
    columns/Dataset Leave One Out Validation Scores/.style={column type={S[round-precision=2, table-format=-1.3, table-number-alignment=center]}, column name=\shortstack{Mean LOOV}},
    every head row/.style={before row=\toprule, after row=\midrule},
    every last row/.style={after row=\bottomrule}
    ]{CSVFiles/Model/datasubsetting_spatial_experiment_all_pollutants_subset_Forecasting_mean.csv}}
    \smallskip
    \caption{{\bfseries Overview of forecasting (test score) and filling missing data spatial (LOOV) performance for a model trained with data that is only available ahead of time (e.g. forecasting ahead of the current date).} When using the dataset families available ahead of time that could be used in a true forecasting mode, e.g. predict ahead of the current date, it can be seen that the model's performance in a forecasting situation is similar to the global model seen in Table \ref{tab:dataSubsetGlobalSpatial}, showing good performance. However, the LOOV is considerably better, with most of the air pollutants having a strong LOOV performance, with only NO and SO$_2$ having negative performance, driven by a small number of bad predictions for particular stations. In this context, the forecasting datasets are road infrastructure, geographic, meteorological and temporal dataset families. The difference between the performance of the forecasting datasets and the global datasets models shows the importance of some datasets to the model's overall performance, such as the road infrastructure is critical for NO$_x$ and NO$_2$ where its inclusion improves performance considerably, particularly for the LOOV results.  }\label{tab:dataSubsetForecastingSpatial}
\end{table}

\clearpage
\section{Research Data Output Summary Statistics}
\label{sec:researchData:summaryStatistics}

\begin{figure}[!htb]
  \hspace*{\fill}   % maximize separation between the subfigures
  \begin{subfigure}{0.95\textwidth}
    \includegraphics[width=\linewidth]{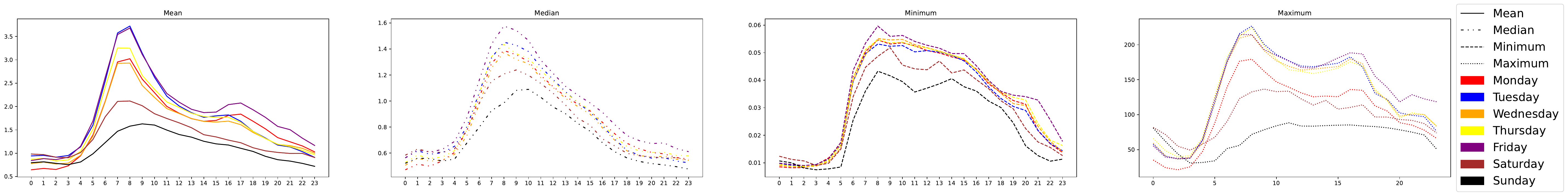}\\
    \includegraphics[width=\linewidth]{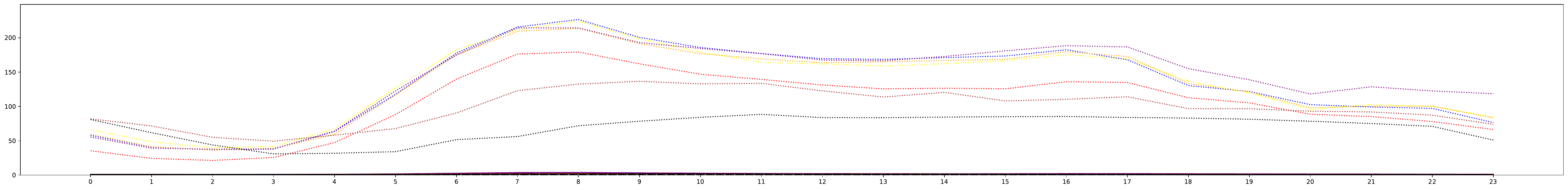}
    \caption{NO}\label{fig:researchData2018SummaryNO}
  \end{subfigure}
  \hspace*{\fill}   % maximize separation between the subfigures
  \\
  \hspace*{\fill}   % maximize separation between the subfigures
  \begin{subfigure}{0.95\textwidth}
    \includegraphics[width=\linewidth]{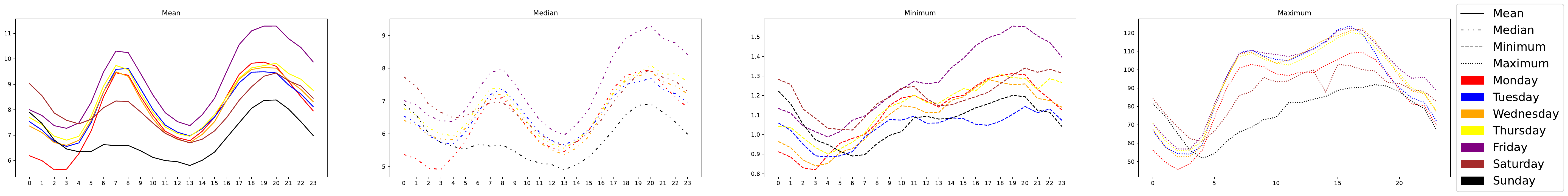}\\
    \includegraphics[width=\linewidth]{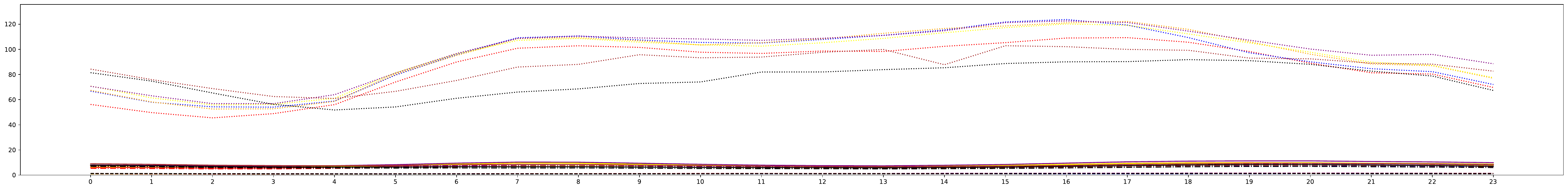}
    \caption{ NO2}\label{fig:researchData2018SummaryNO2}
  \end{subfigure}
  \hspace*{\fill}   % maximize separation between the subfigures
  \\
  \hspace*{\fill}   % maximize separation between the subfigures
  \begin{subfigure}{0.95\textwidth}
    \includegraphics[width=\linewidth]{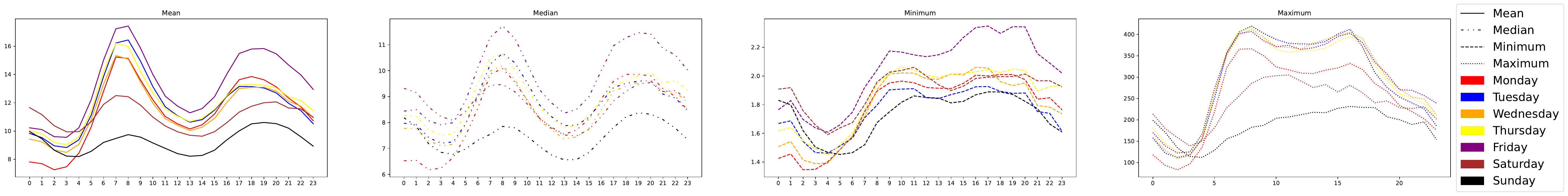}\\
    \includegraphics[width=\linewidth]{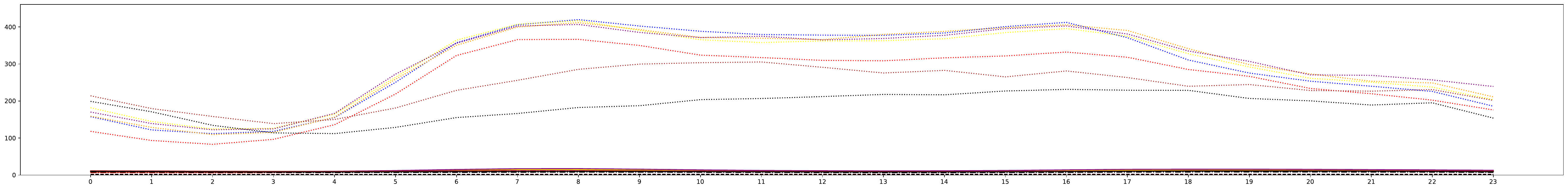}
    \caption{ NOX}\label{fig:researchData2018SummaryNOX}
  \end{subfigure}
  \hspace*{\fill}   % maximize separation between the subfigures
  \\
  \hspace*{\fill}   % maximize separation between the subfigures
  \begin{subfigure}{0.95\textwidth}
    \includegraphics[width=\linewidth]{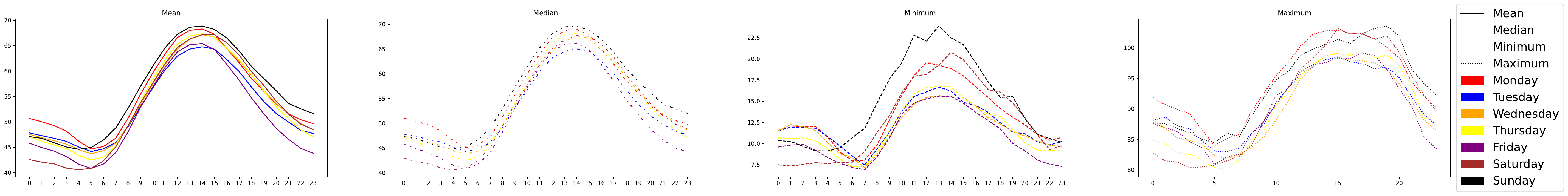}\\
    \includegraphics[width=\linewidth]{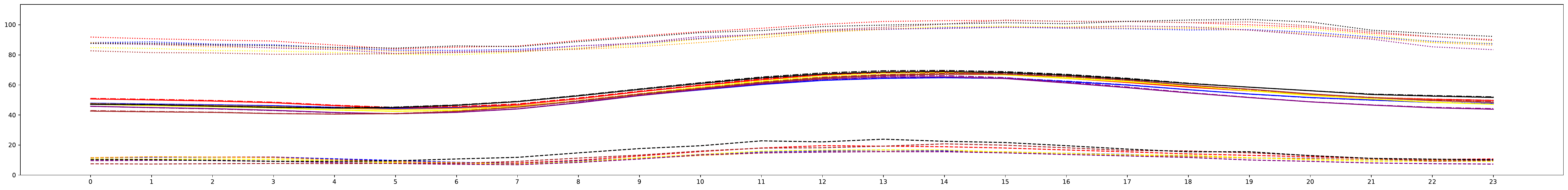}
    \caption{O3}\label{fig:researchData2018SummaryO3}
  \end{subfigure}
  \hspace*{\fill}   % maximize separation between the subfigures
  \caption{{\bfseries Summary of the complete air pollution concentration dataset, spatially and temporal for England at the 1km$^2$ spatial resolution for the air pollutants NO, NO$_2$, NO$_x$, O$_3$, PM$_{10}$, PM$_{2.5}$ and SO$_2$ at the hourly temporal level for 2018. } }
\end{figure}
\begin{figure}[!htb]\ContinuedFloat
  \hspace*{\fill}   % maximize separation between the subfigures
  \begin{subfigure}{0.95\textwidth}
    \includegraphics[width=\linewidth]{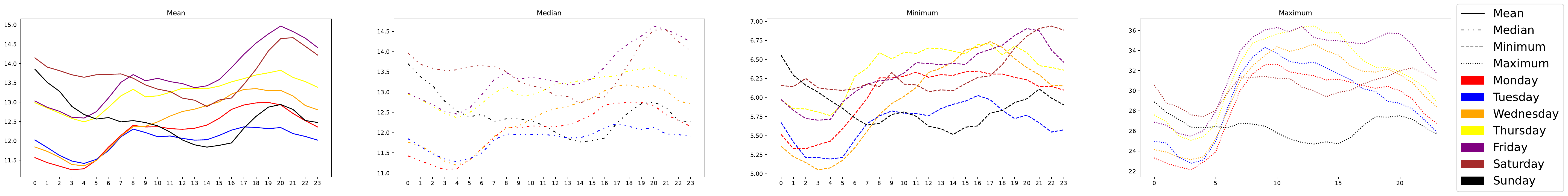}\\
    \includegraphics[width=\linewidth]{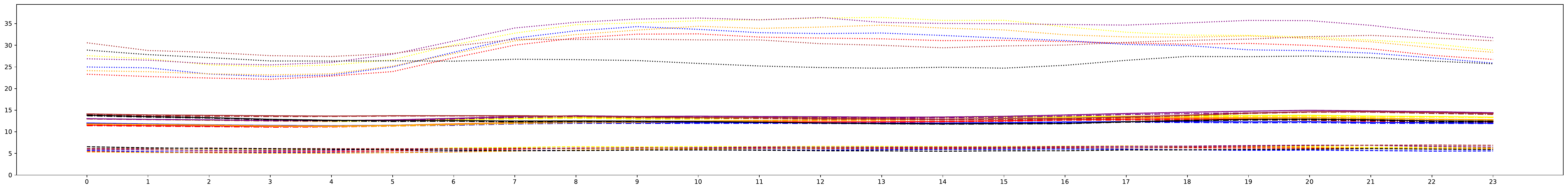}
    \caption{PM10}\label{fig:researchData2018SummarPM10}
  \end{subfigure}
  \hspace*{\fill}   % maximize separation between the subfigures
  \\
  \hspace*{\fill}   % maximize separation between the subfigures
  \begin{subfigure}{0.95\textwidth}
    \includegraphics[width=\linewidth]{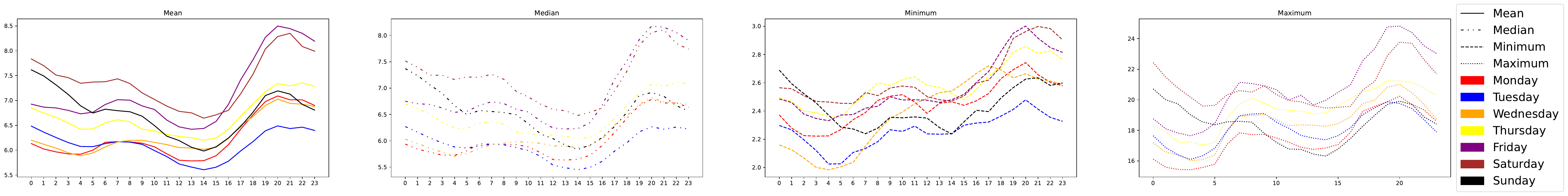}\\
    \includegraphics[width=\linewidth]{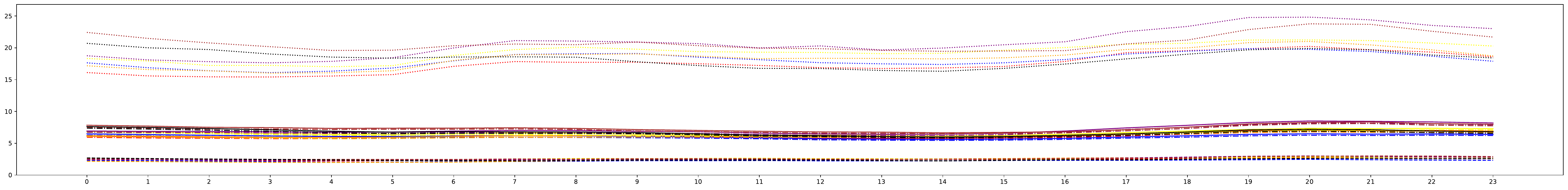}
    \caption{PM$_{2.5}$}\label{fig:researchData2018SummaryPM25}
  \end{subfigure}
  \hspace*{\fill}   % maximize separation between the subfigures
  \\
  \hspace*{\fill}   % maximize separation between the subfigures
  \begin{subfigure}{0.95\textwidth}
    \includegraphics[width=\linewidth]{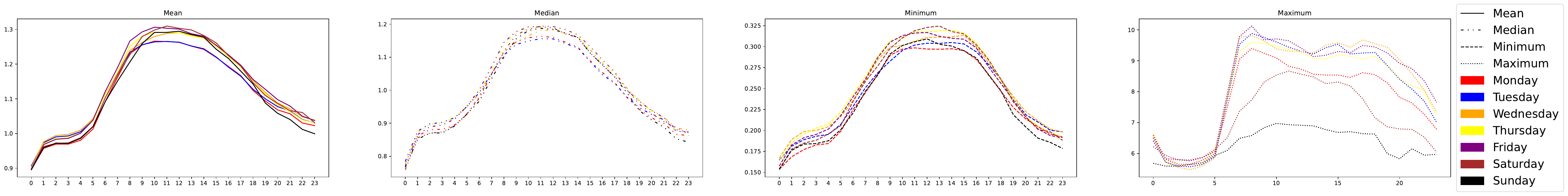}\\
    \includegraphics[width=\linewidth]{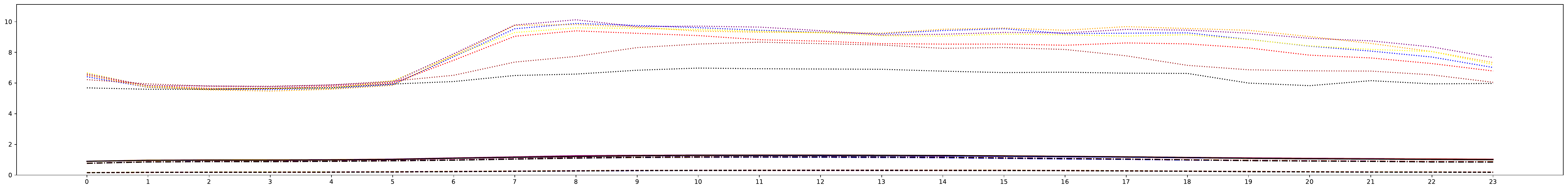}
    \caption{SO2}\label{fig:researchData2018SummarySO2}
  \end{subfigure}
  \hspace*{\fill}   % maximize separation between the subfigures
  \\
\caption{{\bfseries Summary of the complete air pollution concentration dataset, spatially and temporal for England at the 1km$^2$ spatial resolution for the air pollutants NO, NO$_2$, NO$_x$, O$_3$, PM$_{10}$, PM$_{2.5}$ and SO$_2$ at the hourly temporal level for 2018. (cont.)} Summarised in the figures is the dataset made possible by the model developed and presented by this work. The mean, median, minimum and maximum at each hour of the day for each day of the week are shown for each air pollutant, highlighting their overall trends across England. } \label{fig:researchData2018Summary}
\end{figure}

\clearpage
\bibliographystyle{IEEEtran}
\bibliography{bibliography_supplementary}